\pdfoutput=1
\documentclass[11pt]{article}
\usepackage[margin=1in]{geometry}
\usepackage{amsmath,amssymb,amsthm,bm,graphicx,hyperref,authblk,booktabs,microtype,placeins,xcolor,tikz,array}
\usetikzlibrary{arrows.meta,positioning,calc}
\usepackage[labelfont=bf,font=small]{caption}
\usepackage{mdframed}
\hypersetup{hidelinks}
\theoremstyle{plain}
\newtheorem{proposition}{Proposition}
\newtheorem{corollary}{Corollary}
\theoremstyle{remark}
\newtheorem{remark}{Remark}
\newcommand{\bth}{\bm{\theta}}
\newcommand{\bv}{\bm{v}}
\newcommand{\bbeta}{\bm{\beta}}
\newcommand{\bc}{\bm{c}}
\newcommand{\Tr}{\operatorname{Tr}}
\newcommand{\Var}{\operatorname{Var}}
\newcommand{\col}{\operatorname{col}}
\newcommand{\Lc}{\mathcal{L}}
\newcommand{\Jc}{\mathcal{J}}
\newcommand{\BC}{\mathcal{B}_C}
\newcommand{\BQ}{\mathcal{B}_Q}
\newcommand{\GQ}{\bm{G}_Q}
\newcommand{\FQ}{\bm{F}_Q}
\newcommand{\FC}{\bm{F}_C}
\newcommand{\PX}{\Pi_X}
\newcommand{\PXp}{\Pi_X^{\perp}}
\newcommand{\Pir}{\Pi_r}
\newcommand{\lo}{\ell_0}
\newcommand{\lp}{\ell_\perp}
\newcommand{\GamQ}{\Gamma_Q}
\newcommand{\ws}{w^\star}
\newcommand{\normF}[1]{\|#1\|_F}
\newcommand{\normtwo}[1]{\|#1\|_2}
\newcommand{\SLD}{L}
\newcommand{\DOI}{10.5281/zenodo.23272046}

\title{Quantum-Fisher-information-aware optimization yields a quantum reservoir advantage over tuned classical reservoirs}
\author[1]{Phi Hung Nguyen\thanks{Corresponding authors: hpnguye7@arizona.edu (P.H.N.), danielsoh@arizona.edu (D.S.)}}
\author[1]{Daniel Soh\protect\footnotemark[1]}
\affil[1]{James C. Wyant College of Optical Sciences, University of Arizona, Tucson, Arizona 85721, USA}
\date{}

\begin{document}
\maketitle

\begin{abstract}
We show that a quantum reservoir computer optimized with its quantum Fisher information (QFI) has a quantum advantage over a classical reservoir of equal readout cost, provided by a quantum-information advantage: optimizing the quantum parameters with the QFI metric. On Mackey--Glass forecasting, a driven Jaynes--Cummings reservoir with one to three atoms beats a tuned echo-state network at every size from 4 to 30 trained weights, by 25--69\% in held-out error, and is ahead of 99.5\% of 10,000 random networks. After grid and Bayesian searches saturate, the QFI-aware stage lowers the error by 29--51\% at 12--30 weights, beyond equal-budget black-box search or classical-parameter gradients alone, and at 30 weights turns a loss into a win. We then derive a fundamental QFI bound on the first-order rate of change of the retrained loss and factorize the gap to this bound into five link efficiencies that identify the physical mechanism limiting optimization.
\end{abstract}

\section{Introduction}
Reservoir computing replaces a trainable recurrent network by a fixed dynamical system, in physical implementations a photonic, spintronic or mechanical device, and trains only a linear readout \cite{Jaeger2001,Maass2002,Lukosevicius2009,Appeltant2011,Vandoorne2014,Torrejon2017,Tanaka2019}. In quantum reservoir computing the reservoir is a quantum system, the features are expectation values of measured observables, and the linear readout is trained by least squares \cite{Fujii2017,Mujal2021,Nokkala2021,Chen2020,MartinezPena2021,Zhu2025,Senanian2024,Hu2024,Dudas2023,Kobayashi2024}. The simplicity of the readout moves the design problem upstream, to the couplings, dissipation, input encoding and measured observables that determine which physical responses can reach a task \cite{Govia2021,Kalfus2022,Sannia2024,Mujal2023,GrossRieser2026}. Parameter sweeps can find a useful operating point but do not say why a nearby point fails or which part of a device should be changed; the same physical-parameter gradients govern classical physical reservoirs and physical neural networks trained in situ \cite{Wright2022,Momeni2023,Stern2023}.

A weak gradient of a trained loss can arise for different reasons, and, as for barren plateaus in variational circuits \cite{McClean2018}, its magnitude alone does not identify the cause: the state may barely respond, the measurement may hide the response, a readout refit may absorb feature motion inside the current span, or the new motion may point away from the unresolved target. A design theory should therefore separate the available physical response from the fraction that survives measurement, refitting and task projection.

Quantum Fisher information (QFI) is the intrinsic metric for distinguishable state motion \cite{Helstrom1976,BraunsteinCaves1994,Paris2009}, used to study trainability and capacity in parametrized quantum models \cite{Haug2021,Abbas2021,Meyer2021} and to define quantum natural-gradient (QNG) updates \cite{Amari1998,Stokes2020,Koczor2022}. For a quantum reservoir computer it gives a measurement-independent certificate (a direction of zero QFI admits no first-order response for any observable or target), and the symmetric logarithmic derivative (SLD) identifies a measurement that resolves a chosen direction. What is missing is the conversion from this state-level sensitivity to the derivative of a loss whose readout is retrained after every physical update.

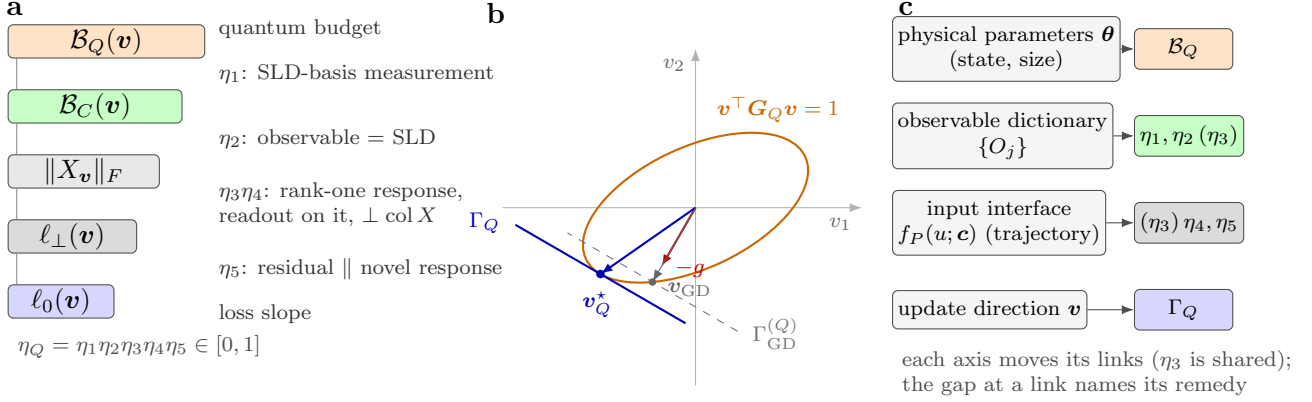
\begin{figure}[p]
\centering
\resizebox{\ifdim\width>\textwidth \textwidth\else\width\fi}{!}{\begin{tikzpicture}[x=1cm,y=1cm,font=\small,>={Latex[length=1.6mm]},
  rung/.style={draw=black!70,rounded corners=2pt,fill=#1,minimum height=0.44cm,inner xsep=4pt,inner ysep=1pt,anchor=west},
  lab/.style={font=\scriptsize,align=left,anchor=west,text=black!80,inner sep=1pt},
  axis/.style={draw=black!70,rounded corners=2pt,fill=gray!8,minimum width=2.2cm,minimum height=0.5cm,align=center,font=\scriptsize,inner xsep=2pt},
  link/.style={draw=black!70,rounded corners=2pt,fill=#1,minimum width=1.3cm,minimum height=0.5cm,align=center,font=\scriptsize,inner xsep=2pt}]
\node[anchor=north west,font=\bfseries] at (-0.15,5.65) {a};
\node[rung=orange!22,minimum width=2.6cm] (r1) at (0,5.0) {$\BQ(\bv)$};
\node[rung=green!22!white,minimum width=2.3cm]   (r2) at (0,4.14) {$\BC(\bv)$};
\node[rung=gray!18,minimum width=2.0cm]   (r3) at (0,3.28) {$\normF{X_{\bv}}$};
\node[rung=black!14,minimum width=1.7cm]   (r4) at (0,2.42) {$\lp(\bv)$};
\node[rung=blue!16,minimum width=1.4cm]   (r5) at (0,1.56) {$\lo(\bv)$};
\node[lab,anchor=south west] at (2.75,5.0) {quantum budget};
\node[lab,anchor=north west] at (2.75,1.56) {loss slope};
\node[lab] at (2.75,4.57) {$\eta_1$: SLD-basis measurement};
\node[lab] at (2.75,3.71) {$\eta_2$: observable $=$ SLD};
\node[lab] at (2.75,2.85) {$\eta_3\eta_4$: rank-one response,\\ readout on it, $\perp\col X$};
\node[lab] at (2.75,1.99) {$\eta_5$: residual $\parallel$ novel response};
\foreach \i/\j in {1/2,2/3,3/4,4/5}{\draw[black!45] ($(r\i.south west)+(0.12,0)$) -- ($(r\j.north west)+(0.12,0)$);}
\node[font=\scriptsize,anchor=north west,text=black!70] at (0,1.25) {$\eta_Q=\eta_1\eta_2\eta_3\eta_4\eta_5\in[0,1]$};
\begin{scope}[shift={(9.1,2.8)}]
\node[anchor=north west,font=\bfseries] at (-2.9,2.85) {b};
\draw[black!30,->] (-2.55,0) -- (2.2,0) node[below left,font=\scriptsize,text=black!60] {$v_1$};
\draw[black!30,->] (0,-2.35) -- (0,2.1) node[below left,font=\scriptsize,text=black!60] {$v_2$};
\draw[thick,orange!80!black,rotate=25] (0,0) ellipse (1.6 and 0.8);
\node[orange!80!black,font=\scriptsize,anchor=west] at (0.15,1.3) {$\bv^\top\GQ\bv=1$};
\draw[blue!70!black,thick] (-0.132,-1.529) -- (-2.383,-0.226);
\draw[black!55,dashed] (0.556,-1.635) -- (-1.694,-0.332);
\draw[->,red!75!black,thick] (0,0) -- (-0.451,-0.779);
\node[red!75!black,font=\scriptsize,anchor=west] at (-0.40,-0.80) {$-g$};
\draw[->,blue!70!black,thick] (0,0) -- (-1.2576,-0.8774);
\fill[blue!70!black] (-1.2576,-0.8774) circle (1.6pt);
\node[blue!70!black,font=\scriptsize,anchor=north] at (-1.30,-0.98) {$\bv_Q^\star$};
\draw[->,black!60] (0,0) -- (-0.5692,-0.9832);
\fill[black!60] (-0.5692,-0.9832) circle (1.4pt);
\node[black!60,font=\scriptsize,anchor=west] at (-0.50,-1.08) {$\bv_{\rm GD}$};
\node[blue!70!black,font=\scriptsize,anchor=east] at (-2.45,-0.20) {$\GamQ$};
\node[black!55,font=\scriptsize,anchor=west] at (0.62,-1.66) {$\Gamma^{(Q)}_{\rm GD}$};
\end{scope}
\begin{scope}[shift={(11.7,0)}]
\node[anchor=north west,font=\bfseries] at (-0.05,5.65) {c};
\node[axis,anchor=west] (a1) at (0,4.9) {physical parameters $\bth$\\ (state, size)};
\node[axis,anchor=west] (a2) at (0,3.75) {observable dictionary\\ $\{O_j\}$};
\node[axis,anchor=west] (a3) at (0,2.6) {input interface\\ $f_P(u;\bc)$ (trajectory)};
\node[axis,anchor=west] (a4) at (0,1.45) {update direction $\bv$};
\node[link=orange!22,anchor=west] (l1) at (3.2,4.9) {$\BQ$};
\node[link=green!22!white,anchor=west]   (l2) at (3.2,3.75) {$\eta_1,\eta_2\,(\eta_3)$};
\node[link=black!14,anchor=west]   (l3) at (3.2,2.6) {$(\eta_3)\,\eta_4,\eta_5$};
\node[link=blue!16,anchor=west]   (l4) at (3.2,1.45) {$\GamQ$};
\foreach \i in {1,2,3,4}{\draw[->,black!70] (a\i.east) -- (l\i.west);}
\node[font=\scriptsize,text=black!70,align=left,anchor=north west] at (0,1.05) {each axis moves its links ($\eta_3$ is shared);\\ the gap at a link names its remedy};
\end{scope}
\end{tikzpicture}}
\caption{\textbf{Quantum Fisher information bounds the retrained-loss gradient through five attainable links.} \textbf{a}, The ladder of Eq.~\eqref{eq:chain} drawn as nested bounds, from the feature-matched quantum budget $\BQ(\bv)$ down to the normalized loss slope $\lo(\bv)$. The five reported factors group the six inequalities; $\eta_3$ combines the two middle norm/readout inequalities, while the remaining factors correspond to single gaps. The measurement link $\eta_1$ closes for a measurement in the SLD eigenbasis, the observable link $\eta_2$ for an observable equal to the SLD, the readout and novelty links $\eta_3\eta_4$ for a rank-one feature response that carries the readout and lies outside the current feature span, and the task link $\eta_5$ for a residual parallel to the novel response. The utilization $\eta_Q$ is the product of the five factors. \textbf{b}, The Rayleigh optimum of Eq.~\eqref{eq:designability} for an illustrative metric. On the unit-QFI ellipse $\bv^\top\GQ\bv=1$ the level line of the first-order loss decrease $-g^\top\bv$ is tangent at the QFI-normalized natural-gradient direction $\bv_Q^\star$, which attains $\GamQ$. The Euclidean steepest-descent direction $\bv_{\rm GD}$ on the same ellipse lies on a lower level line (Supplementary Note~3). \textbf{c}, The design axes and the links they control. The physical parameters set the trajectory and hence the budget $\BQ$, whose weights $\omega_t$ also depend on the deployed dictionary. The observable dictionary sets the measurement factors $\eta_1,\eta_2$. The input interface sets the novelty and task factors $\eta_4,\eta_5$ through the trajectory. The update direction sets $\GamQ$. The readout factor $\eta_3$ is shared, because the dictionary and the input both move it (Fig.~\ref{fig:scan}g). The gap at a link is attributed to the axis that controls it. Panels are schematic and contain no data.}
\label{fig:concept}
\end{figure}

Here we derive that conversion for a quantum reservoir computer with a retrained linear readout (Fig.~\ref{fig:concept}); it reduces to the classical case when the reservoir state is diagonal (Discussion). The retrained-loss gradient is bounded by QFI through a sequence of projections and norms, and five grouped ratios of that chain form an exact decomposition of the gradient into a quantum response budget and five link factors that locate the loss of response at the measurement, readout, novelty or task stage. For a designer, it names the part of a device to change when the device has stopped improving, and how far to push a drive gain.

The main result of this paper is a quantum advantage over a matched classical reservoir, together with its source: a quantum-information advantage of QFI-aware optimization of the quantum parameters. An optimization stage has a \emph{quantum-information advantage} when it acts on information that exists only in the state before measurement (the QFI metric and the SLD eigenbasis) and reaches a reservoir that a search on the measured loss alone does not, with the reservoir family, constraints, readout, data and selection rule held fixed. On Mackey--Glass forecasting a driven Jaynes--Cummings (JC) reservoir optimized this way beats a tuned echo-state network of equal readout cost at every size from $4$ to $30$ trained weights, and a stage test locates the margin in the QFI-aware stage, after grid and Bayesian searches have saturated (Sec.~\ref{sec:matched}). Sections~\ref{sec:bound} to~\ref{sec:advantages} derive the theory behind the limits of this optimization, propose engineering methods and test them in two exactly simulable testbeds: a four-spin transverse-field Ising reservoir over twenty random couplings and a single-atom JC reservoir.

\section{Results}

\subsection{A quantum advantage over a matched echo-state network}
\label{sec:matched}
A driven JC reservoir whose quantum parameters are optimized with the QFI metric beats a tuned echo-state network of equal readout cost at every tested size, and the QFI-aware stage produces the margin (Table~\ref{tab:matched}, Fig.~\ref{fig:advantages}c,d). The reservoir is a cavity coupled to $N=1$--$3$ atoms, driven by the input and read through the cavity quadratures and atomic coherences at $V$ sub-times per input interval, $m=V(2+2N)$ features in all. Every parameter of the device is optimized---detunings, couplings, drive, decay rates, input offset, sampling times and the $m$ gains of the measurement-conditioned input drive, $11$--$44$ parameters in all---under a photon cap $\max_t\langle n\rangle\le3$ and a bounded drive (Methods). The network with $m$ nodes has its spectral radius, leak, input scale and bias tuned over $810$ settings on the training loss averaged over random draws of its matrices, and the selected setting is evaluated over $10{,}000$ draws; both sides train the same ridge readout on the same $2{,}000$ samples and select on training data. The network is used in its original form, with random reservoir and input matrices and no feedback: state feedback is an ad hoc extension of the echo-state network \cite{EhlersNN2025} whose gains are difficult to set without losing the echo-state property, whereas the measurement-conditioned drive of the quantum reservoir is part of how its input is applied (Methods). This follows the prevailing convention: of 38 published comparisons among 66 quantum-reservoir studies, the most common matching criterion is readout count (15 of 38); 30 tune the classical side on at most a small grid, leave it untuned or do not report tuning; none averages more than $1{,}000$ random draws; and none trains every parameter of the classical reservoir (Supplementary Note~16). In an alternative matching at $m=30$ by physical elements, with all $12$--$17$ parameters of four three-node network variants trained, the reservoir's held-out normalized root-mean-square error (NRMSE) is $30$--$64\%$ below that of each variant, and lower on each of $20$ fresh windows (Supplementary Note~15).

Optimization proceeds in three stages (Methods). A grid search and Bayesian optimization (BO), run until they stall, use only the scalar training loss. The QFI-aware stage then takes natural-gradient steps $\bv_Q^\star=-\GQ^+g/\GamQ$ [Eq.~\eqref{eq:designability}] on the quantum parameters, with $\GQ$ computed from the density matrices along the trajectory, alternating with quasi-Newton steps on the offset, sampling times and drive gains. A quasi-random search of $14{,}000$--$18{,}700$ points per configuration over the box ($m\ge12$) ends $1.2$--$1.8$ times above the grid $+$ BO training loss.

\emph{Advantage.} At every $m$ from $4$ to $30$ the optimized reservoir has a lower held-out NRMSE than the network's mean, by $25$--$69\%$, and a lower value than $99.5$--$100\%$ of the individual draws (Table~\ref{tab:matched}).

\emph{Attribution.} The stage test compares each configuration before and after the QFI-aware stage: at $m\le8$ the stage gains $1$--$6\%$; from $m=12$ on it lowers the held-out error of the stalled black-box optimum by $29$--$51\%$ and widens the margin over the network ($m=18$: $28\%\to60\%$; $m=24$: $21\%\to51\%$). At $m=30$ the black-box optimum is $34\%$ \emph{worse} than the network's mean and beats only $3\%$ of its draws; the QFI-aware stage brings it $35\%$ below the mean, ahead of $99.9\%$ of the draws. Two controls from the same starting points isolate the contribution of the quantum parameter optimization block (Methods, Supplementary Note~14): a black-box search continued for the stage's own evaluation count, and exact-gradient descent on the classical parameters alone (drive gains, offset, sampling times) with the quantum parameters frozen. At $m\le8$ the equal-budget search matches or undercuts the stage and the classical-parameter gradient ends within $3\%$ of it; at $m=12$, $16$, $18$ and $30$ the stage ends $13$--$45\%$ below the better control, and at $m=24$, $6\%$ below. Natural-gradient steps on the quantum parameters alone recover a quarter to two thirds of the gain at $m=12$--$18$; the two blocks together exceed either, the quantum step opening the room that the classical step exploits. The gain grows with the dimension $d$ of the design space ($11$ at $m=4$, $44$ at $m=30$), as the grid and the surrogate of BO lose resolution.

The QFI natural-gradient step gives the steepest descent per unit of distinguishable state motion (Sec.~\ref{sec:bound}) and is invariant under reparametrization. The reported margins are obtained from expectation-value simulations; finite-shot costs are analysed in Supplementary Note~14. The optimized closed loops remain contractive along their training trajectories, with \(\mathrm{median}_t|\lambda_1|=0.90\)--\(0.96\), below the dynamical edge identified in Sec.~\ref{sec:feedback}. Robustness to initialization is also verified: from four different initial states and after a 500-sample washout, every optimized design reproduces its held-out NRMSE to within \(0.04\%\) while satisfying the drive and photon-number constraints on every training and held-out sample (Supplementary Note~14).

\begin{table}[!ht]\centering\small\setlength{\tabcolsep}{3pt}
\caption{\textbf{Quantum advantage and quantum-information advantage at equal readout cost $m$ (Mackey--Glass, five-step forecasting, held-out NRMSE).} ESN (echo-state network): mean $\pm$ standard deviation over $10{,}000$ random draws of the selected hyperparameter setting. QRC (quantum reservoir computer): configuration with the lowest training loss among the layouts tested at each $m$ ($N$ atoms, $V$ sub-times), after grid $+$ BO on the training loss (``simple'' optimization) and after the additional QFI-aware stage. \emph{Quantum advantage}: relative margin of the QFI-optimized QRC below the ESN mean, and share of ESN draws above the QRC value. \emph{Quantum-information advantage}: relative reduction of the held-out NRMSE by the QFI-aware stage over the simple optimization of the same reservoir. Stage table with training errors: Supplementary Table~18. At $m=16$ the three-atom reservoir uses $N_c=8$; re-evaluated at $N_c=10$ its held-out NRMSE is $0.0433$ (margin $67\%$, stage gain $34\%$).}
\label{tab:matched}
\begin{tabular}{@{}rlcccccc@{}}
\toprule
& & & \multicolumn{2}{c}{QRC held-out NRMSE} & \multicolumn{2}{c}{Quantum advantage} & Quantum-information\\
$m$ & $N$, $V$ & ESN & grid $+$ BO & $+$ QFI stage & gain over ESN & ESN draws worse & advantage\\
\midrule
4 & 1, 1 & $0.252\pm0.042$ & $0.1781$ & $0.1760$ & $30\%$ & $100\%$ & $1.2\%$\\
6 & 2, 1 & $0.215\pm0.029$ & $0.1673$ & $0.1616$ & $25\%$ & $99.54\%$ & $3.4\%$\\
8 & 1, 2 & $0.192\pm0.024$ & $0.1345$ & $0.1269$ & $34\%$ & $99.97\%$ & $5.7\%$\\
12 & 2, 2 & $0.162\pm0.019$ & $0.0858$ & $0.0608$ & $62\%$ & $100\%$ & $29\%$\\
16 & 3, 2 & $0.131\pm0.022$ & $0.0658$ & $0.0400$ & $69\%$ & $100\%$ & $39\%$\\
18 & 2, 3 & $0.112\pm0.017$ & $0.0806$ & $0.0449$ & $60\%$ & $100\%$ & $44\%$\\
24 & 2, 4 & $0.081\pm0.012$ & $0.0636$ & $0.0394$ & $51\%$ & $100\%$ & $38\%$\\
30 & 2, 5 & $0.061\pm0.009$ & $0.0816$ & $0.0397$ & $35\%$ & $99.93\%$ & $51\%$\\
\bottomrule\end{tabular}
\end{table}

\emph{Testbeds.} The procedures above rely on gradient-based updates, motivating a bound on the attainable first-order loss gradient that we derive and test below. We evaluate the bound in two exactly tractable testbeds: the four-feature single-atom Jaynes--Cummings cavity, the smallest member of the family considered above (held-out NRMSE $0.449$ at the reference anchor and $0.357$ with the conditioned drive), and a four-spin transverse-field Ising reservoir with reset input injection \cite{Fujii2017}. Both systems are small enough that every quantity entering the hierarchy can be computed directly.

\subsection{A retrained-loss gradient is bounded by, and decomposed against, quantum Fisher information}
\label{sec:bound}
The slope of the retrained loss along any physical direction never exceeds the QFI response budget, and the ratio between the two factorizes into five link factors that each name one way the response is lost. Let $\rho_t(\bth,u)$ be the reservoir state at sample $t$ under input $u$ and physical controls $\bth\in\mathbb R^p$. A fixed set of Hermitian observables $\{O_j\}_{j=1}^m$ defines
\begin{equation}
X_{tj}(\bth,u)=\Tr[\rho_t(\bth,u)O_j].
\label{eq:features}
\end{equation}
For a target $y\in\mathbb R^T$, we study the value after retraining the linear readout,
\begin{equation}
\Lc(\bth,u)=\min_w\frac12\|y-Xw\|^2,\qquad \ws=X^+y,\qquad r=y-X\ws.
\label{eq:value}
\end{equation}
On a fixed-rank neighbourhood, $X^\top r=0$. The envelope theorem then gives, for any physical direction $\bv$ \cite{GolubPereyra1973},
\begin{equation}
D_{\bv}\Lc=-r^\top X_{\bv}\ws,\qquad X_{\bv}=D_{\bv}X.
\label{eq:envelope}
\end{equation}
The derivative of the refitted readout does not appear. Define $\PX=XX^+$, $\PXp=I-\PX$ and $\Pir=rr^\top/\|r\|^2$.

For the projective measurement of $O_j$ at time $t$, error propagation and measurement monotonicity give \cite{BraunsteinCaves1994}
\begin{equation}
|D_{\bv}\langle O_j\rangle_t|^2
\le[\bv^\top\FC^{(j)}(t)\bv]\Var_{\rho_t}(O_j)
\le[\bv^\top\FQ(t)\bv]\Var_{\rho_t}(O_j),
\label{eq:fisher_bound}
\end{equation}
where $\FC^{(j)}(t)$ is the classical Fisher information (CFI) of that measurement and $\FQ(t)$ is the QFI of $\rho_t$. Summing over all measured features defines
\begin{align}
\BC^2(\bv)&\equiv\sum_{t,j}[\bv^\top\FC^{(j)}(t)\bv]\Var_{\rho_t}(O_j),\nonumber\\
\BQ^2(\bv)&\equiv\sum_{t,j}[\bv^\top\FQ(t)\bv]\Var_{\rho_t}(O_j)
=\bv^\top\GQ\bv,\qquad
\GQ\equiv\sum_t\omega_t\FQ(t),\quad
\omega_t\equiv\sum_j\Var_{\rho_t}(O_j).
\label{eq:envelopes}
\end{align}
The marginal QFI $\sum_t\FQ(t)$ depends only on the state trajectory; $\GQ$ is feature matched, weighting the QFI at each time by the total variance of the deployed dictionary, and sets the scale of the response envelope for the chosen features without being a joint multi-time QFI.

We normalize the task response by the current residual and readout,
\begin{equation}
\lo(\bv)=\frac{|D_{\bv}\Lc|}{\|r\|\|\ws\|},\qquad
\lp(\bv)=\frac{\|\PXp X_{\bv}\ws\|}{\|\ws\|},\qquad
\eta_Q(\bv)=\frac{\lo(\bv)}{\BQ(\bv)}.
\label{eq:slopes}
\end{equation}
Here $\lp$ is the part of the deployed response that cannot be reproduced by refitting the present feature span.

\begin{mdframed}[linewidth=0.6pt,innertopmargin=6pt,innerbottommargin=8pt,innerleftmargin=8pt,innerrightmargin=8pt,skipabove=8pt,skipbelow=8pt]
\noindent\textbf{Main result.}

\smallskip\noindent\textbf{Bound.} For every physical direction $\bv$,
\begin{equation}
\lo(\bv)\le\lp(\bv)\le\frac{\|X_{\bv}\ws\|}{\|\ws\|}\le\normtwo{X_{\bv}}\le\normF{X_{\bv}}\le\BC(\bv)\le\BQ(\bv).
\label{eq:chain}
\end{equation}
The largest first-order loss decrease within one QFI unit is
\begin{equation}
\GamQ=\max_{\bv^\top\GQ\bv\le1}(-g^\top\bv)=\sqrt{g^\top\GQ^+g}
=\|r\|\|\ws\|\,\eta_Q(\bv_Q^\star),
\label{eq:designability}
\end{equation}
with $g=\nabla_{\bth}\Lc$ and $\bv_Q^\star=-\GQ^+g/\GamQ$.

\smallskip\noindent\textbf{Link-wise attainability.} Each inequality has a separate equality condition. An SLD measurement closes the two measurement links; a rank-one feature tangent and a matching readout close the norm and readout links; $X_{\bv}\ws\perp\col X$ closes the novelty link; $r\parallel\PXp X_{\bv}\ws$ closes the task link; and $\bv_Q^\star$ attains Eq.~\eqref{eq:designability}. These conditions act on different objects and need not hold simultaneously.

\smallskip\noindent\textbf{Exact decomposition.} The five factors group the six inequalities of Eq.~\eqref{eq:chain}; in particular, the readout factor $\eta_3$ combines the two middle norm inequalities. We define
\begin{align*}
\eta_1&=\frac{\BC}{\BQ},&
\eta_2&=\frac{\normF{X_{\bv}}}{\BC},&
\eta_3&=\frac{\|X_{\bv}\ws\|}{\|\ws\|\,\normF{X_{\bv}}},\\
\eta_4&=\frac{\lp}{\|X_{\bv}\ws\|/\|\ws\|},&
\eta_5&=\frac{\lo}{\lp},
\end{align*}
with zero-denominator cases understood by continuity on the nonzero support. Then $\eta_k\in[0,1]$ and
\begin{equation}
|D_{\bv}\Lc|=\|r\|\,\|\ws\|\,\BQ(\bv)\prod_{k=1}^{5}\eta_k,
\qquad
\Delta\log\lo=\Delta\log\BQ+\sum_{k=1}^{5}\Delta\log\eta_k.
\label{eq:budget}
\end{equation}
The decomposition attributes an observed change of the local slope to the links that moved; it does not predict the sign of a finite intervention before the side terms are recomputed (Sec.~\ref{sec:attribution}).
\end{mdframed}

The proof is elementary (Supplementary Note~2). Equation~\eqref{eq:designability} is Cauchy--Schwarz in the metric $\GQ$, so $\bv_Q^\star$ is the QFI-normalized natural-gradient direction \cite{Amari1998,Stokes2020,Koczor2022}, and Euclidean steepest descent rescaled to the same QFI budget achieves $\Gamma^{(Q)}_{\rm GD}=g^\top g/\sqrt{g^\top\GQ g}\le\GamQ$ (Supplementary Note~3); $\bv_Q^\star$ is the update used in the QFI-aware stage of Sec.~\ref{sec:matched}. Table~\ref{tab:glossary} summarizes the factors.

\begin{table}[!ht]\centering\small
\caption{\textbf{Glossary of the loss budget.} Quantities are evaluated at one physical anchor and one direction $\bv$.}
\label{tab:glossary}
\begin{tabular}{@{}lp{0.70\linewidth}@{}}\toprule
Symbol & Meaning and equality condition\\\midrule
$\BQ(\bv)=\sqrt{\bv^\top\GQ\bv}$ & Feature-matched quantum response budget\\
$\sum_t\FQ(t)$ & Marginal trajectory QFI; zero in a direction certifies zero gradient for every later dictionary and task\\
$\BC(\bv)$ & Measured response budget; $\eta_1=\BC/\BQ=1$ for an SLD-optimal measurement\\
$\normF{X_{\bv}}$ & Total feature motion; $\eta_2=\normF{X_{\bv}}/\BC=1$ when the centred observable is proportional to the classical score\\
$\|X_{\bv}\ws\|/\|\ws\|$ & Response used by the fitted predictor; $\eta_3=1$ for a rank-one response aligned with $\ws$\\
$\lp(\bv)$ & Novel response outside $\col X$; $\eta_4=1$ when the deployed response is entirely novel\\
$\lo(\bv)$ & Normalized task slope; $\eta_5=1$ when the residual is parallel to the novel response\\
$\eta_Q(\bv)$ & Utilization $\lo/\BQ=\prod_k\eta_k$\\
$\GamQ$ & Largest first-order loss decrease per QFI unit\\
\bottomrule\end{tabular}
\end{table}

The refit projection $\PXp$ has no analogue in a frozen-readout gradient (Remark~\ref{rem:refit}). The theory is local to a fixed-rank neighbourhood and to separately acquired single-time expectation values. Under a smooth reparametrization of the physical controls, $\GQ$ transforms as a metric and $g$ as a covector, so $\GamQ$, $\eta_Q$ and $\bv_Q^\star$ are invariant, and a common rescaling of all observables leaves every $\eta_k$ unchanged (Supplementary Note~3).

\begin{remark}[Refit projection]\label{rem:refit}
For the retrained loss,
\begin{equation}
\lo(\bv)=\frac{\|\Pir\PXp X_{\bv}\ws\|}{\|\ws\|}.
\label{eq:refit_identity}
\end{equation}
Thus $\PX X_{\bv}\ws$ is the response that a readout refit absorbs, and $\PXp X_{\bv}\ws$ is new response; $\eta_4$ measures this loss.
\end{remark}

\begin{proposition}[Task-link tightness over targets]\label{prop:task}
Fix $X$ and $X_{\bv}$ on a full-column-rank neighbourhood and set $A=\PXp X_{\bv}$.
\begin{enumerate}
\item[(i)] For every readout $w$ with $Aw\neq0$, choose $r=Aw/\|Aw\|$ and $y=Xw+r$. Then the refitted readout is $\ws=w$ and $\lo(\bv)=\lp(\bv)$.
\item[(ii)] Consequently,
\begin{equation}
\sup_y\ \lo(\bv;y)=\normtwo{A}=\normtwo{\PXp X_{\bv}},
\label{eq:sup_targets}
\end{equation}
attained when $w$ is a leading right singular vector of $A$.
\end{enumerate}
\end{proposition}
The proof is given in Supplementary Note~2. Figure~\ref{fig:hierarchy}d realizes part (ii) at the JC QNG direction, and the forecasting targets used elsewhere are independent of this construction. Figure~\ref{fig:hierarchy}a,b checks the chain at the JC reference anchor.

\subsection{Three quantum statements: certificate, centred SLD and incompatible measurements}
\label{sec:tfi}
Zero QFI along a physical direction proves that no measurement, readout or target can produce a gradient there, and finite QFI is resolved in full only by a measurement matched to that direction. Of the six inequalities in Eq.~\eqref{eq:chain}, only $\BC\le\BQ$ is quantum; the others are properties of the classical feature matrix and its derivative. The quantum content consists of three statements, tested in the spin reservoir and, for the second, in the JC reservoir. The spin reservoir is the four-spin transverse-field Ising register \cite{Fujii2017},
$H=J\sum_{i<j}c_{ij}X_iX_j+h\sum_iZ_i$, $U=e^{-i\tau H}$, with physical controls $\bth=(J,h,\tau)$ and base dictionary $\{Z_i\}$. The couplings $c_{ij}$ are drawn from twenty seeds (medians and interquartile ranges, IQR, in Table~\ref{tab:stats}); Fig.~\ref{fig:tfi}a,b,d shows seed 7, the best of the twenty, and every qualitative statement below holds for all of them.

\begin{corollary}[Impossibility certificate]\label{cor:certificate}
If $\bv^\top\FQ(t)\bv=0$ for every sample $t$, then $D_{\bv}\Lc=0$ for every observable dictionary, readout and target. Conversely, if $\bv^\top\FQ(t)\bv>0$ for some $t$, the SLD $\SLD_{\bv}(t)$ has a nonzero feature tangent along $\bv$.
\end{corollary}
The certificate is measurement free, so a zero-QFI physical direction cannot be repaired downstream (proof and sharpness in Supplementary Note~2). A vanishing QFI certifies that no first-order improvement is possible by varying the parameters included in the QFI. In the spin reservoir the dynamics depend only on $J\tau$ and $h\tau$, so the rescaling $(J,h,\tau)\mapsto(\lambda J,\lambda h,\tau/\lambda)$ leaves every state unchanged and its tangent $\bv_{\rm sym}\propto(J,h,-\tau)$ has zero QFI at every time. At $(1,1,1)$ one eigenvalue of $\GQ$ is at the numerical floor for all twenty seeds, every chain quantity along $\bv_{\rm sym}$ falls to the floor with it, and the other tested directions have finite gradients (Fig.~\ref{fig:tfi}a, Table~\ref{tab:stats}). Rotating away from $\bv_{\rm sym}$ by an angle $\alpha$ makes the marginal QFI vanish as $\alpha^2$ and $|D_{\bv}\Lc|$ as $\alpha$ for three dictionaries and two tasks in every seed (Fig.~\ref{fig:tfi}b). Along the finite orbit $\lambda\in[1/2,2]$ the training loss and held-out NRMSE stay constant to better than $3\times10^{-12}$ although the nominal controls change fourfold (Fig.~\ref{fig:tfi}d). A search over twenty seeds, five anchors and eight QFI matrices found no approximately redundant direction (Fig.~\ref{fig:tfi}f, Supplementary Note~9). The certificate distinguishes zero from finite QFI and nothing in between.

\emph{Centred SLD carries no readout weight.} Let $\SLD_{\bv}(t)$ be the SLD defined by $\partial_{\bv}\rho_t=\tfrac12\{\rho_t,\SLD_{\bv}(t)\}$. At the anchor,
\begin{equation}
|D_{\bv}\langle\SLD_{\bv}\rangle_t|^2
=[\bv^\top\FQ(t)\bv]\Var_{\rho_t}(\SLD_{\bv}),
\label{eq:sld_equality}
\end{equation}
and measurement in its eigenbasis attains $\bv^\top\FC\bv=\bv^\top\FQ\bv$ \cite{BraunsteinCaves1994}, so the two measurement links can be closed for any chosen direction. One fixed basis is optimal for all directions only when the multiparameter compatibility conditions are met \cite{Ragy2016}. Closing the links need not improve the task: $\langle\SLD_{\bv}(t)\rangle=0$ at the anchor, so adding the centred SLD creates a zero feature column that raises the visible Fisher response but receives no readout weight to first order. At the JC anchor $(g,\epsilon,\kappa_c)=(30,20,10)$, adding the SLD along $\bv_Q^\star$ raises the measurement factor $\eta_1$ from $0.38$ to $0.55$ and the observable factor $\eta_2$ from $0.83$ to $0.94$, while the readout factor $\eta_3$ falls from $0.21$ to $0.12$ and $\lo$ is unchanged (Fig.~\ref{fig:hierarchy}c). SLD eigenprojectors behave differently because their probability features are nonzero at the anchor and can carry readout weight. At this anchor the two Fisher-leading projectors raise $\eta_1$ to $0.44$, increase $\lo$ by $29\%$ and lower the training loss from $11.16$ to $10.60$.

\emph{Incompatible SLDs.} For a two-outcome observable the centred observable is proportional to the classical score, so $\eta_2=1$ identically in the spin reservoir and the measurement gap is entirely $\eta_1$. The computational basis $Z^{\otimes4}$ captures only $0.10$ of the feature-matched QFI along the physical directions (Fig.~\ref{fig:tfi}c). The SLDs of $(J,h,\tau)$ do not commute, and each SLD eigenbasis attains its own direction but resolves only a fraction of the others (Table~\ref{tab:stats}). A basis optimized independently at each time reaches $0.88$ in both physical directions simultaneously but never unity (Supplementary Note~9). The best-matched single-control basis resolves $\bv_Q^\star$ at $0.90$, and the matching control differs between seeds (Table~\ref{tab:stats}).

\subsection{The decomposition attributes interventions}
\label{sec:attribution}
The smallest link factor points to the intervention that raises the slope about twice as often as the alternative does, but it does not predict the sign of a finite step. For Euclidean-unit parameter directions, the largest raw response budget is
\begin{equation}
\BQ^{\max}=\max_{\|\bv\|=1}\BQ(\bv)=\sqrt{\lambda_{\max}(\GQ)}.
\label{eq:qfi_ceiling}
\end{equation}
With the readout-free task factor $\chi_r(\bv)=\normF{\Pir X_{\bv}}/\normF{X_{\bv}}$ along the dominant-QFI direction, $\chi_r\BQ$ is a task-conditioned ceiling. On the $11\times11$ grid $g,\kappa_c\in\{10,\ldots,20\}$ at $\epsilon=15$, the raw ceiling varies by a factor of $1.6$, $\chi_r\BQ$ by $18$ and the trained loss by $1.5$ (Fig.~\ref{fig:scan}a--c). The largest intrinsic response does not identify the best operating point. Under input shuffling the QFI and its dominant directions are stable while $\chi_r$ halves (Fig.~\ref{fig:scan}d,e).

Along $\bv_Q^\star$, the smallest factor is the novelty factor $\eta_4$ at 76 of 121 grid points and the readout factor $\eta_3$ at the other 45. At every grid point we apply the intervention that acts on the smallest factor (two Fisher-leading SLD eigenprojectors added to the dictionary where $\eta_3$ is smallest, input profiling with $P\le4$ where $\eta_4$ is smallest) and record the change of the local slope $\lo$ along the fixed direction and of the training loss after the finite step (Fig.~\ref{fig:scan}h). Both lower the training loss at all 121 points by construction, so the informative quantity is the slope. The intervention aimed at the smallest factor raises $\lo$ at 65 of 121 points ($53.7\%$), and the other intervention at 35 of 121 ($28.9\%$). The smallest factor favours the matching intervention by roughly two to one but does not determine the outcome; the targeted factor itself rises at most points (Table~\ref{tab:stats}, Supplementary Note~12). At anchor B $=(11,15,15)$, input profiling raises $\lo$, with $89\%$ of the log increase carried by $\eta_4$; at anchor A $=(17,15,15)$, the projectors raise $\eta_3$ by $1.6\times$ and $\lo$ by $42\%$, with side terms $\eta_4$ ($-0.50$) and $\eta_5$ ($+0.30$) of the same size as the targeted term (Table~\ref{tab:budget}, Fig.~\ref{fig:scan}f,g). The upstream budget grows with atom number (Supplementary Note~8); its use depends on the downstream links.

\begin{table}[!ht]\centering\small
\caption{\textbf{Loss budget at two anchors of the parameter scan, before and after the intervention.} Chain quantities of Eq.~\eqref{eq:chain} and the five link factors along $\bv_Q^\star$ (one QFI unit, $\BQ=1$ before the intervention) at anchor A $=(g,\epsilon,\kappa_c)=(17,15,15)$, where the smallest factor is $\eta_3$, and anchor B $=(11,15,15)$, where it is $\eta_4$. The intervention acts on the smallest factor. At A the two Fisher-leading SLD eigenprojectors are added to the dictionary at fixed $\bth$ and trajectory. At B the input is profiled with $P\le4$ at fixed $\bth$. The direction $\bv$ is held fixed. $\BQ$ after the intervention is evaluated with the changed weights $\omega_t$ (A) or on the profiled trajectory (B). The last rows give the log-additive attribution of the change.}
\label{tab:budget}
\begin{tabular}{@{}lcccc@{}}\toprule
 & \multicolumn{2}{c}{Anchor A ($\eta_3$)} & \multicolumn{2}{c}{Anchor B ($\eta_4$)}\\
\cmidrule(lr){2-3}\cmidrule(lr){4-5}
Quantity & before & after & before & after\\\midrule
$\BQ$ & 1 & 1.008 & 1 & 1.007\\
$\BC$ & 0.263 & 0.284 & 0.499 & 0.502\\
$\normF{X_{\bv}}$ & 0.256 & 0.277 & 0.499 & 0.502\\
$\|X_{\bv}\ws\|/\|\ws\|$ & $1.78\times10^{-2}$ & $3.10\times10^{-2}$ & $8.65\times10^{-2}$ & $8.25\times10^{-2}$\\
$\lp$ & $1.83\times10^{-3}$ & $1.93\times10^{-3}$ & $2.25\times10^{-3}$ & $2.50\times10^{-3}$\\
$\lo$ & $6.50\times10^{-4}$ & $9.20\times10^{-4}$ & $6.82\times10^{-4}$ & $8.08\times10^{-4}$\\\midrule
$\eta_1$ & 0.263 & 0.281 & 0.499 & 0.499\\
$\eta_2$ & 0.973 & 0.977 & 1.000 & 1.000\\
$\eta_3$ & \textbf{0.070} & 0.112 & 0.173 & 0.164\\
$\eta_4$ & 0.102 & 0.062 & \textbf{0.026} & 0.030\\
$\eta_5$ & 0.356 & 0.478 & 0.303 & 0.324\\
$\eta_Q$ & $6.50\times10^{-4}$ & $9.13\times10^{-4}$ & $6.82\times10^{-4}$ & $8.03\times10^{-4}$\\\midrule
$\Delta\log\lo$ & -- & $+0.35$ & -- & $+0.17$\\
$\Delta\log\BQ$ & -- & $+0.008$ & -- & $+0.007$\\
$\Delta\log\eta_1,\ \Delta\log\eta_2$ & -- & $+0.07$, $+0.00$ & -- & $0.00$, $0.00$\\
$\Delta\log\eta_3$ & -- & $+0.47$ & -- & $-0.05$\\
$\Delta\log\eta_4$ & -- & $-0.50$ & -- & $+0.15$\\
$\Delta\log\eta_5$ & -- & $+0.30$ & -- & $+0.07$\\
\bottomrule\end{tabular}
\end{table}

\begin{table}[p]\centering\scriptsize\setlength{\tabcolsep}{3pt}
\caption{\textbf{Twenty-seed and 121-point statistics.} Spin reservoir: twenty couplings at $(J,h,\tau)=(1,1,1)$, dictionary $\{Z_i\}$, Mackey--Glass unless stated. JC reservoir: the grid of Fig.~\ref{fig:scan}, thirty drive-gain anchors per task (Methods). Medians with interquartile ranges (IQR) in brackets.}
\label{tab:stats}
\begin{tabular}{@{}p{0.40\linewidth}p{0.21\linewidth}p{0.25\linewidth}p{0.10\linewidth}@{}}\toprule
Quantity & Median [IQR] & Range or count & Source\\\midrule
\multicolumn{4}{@{}l}{\emph{Spin reservoir, twenty seeds}}\\
Training loss & $7.74$ $[6.99,9.24]$ & seed 7 (best): $5.41$ & Fig.~\ref{fig:tfi}e\\
Held-out NRMSE & $0.391$ $[0.359,0.425]$ & seed 7 (best): $0.333$ & Fig.~\ref{fig:tfi}e\\
Nonzero eigenvalues of $\GQ$ & $1.99\times10^4$ $[1.81,2.52]\times10^4$; $9.55\times10^4$ $[7.63,13.2]\times10^4$ & third at the floor, $|\lambda|/\lambda_{\max}\le7\times10^{-15}$ & Fig.~\ref{fig:tfi}a\\
Certificate exponents of QFI and $|D_{\bv}\Lc|$ in $\alpha$ & $2.000$, $1.000$ & every seed & Fig.~\ref{fig:tfi}b\\
Restricted eigenvalue ratio, near-redundant search & -- & minimum $0.0136$ over 800 entries & Fig.~\ref{fig:tfi}f\\
Training-loss change for $|s|\le0.3$ along the best candidate & -- & $-23\%$ to $+92\%$ & SI Note~9\\
Capture by $Z^{\otimes4}$ along physical directions & $0.10$ $[0.08,0.12]$ & $0.05$--$0.16$ & Fig.~\ref{fig:tfi}c\\
Normalized commutator norms of the three SLD pairs & $0.40$, $0.34$, $0.22$; widths $\le0.04$ & -- & SI Note~9\\
Capture by the $J$-SLD basis along $e_h$ & $0.18$ $[0.15,0.23]$ & -- & Fig.~\ref{fig:tfi}c\\
Capture by the $h$-SLD basis along $e_\tau$ & $0.79$ $[0.77,0.82]$ & -- & Fig.~\ref{fig:tfi}c\\
Capture by the $h$-SLD basis along $e_J$ & $0.23$ $[0.14,0.29]$ & -- & Fig.~\ref{fig:tfi}c\\
Per-time optimized basis, both physical directions & $0.88$ $[0.86,0.92]$ & $0.75$--$0.95$ & SI Note~9\\
Best-matched single-control basis along $\bv_Q^\star$ & $0.90$ $[0.78,0.97]$ & $J$ for 10, $\tau$ for 7, $h$ for 3 seeds & Fig.~\ref{fig:tfi}e\\
Task-link closure $\lo=\lp$ at $\bv_Q^\star$ & -- & $2\times10^{-10}$ relative, every seed & SI Note~9\\
Two SLD eigenprojectors added: training loss, held-out NRMSE & $-9\%$, $-0.026$ & lower at 20 of 20 seeds & SI Note~13\\
\midrule
\multicolumn{4}{@{}l}{\emph{JC reservoir, 121 grid points, along $\bv_Q^\star$}}\\
Smallest link factor & -- & $\eta_4$ at 76, $\eta_3$ at 45 points & Fig.~\ref{fig:scan}h\\
Intervention on the smallest factor raises $\lo$ & -- & 65 of 121 ($53.7\%$); 22 of 45 $\eta_3$-limited, 43 of 76 $\eta_4$-limited & Fig.~\ref{fig:scan}h\\
Other intervention raises $\lo$ & -- & 35 of 121 ($28.9\%$) & SI Table~13\\
Targeted factor rises & -- & 30 of 45 ($\eta_3$), 72 of 76 ($\eta_4$) & SI Note~12\\
Spearman correlation, $\Delta\log\lo$ against finite-step loss change & $0.06$ (projectors), $0.84$ (profiling) & -- & SI Note~12\\
Measurement factor $\eta_1$, base dictionary & $0.35$ $[0.29,0.46]$ & -- & Fig.~\ref{fig:advantages}a\\
$\eta_1$ ratio after two SLD eigenprojectors & $\times1.05$ & raised at 121 of 121; $\times1.08$ for the full SLD basis & Fig.~\ref{fig:advantages}a\\
Two SLD eigenprojectors: training loss, held-out NRMSE, $\lo$ along the fixed direction & $-9\%$ (loss) & loss lower at 121 of 121, NRMSE lower at 118 of 121, $\lo$ higher at 55 of 121 & SI Table~16\\
$\GamQ/\Gamma^{(Q)}_{\rm GD}$ & $1.13$ & up to $3.3$; above one at 121 of 121 & SI Note~3\\
\midrule
\multicolumn{4}{@{}l}{\emph{Measurement-conditioned drive, thirty Mackey--Glass anchors}}\\
$\eta_5(\beta_{\rm slope})/\eta_5(0)$ & $0.053$ $[0.023,0.079]$ & medians $0.40\to0.020$ & SI Table~10\\
$\BQ(e_\beta)$ at $\beta_{\rm slope}$ against $\beta=0$ & $48$ against $82$ (ratio $0.61$) & -- & SI Note~11\\
$\mathrm{median}_t|\lambda_1|$ at $\beta_{\rm slope}$ & $0.953$ $[0.950,0.956]$ & below one at 30 of 30 & SI Note~11\\
$\beta_{\rm slope}-\beta^{\rm test}_{\rm opt}$ & $-0.035$ $[-0.038,-0.032]$ & $-0.054$ to $-0.016$; negative at 30 of 30 & Fig.~\ref{fig:feedback}f\\
Held-out regret at $\beta_{\rm slope}$ & $0.36\%$ & maximum $0.85\%$ & SI Table~10\\
$\beta_{\rm edge}-\beta^{\rm test}_{\rm opt}$ & $0.155$ $[0.119,0.177]$ & minimum $0.091$; positive at 30 of 30 & Fig.~\ref{fig:feedback}f\\
Evaluations to $\beta_{\rm slope}$, safeguarded root search & $9$ & $5$--$9$ (replay, 30 anchors), $7$--$11$ (new trajectories, 10 anchors); sweep $26$ to the edge, $48$ in all; golden section $13$ & Fig.~\ref{fig:advantages}b\\
\midrule
\multicolumn{4}{@{}l}{\emph{Measurement-conditioned drive, thirty NARMA10 anchors}}\\
$\BQ(e_\beta)$ growth factor to $\beta_{\rm slope}$; $\eta_5$ at $\beta_{\rm slope}$ & $2.8$; $0.011$ & 21 anchors in the converged range & SI Table~11\\
Truncation limit $\beta_{\rm nc}$; $\beta_{\rm slope}$ within $0.05$ of the held-out optimum & $-0.90$ & 15 of the 21 with $\beta_{\rm slope}$ before $\beta_{\rm nc}$; no zero of $D_\beta\Lc$ before $\beta_{\rm nc}$ at 9 of 30 & SI Table~11\\
\bottomrule\end{tabular}
\end{table}

\subsection{Measurement-conditioned drive: a diagnosis of where the gain stops helping}
\label{sec:feedback}
Conditioning the drive on a measured readout stops helping where the task link closes, before the dynamical edge and while the QFI budget is still large, and the zero of the exact slope at the training minimum locates that gain from training data alone. We drive the JC reservoir with $\tilde u_t=u_t+\beta\langle p\rangle_t$, the measurement-conditioned drive introduced as measured feedback in Refs.~\cite{Kobayashi2024,Zhu2025npj}. The reference anchor is $(30,20,10)$ (Fig.~\ref{fig:feedback}a--e); the statistics use thirty anchors per task (Methods).

From $\beta=0$ to $0.75$ at the reference anchor, along $e_g$, $\Delta\log\lo=1.44$ with $\Delta\log\BQ=0.55$ and $\Delta\log\eta_{1..5}=(0.06,0.29,-0.35,1.82,-0.93)$ (Fig.~\ref{fig:feedback}d). The novelty factor $\eta_4$ accounts for $126\%$ of the net increase, rising from $0.12$ to $0.76$, and negative side terms partly cancel its gain. The rise in novelty comes from retention, the only memory ratio that changes strongly ($5.5\times$ at lag ten; Supplementary Note~11).

The diagnosis follows from the chain along $e_\beta$. Over the stable branch, $\BQ(e_\beta)$ remains large, falling smoothly from $187$ to $47$ at the reference anchor (Fig.~\ref{fig:feedback}c). The task factor $\eta_5$ closes. It falls from $0.32$ to $0.032$ near $\beta=0.70$, where the exact slope $D_\beta\Lc$ crosses zero at $\beta_{\rm slope}=0.684$ and the training loss is minimal ($\beta^{\rm train}_{\rm opt}=0.682$), while the held-out NRMSE falls from $0.449$ to $0.357$ (Fig.~\ref{fig:feedback}a,b). The training optimum also precedes the dynamical edge: the slow mode $\mathrm{median}_t|\lambda_1|$ of the closed-loop tangent map, Eq.~\eqref{eq:feedback_tangent}, reaches one only at $\beta_{\rm edge}=0.803$ (Fig.~\ref{fig:feedback}b). Beyond the edge ($\beta=0.9$), $\BQ(e_\beta)$ grows by two orders of magnitude while $\eta_3=0.05$, $\eta_5=0.03$ and $\GamQ=0.012$ (Fig.~\ref{fig:feedback}c,e): the gain with the largest QFI has among the smallest QFI-normalized slopes.

The thirty Mackey--Glass anchors show the same picture (Table~\ref{tab:stats}): at the gain nearest $\beta_{\rm slope}$, the task factor has fallen from $0.40$ to $0.020$, $\BQ$ remains at $61\%$ of its $\beta=0$ value, the novel response $\lp$ doubles and the loop is still contracting ($\mathrm{median}_t|\lambda_1|=0.953$). Geometrically, $\eta_5=|\cos\theta_{\rm task}|$ with $\theta_{\rm task}$ the angle between the residual and the novel response: the drive keeps producing new motion, but both vectors turn until that motion is perpendicular to what the task still needs (Supplementary Note~11, Supplementary Fig.~8). At every one of the thirty anchors, $\beta_{\rm slope}$ lies below the held-out optimum $\beta^{\rm test}_{\rm opt}$, by $0.035$ at the median, and the held-out NRMSE at $\beta_{\rm slope}$ is within $0.36\%$ of the best held-out value in the converged range (median regret; Fig.~\ref{fig:feedback}f).

Two caveats fix the wording of the rule. It refers to the zero of the exact slope at the training minimum, which need not be the first zero: in a two-atom JC reservoir at the same anchor the first zero ($\beta=0.249$) is a shallow local minimum, whereas the held-out optimum ($0.783$, NRMSE $0.299$) coincides with the zero at the pre-collapse training minimum ($0.779$), and stopping at the first zero would forfeit $13\%$ in held-out NRMSE (Fig.~\ref{fig:feedback}g). NARMA10 shows the same diagnosis in a different regime: the descent direction is $\beta<0$, $\BQ$ grows by a median factor of $2.8$ while $\eta_5$ again collapses to $0.011$, and validity is limited by photon-number growth beyond a gain $\beta_{\rm nc}$ at which the truncation stops converging; $\beta_{\rm slope}$ is within $0.05$ of the held-out optimum at 15 of the 21 anchors where it lies in the converged range (Table~\ref{tab:stats}, Supplementary Note~11, Supplementary Fig.~7).

\subsection{Two further design gains from pre-measurement information}
\label{sec:advantages}
The same pre-measurement information improves two further design decisions in the small JC reservoir (Fig.~\ref{fig:advantages}a,b, Table~\ref{tab:stats}).

\emph{Observables.} Across the 121 grid points of Fig.~\ref{fig:scan} the deployed dictionary certifies only $0.35$ of $\BQ$ along $\bv_Q^\star$. Adding the two Fisher-leading SLD eigenprojectors raises the measurement factor $\eta_1$ at every point, lowers the training loss at 121 of 121 points ($-9\%$ median) and the held-out NRMSE at 118, and lowers both at 20 of 20 spin seeds (Fig.~\ref{fig:advantages}a). The added observables are chosen from the SLD of the state without reference to the task. The slope along the fixed direction rises at only 55 of 121 points; the gain of a matched measurement lies in the fit, not in the shot cost of the gradient (Supplementary Note~13).

\emph{Drive gain.} A safeguarded root search on the exact slope $D_\beta\Lc$, replayed on the stored scans, returns $\beta_{\rm slope}$ at all thirty anchors in $5$--$9$ evaluations, against a median of $26$ gains of a sweep to the edge and $48$ in all, at the same held-out regret (Fig.~\ref{fig:advantages}b); with new trajectories at ten anchors it takes $7$--$11$ (Supplementary Note~13). A golden-section search on the loss alone needs $13$ evaluations at the median on the replay and $16$ with new trajectories, but less wall time ($107$ against $153$~s), because each slope evaluation adds a complex-step derivative; the chain diagnosis adds $16\%$ to an evaluation.

\section{Discussion}
The advantage of Sec.~\ref{sec:matched} is an advantage of physical information. The grid, the Bayesian search and the network's hyperparameter search see nothing but the training loss; the QFI-aware stage sees the metric $\GQ$, a property of the state before measurement that no search on the measured loss can use. The stage is most effective where black-box search is weakest, in $23$--$44$-dimensional design spaces whose parameters differ by orders of magnitude in sensitivity (the eigenvalues of $\GQ$ over the physical parameters span $3.5$--$5.5$ decades at the optima). The comparison is at matched readout cost: the network is tuned in its hyperparameters and averaged over random reservoirs, as is standard, whereas every parameter of the quantum device is optimized, the design freedom that the QFI metric organizes. The errors are expectation values; holding the margin at finite shot number requires $7\times10^{6}$ ($m=16$) to $2\times10^{10}$ ($m=8$) shots per feature and time sample with a refitted readout (Supplementary Note~14).

Equation~\eqref{eq:budget} separates a physical learning gradient into an upstream response budget and five conversion factors, one for each way the same small gradient can arise: the certificate (Corollary~\ref{cor:certificate}) settles whether a direction can be trained at all, the measurement factors whether the measurement is the bottleneck, $\eta_4$ how much response a readout refit absorbs, and $\eta_5$ whether the surviving response points at the unresolved target. The bound itself is loose by two to six orders of magnitude at every evaluated point (two to five along $\bv_Q^\star$), because the six equality conditions are not met simultaneously, while Eq.~\eqref{eq:envelope} supplies the gradient exactly. The advantage of Sec.~\ref{sec:matched} and the design gains of Sec.~\ref{sec:advantages} spend the same surplus. The pre-measurement state carries a factor of $1/\eta_1\approx2$--$5$ more Fisher response than the deployed readout returns, and the QFI metric, the SLD eigenbasis and the exact slope direct that surplus to the parameter update, the observables and the drive gain.

Conditioning the input drive on measured readouts is usually explained through memory \cite{Kobayashi2024,Zhu2025npj}; in the JC reservoir it helps by creating novel response through retention, and its useful regime ends when the task link closes. These statements differ from capacity measures \cite{Dambre2012,Innocenti2023} and from Fisher analyses of frozen variational predictors \cite{Haug2021,Abbas2021,Meyer2021}: here the readout is refitted after every physical update, and both the refit and the residual projection are essential. For a classical stochastic reservoir \cite{Ehlers2025}, whose features are expectation values of a diagonal state, the deployed measurement attains $\FC=\FQ$ and the chain reduces to its four downstream links.

The response budget $\BQ$, the SLD eigenbases and $\eta_1$ require the density matrix, so the decomposition and the QFI-aware stage are design-time tools; the decomposition concerns the local slope, and a finite intervention also moves the side terms (Sec.~\ref{sec:attribution}, Supplementary Note~12).

Two experimental routes narrow the density-matrix requirement. The CFI of the deployed dictionary follows from the same outcome statistics as the features and gives $\BC$ directly; over the JC grid it lies a factor of $2$--$5$ below $\BQ$ ($\eta_1=0.51$ along the stable branch of the conditioned drive), while the four downstream links involve only measured features (Supplementary Note~12). Randomized-measurement protocols estimate the QFI without tomography \cite{Rath2021,Yu2021}, at a cost that grows with dimension and, because the chain uses single-time expectation values, with a restart from washout per sample.

\section{Methods}
\subsection{Jaynes--Cummings reservoir}
The $N$-atom cavity state obeys
\begin{align}
\dot\rho={}&-i[H_t,\rho]+\tfrac{\kappa_c}{2}\mathcal D[a]\rho+\gamma\sum_{i=1}^N\mathcal D[\sigma_i^-]\rho,\nonumber\\
H_t={}&\omega_ca^\dagger a+\omega_a\sum_{i=1}^N\sigma_i^+\sigma_i^-
+\sum_{i=1}^Ng_i(a^\dagger\sigma_i^-+a\sigma_i^+)
+i\epsilon u_t(a^\dagger-a),
\label{eq:jc}
\end{align}
with $\mathcal D[L]\rho=L\rho L^\dagger-\tfrac12\{L^\dagger L,\rho\}$. We work in the drive-rotating frame, with all rates expressed in a common reference unit. Inputs are piecewise constant over intervals $\Delta t=1/30$, and the state is propagated by the exact matrix exponential of the Liouvillian. Feature row $t$ uses the state at the beginning of interval $t$, after $u_0,\ldots,u_{t-1}$. Unless stated otherwise, $N_c=10$, $\omega_c=40$, $\omega_a=20$, and 100 samples are discarded as washout. For the single-atom data the atomic decay is tied to the cavity decay, $\gamma=\kappa_c/2$, so the controls are $\bth=(g,\epsilon,\kappa_c)$. The four measured features are $x=a+a^\dagger$, $p=-i(a-a^\dagger)$, $\sigma_x$ and $\sigma_y$, with a separate bias column and no virtual nodes. In the matched-cost comparison (Sec.~\ref{sec:methods_matched}) the rotating-frame frequencies $\omega_c\to\Delta_c$ and $\omega_a\to\Delta_{a,i}$, the atomic decay rates $\gamma\to\gamma_i$ and the couplings $g_i$ are individually free, and $\sigma^x_i,\sigma^y_i$ are measured for every atom.

Cutoff convergence is checked from the photon occupation and repeated calculations at adjacent $N_c$. At $(30,20,10)$, $(15,15,15)$ and $(11,15,15)$, the training loss, $\BQ^{\max}$ and $\GamQ$ change by less than $10^{-4}$ in relative terms under $N_c\to N_c+2$; $\BC$, $\eta_1$ and $\eta_2$ change by less than $6\times10^{-5}$ for the directions reported in Fig.~\ref{fig:hierarchy}. The quadrature CFI is computed from the projective measurement in the eigenbasis of the truncated $x$ or $p$ operator, the finite-dimensional representation of homodyne detection.

\subsection{Measurement-conditioned input drive}
\label{sec:methods-feedback}
The input enters the reservoir through a drive that is conditioned on the reservoir's own measured readouts. This is part of how the reservoir is built, the counterpart of the input encoding of an echo-state network; the construction was introduced as measured feedback in Refs.~\cite{Kobayashi2024,Zhu2025npj}. Both studies of the drive (Secs.~\ref{sec:matched} and~\ref{sec:feedback}) use one law, with an input offset and a gain on every feature,
\begin{equation}
\tilde u_t=u_t+b+\bbeta^{\top}X_t=u_t+b+\sum_{s=1}^{V}\Tr\bigl[\rho_{t-1,s}\,O_{\beta,s}\bigr],\qquad O_{\beta,s}\equiv\sum_j\beta_{js}O_j .
\label{eq:fbdrive}
\end{equation}
Here $X_t\in\mathbb{R}^m$ holds the $V$ sub-time samples of interval $t-1$, $\rho_{t-1,s}$ being the state at its $s$-th sampling time, the last at the end of the interval ($\rho_{t-1,V}=\rho_t$). All samples are taken before interval $t$ starts, and $\tilde u_t$ drives the evolution from $\rho_t$ to $\rho_{t+1}$, so the loop is causal. Section~\ref{sec:feedback} uses the special case $V=1$, $b=0$, $\bbeta=\beta e_p$, that is $\tilde u_t=u_t+\beta\langle p\rangle_t$ with $\langle p\rangle_t=\Tr[\rho_tO_p]$, one scalar gain on one observable; Sec.~\ref{sec:matched} optimizes $b$ and all $m$ components of $\bbeta$. We use ensemble expectation values, corresponding to the infinite-trajectory limit of the continuous-measurement protocol of Ref.~\cite{Zhu2025npj}; our $p=i(a^\dagger-a)$ and $\sigma_y$ have the opposite sign to that reference, which only flips the sign of the corresponding gains. Because the fed-back observables are fixed, the trajectory depends on $(\bth,b,\bbeta)$ but not on $\ws$, so Eq.~\eqref{eq:envelope} also holds for $\mu=\beta$ and for every component of $\bbeta$. Derivatives with respect to $\bth$, $\beta$ and past inputs are propagated through the loop by complex-step differentiation, and derivatives with respect to $(b,\bbeta)$ in Sec.~\ref{sec:matched} by the exact adjoint.

Let $\Phi_t[A]=e^{\Lc(\tilde u_t)\Delta t}[A]$ and
$c_t=\partial_u\Phi(u)[\rho_t]|_{u=\tilde u_t}$. For the single-channel loop, a perturbation $A$ changes the conditioning signal by $\beta\Tr[AO_p]$, and the closed-loop tangent superoperator is
\begin{equation}
\delta\rho_{t+1}=\Jc_t[\delta\rho_t]+c_t\,\delta u_t,
\qquad
\Jc_t[A]=\Phi_t[A]+\beta c_t\Tr[AO_p].
\label{eq:feedback_tangent}
\end{equation}
The second term is the measurement--drive--state return path. Trace preservation makes the traceless perturbation subspace invariant; $\lambda_1$ is the largest-magnitude eigenvalue of $\Jc_t$ on that subspace. For the general loop of Eq.~\eqref{eq:fbdrive} the state of the loop is $(\rho_t,s_t)$ with $s_t=\bbeta^\top X_t$; one interval maps $\delta\rho_{t,s}=\Phi_{t,s}[\delta\rho_{t,s-1}]+c_{t,s}\,\delta s_t$ over its $V$ sub-steps ($\delta\rho_{t,0}=\delta\rho_t$, $\delta\rho_{t,V}=\delta\rho_{t+1}$, $\Phi_{t,s}$ and $c_{t,s}$ defined as above for sub-step $s$) and $\delta s_{t+1}=\sum_s\Tr[\delta\rho_{t,s}O_{\beta,s}]$, which reduces to Eq.~\eqref{eq:feedback_tangent} for $V=1$, $b=0$, $\bbeta=\beta e_p$, and $\lambda_1$ is taken on traceless $\delta\rho$ together with $\delta s$. The open-loop frozen-map estimate is retained as a separate diagnostic in Supplementary Note~11. Runs with $|\operatorname{Re}\tilde u_t|>10$ are classified as unstable in Sec.~\ref{sec:feedback} and inadmissible in Sec.~\ref{sec:matched}. The photon number is controlled differently in the two studies: Sec.~\ref{sec:feedback} scans gains until the truncation stops converging ($\beta_{\rm nc}$ below); Sec.~\ref{sec:matched} imposes $\max_t\langle n\rangle\le3$ during the search and re-evaluates every end point at $N_c+2$. The input-memory ratios are averaged over lags 1--10 and five seeds; the effective rank of the lag response is the participation ratio of the singular values of the $4\times10$ matrix $\partial\langle O_j\rangle_t/\partial u_{t-k}$ after column normalization.

\emph{Stopping-rule quantities.} At each anchor the gain is scanned from $\beta=0$ in the descent direction of the training loss (the sign of $-D_\beta\Lc$ at $\beta=0$: $+\beta$ for Mackey--Glass, $-\beta$ for NARMA10), the readout is refitted at every gain, and the following are extracted with fixed conventions. $\beta_{\rm slope}$ is the zero of the exact slope $D_\beta\Lc$ (envelope identity, complex step through the loop) at the training minimum, found by linear interpolation between grid points; when $D_\beta\Lc$ has several zeros before the edge, the zero at the global training minimum of the converged range is used and the first zero is reported separately (Supplementary Note~11). $\beta^{\rm train}_{\rm opt}$ is the vertex of the parabola through the first local minimum of the sampled training loss along the descent coordinate and its two neighbours; $\beta^{\rm test}_{\rm opt}$ is the same rule applied to the held-out NRMSE, used only as a check on the training-based rule and never for selection. $\beta_{\rm edge}$ is the first crossing of $\operatorname{median}_t|\lambda_1(t,\beta)|=1$ over $n_{\rm samp}$ sampled training states ($n_{\rm samp}=10$ for the reference-anchor data of Fig.~\ref{fig:feedback}a--e and $6$ for the thirty-anchor scans and the two-atom sweep; at the reference anchor the two scans give $\beta_{\rm edge}=0.803$ and $0.825$ and $\beta_{\rm slope}=0.684$ and $0.689$; $D_\beta\Lc$ and the loss do not depend on $n_{\rm samp}$, so the $\beta_{\rm slope}$ difference reflects the gain spacing, $0.05$ against $0.025$). $\beta^{\max}_{\rm edge}$ uses $\max_t|\lambda_1|$ instead. $\beta_{\rm nc}$ marks the first gain at which the maximum photon number exceeds twice its $\beta=0$ value. Beyond it, the $N_c=10$ truncation is not converged and no number is quoted. The regret of $\beta_{\rm slope}$ is the relative excess of the held-out NRMSE at $\beta_{\rm slope}$ over the best held-out value on the grid before $\min(\beta_{\rm edge},\beta_{\rm nc})$.

\emph{Anchor set.} The one-atom statistics use thirty anchors per task: the $5\times5$ sub-grid $g,\kappa_c\in\{10,12,15,18,20\}$ at $\epsilon=15$, the four additional grid anchors $(12,15,11)$, $(15,15,11)$, $(18,15,11)$ and $(18,15,19)$, and the reference anchor $(30,20,10)$. Gains run from $0$ to $1$ in steps of $0.025$, then more coarsely to $2$, for Mackey--Glass, and from $+0.1$ to $-1$ in steps of $0.025$, then more coarsely to $-3$, for NARMA10. The two-atom reservoir uses $N=2$ with distinct couplings $g_i=g\,(1,\tfrac12)$, six features ($x$, $p$, $\sigma_{x,i}$, $\sigma_{y,i}$), the same anchor and channel, and gains in steps of $0.05$. The scan and statistics scripts are part of the deposited code.

\subsection{Transverse-field Ising reservoir}
The spin reservoir has four qubits and
$H=J\sum_{i<j}c_{ij}X_iX_j+h\sum_iZ_i$, with fixed $c_{ij}\sim\mathcal U(-1,1)$. At each sample, qubit 1 is reset to $|\psi_u\rangle=\sqrt{1-u_t}|0\rangle+\sqrt{u_t}|1\rangle$ and the register evolves under $U=e^{-i\tau H}$. The initial state is $|0\rangle^{\otimes4}$ and the controls are $(J,h,\tau)$ at the anchor $(1,1,1)$. Mackey--Glass values are mapped affinely to $[0,1]$ for the reset amplitude; NARMA10 uses $u/0.5$. The base dictionary is $\{Z_i\}_{i=1}^4$ plus a bias; enlarged dictionaries add $\{X_i\}$ or $\{Z_iZ_j\}_{i<j}$. Complex-step state derivatives use $h=10^{-6}$ and agree with centred feature differences to the expected truncation error (relative $1.5\times10^{-4}$ median over seeds, worst $3.3\times10^{-3}$). SLD eigenprojectors with eigenvalues closer than $10^{-7}$ are merged when computing the measurement families of Fig.~\ref{fig:tfi}c.

\emph{Seeds and statistics.} The couplings $c_{ij}$ are drawn with the NumPy generator seeded by $s\in\{0,\ldots,19\}$; the single-seed panels use seed 7, shown in Fig.~\ref{fig:tfi}a,b,d. For every seed the full chain along the six reported directions, the certificate sweep, the symmetry orbit, the measurement-family ratios and the per-time optimized basis (Supplementary Note~9) are recomputed at $(1,1,1)$ for the three dictionaries and both tasks. The recomputation for seed 7 reproduces the single-seed values of Fig.~\ref{fig:tfi}a,b,d to better than $2.5\times10^{-4}$ relative for every quantity above the numerical floor. We report the median and the interquartile range over the twenty seeds; a quantity is ``at the floor'' when its magnitude relative to the largest eigenvalue of $\GQ$ is below $10^{-8}$ (observed $\le7\times10^{-15}$), and floor values are not quoted as numbers because they change between runs. The near-zero-QFI search (Fig.~\ref{fig:tfi}f) diagonalizes, for each seed, the marginal QFI $\sum_t\FQ(t)$ of each input drive and $\GQ$ of each dictionary and task at the five anchors $(1,1,1)$, $(0.5,1,1)$, $(1,0.5,1)$, $(1,1,0.5)$ and $(2,1,1)$ on the plane orthogonal to $\bv_{\rm sym}$, and reports the ratio of the smaller to the larger restricted eigenvalue; the best candidate is then evaluated along the chain and by finite moves of $|s|\le0.3$.

\subsection{Tasks}
Mackey--Glass forecasting uses $\dot x=0.2x(t-17)/[1+x(t-17)^{10}]-0.1x$, integrated by Euler steps of one from $x(0)=1$ after discarding 1,000 transient steps, with a five-step-ahead target. The 3,000-sample sequence uses 100 washout samples, 2,000 training samples and the remaining 895 samples for testing after the target shift. NARMA10 follows $y_{t+1}=0.3y_t+0.05y_t\sum_{i=0}^{9}y_{t-i}+1.5u_{t-9}u_t+0.1$, with $u_t\sim\mathcal U(0,0.5)$ and the same split. Held-out data are never used to select directions, input maps, model orders or stopping rules.

\subsection{Readout loss and forecasting protocol}
The theory is written for minimum-norm ordinary least squares (OLS). Numerical fits include a bias and a ridge stabilizer $\delta=10^{-10}$,
\begin{equation}
\Lc_\delta=\frac12\|y-Xw\|^2+\frac\delta2\|w\|^2.
\end{equation}
The envelope identity remains exact for this optimized value. The OLS orthogonality used in the theorem is violated only by $X^\top r=\delta\ws$, below the reported precision of the chain. Physical and input updates are selected on the training loss; held-out NRMSE is the root-mean-square error divided by the standard deviation of the held-out target. Sample-size checks are given in Supplementary Note~10.

\subsection{Trajectory QFI, SLD and QFI-normalized direction}
For $\rho=\sum_p\lambda_p|p\rangle\langle p|$,
\begin{equation}
[\FQ]_{\mu\nu}=2\sum_{p,q:\lambda_p+\lambda_q>0}
\frac{\operatorname{Re}[\langle p|\partial_\mu\rho|q\rangle\langle q|\partial_\nu\rho|p\rangle]}{\lambda_p+\lambda_q},\qquad
\SLD_{\bv}=2\sum_{p,q:\lambda_p+\lambda_q>0}\frac{\langle p|\partial_{\bv}\rho|q\rangle}{\lambda_p+\lambda_q}|p\rangle\langle q|.
\label{eq:qfi}
\end{equation}
Terms with $\lambda_p+\lambda_q\le10^{-12}\lambda_{\max}$ are discarded. Physical state derivatives use a complex-step rule with $h=10^{-4}$; feature derivatives follow by linearity. QFI matrices are symmetrized and inverted on a common relative spectral support of $10^{-10}$. The feature-matched metric is $\GQ=\sum_t\omega_t\FQ(t)$ with $\omega_t=\sum_j\Var_{\rho_t}(O_j)$. At each anchor, $\bv_Q^\star=-\GQ^+g/\GamQ$ is normalized to one QFI unit. For the SLD closure test, this direction and QFI unit are defined by the base dictionary and then held fixed while the augmented dictionary is evaluated. The matched numerical controls are listed in Supplementary Note~6.

\subsection{Monotone input family and profiling}
Normalize the input as $z(u)=(u-u_{\min})/(u_{\max}-u_{\min})$ and define
\begin{equation}
f_P(u;\bc)=u_{\min}+(u_{\max}-u_{\min})
\frac{\int_0^{z(u)}\exp[\sum_{k=1}^Pc_kP_k(2s-1)]\,ds}
{\int_0^1\exp[\sum_{k=1}^Pc_kP_k(2s-1)]\,ds}.
\label{eq:inputmap}
\end{equation}
The map is strictly increasing, fixes the endpoints, and reduces to the identity at $\bc=0$. Centred coefficient probes ($\delta=0.05$) form a prediction-response bank $D_P$; updates $\Delta\bc\propto D_P^+r$ are clipped to norm $0.5$ and backtracked on the exact nonlinear training loss. An order is saturated when an accepted step lowers the loss by less than $10^{-4}$ in relative terms or after the declared iteration cap. The full algorithm, nesting proof and inversion checks are in Supplementary Note~4.

\subsection{Sign-agreement scan}
At each of the 121 points of the $(g,\kappa_c)$ grid at $\epsilon=15$ (Mackey--Glass task), the chain along $\bv_Q^\star$ is evaluated as in Table~\ref{tab:budget} and both interventions are applied at fixed $\bth$: the two Fisher-leading SLD eigenprojectors of $\bv_Q^\star$ added to the dictionary (same per-time projector rule applied at held-out times) and input profiling with $P\le4$. For each intervention we record $\Delta\log\lo$, its decomposition of Eq.~\eqref{eq:budget}, the training loss and the held-out NRMSE before and after. The intervention ``on the smallest factor'' is the dictionary intervention where $\eta_3<\eta_4$ and profiling otherwise. The quadrant fractions of Fig.~\ref{fig:scan}h count grid points by the signs of $\Delta\log\lo$ and of $\Delta\log$ (training loss); held-out values are reported but enter no selection. Anchor A of Table~\ref{tab:budget} is $(17,15,15)$: among the 22 $\eta_3$-limited grid points at which the dictionary intervention raises $\lo$, it lies near $(15,15,15)$ and has $\Delta\log\eta_3$ as the largest term of the rise of $\lo$; the nearest such point, $(16,15,15)$, gives the same picture with a smaller rise (Supplementary Note~12). Its values are those of the scan at full settings. Anchor B was evaluated independently of the scan, and the two evaluations agree exactly. The counterexample anchor $(15,15,15)$ is documented in Supplementary Note~12. The CFI proxy of the Discussion uses the $3\times3$ matrices $\GQ$ and $\bm G_C=\sum_t\sum_j\Var_{\rho_t}(O_j)\FC^{(j)}(t)$ stored at every grid point, with $\eta_1^2=\bv^\top\bm G_C\bv/\bv^\top\GQ\bv$.

\subsection{Matched-cost comparison and stage test}
\label{sec:methods_matched}
\emph{Quantum reservoir.} Readout cost is the number of trained weights $m$, plus a bias. The reservoir is the driven JC model of Eq.~\eqref{eq:jc} with $N=1$, $2$ or $3$ atoms (individually coupled with $g_i$, detuned by $\Delta_{a,i}$ from the drive and decaying at $\gamma_i$), a cavity detuning $\Delta_c$, drive $\epsilon$ and cavity decay $\kappa_c$, and $N_c=10$ ($8$ for three atoms). The input enters through the drive law of Eq.~\eqref{eq:fbdrive}, $\tilde u_t=u_t+b+\bbeta^\top X_t$, with an offset $b$ and a gain vector $\bbeta$ over all $m$ features; the features $\langle x\rangle,\langle p\rangle,\langle\sigma^x_i\rangle,\langle\sigma^y_i\rangle$ are sampled at $V$ sub-times per input interval, the $V-1$ interior sub-times being free, so that $m=V(2+2N)$. The layouts are $(N,V)=(1,1)$, $(2,1)$, $(1,2)$, $(1,3)$ and $(2,2)$, $(1,4)$ and $(3,2)$, $(2,3)$, $(2,4)$, $(2,5)$ for $m=4$, $6$, $8$, $12$, $16$, $18$, $24$, $30$; where two layouts share an $m$, the one with the lower training loss after the QFI stage is reported and the other is listed in Supplementary Note~14. Every parameter is optimized: $\bth=(\Delta_c,\Delta_{a,i},g_i,\epsilon,\kappa_c,\gamma_i)$, $b$, the interior sub-times and $\bbeta$, in total $11$--$44$ parameters, inside the box $g_i\in[2,64]$, $\Delta_c\in[1,80]$, $\Delta_{a,i}\in[-80,80]$, $\epsilon\in[2,48]$, $\kappa_c\in[1,48]$, $\gamma_i\in[0.1,24]$, $|b|\le\sigma_u$, under the constraints $|\operatorname{Re}\tilde u_t|\le10$ and $\max_t\langle n\rangle\le3$; a bound that an optimum reaches is doubled (detunings: extended by $40$) and the search repeated until the gain falls below $10^{-3}$. A sensitivity screen kept every parameter, and the time step, drive frequency, sign conventions, quadrature rotation and initial state were verified to be redundant with the listed parameters (Supplementary Note~14).

\emph{Stages.} A grid with three levels per parameter around the reference anchor ($2{,}000$--$4{,}000$ points) is followed by Bayesian optimization (Gaussian-process expected improvement, then a trust region) from the best grid point, stopped when the relative improvement over $40$ evaluations falls below $10^{-3}$ or at $600$ evaluations; both use only the training loss. The QFI-aware stage alternates rounds of natural-gradient steps $\bv_Q^\star$ on $\bth$, with $\GQ$ formed from complex-step state derivatives along the training trajectory and the step halved until the training loss decreases, with BFGS steps on $(b,\tau,\bbeta)$ (exact adjoint gradient for $b$ and $\bbeta$, complex step for $\tau$, Armijo backtracking), for $6$ h ($m\le18$) or $12$ h ($m\ge24$) of wall time, extended by one budget while the last round improved the loss by more than $10^{-3}$ in relative terms or the end point sat on a box face with an outward descent direction (at most $36$ h); the three-atom run at $m=16$ was stopped at $18$ h with a last-round improvement of $6\times10^{-3}$. At $m=24$ and $30$ the grid and BO stages searched the drive vector with the controls $(g_i,\epsilon,\kappa_c)$ held at $\bth^\star$, the training-selected end point of earlier natural-gradient runs from local starts; the QFI-aware stage then comprises a natural-gradient stage on the controls alternating with BFGS on the drive vector (the controls-and-drive-vector search), followed by the all-parameter protocol (grid, BO and QFI-aware stage on all parameters) from its end point (Supplementary Note~14). A quasi-random (Sobol) search of $13{,}872$--$18{,}698$ points over the full box at $m\ge12$ serves as a control on the black-box stages. Each end point was re-evaluated at $N_c+2$. Training NRMSE is the root-mean-square residual of the ridge fit on the $2{,}000$ training rows divided by the standard deviation of the training target; held-out NRMSE uses the $895$ held-out rows and is reported only after all selection.

\emph{Controls for the stage test.} From the point at which the QFI-aware stage starts (the grid $+$ BO optimum; at $m=24$ and $30$ also the BO optimum of the controls-and-drive-vector search), three arms run under the stage's constraints, accounting and budgets: BFGS with the exact adjoint gradient on $(b,\tau,\bbeta)$ with the quantum parameters frozen; natural-gradient rounds on $\bth$ with $(b,\tau,\bbeta)$ frozen; and trust-region BO continued for the number of trajectory evaluations the stage used. Each end point is evaluated on the held-out rows once (Supplementary Note~14). A ridge chosen by training-internal validation ($\delta\in[10^{-10},1]$), a moving-block bootstrap of the held-out residuals and five further Mackey--Glass windows with the readout refit were used to check the sensitivity of Table~\ref{tab:matched} (Supplementary Note~14).

\emph{Echo-state network.} The network with $m$ nodes updates $x_{t+1}=(1-a)x_t+a\tanh(Wx_t+W_{\rm in}u_t+b)$ from $x_0=0$, with $W$ a dense Gaussian matrix rescaled to spectral radius $\rho\in\{0.1,0.2,0.3,0.5,0.7,0.9,0.95,0.99,1.1\}$, leak $a\in\{0.1,0.3,0.5,0.7,1\}$, $W_{\rm in}\sim\mathcal U(-s,s)$ with $s\in\{0.005,0.01,0.02,0.05,0.1,0.3,0.5,1,2\}$ and bias $b\sim\mathcal U(-0.1,0.1)$ or $0$; the state before the input of interval $t$ is the feature row, with the same washout, split, ridge and bias column as the quantum reservoir and no direct input term. Its four hyperparameters are chosen among these $810$ settings on the mean training loss over $1{,}000$ random draws of $W$, $W_{\rm in}$ and $b$; the five best settings were re-evaluated over $10{,}000$ draws and the one with the lowest mean training loss selected. The reported statistics are the mean, standard deviation and empirical distribution of the held-out NRMSE over the $10{,}000$ draws of that setting. NARMA10 was not part of this comparison.

\emph{Design gains.} The SLD-eigenprojector augmentation and the root search for $\beta_{\rm slope}$ of Sec.~\ref{sec:advantages} and Supplementary Note~13 use the conventions above. The root search brackets the zero of $D_\beta\Lc$ by doubling from $\beta=0$ with step $0.1$, closes the bracket when the training loss rises, and bisects to $0.025$ (replay) or $0.01$ (new trajectories). Finite-shot estimates add Gaussian noise of variance $\Var_t(O_j)/n$ to every feature, refit the readout and take the median over ten noise seeds; the shot count $n^\star$ at which the reservoir's margin over the network's mean vanishes is computed for the refitted and for the fixed readout (Supplementary Note~14).

\subsection{Finite-shot cost}
For a centred difference with step $h$ and $n$ shots per side, the number of shots required to estimate $D_{\bv}\Lc$ to relative precision $\epsilon_{\rm rel}$ is
\begin{equation}
n_{\rm req}=\frac{\sigma_{\bv}^2}{2h^2\epsilon_{\rm rel}^2(D_{\bv}\Lc)^2}
=\frac{\sigma_{\bv}^2}{2h^2\epsilon_{\rm rel}^2\|r\|^2\|\ws\|^2\BQ^2(\bv)\eta_Q^2(\bv)}.
\label{eq:shots}
\end{equation}
Here $\sigma_{\bv}^2$ is the single-shot variance of the loss estimate at $\bth\pm h\bv$. The denominator is the useful physical response and the numerator is the noise of the measured representation. Reducing the numerator while preserving the derivative requires a weak representation fibre. The four-feature JC dictionary has none. Its derivative-preserving weak rank is zero, so a representation redesign at fixed readout saves at most $3\%$ of the shots, and the $9$--$56\%$ saving of a joint redesign comes with derivative ratios between $0.4$ and $14$ (Supplementary Notes~5 and~7).

\subsection{Atom-number controls}
The atom scan uses $N=1,2,3,4$, $N_c=8$ and eight paired log-parameter samples around $(G\ \mathrm{or}\ g_0,\epsilon,\kappa_c,\gamma)=(10,10,5,5)$ with log-standard deviation $0.12$. Cavity and atomic decay are independent in this scan, with cavity dissipator $\kappa_c\mathcal D[a]$. QFI with respect to the four logarithmic controls is evaluated by centred differences of relative step $10^{-3}$ over a common 550-sample post-washout window. Fixed collective coupling uses $g_i=G/\sqrt N$; fixed individual coupling uses $g_i=g_0$. We report medians and interquartile ranges of the dominant QFI eigenvalue and trace; raw ranks, photon occupations and convergence checks are included in the deposited data.

\subsection{Use of AI tools}
A large language model (Claude, Anthropic) was used to assist with developing and reviewing analysis code, verifying numerical values reported in the manuscript against the deposited data, organizing the code and data for public release, and improving the clarity and presentation of the manuscript. All research questions, theoretical derivations, methodological decisions, and scientific interpretations were developed by the authors. The authors independently reviewed and validated the AI-assisted outputs and take full responsibility for the accuracy and integrity of the published work.

\section*{Data availability}
The numerical arrays underlying every plotted point, including the chain quantities and link factors of both reservoirs, the certificate sweep and measurement-family ratios of the spin reservoir, input-order traces, paired QNG-path evaluations, atom-number QFI samples, derivative-sampling estimates, the drive-gain scans, grid tests and finite-shot runs, the adversarial tests of the spin reservoir, the twenty-seed spin-reservoir results and near-zero-QFI search, the thirty-anchor drive-gain scans for both tasks and the two-atom sweep, the sign-agreement scan and the CFI-proxy check, and the matched-cost comparison (per-configuration stage records, optimized parameters and stored features, the $10{,}000$-draw echo-state-network distributions, the aggregate results of the quasi-random control search and the finite-shot tables), are deposited in a versioned Zenodo record (\DOI).

\section*{Code availability}
Simulation and analysis code, exact run configurations, software environments and random seeds for every panel are deposited in the same Zenodo record (\DOI).

\section*{Acknowledgements}
D.S. acknowledges support from the U.S. Department of Energy, Office of Science, under Award No. DE-SC0025910, and from the National Science Foundation under Award No. 2529700. The funders played no role in study design, data collection, analysis and interpretation of data, or the writing of this manuscript.

\section*{Author contributions}
P.H.N. conceived the study, developed the theory and performed the numerical work. D.S. supervised the project and contributed to interpretation and manuscript preparation. Both authors reviewed the manuscript.

\section*{Competing interests}
The authors declare no financial or non-financial competing interests.

\clearpage
\newgeometry{left=0.6in,right=0.6in,top=0.6in,bottom=0.7in}
\captionsetup{font=small}

\clearpage
\begin{figure}[p]
\centering
\includegraphics[width=\textwidth,height=0.84\textheight,keepaspectratio]{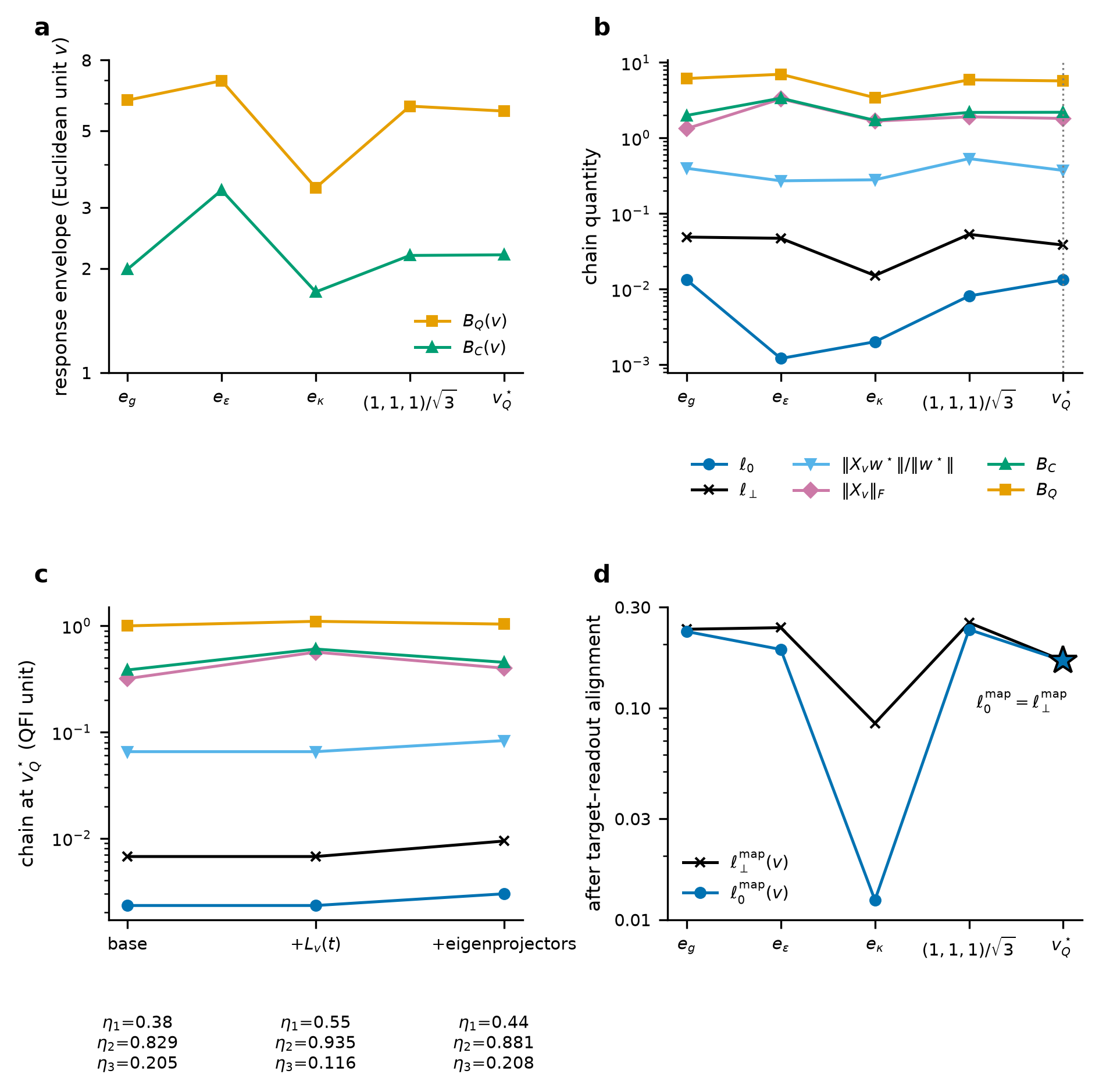}
\caption{\textbf{Every link of the bound is checked and attained in a single-atom Jaynes--Cummings reservoir.} Anchor $(g,\epsilon,\kappa_c)=(30,20,10)$, five-step Mackey--Glass forecasting, four Euclidean-unit directions ($e_g$, $e_\epsilon$, $e_\kappa$ and $(1,1,1)/\sqrt3$) and the QNG direction $\bv_Q^\star$ (one QFI unit). \textbf{a}, Measurement and quantum response envelopes $\BC(\bv)$ and $\BQ(\bv)$. \textbf{b}, The full chain of Eq.~\eqref{eq:chain} for the same directions, from $\BQ$ down to $\lo$. The five factors quantify the grouped gaps; $\eta_3$ combines the two middle norm/readout inequalities. \textbf{c}, Closure of the measurement links at $\bv_Q^\star$. Adding the centred SLD $\SLD_{\bv}(t)$ to the dictionary moves $\BC$ towards $\BQ$ and $\normF{X_{\bv}}$ towards $\BC$ while $\lo$ is unchanged, so the gap moves to the readout factor $\eta_3$. Adding the two Fisher-leading eigenprojectors of $\SLD_{\bv}(t)$ instead raises $\lo$, because the projector features carry readout weight. The values of $\eta_1$, $\eta_2$ and $\eta_3$ are printed under each dictionary. \textbf{d}, Closure of the task link by Proposition~\ref{prop:task}(ii). The target $y=Xw+r$, with $w$ and $r$ the leading singular pair of $\PXp X_{\bv}$, gives $\lo=\lp$ at the QNG direction (star). For the other directions $\lo^{\rm map}\le\lp^{\rm map}$, as required. The forecasting targets used elsewhere are fixed independently of this construction.}
\label{fig:hierarchy}
\end{figure}

\clearpage
\begin{figure}[p]\centering
\includegraphics[width=\textwidth,height=0.775\textheight,keepaspectratio]{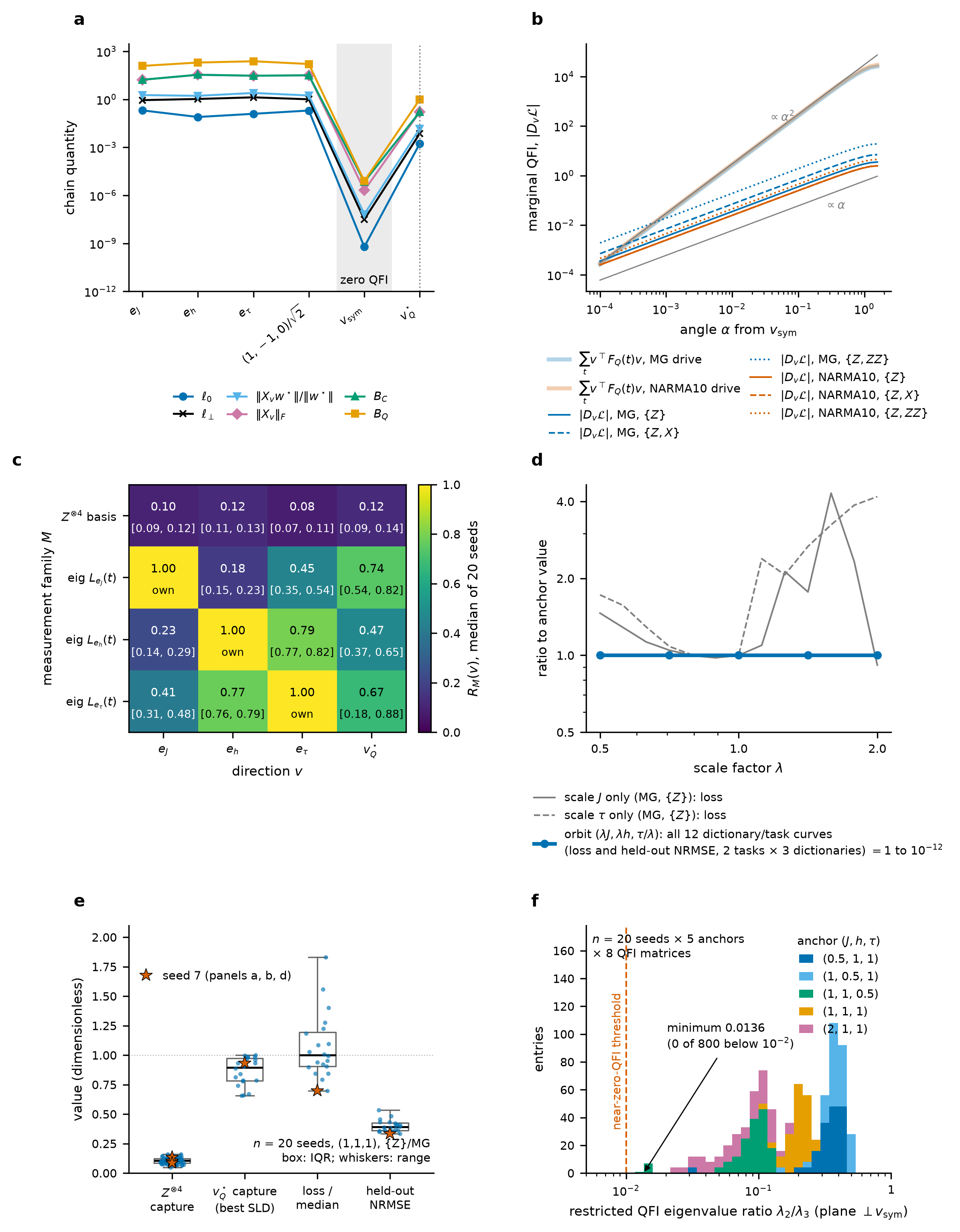}
\caption{\textbf{Chain, certificate and measurement links across twenty random four-spin Ising reservoirs.} Anchor $(J,h,\tau)=(1,1,1)$, dictionary $\{Z_i\}$ and five-step Mackey--Glass forecasting unless stated. Panels a, b and d show seed 7; panels c, e and f pool twenty seeds. \textbf{a}, Chain of Eq.~\eqref{eq:chain} for four Euclidean-unit directions, the symmetry direction $\bv_{\rm sym}\propto(J,h,-\tau)$ and the QNG direction $\bv_Q^\star$ (one QFI unit). Along $\bv_{\rm sym}$ every quantity is at the numerical floor. \textbf{b}, Certificate test along $\bv(\alpha)=\cos\alpha\,\bv_{\rm sym}+\sin\alpha\,\bv_\perp$: marginal QFI vanishes as $\alpha^2$ and $|D_{\bv}\Lc|$ as $\alpha$ for every tested dictionary and task. \textbf{c}, Fraction $R_M(\bv)$ of feature-matched QFI captured by the computational basis or by the SLD eigenbasis of one control. Each SLD basis captures its own direction fully but only part of the others because the SLDs do not commute. \textbf{d}, Training loss and held-out NRMSE along the exact symmetry orbit $(\lambda J,\lambda h,\tau/\lambda)$ remain at the anchor value; grey curves vary $J$ or $\tau$ alone. \textbf{e}, Seed-to-seed spread of computational-basis capture, best single-control SLD capture of $\bv_Q^\star$, training loss and held-out NRMSE; seed 7 is highlighted. \textbf{f}, Restricted-QFI eigenvalue-ratio search over twenty seeds, five anchors and eight QFI matrices (800 entries). No entry lies below $10^{-2}$. Results use $N=4$, $T=3{,}000$ samples and a 100-sample washout.}
\label{fig:tfi}
\end{figure}

\clearpage
\begin{figure}[p]\centering
\includegraphics[width=\textwidth,height=0.69\textheight,keepaspectratio]{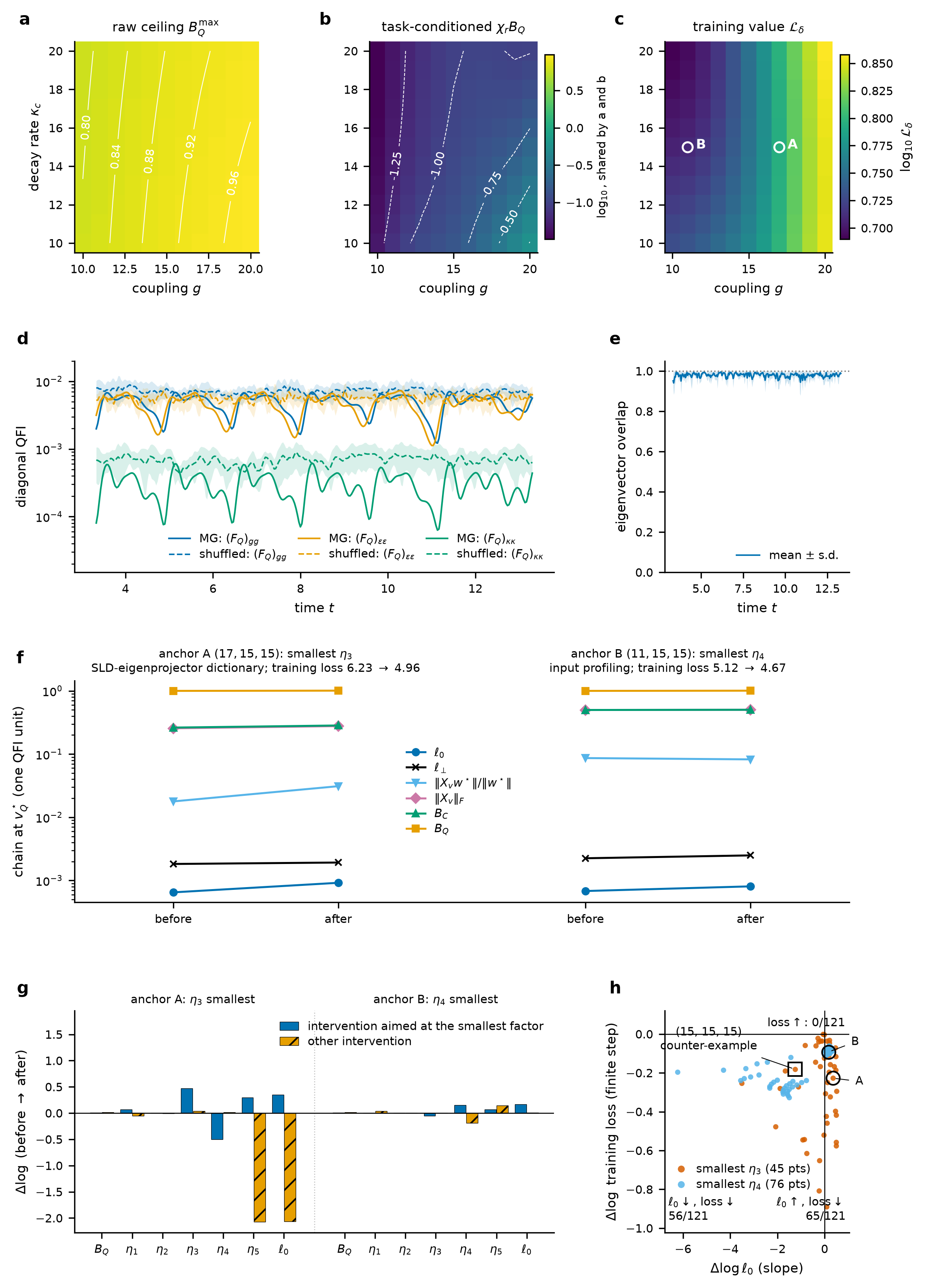}
\caption{\textbf{The decomposition attributes each intervention; the sign scan tests whether the slope predicts the step.} \textbf{a}, Raw task-independent ceiling $\BQ^{\max}=\sqrt{\lambda_{\max}(\GQ)}$ across the atom--cavity coupling $g$ and the decay rate $\kappa_c$ (shared equally by cavity and atom, Methods) at fixed drive $\epsilon=15$. \textbf{b}, Mackey--Glass task-conditioned ceiling $\chi_r\BQ$ along the dominant-QFI direction, with the readout-free task factor $\chi_r=\normF{\Pir X_{\bv}}/\normF{X_{\bv}}$. Panels a and b share a logarithmic colour scale. \textbf{c}, Training value $\Lc_\delta$ on a logarithmic scale, with the anchors A $=(17,15,15)$ and B $=(11,15,15)$ of panels f and g marked. \textbf{d}, Time-resolved diagonal elements of $\FQ(t)$ for $\bth=(g,\epsilon,\kappa_c)$ at $(15,15,15)$ over the first 300 training samples. Solid curves show the original Mackey--Glass input. Dashed curves and bands show the mean and standard deviation over ten independent permutations of the same values. \textbf{e}, Absolute overlap between the instantaneous dominant-QFI directions for the original and shuffled inputs. Points are retained only when both normalized eigengaps exceed $0.05$. \textbf{f}, Loss budget of Eq.~\eqref{eq:budget} at anchors A and B along $\bv_Q^\star$, before and after the intervention that acts on the smallest factor (A, SLD-eigenprojector dictionary; B, input profiling). The link factors are listed in Table~\ref{tab:budget}. \textbf{g}, Attribution. Decomposition of $\Delta\log\lo$ into $\Delta\log\BQ$ and $\Delta\log\eta_1,\dots,\Delta\log\eta_5$ for the intervention aimed at the smallest factor (solid) and for the other one (hatched), at A (left group) and B (right group). \textbf{h}, Sign-agreement scan. Change of the local slope, $\Delta\log\lo$, against the change of the training loss after the finite step, $\Delta\log\Lc_\delta$, for the intervention on the smallest factor at all 121 grid points. Colour marks the smallest factor ($\eta_3$ or $\eta_4$) and the number of points in each quadrant is printed. The anchor $(15,15,15)$, where the slope falls and the loss falls (Supplementary Note~12), and anchor A are marked. Results use one atom, $N_c=10$, $\omega_c=40$, $\omega_a=20$, $\Delta t=1/30$ and a 100-sample washout.}
\label{fig:scan}
\end{figure}

\clearpage
\begin{figure}[p]
\centering
\includegraphics[width=0.92\textwidth,height=0.84\textheight,keepaspectratio]{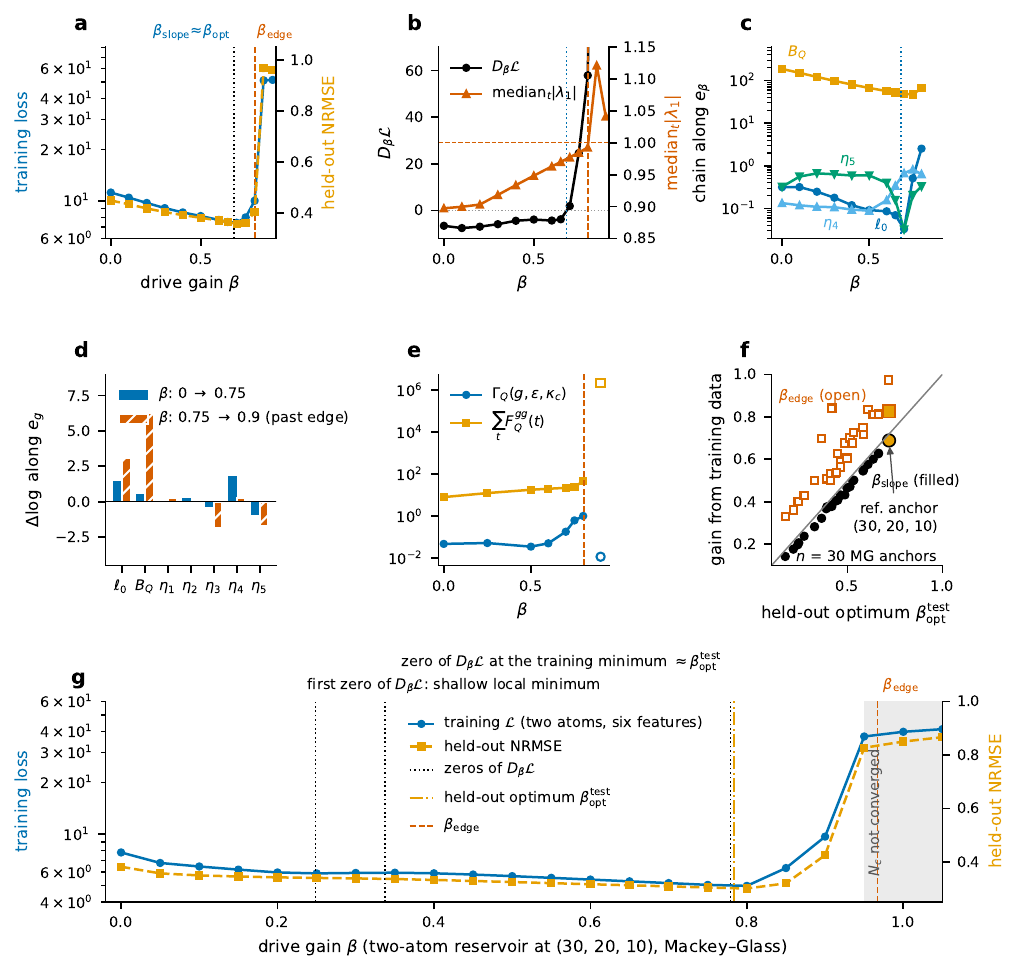}
\caption{\textbf{The exact slope locates the drive gain at which the loop stops helping.} Driven JC reservoir with $\tilde u_t=u_t+\beta\langle p\rangle_t$, five-step Mackey--Glass forecasting. Panels a--e use the reference anchor $(30,20,10)$. \textbf{a}, Training loss and held-out NRMSE versus gain, with the slope zero $\beta_{\rm slope}=0.684$, training minimum $0.682$ and dynamical edge $0.803$. \textbf{b}, Exact slope $D_\beta\Lc$ and slow-mode gain $\mathrm{median}_t|\lambda_1|$ of the closed-loop tangent map. The slope crosses zero while the loop is still contracting. \textbf{c}, Chain along $e_\beta$ on the stable branch: $\BQ$ remains large while the task factor $\eta_5$ collapses and the novelty factor $\eta_4$ rises. \textbf{d}, Log-additive attribution along $e_g$: from $\beta=0$ to $0.75$ the increase is carried mainly by $\eta_4$; beyond the edge the rising budget is removed by $\eta_3$ and $\eta_5$. \textbf{e}, QFI-normalized slope $\GamQ$ and marginal QFI along $g$; open symbols are beyond the converged stable branch. \textbf{f}, Thirty-anchor held-out check. The training-based $\beta_{\rm slope}$ and the dynamical edge are plotted against the held-out optimum; the held-out set is used only for this check. \textbf{g}, Two-atom reservoir: the first slope zero at $0.249$ is a shallow local minimum, whereas the held-out optimum $0.783$ coincides with the zero at the pre-collapse training minimum $0.779$.}
\label{fig:feedback}
\end{figure}

\clearpage
\begin{figure}[p]
\centering
\includegraphics[width=\textwidth,height=0.84\textheight,keepaspectratio]{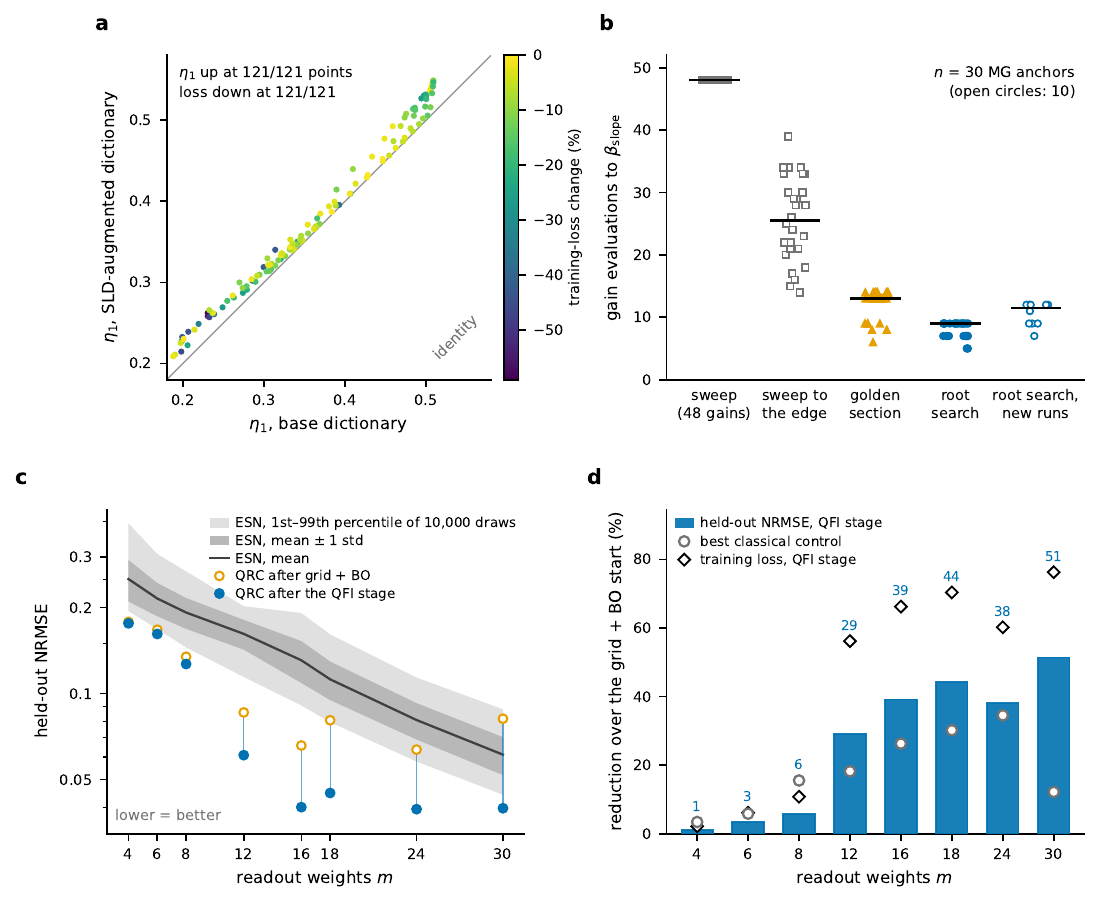}
\caption{\textbf{Quantum advantage over a matched echo-state network, its attribution to the QFI-aware stage, and two further design gains.} \textbf{a}--\textbf{b}, Design gains in the four-feature JC reservoir. \textbf{a}, Measurement link $\eta_1$ along $\bv_Q^\star$ at the 121 grid points of Fig.~\ref{fig:scan}, for the base dictionary and for the dictionary augmented by the two Fisher-leading eigenprojectors of the SLD of $\bv_Q^\star$ at fixed $\bth$ and trajectory. Colour gives the relative change of the training loss. Every point lies above the identity line and every training loss falls. \textbf{b}, Gain evaluations needed to locate the stopping gain $\beta_{\rm slope}$ of Fig.~\ref{fig:feedback}, one marker per anchor and a black tick at the median. The stored sweep of 48 gains and the sweep up to the dynamical edge are compared with a golden-section search on the training loss alone and with the safeguarded root search on the exact slope $D_\beta\Lc$, both replayed on the stored gains of the thirty Mackey--Glass anchors. Open circles show the secant variant of the root search with new trajectories at ten anchors. \textbf{c}, Held-out NRMSE against the number of trained weights $m$ on five-step Mackey--Glass forecasting: the tuned echo-state network (line: mean over $10{,}000$ random draws of the selected hyperparameter setting; dark band: $\pm1$ standard deviation; light band: 1st--99th percentile) and the quantum reservoir configuration of Table~\ref{tab:matched} after grid $+$ Bayesian optimization on the training loss (open markers) and after the additional QFI-aware stage (filled markers). \textbf{d}, Stage test: relative reduction of the held-out NRMSE (bars) and of the training loss (diamonds) produced by the QFI-aware stage over the stalled grid $+$ BO optimum, per $m$; open circles, the held-out reduction of the better classical control from the same start (exact-gradient descent on the classical parameters alone, or black-box search for the stage's evaluation count; at $m=24$ and $30$ from the BO optimum of the controls-and-drive-vector search). Expectation-value results; finite-shot cost in Supplementary Note~14.}
\label{fig:advantages}
\end{figure}
\restoregeometry

\end{document}


\maketitle

\section{Notation and scope}
This note fixes the notation used throughout the Supplementary Notes and states the scope of every theoretical claim. It supports main-text Sec.~2.2, which introduces the chain quantities, and the testbed paragraph of Sec.~2.1.

The main text defines the chain quantities $\BQ$, $\BC$, $\normF{X_{\bv}}$, $\|X_{\bv}\ws\|/\|\ws\|$, $\lp$, $\lo$, the factors $\eta_1,\ldots,\eta_5$, the utilization $\eta_Q$, the QFI-normalized slope $\GamQ$ and the readout-free task factor $\chi_r$. Supplementary Table~\ref{tab:snotation} lists only the additional notation used below.

\begin{table}[!ht]\centering\small
\caption{Additional notation used in the Supplementary Notes.}
\label{tab:snotation}
\begin{tabular}{ll}\toprule
Symbol & Definition\\\midrule
$\rho_t(\bth,u)$ & Reservoir state at sample $t$\\
$X,X_{\bv}$ & Feature matrix and directional feature derivative\\
$y,\ws,r$ & Target, fitted readout and residual\\
$\Pir,\PX,\PXp$ & Residual projector, feature-span projector and its complement\\
$\FC^{(j)}(t),\FQ(t)$ & CFI of the measurement of $O_j$ and QFI of $\rho_t$\\
$\SLD_{\bv}(t)$ & SLD along $\bv$, $\partial_{\bv}\rho_t=\tfrac12\{\rho_t,\SLD_{\bv}(t)\}$\\
$A$ & Novel-response operator $\PXp X_{\bv}$\\
$g,\GQ$ & Loss gradient and feature-matched trajectory metric\\
$f_P(u;\bc),D_P$ & Monotone input map and its prediction-response bank\\
$Z,C,N_h,\Sigma$ & Accessible dictionary, representation map, weak kernel and covariance weight\\
\bottomrule\end{tabular}
\end{table}

All theoretical statements are local to a fixed-rank neighbourhood of one physical anchor. Features are separately acquired single-time expectation values, so $\GQ$ is a feature-matched marginal metric rather than a process QFI. Fisher envelopes are compared only at matched trajectory length, sample weights, observable normalization and measurement protocol. The QFI-normalized natural-gradient direction $\bv_Q^\star$ enters as the maximizer of the QFI-normalized first-order slope. Its use implies no convergence claim. The numerical results are finite-size checks of the stated reservoirs. No asymptotic scaling or complexity-theoretic claim is made; the quantum advantage of main-text Sec.~2.1 is a comparison of expectation-value performance at matched readout cost. The two design gains of \ref{sec:advantage} are gains over the same reservoir designed from measured features alone; the matched-cost comparison with a classical learner and the stage test behind the main-text advantage claim are reported in \ref{sec:matched}.

\section{Proof of the chain and of every tightness condition}
\label{sec:chain}
This note proves the chain of inequalities of the main text and the equality condition of every link. It supports the bound and decomposition of main-text Sec.~2.2, Proposition~1 (task-link tightness) and the impossibility certificate of Sec.~2.3.

\subsection*{Envelope identity}
\begin{slemma}[Envelope identity]\label{lem:envelope}
Let $\Lc(\bth)=\min_w\frac12\|y-X(\bth)w\|^2$ with $\ws=X^+y$ and $r=y-X\ws$. On a fixed-rank neighbourhood,
\begin{equation}
D_{\bv}\Lc=-r^\top X_{\bv}\ws .
\label{eq:s-envelope}
\end{equation}
\end{slemma}
\begin{proof}
Write $f(\bth,w)=\frac12\|y-X(\bth)w\|^2$. The normal equations $X^\top(y-X\ws)=0$ characterize every minimizer, and on a fixed-rank neighbourhood the minimum-norm minimizer $\ws(\bth)$ is differentiable \cite{S3}. Differentiating $\Lc(\bth)=f(\bth,\ws(\bth))$ gives $D_{\bv}\Lc=\partial_{\bth}f\cdot\bv+\partial_wf\cdot D_{\bv}\ws$. The second term vanishes because $\partial_wf=-X^\top r=0$ at the minimizer, and the first term is $-r^\top X_{\bv}\ws$. No derivative of the refitted readout appears. The same argument applies to the ridge value $\Lc_\delta$ of the main text, whose minimizer satisfies $X^\top r=\delta\ws$. The identity Eq.~\eqref{eq:s-envelope} is unchanged, and the orthogonality $r\perp\col X$ used below holds up to $O(\delta)$.
\end{proof}

\subsection*{Measurement lemmas}
\begin{slemma}[Classical response bound]\label{lem:classical}
For an observable $O=\sum_mo_mP_m$ measured projectively on $\rho$, with outcome probabilities $p_m(\bth)=\Tr(\rho P_m)$,
\begin{equation}
|D_{\bv}\langle O\rangle|^2\le(\bv^\top\FC\bv)\Var_\rho(O),
\qquad
\bv^\top\FC\bv=\sum_{m:p_m>0}\frac{(D_{\bv}p_m)^2}{p_m},
\label{eq:s-classical}
\end{equation}
with equality if and only if $o_m-\langle O\rangle=\kappa\,D_{\bv}p_m/p_m$ on the support for some constant $\kappa$, that is, when the centred observable is proportional to the classical score.
\end{slemma}
\begin{proof}
Since $\sum_mD_{\bv}p_m=0$, $D_{\bv}\langle O\rangle=\sum_m(o_m-\langle O\rangle)D_{\bv}p_m=\sum_m\big[(o_m-\langle O\rangle)\sqrt{p_m}\big]\big[D_{\bv}p_m/\sqrt{p_m}\big]$. Cauchy--Schwarz gives the bound, with equality if and only if the two bracketed vectors are proportional.
\end{proof}

\begin{slemma}[Measurement monotonicity]\label{lem:bc}
For every measurement, $\FC\preceq\FQ$ \cite{S1}. For a fixed direction $\bv$, the projective measurement in the eigenbasis of $\SLD_{\bv}$ attains $\bv^\top\FC\bv=\bv^\top\FQ\bv$.
\end{slemma}
This is the Braunstein--Caves theorem. Summing Eqs.~\eqref{eq:s-classical} and the monotonicity over matched times and observables gives
\begin{equation}
\normF{X_{\bv}}^2=\sum_{t,j}|D_{\bv}\langle O_j\rangle_t|^2\le\BC^2(\bv)\le\BQ^2(\bv).
\label{eq:s-measurement}
\end{equation}

\subsection*{The chain}
\begin{sprop}[Chain]\label{prop:chain}
For every direction $\bv$,
\begin{equation}
\lo(\bv)\le\lp(\bv)\le\frac{\|X_{\bv}\ws\|}{\|\ws\|}\le\normtwo{X_{\bv}}\le\normF{X_{\bv}}\le\BC(\bv)\le\BQ(\bv),
\label{eq:s-chain}
\end{equation}
with the following equality conditions, link by link: (i) $\lo=\lp$ if and only if $r\parallel\PXp X_{\bv}\ws$; (ii) $\lp=\|X_{\bv}\ws\|/\|\ws\|$ if and only if $X_{\bv}\ws\perp\col X$; (iii) $\|X_{\bv}\ws\|=\normtwo{X_{\bv}}\|\ws\|$ if and only if $\ws$ is a leading right singular vector of $X_{\bv}$; (iv) $\normtwo{X_{\bv}}=\normF{X_{\bv}}$ if and only if $X_{\bv}$ has rank one; (v) $\normF{X_{\bv}}=\BC$ if and only if every centred observable is proportional to the classical score of its own measurement at every time; (vi) $\BC=\BQ$ if and only if every deployed measurement attains the QFI along $\bv$ at every time.
\end{sprop}
\begin{proof}
By Eq.~\eqref{eq:s-envelope}, $|D_{\bv}\Lc|=|r^\top X_{\bv}\ws|=\|r\|\,\|\Pir X_{\bv}\ws\|$, so $\lo=\|\Pir X_{\bv}\ws\|/\|\ws\|$ exactly. Because $r\perp\col X$, $\Pir\PX=0$ and hence $\Pir=\Pir\PXp$, which gives $\lo=\|\Pir\PXp X_{\bv}\ws\|/\|\ws\|\le\|\PXp X_{\bv}\ws\|/\|\ws\|=\lp$ with equality if and only if $\PXp X_{\bv}\ws$ lies in the span of $r$ (i). The second inequality is $\|\PXp z\|\le\|z\|$ for $z=X_{\bv}\ws$, with equality if and only if $\PX z=0$ (ii). The third and fourth are the definitions of the spectral norm and of $\normtwo{\cdot}\le\normF{\cdot}$, with the standard equality conditions (iii) and (iv). The last two are Eq.~\eqref{eq:s-measurement}, and their equality conditions are those of Lemmas~\ref{lem:classical} and~\ref{lem:bc} applied to every $(t,j)$ term (v), (vi).
\end{proof}
The identity $\lo=\|\Pir\PXp X_{\bv}\ws\|/\|\ws\|$ in the proof is Remark~1 of the main text. The factor $\PXp$ is available because the refit keeps $r\perp\col X$ at every anchor. For a predictor whose readout $w$ is not refitted, the derivative of $\frac12\|y-Xw\|^2$ at fixed $w$ is $-(y-Xw)^\top X_{\bv}w$ with $y-Xw\not\perp\col X$ in general, so the corresponding bound $|D_{\bv}\Lc_{\rm frozen}|\le\|y-Xw\|\,\|X_{\bv}w\|$ contains no projection onto $\col X^\perp$ and cannot separate response that a refit would reproduce from response that is new.

\subsection*{Closing the measurement links with the symmetric logarithmic derivative}
\begin{sprop}[SLD closure]\label{prop:sld}
Fix $t$ and $\bv$, let $\SLD\equiv\SLD_{\bv}(t)$ be the SLD of $\rho\equiv\rho_t$ along $\bv$, and take $O=\SLD$ as the observable at time $t$, held fixed at the anchor. Then
\begin{equation}
\langle\SLD\rangle=0,\qquad D_{\bv}\langle\SLD\rangle=\bv^\top\FQ(t)\bv,\qquad \Var_\rho(\SLD)=\bv^\top\FQ(t)\bv,
\label{eq:s-sld}
\end{equation}
so that $|D_{\bv}\langle\SLD\rangle|^2=[\bv^\top\FQ(t)\bv]\Var_\rho(\SLD)$, and the projective measurement in the eigenbasis of $\SLD$ has $\bv^\top\FC\bv=\bv^\top\FQ(t)\bv$. Both measurement links of Eq.~\eqref{eq:s-chain} are equalities for this observable.
\end{sprop}
\begin{proof}
From $\partial_{\bv}\rho=\tfrac12\{\rho,\SLD\}$ and $\Tr\partial_{\bv}\rho=0$, $\langle\SLD\rangle=\Tr(\rho\SLD)=\Tr(\partial_{\bv}\rho)=0$. With the operator held fixed, $D_{\bv}\langle\SLD\rangle=\Tr(\partial_{\bv}\rho\,\SLD)=\tfrac12\Tr(\{\rho,\SLD\}\SLD)=\Tr(\rho\SLD^2)=\bv^\top\FQ(t)\bv$, and $\Var_\rho(\SLD)=\Tr(\rho\SLD^2)-\langle\SLD\rangle^2=\bv^\top\FQ(t)\bv$. For the projective measurement in the eigenbasis $\SLD=\sum_m\lambda_mP_m$, the score is $D_{\bv}p_m/p_m=\Tr(\partial_{\bv}\rho P_m)/p_m=\tfrac12\Tr(\{\rho,\SLD\}P_m)/p_m=\lambda_m$, because $P_m$ commutes with $\SLD$. Hence $o_m-\langle O\rangle=\lambda_m$ equals the score, which is the equality condition of Lemma~\ref{lem:classical} with $\kappa=1$. Finally $\bv^\top\FC\bv=\sum_m\lambda_m^2p_m=\Tr(\rho\SLD^2)=\bv^\top\FQ(t)\bv$, which is the equality case of Lemma~\ref{lem:bc}.
\end{proof}
\emph{Remark (centred SLD).} Since $\langle\SLD_{\bv}(t)\rangle=0$, the column $\Tr(\rho_t\SLD_{\bv}(t))$ of $X$ vanishes at the anchor. Appending a zero column leaves $X^+y$ unchanged apart from a zero entry (and likewise the ridge readout with $\delta=10^{-10}$), so $\ws$, $r$, $\lp$ and $\lo$ are unchanged, while $\normF{X_{\bv}}^2$, $\BC^2$ and $\BQ^2$ each increase by $\sum_t[\bv^\top\FQ(t)\bv]^2$. The measurement links close and the readout link opens, with $\BQ\eta_1\eta_2\eta_3=\|X_{\bv}\ws\|/\|\ws\|$, so the loss-budget product of the main text is preserved. The SLD makes the state motion measurable but contributes to the loss only at second order in the update. The SLD eigenprojectors $P_m(t)$ behave differently. They are not centred (their features are the outcome probabilities $p_m$), and they can carry readout weight. In the envelope $\BC$ each projector enters as its own two-outcome measurement $\{P_m,I-P_m\}$, whose centred observable is proportional to its classical score, so its observable link closes. Its classical Fisher information is $(D_{\bv}p_m)^2/[p_m(1-p_m)]=p_m\lambda_m^2/(1-p_m)$, which reaches $\bv^\top\FQ\bv=\sum_kp_k\lambda_k^2$ only in special cases (for instance a two-outcome spectrum), so the projectors raise $\eta_1$ only partially; the full projective measurement in the SLD eigenbasis attains $\FC=\FQ$, but its outcomes enter the features one projector at a time.

$\SLD_{\bv}(t)$ depends on $t$ through $\rho_t$ and on $\bv$ linearly, $\SLD_{\bv}=\sum_\mu v_\mu\SLD_\mu$. A single fixed measurement therefore attains the ceiling for every direction simultaneously only when the CFI matrix of that measurement equals $\FQ(t)$, which is the compatibility question of multiparameter metrology \cite{S4}. Commuting SLDs $\{\SLD_\mu\}$ suffice, because the joint eigenbasis is then optimal for every $\bv$. When the SLD is added to a dictionary that already contains other observables, only its own $(t,j)$ terms in Eq.~\eqref{eq:s-measurement} are equalities. The ratios $\eta_1$ and $\eta_2$ reported in the main text quantify the closure of the summed envelopes. In the JC reservoir at $(30,20,10)$ and $\bv_Q^\star$, adding $\SLD_{\bv}(t)$ raises $\eta_1$ from $0.38$ to $0.55$ and $\eta_2$ from $0.83$ to $0.94$ with $\lo$ unchanged. Adding its two Fisher-leading eigenprojectors instead raises them to $0.44$ and $0.88$ (main-text Fig.~2c).

\subsection*{Closing the task link: tightness over targets}
\begin{sprop}[Task-link tightness over targets]\label{prop:align}
Let $X$ have full column rank on the fixed-rank neighbourhood and let $A=\PXp X_{\bv}$ be the novel-response operator.
\begin{enumerate}[nosep,leftmargin=2em]
\item[(i)] For every readout $w$ with $Aw\neq0$, the target $y=Xw+r$ with $r=Aw/\|Aw\|$ has minimum-norm readout $\ws=w$ and residual $r$, and for it $\lo(\bv)=\lp(\bv)=\|Aw\|/\|w\|$.
\item[(ii)] Over all targets,
\begin{equation}
\sup_y\ \lo(\bv;y)=\normtwo{A},
\label{eq:s-align}
\end{equation}
attained by the target of (i) with $w$ a leading right singular vector of $A$ (and then $r$ the matching left singular vector).
\end{enumerate}
\end{sprop}
\begin{proof}
(i) The columns of $A$ lie in $\col X^\perp$, so $r\perp\col X$ and $X^+r=0$; with full column rank, $X^+X=I$, so $\ws=X^+y=X^+Xw+X^+r=w$ and $y-X\ws=r$. Then $|D_{\bv}\Lc|=|r^\top X_{\bv}w|=|r^\top Aw|=\|Aw\|$ (using $r^\top\PXp=r^\top$ and $r=Aw/\|Aw\|$), so $\lo=\|Aw\|/(\|r\|\|w\|)=\|Aw\|/\|w\|$, and $\lp=\|Aw\|/\|w\|$ by definition. (ii) For every target, Eq.~\eqref{eq:s-chain} gives $\lo\le\lp=\|A\ws\|/\|\ws\|\le\normtwo{A}$; the target of (i) with $w$ a leading right singular vector gives $\|Aw\|/\|w\|=\normtwo{A}$, so the supremum is attained. For that $w$, $Aw/\|Aw\|$ is the matching left singular vector.
\end{proof}
Part (i) shows that the task link can be closed for any readout, so no bound on $\lo$ that depends on the target only through $\ws$ can be smaller than $\lp$. Part (ii) shows that no target-independent bound can be smaller than $\normtwo{\PXp X_{\bv}}$. Neither part asserts that a performance-selected target or input map attains the preceding links. Those links are closed by their own conditions in Supplementary Proposition~\ref{prop:chain}, and the conditions involve different objects that are not claimed to hold simultaneously. The construction of (ii) is used in the main text only to attain the task link at a chosen direction (Fig.~2d). The forecasting targets used in all other panels are fixed independently of it.

\subsection*{Readout-free task factor and the impossibility certificate}
For the parameter scan of the main text, the readout-free task factor $\chi_r(\bv)=\normF{\Pir X_{\bv}}/\normF{X_{\bv}}\le1$ is used. It satisfies $\lo=\|\Pir X_{\bv}\ws\|/\|\ws\|\le\normtwo{\Pir X_{\bv}}=\normF{\Pir X_{\bv}}=\chi_r\normF{X_{\bv}}$ (the two norms coincide because $\Pir X_{\bv}$ has rank at most one), so $\chi_r\BQ$ is a task-conditioned ceiling that does not depend on the readout. It is a diagnostic for the scan and is not part of the chain.

\begin{scor}[Impossibility certificate]\label{cor:cert}
If $\bv^\top\FQ(t)\bv=0$ for every $t$ in the window, then $D_{\bv}\Lc=0$ for every dictionary, readout and target. Conversely, if $\bv^\top\FQ(t)\bv>0$ for some $t$, the SLD observable of Supplementary Proposition~\ref{prop:sld} has a nonzero feature tangent along $\bv$.
\end{scor}
\begin{proof}
$\bv^\top\GQ\bv=\sum_t\omega_t\bv^\top\FQ(t)\bv=0$ for every choice of the nonnegative weights $\omega_t$, so $\BQ(\bv)=0$ and Eq.~\eqref{eq:s-chain} forces $\lo=0$. The converse is Eq.~\eqref{eq:s-sld}.
\end{proof}

\section{Rayleigh optimum, invariance, Euclidean comparison and the anchor sequence}
\label{sec:rayleigh}
This note derives the QFI-normalized natural-gradient direction $\bv_Q^\star$, its invariances and its comparison with Euclidean steepest descent, and defines the common anchor sequence used by later notes. It supports the QFI-normalized slope $\GamQ$ and the invariance statements of main-text Sec.~2.2 and Fig.~1b.

Let $g=\nabla_{\bth}\Lc$ and $\GQ=\sum_t\omega_t\FQ(t)$ with positive weights on the retained window. If $\bv^\top\GQ\bv=0$ then every directional QFI contribution vanishes, so by Supplementary Corollary~\ref{cor:cert} $D_{\bv}\Lc=g^\top\bv=0$. Hence $g\in\operatorname{range}(\GQ)$ on the retained support, and $\GQ^+$ acts on $g$ as a true inverse.

\begin{sprop}[Rayleigh optimum]\label{prop:rayleigh}
\begin{equation}
\GamQ\equiv\max_{\bv^\top\GQ\bv\le1}(-g^\top\bv)=\sqrt{g^\top\GQ^+g},\qquad
\bv_Q^\star=-\frac{\GQ^+g}{\sqrt{g^\top\GQ^+g}},\qquad
\GamQ=\|r\|\,\|\ws\|\,\eta_Q(\bv_Q^\star).
\label{eq:s-rayleigh}
\end{equation}
\end{sprop}
\begin{proof}
For $\bv$ on the support set $z=\GQ^{1/2}\bv$; then $g^\top\bv=(\GQ^{+1/2}g)^\top z$ and Cauchy--Schwarz gives $(g^\top\bv)^2\le(g^\top\GQ^+g)(\bv^\top\GQ\bv)$, with equality if and only if $z\propto\GQ^{+1/2}g$. Maximizing $-g^\top\bv$ over $\bv^\top\GQ\bv\le1$ therefore gives $\sqrt{g^\top\GQ^+g}$, attained at $\bv_Q^\star$, which satisfies $\bv_Q^{\star\top}\GQ\bv_Q^\star=1$ and $-g^\top\bv_Q^\star=\GamQ$. For the identity, $\BQ(\bv_Q^\star)=1$ and $|D_{\bv_Q^\star}\Lc|=-g^\top\bv_Q^\star=\GamQ$, so $\eta_Q(\bv_Q^\star)=\lo(\bv_Q^\star)/\BQ(\bv_Q^\star)=\GamQ/(\|r\|\|\ws\|)$. Since $\lo\le\BQ$ by Eq.~\eqref{eq:s-chain}, $\eta_Q\in[0,1]$, so $\GamQ\le\|r\|\|\ws\|$.
\end{proof}
The generalized Rayleigh quotient $(g^\top\bv)^2/(\bv^\top\GQ\bv)$ is maximized by the dominant vector of the pencil $(gg^\top,\GQ)$, which is how $\bv_Q^\star$ is computed numerically (main-text Methods). Equation~\eqref{eq:s-rayleigh} is the natural-gradient optimum of Amari \cite{S5} for the metric $\GQ$. The contribution here is the identification of the metric appropriate to the retrained loss and the identity with $\eta_Q$.

\subsection*{Invariance}
\begin{slemma}[Reparameterization]\label{lem:reparam}
Under a local reparameterization $\bth=\bth(\bm\phi)$ with Jacobian $J$ of full rank, $g_\phi=J^\top g_\theta$ and $\GQ^{(\phi)}=J^\top\GQ J$, so $g_\phi^\top(\GQ^{(\phi)})^+g_\phi=g_\theta^\top\GQ^+g_\theta$ on the common retained support. Hence $\GamQ$ is invariant, and so are $\eta_Q(\bv_Q^\star)=\GamQ/(\|r\|\|\ws\|)$ and $\eta_Q(\bv)$ for every direction transported as $\bv_\phi=J^{-1}\bv_\theta$, because $D_{\bv}\Lc$ and $\bv^\top\GQ\bv$ are both scalars under transport.
\end{slemma}
No invariance is claimed across a change of numerical support.

\begin{slemma}[Observable rescaling]\label{lem:scaling}
Let $X$ have full column rank and let the observables be rescaled, $O_j\to s_jO_j$ with $S=\operatorname{diag}(s_1,\dots,s_m)$, $s_j>0$. Then $X\to XS$, $X_{\bv}\to X_{\bv}S$, $\ws\to S^{-1}\ws$, and $r$, $\Lc$ and $D_{\bv}\Lc$ are unchanged. Under a common rescaling $S=sI$, every chain quantity in Eq.~\eqref{eq:s-chain} is multiplied by $s$, so $\eta_1,\dots,\eta_5$ and $\eta_Q(\bv)$ are invariant. Under a non-uniform $S$ the chain quantities and the link factors change individually.
\end{slemma}
\begin{proof}
With full column rank, $(XS)^+=S^{-1}X^+$, so the readout becomes $S^{-1}\ws$, the prediction $XS\,S^{-1}\ws=X\ws$, the residual and the loss are unchanged, and by Eq.~\eqref{eq:s-envelope} $D_{\bv}\Lc\to-r^\top X_{\bv}S\,S^{-1}\ws$ is unchanged. For $S=sI$, $\|\ws\|\to\|\ws\|/s$ gives $\lo\to s\lo$. The vectors $\PXp X_{\bv}\ws$ and $X_{\bv}\ws$ are unchanged while $\|\ws\|$ scales by $s^{-1}$, so $\lp$ and $\|X_{\bv}\ws\|/\|\ws\|$ scale by $s$. The norms $\normtwo{X_{\bv}}$ and $\normF{X_{\bv}}$ scale by $s$, and $\Var(sO_j)=s^2\Var(O_j)$ gives $\BC\to s\BC$, $\BQ\to s\BQ$. Each ratio used in the five-factor definition is therefore invariant. For non-uniform $S$ the weights $\omega_t=\sum_js_j^2\Var(O_j)$ and the norm $\|S^{-1}\ws\|$ change by direction-dependent factors, and the ratios are not preserved.
\end{proof}
Because of Lemma~\ref{lem:scaling}, every comparison in the paper holds the observable normalization fixed. The numerical values of the individual link factors depend on the normalization of the observable channels. In all numerical results we use the observables in their native normalization, without additional variance rescaling.

\begin{slemma}[Dictionary dependence of the budget]\label{lem:dict}
Write $q_t(\bv)=\bv^\top\FQ(t)\bv$, which is measurement free, and $M(\bv)=\sum_tq_t(\bv)$ for the marginal QFI along $\bv$. With $\omega_t=\sum_j\Var_{\rho_t}(O_j)$ and $\Delta_j$ the spectral spread (largest minus smallest eigenvalue) of $O_j$,
\begin{equation}
\Big(\min_t\omega_t\Big)M(\bv)\;\le\;\BQ^2(\bv)\;\le\;\Big(\max_t\omega_t\Big)M(\bv)\;\le\;\Big(\tfrac14\sum_j\Delta_j^2\Big)M(\bv),
\label{eq:s-dict}
\end{equation}
and $\BQ^2(\bv)=\bar\omega\,M(\bv)+T\,\mathrm{cov}_t(\omega_t,q_t(\bv))$ with $\bar\omega$ the time average of $\omega_t$.
\end{slemma}
\begin{proof}
$\BQ^2=\sum_t\omega_tq_t$ with $q_t\ge0$ gives the first two inequalities and the covariance identity. The last is Popoviciu's inequality $\Var_\rho(O)\le\Delta^2/4$, applied to each observable at each time.
\end{proof}
Thus the dictionary changes $\BQ$ through the variance weights $\omega_t$, whereas its measurement content enters through $\eta_1$ and $\eta_2$. The zero set is unchanged, since a vanishing marginal QFI $M(\bv)$ makes $\BQ$ vanish for every dictionary. At the Ising anchor, $\omega_t\in[2.71,4.00]$ and $\BQ^2$ is $92$--$93\%$ of the Pauli upper envelope $4M(\bv)$ for the reported directions. At the JC anchor, $\omega_t\in[4.32,4.85]$, and $\BQ^2$ differs from $\bar\omega M(\bv)$ by less than $0.5\%$ for the directions of main-text Fig.~2. The data and scripts are part of the deposited code and data.

\subsection*{Euclidean comparison}
\begin{sprop}[Euclidean comparison]\label{prop:gd}
The Euclidean steepest-descent direction normalized to the same QFI budget, $\bv_{\rm GD}=-g/\sqrt{g^\top\GQ g}$, achieves
\begin{equation}
\Gamma_{\rm GD}^{(Q)}=-g^\top\bv_{\rm GD}=\frac{g^\top g}{\sqrt{g^\top\GQ g}}\le\sqrt{g^\top\GQ^+g}=\GamQ,
\label{eq:s-gd}
\end{equation}
with equality if and only if $g$ is an eigenvector of $\GQ$ on the retained support.
\end{sprop}
\begin{proof}
$g^\top g=(\GQ^{1/2}g)^\top(\GQ^{+1/2}g)\le\|\GQ^{1/2}g\|\,\|\GQ^{+1/2}g\|=\sqrt{g^\top\GQ g}\sqrt{g^\top\GQ^+g}$, with equality if and only if $\GQ^{1/2}g\propto\GQ^{+1/2}g$, that is, $\GQ g\propto g$.
\end{proof}
Supplementary Proposition~\ref{prop:gd} concerns the metric rather than an optimizer. The QFI unit ranks directions by first-order loss decrease per unit of distinguishable state motion. No claim about convergence rate or final loss is made or tested here. The design protocol of \ref{sec:matched} uses this direction on the physical parameters in its final stage; its effect is measured there by the stage test.

On the $11\times11$ JC grid of main-text Fig.~4, the metric advantage is present at every point: $\GamQ/\Gamma_{\rm GD}^{(Q)}$ has median $1.13$, ranges from $1.049$ to $3.308$, and exceeds one at all 121 points. These values are computed directly from the stored $\GQ$ and loss gradients at each grid point and compare first-order directions under the same QFI budget; they are not a claim about optimizer convergence.

\subsection*{The anchor sequence}
Several checks use a common sequence of 31 JC anchors generated from $(30,20,10)$. At each point, $\bv_Q^\star$ is normalized to one QFI unit and a nominal per-sample-QFI step of $0.1$ is backtracked until the retrained loss decreases. All states, derivatives, readouts and QFI matrices are recomputed after every accepted step. The sequence spans training loss $11.16\to6.67$, held-out NRMSE $0.449\to0.349$, and $\GamQ=4.9\times10^{-2}\to6\times10^{-8}$. The maximum photon occupation is $1.32$. The sequence serves only as a reproducible set of operating points.

As a numerical check of Supplementary Proposition~\ref{prop:gd}, $\GamQ\ge\Gamma^{(Q)}_{\rm GD}$ holds at all 31 anchors, with $\GamQ/\Gamma^{(Q)}_{\rm GD}=1.02$ at the initial anchor, $1.05$, $1.08$, $1.12$, $1.13$, $1.13$, $1.12$, $1.34$, $1.25$, $1.13$ and $1.16$ at updates $1$--$10$, a maximum of $1.45$ at update $12$, and values between $1.06$ and $1.30$ thereafter. The per-update values are included in the deposited data.

\section{Input profiling: nested maps and paired checks}
\label{sec:input}
This note gives the nested input-map family, the profiling algorithm and the paired checks behind the input intervention of main-text Table~3 and the sign-agreement scan of Sec.~2.4.

The input interface changes the trajectory and therefore the downstream links of the loss budget. We profile the training loss over a nested family of monotone pointwise bijections $f_P(u;\bc)$ \cite{Kuriki2018,Penkovsky2018,Govia2022,Picco2025}. Normalize $u$ to $z(u)=(u-u_{\min})/(u_{\max}-u_{\min})$ and define
\begin{equation}
f_P(u;\bc)=u_{\min}+(u_{\max}-u_{\min})
\frac{\int_0^{z(u)}\exp[\sum_{k=1}^Pc_kP_k(2s-1)]\,ds}
{\int_0^1\exp[\sum_{k=1}^Pc_kP_k(2s-1)]\,ds}.
\label{eq:s-map}
\end{equation}
The map is strictly increasing, fixes both endpoints and reduces to the identity at $\bc=0$. Appending a zero coefficient embeds order $P$ in order $P+1$. Hence
\begin{equation}
\widetilde\Lc_P(\bth)=\min_{\bc\in\mathbb R^P}\Lc(\bth,f_P(u;\bc)),\qquad
\widetilde\Lc_{P+1}(\bth)\le\widetilde\Lc_P(\bth)\le\Lc(\bth,u).
\label{eq:s-nesting}
\end{equation}
For a threshold $L_\star$, the acceptable fraction
\begin{equation}
Y_P(L_\star)=\frac{|\{\bth\in\mathcal G:\widetilde\Lc_P(\bth)\le L_\star\}|}{|\mathcal G|}
\label{eq:s-yield}
\end{equation}
cannot decrease with $P$ on the training set. This nesting statement is exact. Held-out improvement remains empirical.

\subsection*{Profiling algorithm}
At fixed $\bth$, start from the identity map and form a prediction-response bank $D_P=[d_1\cdots d_P]$ from centred coefficient probes. Propose $\Delta\bc\propto D_P^+r$, clip its norm, evaluate the full nonlinear reservoir, and backtrack until training loss decreases. Recompute $D_P$ after every accepted update. When an order saturates, append a zero coefficient and continue at $P+1$. The procedure is a local profile and does not certify a global minimum. Numerical inversion of every accepted map is checked before held-out evaluation. When the inner optimum is locally regular, the envelope theorem gives
\begin{equation}
\nabla_{\bth}\widetilde\Lc_P(\bth)
=\partial_{\bth}\Lc(\bth,f_P(u;\bc))\big|_{\bc=\bc^\star(\bth)},
\end{equation}
so the physical derivative is recomputed on the redesigned trajectory without differentiating $\bc^\star$.

\subsection*{Results at the JC anchor and on the parameter grid}
\begin{figure}[!ht]
\centering
\includegraphics[width=\textwidth,height=0.6\textheight,keepaspectratio]{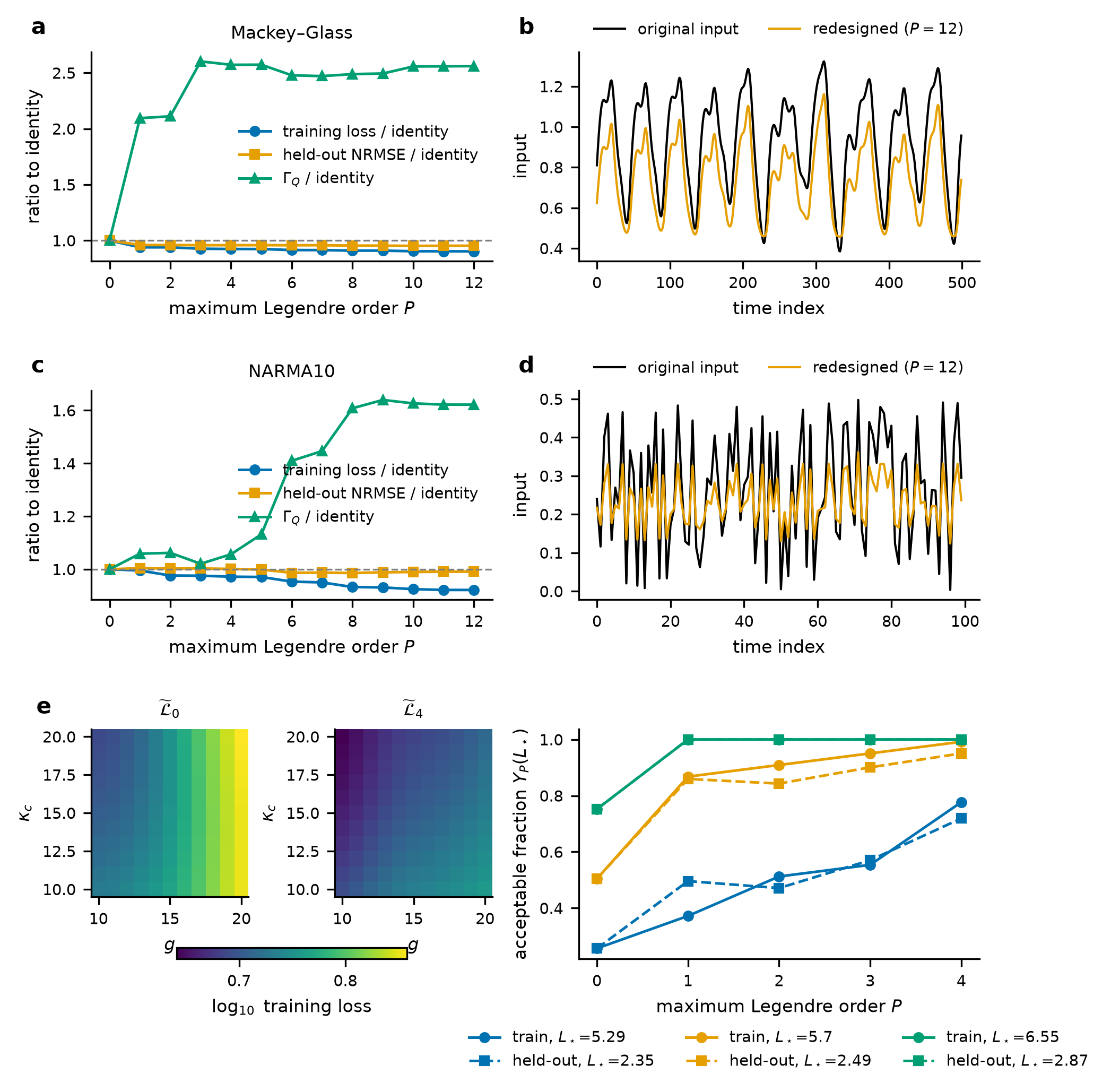}
\caption{\textbf{Nested input profiling.} \textbf{a},\textbf{c}, Training loss, held-out NRMSE and $\GamQ$ at $(30,20,10)$ against the maximum Legendre order for Mackey--Glass and NARMA10. \textbf{b},\textbf{d}, Representative original and redesigned inputs. \textbf{e}, Profiled-loss landscape on the main-text $(g,\kappa_c)$ grid and acceptable fraction $Y_P(L_\star)$ for three identity-loss thresholds.}
\label{fig:profiled}
\end{figure}

At $(30,20,10)$, Mackey--Glass training loss falls by $8\%$ at $P=4$ and $10\%$ at $P=12$ (Supplementary Fig.~\ref{fig:profiled}a,b). Held-out NRMSE falls from $0.449$ to $0.430$, and $\GamQ$ increases by a factor $2.6$. For NARMA10, training loss falls by $8\%$, held-out NRMSE changes by $1\%$, and $\GamQ$ rises by a factor $1.6$ (Supplementary Fig.~\ref{fig:profiled}c,d). Most of the Mackey--Glass gain appears at low order. The accepted Mackey--Glass map emphasizes the peaks of the series, whereas the NARMA10 map compresses the bulk of the uniform input.

On the $11\times11$ grid of the main-text scan, profiling lowers the training and held-out loss at every point (Supplementary Fig.~\ref{fig:profiled}e). At the lower-quartile identity-loss threshold, the acceptable fraction rises from $0.26$ to $0.78$ by $P=4$ ($0.72$ held out). At the median threshold it rises from $0.50$ to $0.99$ ($0.95$ held out). These numbers illustrate the nesting result rather than a fabrication yield.

\subsection*{Scope of the testbeds}
The JC and Ising reservoirs are minimal testbeds chosen for exact simulation of every chain quantity, and at the reference anchor with the four-feature dictionary they are not competitive learners (held-out NRMSE $0.449$, and $0.357$ with the measurement-conditioned drive). The multiplexed reservoirs designed under the all-parameter protocol and their matched-cost comparison with a tuned echo-state network are reported in \ref{sec:matched}.

\subsection*{Paired profiling along the anchor sequence}
\begin{figure}[!ht]
\centering
\includegraphics[width=\textwidth,height=0.6\textheight,keepaspectratio]{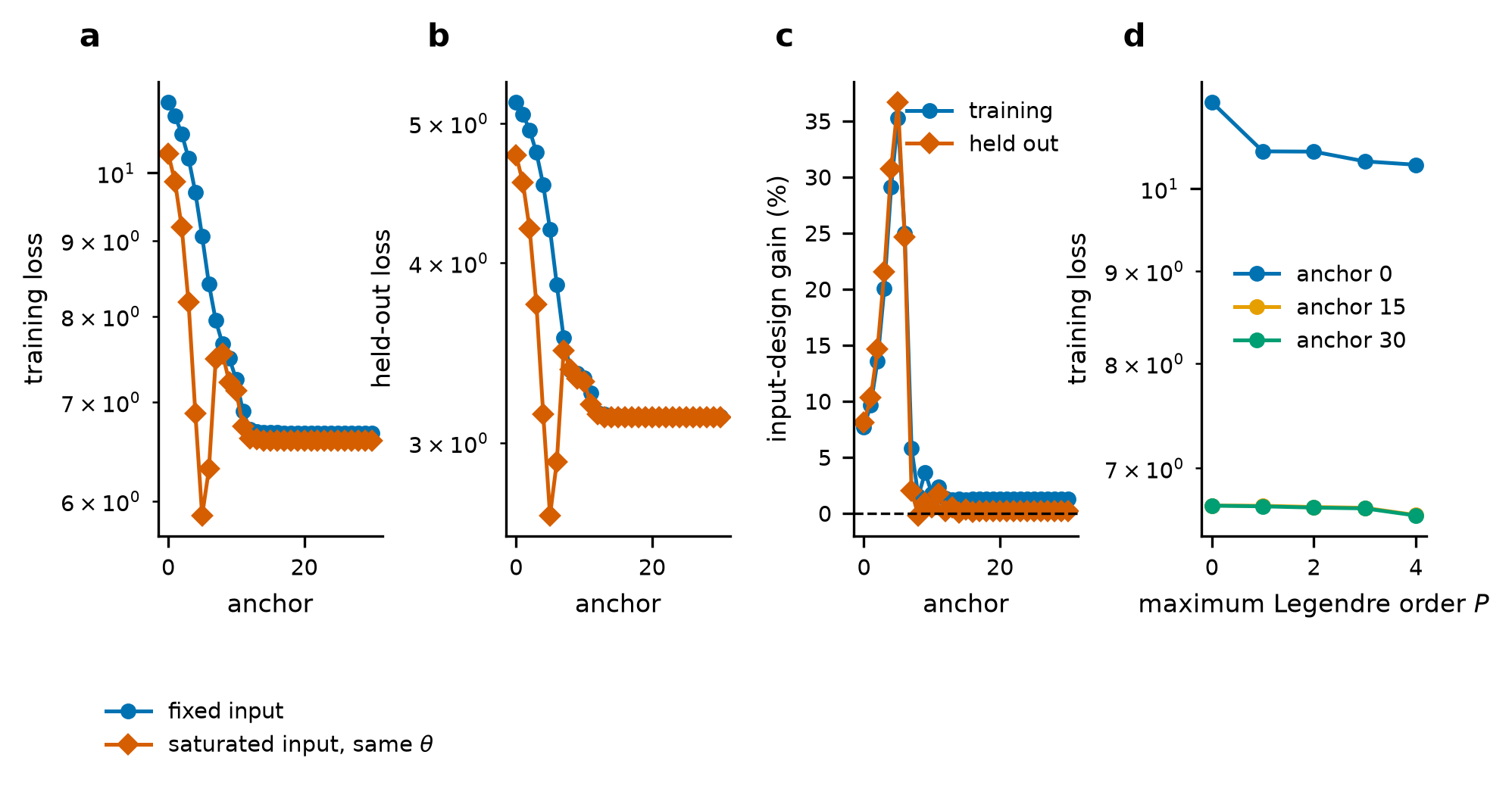}
\caption{\textbf{Input profiling along the common anchor sequence.} \textbf{a}, Training loss for identity and redesigned inputs. \textbf{b}, Held-out loss after the map is fixed. \textbf{c}, Paired training-loss gain. \textbf{d}, Order-by-order traces at three representative anchors. Each redesign starts from the identity map and uses training data only.}
\label{fig:paired}
\end{figure}

At each of the 31 anchors of \ref{sec:rayleigh}, the input map is restarted from identity and optimized on training data only with $P\le4$ (Supplementary Fig.~\ref{fig:paired}). Training loss decreases at every anchor by $1.2$--$35\%$. Held-out changes track the same trend except at one anchor. The benefit is largest in the middle of the sequence and falls to about $1\%$ near its end. The evaluation is paired at fixed physical parameters, and no optimizer is compared.

\section{Shot cost and representation freedom}
\label{sec:shots}
This note derives the finite-shot cost of estimating the gradient and tests whether representation freedom can lower it in the JC reservoir. It supports the finite-shot equation of the main text (Eq.~\eqref{eq:s-shots} below) and the negative result quoted with it.

For an independent centred-difference estimate at $\bth\pm h\bv$ using $n$ trajectories per side,
\begin{equation}
\Var(\widehat{D_{\bv}\Lc})\simeq\frac{\sigma_{\bv}^2}{2nh^2},\qquad
n_{\rm req}=\frac{\sigma_{\bv}^2}{2h^2\epsilon_{\rm rel}^2(D_{\bv}\Lc)^2}
=\frac{\sigma_{\bv}^2}{2h^2\epsilon_{\rm rel}^2\|r\|^2\|\ws\|^2\BQ^2(\bv)\eta_Q^2(\bv)}.
\label{eq:s-shots}
\end{equation}
The denominator is the useful derivative set by the loss budget. The numerator is the propagated noise of the measured representation. The question is whether the representation can reduce $\sigma_{\bv}^2$ without changing the calibrated prediction or the derivative being estimated.

\subsection*{Mean-equivalence fibre}
Let $Z_{tk}=\Tr(\rho_tB_k)$ be an accessible operator dictionary and $X=ZC_0$. If $N_h$ spans a common weak kernel of the pilot dictionaries, then
\begin{equation}
C(A)=C_0+N_hA
\label{eq:s-fiber}
\end{equation}
preserves the pilot feature means while changing the shot covariance. For one deployed channel $c=c_0+N_ha$ with covariance weight $\Sigma$, the minimum-variance representative is
\begin{equation}
a^\star=-(N_h^\top\Sigma N_h)^+N_h^\top\Sigma c_0.
\label{eq:s-variance_design}
\end{equation}
A strict gain requires covariance along the weak subspace. Weak rank alone is not enough. This is a control-variate construction \cite{S2}, distinct from optimizing the observable for prediction \cite{S6}.

We consider two variants. In the $C$-only design the trained readout is fixed and $C$ moves within the mean-equivalence fibre. In the joint design, the readout may also move within a $2\%$ predictor-drift tolerance, enlarging the admissible weak subspace. Because the physical derivative is cancellation dominated, a representation that nearly preserves the prediction can still change $D_{\bv}\Lc$ appreciably. We therefore report both the variance reduction and the derivative ratio $(r^\top Z_{\bv}c)^2/(r^\top X_{\bv}\ws)^2$. A reduction in shot noise is useful only when this ratio remains close to one.

\begin{figure}[!ht]
\centering
\includegraphics[width=\textwidth,height=0.6\textheight,keepaspectratio]{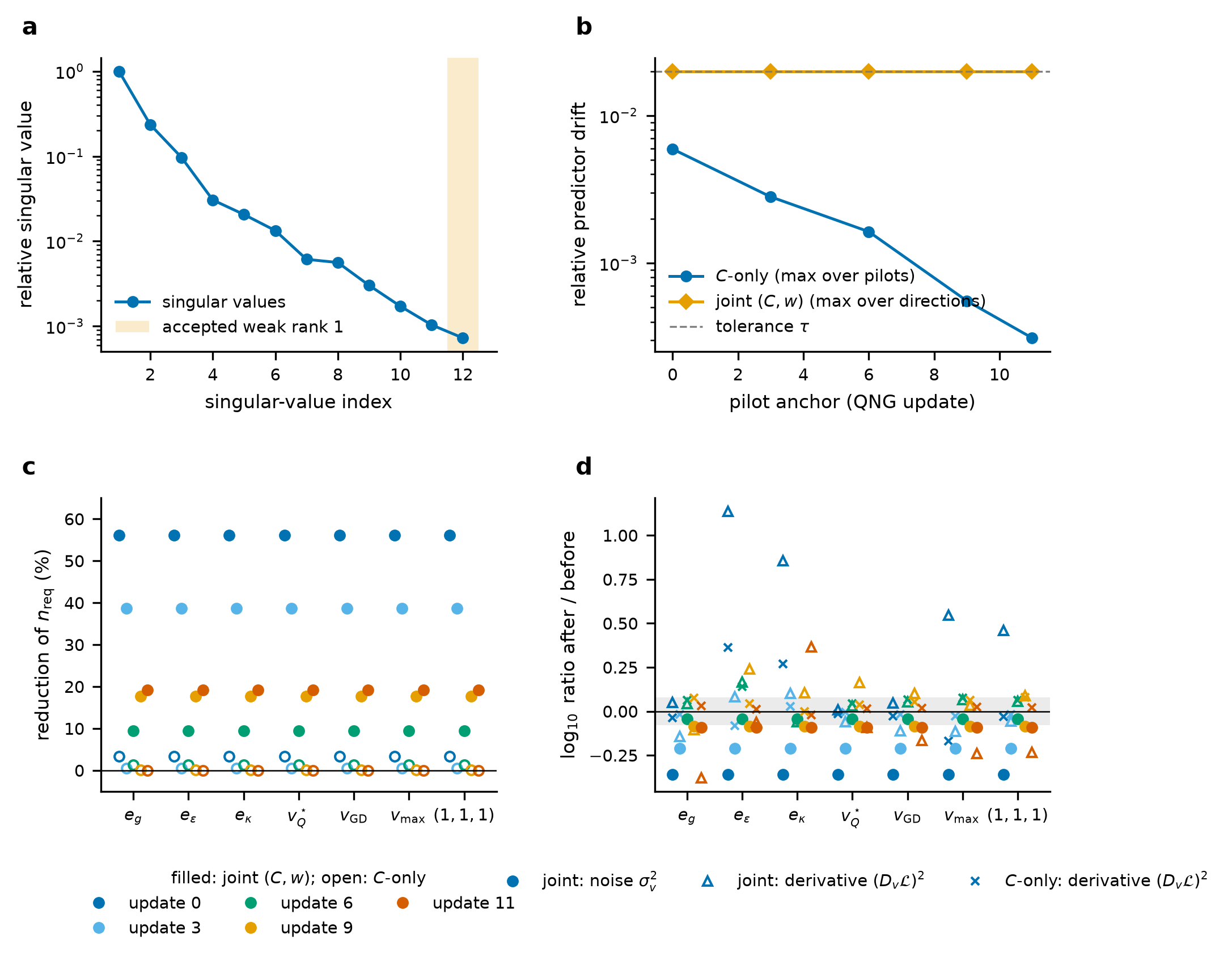}
\caption{\textbf{Shot-cost redesign in the JC reservoir.} \textbf{a}, Relative singular spectrum of the stacked pilot dictionaries and accepted weak rank. \textbf{b}, Predictor drift for the $C$-only and joint designs. \textbf{c}, Change in the required shots of Eq.~\eqref{eq:s-shots} across the declared physical directions. \textbf{d}, Corresponding variance and derivative ratios; the latter quantifies bias introduced when an approximate fibre changes the useful derivative.}
\label{fig:jcshots}
\end{figure}

For quantum features, covariances are assembled within compatible measurement groups and independent groups receive independent shot allocations. In the JC test the accessible dictionary contains twelve outcome functions from the compatible groups $\{x,\sigma_x\}$ and $\{p,\sigma_y\}$. Five pilot anchors from the common anchor sequence and their centred-difference probes are used with $10^3$ pilot shots per sample and group. The stacked dictionary has accepted weak rank one and strict derivative-preserving rank zero (Supplementary Fig.~\ref{fig:jcshots}a). Consequently the $C$-only design reduces the variance by at most $3\%$, while the joint design reduces it by $9$--$56\%$ but moves the derivative substantially (derivative ratios between $0.4$ and $14$ across the declared directions; Supplementary Fig.~\ref{fig:jcshots}b--d). Direct multinomial sampling verifies Eq.~\eqref{eq:s-shots} to within $5\%$. The result is negative. The four-feature JC representation does not contain enough derivative-preserving redundancy for a large shot saving. A 100-node stochastic ESN with larger weak rank is reported in \ref{sec:esn}.

\section{Numerical controls and reporting protocol}
\label{sec:controls}
This note lists the matched controls applied in every numerical comparison of the paper and the reporting protocol for the deposited data. It supports the Methods of the main text, which refer here for the additional numerical conventions.

Supplementary Table~\ref{tab:controls} gives the controls experiment by experiment.
\begin{table}[!ht]\centering\small
\caption{Matched controls used in the numerical comparisons.}
\label{tab:controls}
\begin{tabular}{p{0.25\linewidth}p{0.67\linewidth}}\toprule
Experiment & Controls\\\midrule
Chain and SLD checks & Common trajectory window, QFI support and observable normalization; QFI direction defined from the base dictionary and held fixed during the closure test; cutoff convergence checked at adjacent $N_c$.\\
Task-link construction & Target constructed only for the dedicated tightness panel; forecasting targets elsewhere are unchanged.\\
Parameter scan and attribution & Identical physical grid and direction convention; chain quantities and factors reported before and after each intervention.\\
Input profiling & Identity included at every order; optimization uses training loss only; each grid point or anchor is restarted from identity; held-out data are evaluated after the map is fixed.\\
Shot-cost redesign & Pilot cost separated from repeated-use savings; predictor drift and derivative ratio reported with every variance reduction; compatible measurement groups treated independently.\\
Atom-number scan & Paired log-parameter samples, matched QFI window and cutoff checks; both fixed-individual and fixed-collective coupling reported.\\
Spin reservoir & Same data split and readout conventions as the JC reservoir; symmetry and complex-step derivatives checked independently.\\
Drive-gain stopping & Readout refitted at every gain; slope zero, training optimum, closed-loop edge and $N_c$ marker extracted with fixed interpolation conventions; the held-out optimum is computed after the fact as a check and enters no selection; cutoff checks repeated on the stable and post-edge branches.\\
\bottomrule\end{tabular}
\end{table}
All random seeds, rejected updates, grid failures and figure-level numerical arrays are retained in the deposited data.

\section{Weak-rank design in a stochastic echo-state network}
\label{sec:esn}
This note repeats the weak-rank shot-cost design of \ref{sec:shots} in a 100-node stochastic echo-state network, where a high-dimensional weak fibre is accessible. It supports the finite-shot paragraph of the main text, which quotes the derivative ratios of this classical case alongside those of the JC reservoir.

The mean-equivalence construction of \ref{sec:shots} is platform independent, and a 100-node stochastic echo-state network (ESN) is the setting in which a high-dimensional weak fibre is accessible and repeated-trajectory statistics are stable. It is also the classical special case of the chain. The sampled node distribution is the declared channel, $\FQ(t)$ is its Fisher information, and $\BC=\BQ$ for the full distribution.

\subsection*{Model}
For stochastic realization $s$, the leaky ESN state is
\begin{equation}x_{t+1}^{(s)}=(1-\alpha)x_t^{(s)}+\alpha\tanh\!\left(Wx_t^{(s)}+W_{\rm in}u_t+b+\xi_t^{(s)}\right),\qquad \xi_t^{(s)}\sim\mathcal N(0,\sigma_{\rm dyn}^2I),\label{eq:s-esn}\end{equation}
with feature mean $X_{tj}=S^{-1}\sum_sx_{t,j}^{(s)}$. The fixed reservoir realization has 100 nodes, leak rate $\alpha=0.3$, $W$ Gaussian and rescaled to spectral radius $0.9$, $W_{\rm in}\sim\mathcal U(-0.5,0.5)$ (input scaling $0.5$), $b\sim\mathcal U(-0.1,0.1)$, dynamical noise $\sigma_{\rm dyn}=0.01$ and $S=100$ realizations with common random numbers across parameter probes. The task and data split are those of the Mackey--Glass protocol of the main text (training loss $1.27$, held-out NRMSE $0.174$ at the anchor). Physical directions are unit vectors in the flattened parameter vector $(W,W_{\rm in},\alpha)$, namely a random direction in $W$, the leak rate, a random direction in $W_{\rm in}$, three fixed random combinations of the blocks, and the negative loss gradient projected on their span. Centred differences use $h=0.1$, the pilot set is the anchor with its $\pm h$ probes, and the required-trajectory count uses $\epsilon_{\rm rel}=0.1$.

\subsection*{Results}
\begin{figure}[!ht]
\centering
\includegraphics[width=\textwidth,height=0.6\textheight,keepaspectratio]{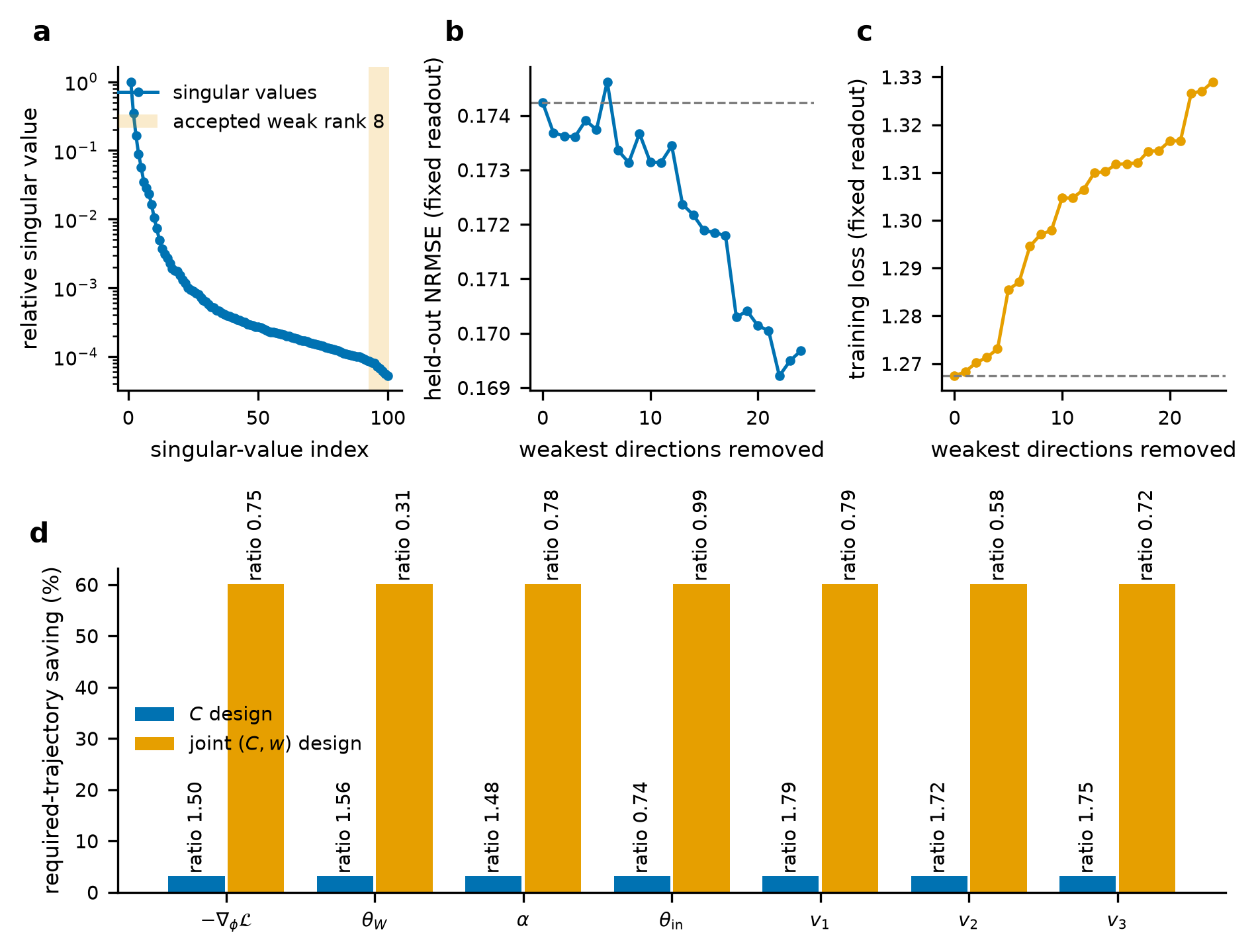}
\caption{\textbf{Weak-rank representation freedom in a stochastic echo-state network.} \textbf{a}, Relative singular spectrum of the stacked pilot feature matrices of the 100-node network and the accepted weak rank (shaded). \textbf{b},\textbf{c}, Held-out NRMSE (b) and training loss (c) after naively deleting increasing numbers of weak directions while holding the readout fixed; both expose predictor drift. \textbf{d}, Reduction in the number of repeated stochastic trajectories required to estimate centred physical derivatives at relative precision $\epsilon_{\rm rel}=0.1$, for the seven declared directions ($-\nabla_\phi\Lc$ projected on the declared span, $\theta_W$, $\alpha$, $\theta_{\rm in}$ and three fixed random combinations $v_1$--$v_3$). Blue bars redesign the representation coefficients $C$ within the accepted mean-equivalence fibre with the readout fixed; orange bars jointly redesign $C$ and the readout $w$ within the drift tolerance $\tau=0.02$. The label above each bar is the derivative ratio $(r^\top Z_{\bv}c)^2/(r^\top X_{\bv}\ws)^2$ of \ref{sec:shots}, the bias that accompanies the saving.}
\label{fig:esn}
\end{figure}

The stacked ESN pilot matrices have accepted weak rank eight and strict derivative-preserving rank two (Supplementary Fig.~\ref{fig:esn}a). Naively deleting weak directions changes the predictor (Supplementary Fig.~\ref{fig:esn}b,c). Moving within the accepted fibre with the readout fixed reduces the repeated-trajectory requirement by only $3\%$, with derivative ratios $0.74$--$1.79$. Allowing a $2\%$ predictor drift reduces the requirement by about $60\%$ but gives derivative ratios $0.31$--$0.99$ (Supplementary Fig.~\ref{fig:esn}d). The ESN therefore shows the same tradeoff as the JC test at larger weak rank. Representation freedom can reduce sampling noise, but only derivative-preserving freedom gives an unbiased saving.

\section{Atom-number scan of the upstream budget}
\label{sec:atoms}
This note reports the atom-number scan of the upstream QFI budget under two coupling controls. It supports the statement of main-text Sec.~2.4 that the upstream budget grows with atom number while its use is decided by the downstream links.

\begin{figure}[!ht]
\centering
\includegraphics[width=\textwidth,height=0.6\textheight,keepaspectratio]{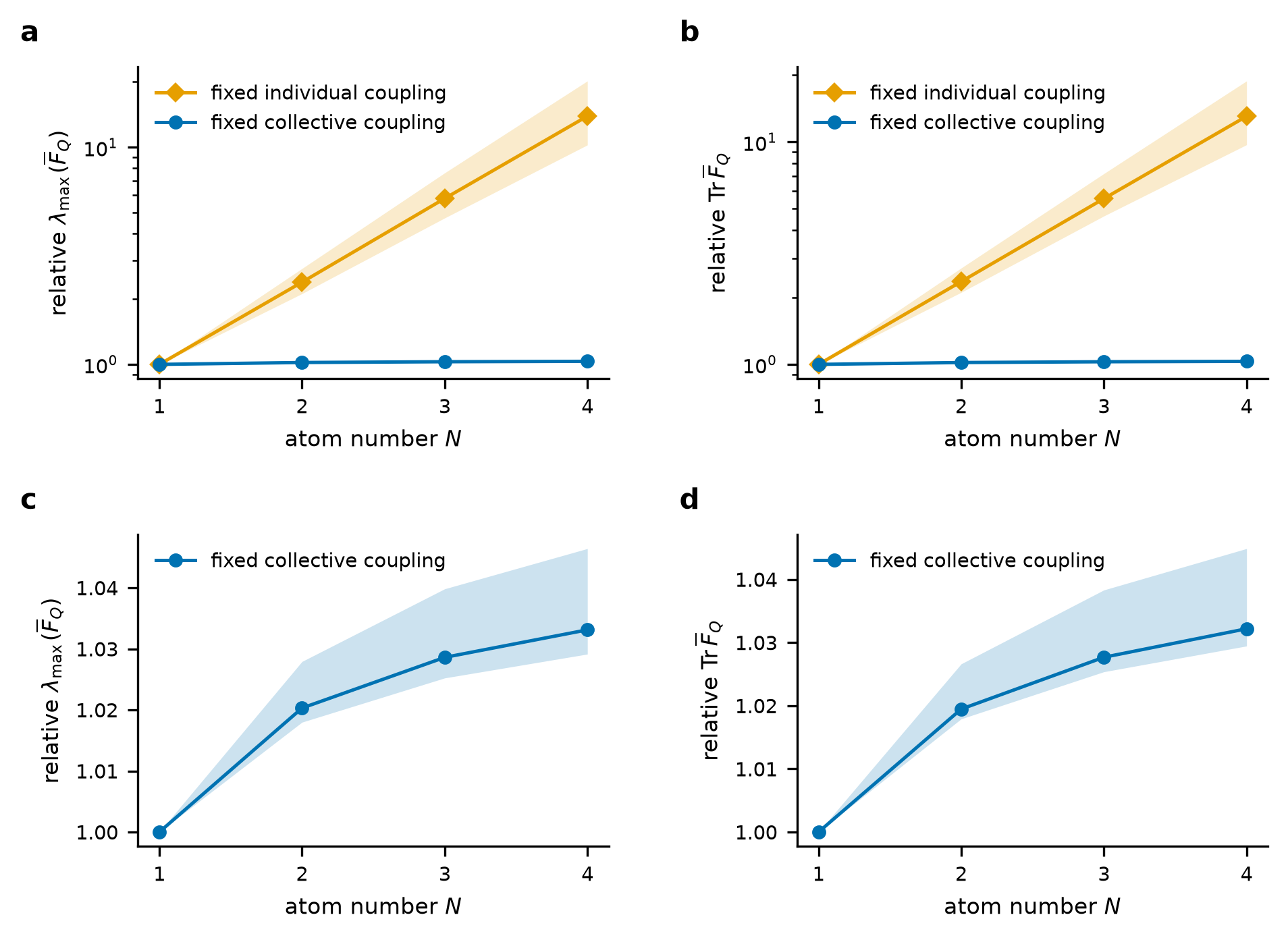}
\caption{\textbf{Atom number enlarges the upstream QFI budget under matched coupling controls.} The mean marginal QFI is $\overline{\FQ}=T^{-1}\sum_{t=1}^{T}\FQ(t)$ over $T=550$ post-washout samples of a driven multi-atom JC reservoir, with respect to the logarithmic controls $(\log G\ \text{or}\ \log g_0,\log\epsilon,\log\kappa_c,\log\gamma)$. \textbf{a}, Relative dominant QFI $\lambda_{\max}(\overline{\FQ})$ (logarithmic axis). \textbf{b}, Relative total QFI $\Tr\overline{\FQ}$. Orange diamonds use fixed individual coupling $g_i=g_0$; blue circles use fixed collective coupling $g_i=G/\sqrt N$, for which $\sqrt{\sum_ig_i^2}=G$. \textbf{c},\textbf{d}, Expanded fixed-collective results. Symbols and bands are medians and interquartile ranges over eight paired log-parameter samples around $(G\ \text{or}\ g_0,\epsilon,\kappa_c,\gamma)=(10,10,5,5)$ with standard deviation $0.12$. Results use $N=1,2,3,4$, $N_c=8$, $\omega_c=40$, $\omega_a=20$, $\Delta t=1/30$ and a 100-sample washout. Values are normalized to the paired $N=1$ result.}
\label{fig:atoms}
\end{figure}

We separate the system-size effect from the growth of the total coupling by comparing $N=1,2,3,4$ atoms under two controls: fixed individual coupling $g_i=g_0$ and fixed collective coupling $g_i=G/\sqrt N$, for which $\sqrt{\sum_i g_i^2}=G$ is constant. The scan uses eight paired log-parameter samples, a common 550-sample QFI window and the same numerical checks for every $N$. Under fixed collective coupling, the median dominant eigenvalue and trace of the mean trajectory QFI at $N=4$ are both about $3\%$ above their paired $N=1$ values, with an increase in every sample (Supplementary Fig.~\ref{fig:atoms}). Under fixed individual coupling they grow by roughly factors $14$ and $13$, respectively, largely because the total coupling itself grows. The fixed-$G$ result is therefore a small but reproducible enlargement of the upstream state-sensitivity budget at matched collective coupling. No asymptotic scaling or task-performance conclusion is drawn. Downstream usefulness was not evaluated as a function of $N$.

\section{Transverse-field Ising reservoir: chain values and checks}
\label{sec:tfi-si}
This note collects the chain values, implementation checks, twenty-seed statistics and adversarial tests for the four-spin transverse-field Ising reservoir. It supports main-text Fig.~3 and the three quantum statements of Sec.~2.3 (certificate, centred SLD and incompatible SLDs), including the disclosure that seed 7 is the best of the twenty seeds.

Supplementary Table~\ref{tab:tfi} gives the chain values behind main-text Fig.~3a for coupling seed 7. The twenty-seed statistics that the main text quotes are collected in Supplementary Tables~\ref{tab:tfi-seeds} and~\ref{tab:tfi-seeds-dirs} below, and the negative search for a second near-redundant direction in Supplementary Tables~\ref{tab:neardeg} and~\ref{tab:neardeg-cand}. Three independent checks confirm the implementation for seed 7. The envelope derivative agrees with centred differences to four digits for every nonzero-QFI direction, and both vanish on $\bv_{\rm sym}$. The SLD equality holds term by term to relative $3\times10^{-15}$. The constructed task target at $\bv_Q^\star$ gives $\lo=\lp=\normtwo{A}=0.057712644$ with the intended readout recovered to $1.5\times10^{-9}$.

\begin{table}[!ht]\centering\small
\caption{Chain quantities and link factors in the four-spin transverse-field Ising reservoir at $(J,h,\tau)=(1,1,1)$, coupling seed $7$, dictionary $\{Z_i\}$, five-step Mackey--Glass forecasting. Euclidean-unit directions except $\bv_Q^\star$ (one QFI unit). $\eta_2=1$ for every direction because every observable has two outcomes. The row $\bv_{\rm sym}$ is at the numerical floor and its factors are not meaningful.}
\label{tab:tfi}
\resizebox{\textwidth}{!}{\begin{tabular}{@{}lccccccc@{}}\toprule
Direction & $\BQ$ & $\BC=\normF{X_{\bv}}$ & $\normtwo{X_{\bv}}$ & $\|X_{\bv}\ws\|/\|\ws\|$ & $\lp$ & $\lo$ & $(\eta_1,\eta_3,\eta_4,\eta_5)$\\\midrule
$e_J$ & 123 & 16.8 & 14.1 & 1.85 & 0.894 & 0.204 & (0.136, 0.110, 0.483, 0.228)\\
$e_h$ & 200 & 34.4 & 33.7 & 1.66 & 1.06 & 0.079 & (0.172, 0.048, 0.636, 0.075)\\
$e_\tau$ & 240 & 30.0 & 28.9 & 2.51 & 1.34 & 0.125 & (0.125, 0.084, 0.534, 0.093)\\
$(1,-1,0)/\sqrt2$ & 162 & 31.9 & 29.8 & 1.74 & 1.01 & 0.200 & (0.197, 0.055, 0.579, 0.198)\\
$(1,1,1)/\sqrt3$ & 278 & 34.6 & 33.4 & 2.89 & 1.55 & 0.144 & (0.125, 0.084, 0.534, 0.093)\\
$\bv_{\rm sym}$ & $8\times10^{-6}$ & $2\times10^{-6}$ & $2\times10^{-6}$ & $7\times10^{-8}$ & $3\times10^{-8}$ & $6\times10^{-10}$ & --\\
$\bv_Q^\star$ & 1 & 0.163 & 0.136 & $1.48\times10^{-2}$ & $7.32\times10^{-3}$ & $1.73\times10^{-3}$ & (0.163, 0.091, 0.495, 0.236)\\
\bottomrule\end{tabular}}
\end{table}

The direction $(1,1,1)/\sqrt3$ equals $(2/\sqrt3)e_\tau$ plus a multiple of $\bv_{\rm sym}$ at this anchor, so its feature tangent is proportional to that of $e_\tau$ and its four nontrivial factors coincide with those of $e_\tau$. Only $\BQ$ and the absolute chain values differ, by the factor $2/\sqrt3$. A symmetry component therefore adds nothing to any link, as the certificate requires.

\emph{Symmetry orbit and tangent line.} Along the orbit $\bth(\lambda)=(\lambda J,\lambda h,\tau/\lambda)$, $\lambda\in[\tfrac12,2]$ (13 points), the training loss of the three dictionaries and two tasks is constant to a relative $7\times10^{-15}$ and the held-out NRMSE to $6\times10^{-13}$ (main-text Fig.~3d). Scaling $J$ alone changes the training loss by a factor between $0.67$ and $4.3$, and scaling $\tau$ alone between $0.59$ and $4.2$ over the same range. Along the straight tangent line $\bth+s\bv_{\rm sym}$ the unitary is $e^{-i(1-s^2)\tau H}$, so the loss is stationary at $s=0$ and changes at second order only, by at most $3\%$ for $|s|\le0.1$ and $21\%$ for $|s|\le0.3$ across the six task--dictionary combinations, symmetric in $s$. The certificate is the first-order statement. The orbit is its finite version for this symmetry.

\emph{Measurement families.} For $R_M(\bv)=\sum_t\omega_t\,\bv^\top\FC^{M(t)}\bv/\bv^\top\GQ\bv$ with $\omega_t=\sum_j\Var_{\rho_t}(Z_j)$, the values of main-text Fig.~3c are: $Z^{\otimes4}$ basis $(0.104,0.133,0.085,0.131)$; eigenbasis of $\SLD_{e_J}(t)$: $(1.000,0.153,0.416,0.932)$; of $\SLD_{e_h}(t)$: $(0.148,1.000,0.776,0.190)$; of $\SLD_{e_\tau}(t)$: $(0.365,0.758,1.000,0.161)$, in the order $(e_J,e_h,e_\tau,\bv_Q^\star)$. Normalized commutator norms $\|[\SLD_a,\SLD_b]\|_F/(\|\SLD_a\|_F\|\SLD_b\|_F)$ over the training window have medians $0.405$, $0.346$ and $0.236$ and minima $0.381$, $0.295$ and $0.192$ for the $(J,h)$, $(J,\tau)$ and $(h,\tau)$ pairs. $R_M$ compares Fisher informations and is therefore the square of an envelope ratio. For the deployed two-outcome measurements of $\{Z_i\}$ taken separately, $\eta_1^2\le R_{Z^{\otimes4}}$, consistent with $\eta_1=0.136$ and $R=0.104$ along $e_J$.

\subsection*{Adversarial tests}
\emph{Bound.} At the anchor $(1,1,1)$ with the Mackey--Glass trajectory, 600 random dictionaries (between 1 and 16 random Hermitian observables on the full sixteen-dimensional space, each normalized to unit mean variance), random unit directions in $(J,h,\tau)$ and random targets (the task target, Gaussian noise and three random nonlinear functions of the input history) give $|D_{\bv}\Lc|/(\|r\|\|\ws\|\BQ(\bv))\le0.050$ (median $7\times10^{-4}$) and no violation. Because a random target is a weak adversary, 400 further dictionaries were scored against the worst-case target of Proposition~1, for which $\lo=\normtwo{\PXp X_{\bv}}$. The largest value of $\lo/\BQ$ is $0.097$ (median $0.012$), again with no violation. It falls with dictionary size (maximum $0.097$ for $m\le4$ and $0.013$ for $m\ge13$) because the refit absorbs more of the response.

\emph{Certificate.} For 200 random dictionaries and targets the retrained loss along the finite orbit $(\lambda J,\lambda h,\tau/\lambda)$, $\lambda=1.05^{\pm1}$, changes by at most $6\times10^{-13}$ relative, and the states are bitwise identical. The complex-step derivative along $\bv_{\rm sym}$ is at most $10^{-5}$ of that along $e_J$ for the same dictionary. This residual is a cancellation among three $O(10^2)$ derivatives and is uncorrelated with the readout norm (correlation $0.01$). It is a round-off residual and carries no gradient information.

\emph{Incompatibility.} At three times in the training window the minimum over directions in the two-dimensional QFI support of $\bv^\top\FC\bv/\bv^\top\FQ\bv$ was maximized over all sixteen-outcome projective bases, parametrized as $U=U_0e^{iH}$ with $H$ Hermitian and started from the SLD eigenbases, the computational basis and random unitaries. The maximum is $0.919$, $0.863$ and $0.860$ (converged to $0.925$ at the first time under a sharper smooth-minimum objective), always reached from an SLD start and with both eigenvalues nearly equal, against $0.03$--$0.21$ for the SLD bases themselves and $0.02$--$0.06$ for $Z^{\otimes4}$. The weak commutator in QFI-orthonormal coordinates is $|\Tr\rho[\SLD_1,\SLD_2]|=0.046$, the strong one $\|[\SLD_1,\SLD_2]\|/(\|\SLD_1\|\|\SLD_2\|)=0.39$, and $\rho_t$ has rank eight. The kernel lets an optimized basis recover most of what the SLD bases lose, and the nonzero weak commutator bars the rest. For an optimized measurement the incompatibility cost is of order ten percent. Over the twenty seeds (three times each, same starts) the maximum is $0.88$ $[0.86,0.92]$, range $0.75$--$0.95$, never unity. The optimum is reached from an SLD start in 55 of 60 cases, and $\rho_t$ has rank eight in all of them.

\subsection*{Twenty random-coupling seeds}
Every quantity of this note and of main-text Fig.~3 was recomputed for the coupling seeds $s=0,\ldots,19$ at the anchor $(1,1,1)$. The recomputation for seed 7 reproduces the values of Supplementary Table~\ref{tab:tfi} to better than $2.5\times10^{-4}$ relative for every quantity above the numerical floor. Every qualitative statement of the main text holds for all twenty seeds: one eigenvalue of $\GQ$ at the floor ($|\lambda|/\lambda_{\max}\le7\times10^{-15}$), certificate exponents $2$ and $1$ for every seed, dictionary and task (fitted $2.000$ and $1.000$, worst $0.9987$), the orbit flat to $\le1.6\times10^{-13}$ (loss) and $\le2.4\times10^{-12}$ (NRMSE), the task link closed at $\bv_Q^\star$ ($|\lo-\lp|/\lp\le1.6\times10^{-10}$), and $D_{\bv_Q^\star}\Lc=-\GamQ$ to machine precision. Two quantities are seed specific, and the main text quotes their distributions from Supplementary Table~\ref{tab:tfi-seeds}. The first is the resolution of $\bv_Q^\star$ by the $J$-SLD basis ($0.93$ for seed 7; $0.74$ $[0.54,0.82]$ over seeds, with the best-matched single-control basis at $0.90$ $[0.78,0.97]$ and the best control $J$, $\tau$ or $h$ for 10, 7 and 3 seeds). The second is the performance of seed 7, which is the best of the twenty on both training loss and held-out NRMSE for $\{Z\}$/Mackey--Glass. The upstream budget varies little across seeds (interquartile range of $\BQ$ along the Euclidean directions between $-12\%$ and $+20\%$ of the median), whereas $\lo$ varies by an order of magnitude and $D_{\bv}\Lc$ changes sign between seeds along $e_J$, $e_h$ and $e_\tau$ (Supplementary Table~\ref{tab:tfi-seeds-dirs}). The budget is therefore a property of the model class, whereas the gradient depends on the particular couplings.

\begin{table}[!ht]\centering\small
\caption{Twenty-seed distribution of the quantities quoted in the main text (median [interquartile range]; min--max over seeds), anchor $(1,1,1)$, dictionary $\{Z_i\}$, Mackey--Glass unless stated. ``Floor'' denotes a round-off value ($|\lambda|/\lambda_{\max}<10^{-8}$) that changes between runs and is not quoted as a number. Capture fractions are $R_M(\bv)$ of main-text Fig.~3c.}
\label{tab:tfi-seeds}
\resizebox{\textwidth}{!}{\begin{tabular}{@{}lccc@{}}\toprule
Quantity & Seed 7 & 20 seeds: median [IQR] & min--max\\\midrule
$\GQ$ eigenvalues & floor, $1.9\times10^4$, $9.3\times10^4$ & floor (20/20), $1.99\times10^4$ $[1.81,2.52]$, $9.55\times10^4$ $[7.63,13.2]$ & $1.16$--$3.0\times10^4$; $5.4$--$26.9\times10^4$\\
Certificate exponents (QFI, $|D_{\bv}\Lc|$) & 2, 1 & $2.000$ $[2.000,2.000]$, $1.000$ $[1.000,1.000]$ & worst $0.9987$\\
Orbit: $\max|\Lc/\Lc_0-1|$, $\max|{\rm NRMSE}/{\rm NRMSE}_0-1|$ & $7\times10^{-15}$, $6\times10^{-13}$ & $1.4\times10^{-14}$, $2.7\times10^{-13}$ & $\le1.6\times10^{-13}$, $\le2.4\times10^{-12}$\\
$Z^{\otimes4}$ capture, pooled over $e_J,e_h,e_\tau$ & $0.09$--$0.13$ & $0.104$ $[0.084,0.124]$ & $0.048$--$0.161$\\
\quad along $e_J$ / $e_h$ / $e_\tau$ & $0.104$ / $0.133$ / $0.085$ & $0.100$ $[0.089,0.122]$ / $0.121$ $[0.108,0.131]$ / $0.083$ $[0.073,0.106]$ & \\
Commutator norms $(J,h)$, $(J,\tau)$, $(h,\tau)$ & $0.40$, $0.35$, $0.24$ & $0.40$ $[0.40,0.41]$, $0.34$ $[0.32,0.36]$, $0.22$ $[0.22,0.23]$ & $0.37$--$0.42$, $0.26$--$0.39$, $0.17$--$0.25$\\
$J$-SLD basis along $e_h$ & $0.15$ & $0.18$ $[0.15,0.23]$ & $0.12$--$0.41$\\
$h$-SLD basis along $e_\tau$ / $e_J$ & $0.78$ / $0.15$ & $0.79$ $[0.77,0.82]$ / $0.23$ $[0.14,0.29]$ & $0.73$--$0.88$ / $0.10$--$0.43$\\
Per-time optimized basis (min over directions) & $0.86$--$0.92$ & $0.88$ $[0.86,0.92]$ (60 seed--time pairs) & $0.75$--$0.95$\\
$\bv_Q^\star$ resolved by the $J$-SLD basis & $0.93$ & $0.74$ $[0.54,0.82]$ & $0.18$--$0.98$\\
$\bv_Q^\star$ resolved by the best single-control SLD basis & $0.93$ ($J$) & $0.90$ $[0.78,0.97]$; best control $J$/$h$/$\tau$ in 10/3/7 seeds & $0.66$--$1.00$\\
Task link at $\bv_Q^\star$: $\lo=\lp$ & $0.0577$ & $0.029$ $[0.023,0.044]$ & $0.012$--$0.062$\\
Training loss & $5.41$ & $7.74$ $[6.99,9.24]$ & $5.41$--$14.2$\\
Held-out NRMSE & $0.333$ & $0.391$ $[0.359,0.425]$ & $0.333$--$0.533$\\
$\GamQ$ & $0.0712$ & $0.100$ $[0.029,0.200]$ & $0.015$--$0.38$\\
$\eta_Q(\bv_Q^\star)$ & $1.73\times10^{-3}$ & $3.0\times10^{-3}$ $[0.8,5.6]$ & $1.7\times10^{-4}$--$1.0\times10^{-2}$\\
Complex step vs.\ central difference (relative) & $7.7\times10^{-5}$ & $1.5\times10^{-4}$ $[0.8,2.5]$ & $5.6\times10^{-5}$--$3.3\times10^{-3}$\\
\bottomrule\end{tabular}}
\end{table}

\begin{table}[!ht]\centering\small
\caption{Chain quantities along the reported directions over twenty seeds, dictionary $\{Z_i\}$, Mackey--Glass (median [IQR]; seed-7 value in parentheses). $\bv_{\rm sym}$ is at the floor in every seed ($\BQ\le2\times10^{-5}$, $\lo\le2\times10^{-9}$) and is omitted.}
\label{tab:tfi-seeds-dirs}
\resizebox{\textwidth}{!}{\begin{tabular}{@{}lccccc@{}}\toprule
 & $e_J$ & $e_h$ & $e_\tau$ & $(1,-1,0)/\sqrt2$ & $\bv_Q^\star$\\\midrule
$\BQ$ & $126$ $[115,148]$ ($123$) & $204$ $[187,227]$ ($200$) & $243$ $[215,291]$ ($240$) & $164$ $[157,175]$ ($162$) & $1$\\
$\BC=\normF{X_{\bv}}$ & $16.5$ $[14.1,18.4]$ ($16.8$) & $34.2$ $[30.8,35.7]$ ($34.4$) & $31.5$ $[27.4,32.4]$ ($30.0$) & $29.1$ $[27.9,32.3]$ ($31.9$) & $0.157$ $[0.120,0.175]$ ($0.163$)\\
$\|X_{\bv}\ws\|/\|\ws\|$ & $3.29$ $[1.90,6.78]$ ($1.85$) & $3.77$ $[1.84,4.50]$ ($1.66$) & $4.82$ $[3.00,8.23]$ ($2.51$) & $3.32$ $[2.02,6.81]$ ($1.74$) & $0.019$ $[0.014,0.038]$ ($0.0148$)\\
$\lp$ & $1.12$ $[0.76,1.72]$ ($0.894$) & $1.50$ $[1.20,1.96]$ ($1.06$) & $1.96$ $[1.24,2.96]$ ($1.34$) & $1.02$ $[0.63,1.61]$ ($1.01$) & $7.8\times10^{-3}$ $[6.6,13.9]$ ($7.32\times10^{-3}$)\\
$\lo$ & $0.197$ $[0.112,0.674]$ ($0.204$) & $0.47$ $[0.07,0.95]$ ($0.079$) & $0.28$ $[0.10,1.02]$ ($0.125$) & $0.118$ $[0.075,0.269]$ ($0.200$) & $3.0\times10^{-3}$ $[0.8,5.6]$ ($1.73\times10^{-3}$)\\
$D_{\bv}\Lc$ & $1.3$ $[-9.1,5.3]$ ($-8.41$) & $2.7$ $[-2.8,10.5]$ ($3.26$) & $0.7$ $[-6.7,13.9]$ ($-5.15$) & $-0.4$ $[-4.8,3.8]$ ($-8.25$) & $-0.100$ $[-0.200,-0.029]$ ($-0.0712$)\\
\bottomrule\end{tabular}}
\end{table}

\subsection*{Negative search for a second near-redundant direction}
The symmetry direction is visible from $U=e^{-i\tau H}$. To test whether the certificate also flags a direction that is only approximately redundant, we diagonalized, for the twenty seeds and the five anchors $(1,1,1)$, $(0.5,1,1)$, $(1,0.5,1)$, $(1,1,0.5)$ and $(2,1,1)$, eight $3\times3$ QFI matrices (the marginal $\sum_t\FQ(t)$ for each input drive and $\GQ$ for each task and dictionary) on the Euclidean plane orthogonal to $\bv_{\rm sym}$, and recorded the ratio of the smaller to the larger restricted eigenvalue. Supplementary Table~\ref{tab:neardeg} summarizes the 800 entries. None has a ratio below $10^{-2}$. The minimum is $0.0136$ (seed 6, anchor $(1,1,0.5)$, NARMA10 drive; $0.0150$ for $\GQ$ of $\{Z\}$/Mackey--Glass), along $\bv_c=(0.734,-0.665,0.139)$, an antisymmetric $J$--$h$ direction with a small $\tau$ component. The ratio is independent of the dictionary (the three $\GQ$ agree to $<1\%$) and of the drive (within a few percent). The weak direction is therefore set by the Hamiltonian and the anchor, and the readout and the task do not enter. Per time step the restricted ratio of $\FQ(t)$ never falls below $4.7\times10^{-3}$ and is below $10^{-2}$ at no more than $6\%$ of the training times of any entry. The per-time weakest direction is stable in time (median $|$overlap$|$ with the summed candidate $0.998$), so no intermittently silent direction hides in the time average. $\bv_{\rm sym}$ itself is at the floor (relative marginal QFI $\le1.1\times10^{-14}$) in all 800 entries. The ratio decreases with $\tau$ ($0.20$ at $\tau=1$, $0.095$ at $\tau=0.5$), which suggests that a weak direction emerges perturbatively as the dynamics approach the identity. That regime was not scanned.

\begin{table}[!ht]\centering\small
\caption{Near-zero-QFI search: ratio of the smaller to the larger eigenvalue of the QFI matrices restricted to the plane orthogonal to $\bv_{\rm sym}$, twenty seeds $\times$ eight matrices per anchor (median [IQR]; min--max). Entries with ratio $<10^{-2}$: 0 of 800; with ratio $<10^{-1}$: 200 of 800.}
\label{tab:neardeg}
\begin{tabular}{@{}lccc@{}}\toprule
Anchor $(J,h,\tau)$ & ratio, median [IQR] & min--max & entries\\\midrule
$(0.5,1,1)$ & $0.351$ $[0.324,0.419]$ & $0.031$--$0.462$ & 160\\
$(1,0.5,1)$ & $0.395$ $[0.364,0.449]$ & $0.146$--$0.529$ & 160\\
$(1,1,0.5)$ & $0.095$ $[0.081,0.112]$ & $0.0136$--$0.123$ & 160\\
$(1,1,1)$ & $0.203$ $[0.182,0.220]$ & $0.108$--$0.272$ & 160\\
$(2,1,1)$ & $0.083$ $[0.051,0.113]$ & $0.024$--$0.181$ & 160\\\midrule
Per time step: $\min_t$ ratio & $0.088$ $[0.047,0.152]$ & $0.0047$--$0.315$ & 200 (seed, anchor, task)\\
Fraction of training times with ratio $<10^{-2}$ & $0$ $[0,0]$ & $0$--$0.061$ & 200\\
\bottomrule\end{tabular}
\end{table}

\emph{Continuity of the certificate at the best candidate.} Supplementary Table~\ref{tab:neardeg-cand} evaluates the chain along $\bv_c$ (seed 6, anchor $(1,1,0.5)$; marginal QFI $1.5\%$ of the dominant restricted eigenvalue, hence $\BQ(\bv_c)$ $12\%$ of $\BQ(\bv_{\rm dom})$) against the dominant direction $\bv_{\rm dom}$ and $\bv_{\rm sym}$. The bound holds in all six task--dictionary combinations ($|D_{\bv}\Lc|/(\|r\|\|\ws\|\BQ)$ between $1.4\times10^{-4}$ and $6.4\times10^{-3}$). Along $\bv_c$, $|D_{\bv}\Lc|$ ($6$--$21$) is of the same order as along $\bv_{\rm dom}$ ($9$--$50$) although the QFI is $66\times$ smaller, so $\eta_Q(\bv_c)$ exceeds $\eta_Q(\bv_{\rm dom})$ by $3$--$15\times$. A finite move $\bth+s\bv_c$, $|s|\le0.3$, changes the training loss by factors between $0.77$ and $1.92$ and the held-out NRMSE by up to $+31\%$, whereas the exact orbit stays flat to $10^{-13}$ and the straight $\bv_{\rm sym}$ line changes the loss only at second order ($\le+9.5\%$). The complex-step $D_{\bv_c}\Lc$ agrees with a central difference ($s=10^{-3}$) to $0.1$--$0.7\%$. The loss is strongly curved along $\bv_c$, and the linear regime is narrower than $|s|\approx0.01$. A direction retaining about one percent of the QFI budget is therefore an ordinary trainable direction. The certificate distinguishes zero QFI from finite QFI, and for finite QFI the quantitative bound of the chain applies. The $\alpha$-sweep $\bv(\alpha)=\cos\alpha\,\bv_c+\sin\alpha\,\bv_{\rm dom}$ confirms this. Nothing vanishes at $\alpha=0$ (contrast main-text Fig.~3b), the marginal QFI interpolates between the two restricted eigenvalues, and $|D_{\bv}\Lc|$ stays of order ten without being monotone in $\alpha$.

\begin{table}[!ht]\centering\small
\caption{Best near-degenerate candidate: seed 6, anchor $(1,1,0.5)$, $\bv_c=(0.734,-0.665,0.139)$ (restricted ratio $0.0150$ for $\GQ$ of $\{Z\}$/Mackey--Glass), $\bv_{\rm dom}=(0.129,0.337,0.933)$. Chain quantities along $\bv_c$ and $\bv_{\rm dom}$ for the base dictionary and both tasks, and the range of the training-loss ratio $\Lc(\bth+s\bv)/\Lc_0$ over $|s|\le0.3$ (13 points); the exact orbit changes the loss by $<10^{-13}$.}
\label{tab:neardeg-cand}
\begin{tabular}{@{}llcccccc@{}}\toprule
Task & Direction & $\sum_t\bv^\top\FQ\bv$ & $\BQ$ & $|D_{\bv}\Lc|$ & $\|r\|\|\ws\|\BQ$ & $|D_{\bv}\Lc|/$bound & loss ratio, $|s|\le0.3$\\\midrule
MG & $\bv_c$ & $1.59\times10^4$ & $240$ & $20.9$ & $7.84\times10^3$ & $2.7\times10^{-3}$ & $[0.87,1.71]$\\
 & $\bv_{\rm dom}$ & $1.07\times10^6$ & $1.96\times10^3$ & $49.9$ & $6.41\times10^4$ & $7.8\times10^{-4}$ & $[0.49,2.13]$\\
 & $\bv_{\rm sym}$ & $5.6\times10^{-9}$ & $1.5\times10^{-4}$ & $1.6\times10^{-7}$ & $5.0\times10^{-3}$ & $3.2\times10^{-5}$ & $[1.000,1.095]$\\
NARMA10 & $\bv_c$ & $7.00\times10^3$ & $159$ & $17.5$ & $2.75\times10^3$ & $6.4\times10^{-3}$ & $[0.78,1.30]$\\
 & $\bv_{\rm dom}$ & $5.03\times10^5$ & $1.34\times10^3$ & $13.5$ & $2.31\times10^4$ & $5.9\times10^{-4}$ & $[0.66,1.45]$\\
 & $\bv_{\rm sym}$ & $1.3\times10^{-9}$ & $6.3\times10^{-5}$ & $1.7\times10^{-7}$ & $1.1\times10^{-3}$ & $1.6\times10^{-4}$ & $[0.998,1.048]$\\
\bottomrule\end{tabular}
\end{table}

\section{Dependence on the number of samples}
\label{sec:tscale}
This note tests whether the small utilizations $\eta_Q$ reported in the main text depend on the length of the training window. It supports the attribution of small $\eta_Q$ to the measurement, readout and novelty links in main-text Secs.~2.2 and~2.3 and the sample-size statement in Methods.

The small utilizations in the main text are not an artefact of the training-window length. Most links are ratios of quantities with the same extensive scaling in $T$. Only the task alignment has a $T$-dependent chance value.

\begin{sprop}[Intensive link factors and the chance value of the task link]\label{prop:tscale}
Suppose the features, feature tangents, per-sample QFI and variances are stationary over the training window, so that $\BQ^2$, $\BC^2$, $\normF{X_{\bv}}^2$ and $\|X_{\bv}\ws\|^2$ are each $T$ times a per-sample mean. Then (i) $\eta_1$, $\eta_2$ and $\eta_3$ are ratios of such quantities and have $T$-independent limits; (ii) $\eta_4=\|\PXp z\|/\|z\|$ with $z=X_{\bv}\ws$ has a $T$-independent limit whenever the part of $z$ outside the fixed-dimensional span $\col X$ has a nonvanishing per-sample mean square; (iii) $\eta_5=|\langle\hat r,\hat a\rangle|$ with $\hat a=\PXp X_{\bv}\ws/\|\PXp X_{\bv}\ws\|$ and $\hat r=r/\|r\|$, both in $\col X^\perp$ of dimension $n=T-\operatorname{rank}X$. If $\hat r$ were uniformly distributed on the unit sphere of $\col X^\perp$, independent of the dynamics, then
\begin{equation}
\mathbb E\,\eta_5^2=\frac1n,\qquad \eta_5^{\rm chance}=\frac{1}{\sqrt{T-\operatorname{rank}X}},
\label{eq:s-chance}
\end{equation}
and $\eta_Q$ would decrease as $T^{-1/2}$. A residual that is coherent with the novel response instead gives a $T$-independent $\eta_5$, and $\eta_5\sqrt{T-\operatorname{rank}X}$ grows as $\sqrt T$.
\end{sprop}
\begin{proof}
(i) and (ii) follow from writing each norm as the square root of $T$ times a per-sample mean. For (iii), $r\perp\col X$ by the refit and $\PXp X_{\bv}\ws\in\col X^\perp$, so both unit vectors lie in the $n$-dimensional subspace $\col X^\perp$ and $\eta_5=\lo/\lp$ is the modulus of their inner product (Supplementary Proposition~\ref{prop:chain}). For $\hat r$ uniform on the unit sphere $S^{n-1}$ of that subspace and any fixed unit $\hat a$, symmetry gives $\mathbb E\langle\hat r,e_i\rangle^2$ equal for every element $e_i$ of an orthonormal basis, and these sum to one, so $\mathbb E\langle\hat r,\hat a\rangle^2=1/n$.
\end{proof}

The normalized quantity $\eta_5\sqrt{T-\operatorname{rank}X}$ is therefore the task alignment in units of chance. Supplementary Table~\ref{tab:tscale} evaluates the chain on nested training windows (the first $T$ samples of the training window, readout refitted on each) at the JC anchor $(30,20,10)$ and the Ising anchor $(1,1,1)$ with the Mackey--Glass task and the base dictionaries ($\operatorname{rank}X=5$ including the bias). $\BQ/\sqrt T$ along a fixed Euclidean direction varies by less than $3\%$ over $T=125$--$2000$, confirming the extensive scaling of the envelopes. For $T\ge250$, $\eta_Q$ stays within $(2.2$--$4.2)\times10^{-3}$ for the JC reservoir and $(1.5$--$1.9)\times10^{-3}$ for the Ising reservoir, with no trend in $T$. Over the same windows $\eta_5$ stays at $0.19$--$0.36$, while its chance value falls from $0.064$ to $0.022$. At $T=2000$ the measured task alignment is $10$--$15$ times chance and grows as $\sqrt T$, which identifies it as coherent. The shortest windows ($T=125$, a few quasi-periods of the Mackey--Glass series) fluctuate, including one value below chance. The smallness of $\eta_Q$ at these anchors is thus due to the measurement, readout and novelty links, as the attribution of the main text states. The number of samples does not cause it.

Equation~\eqref{eq:s-chance} is also a prediction. When the residual is dominated by variation that the dynamics cannot reach, for instance measurement noise on the target or an unpredictable forecasting horizon, $\eta_5$ falls to its chance value and $\eta_Q\propto T^{-1/2}$. The bound then reports correctly that fitting that residual is not learnable through the physical parameters. Reporting $\eta_5\sqrt{T-\operatorname{rank}X}$ alongside $\eta_5$ distinguishes the two regimes.

\begin{table}[!ht]\centering\small
\caption{Link factors on nested training windows of length $T$. JC reservoir at $(g,\epsilon,\kappa_c)=(30,20,10)$ and Ising reservoir at $(J,h,\tau)=(1,1,1)$, five-step Mackey--Glass forecasting, base dictionaries, readout refitted on each window; $\bv_Q^\star$ recomputed on each window. $\eta_5\sqrt{T-5}$ is the task alignment in units of its chance value, Eq.~\eqref{eq:s-chance}.}
\label{tab:tscale}
\begin{tabular}{@{}llccccc@{}}\toprule
Reservoir, direction & & $T=125$ & $250$ & $500$ & $1000$ & $2000$\\\midrule
JC, $e_g$ & $\eta_Q$ & $2.5\times10^{-3}$ & $3.0\times10^{-3}$ & $2.2\times10^{-3}$ & $2.4\times10^{-3}$ & $2.2\times10^{-3}$\\
 & $\eta_5$ & 0.31 & 0.30 & 0.26 & 0.29 & 0.27\\
 & $\eta_5\sqrt{T-5}$ & 3.4 & 4.8 & 5.8 & 9.0 & 12.1\\[2pt]
JC, $\bv_Q^\star$ & $\eta_Q$ & $5.5\times10^{-3}$ & $4.2\times10^{-3}$ & $2.5\times10^{-3}$ & $2.6\times10^{-3}$ & $2.3\times10^{-3}$\\
 & $\eta_5$ & 0.54 & 0.36 & 0.34 & 0.34 & 0.35\\
 & $\eta_5\sqrt{T-5}$ & 5.9 & 5.7 & 7.6 & 10.9 & 15.4\\[2pt]
Ising, $e_J$ & $\eta_Q$ & $1.9\times10^{-4}$ & $1.5\times10^{-3}$ & $1.9\times10^{-3}$ & $1.5\times10^{-3}$ & $1.7\times10^{-3}$\\
 & $\eta_5$ & 0.03 & 0.19 & 0.25 & 0.21 & 0.23\\
 & $\eta_5\sqrt{T-5}$ & 0.3 & 3.0 & 5.6 & 6.7 & 10.2\\[2pt]
Ising, $\bv_Q^\star$ & $\eta_Q$ & $1.3\times10^{-3}$ & $1.5\times10^{-3}$ & $1.9\times10^{-3}$ & $1.6\times10^{-3}$ & $1.7\times10^{-3}$\\
 & $\eta_5$ & 0.44 & 0.20 & 0.25 & 0.23 & 0.24\\
 & $\eta_5\sqrt{T-5}$ & 4.8 & 3.1 & 5.6 & 7.1 & 10.5\\[2pt]
\bottomrule\end{tabular}
\end{table}

\section{Measurement-conditioned drive: memory, stopping rule and checks}
\label{sec:feedback-si}
This note documents the memory diagnostics, the stopping rule and its checks over thirty anchors per task, and the two-atom reservoir behind the results on the measurement-conditioned drive. It supports main-text Sec.~2.5 and Fig.~5.

The reference system is the JC anchor $(30,20,10)$ with five-step Mackey--Glass forecasting and $\tilde u_t=u_t+\beta\langle p\rangle_t$, the case $V=1$, $b=0$, $\bbeta=\beta e_p$ of the drive law of main-text Eq.~(15); Supplementary Note~14 uses the same law with a gain on every feature and an offset. The readout is refitted at every gain.

\subsection*{Choice of drive channel}
Among the four base observables, $p$ gives the clearest stable improvement. Conditioning on $x$, $\sigma_x$ or $\sigma_y$ changes the held-out NRMSE by at most about $10\%$ over the stable small-gain range, while sufficiently large gains can leave the low-excitation branch. The main text therefore uses $\bbeta=\beta e_p$. The full channel-scan data are part of the deposited data.

\subsection*{Memory ratios}
\begin{figure}[!ht]
\centering
\includegraphics[width=\textwidth,height=0.6\textheight,keepaspectratio]{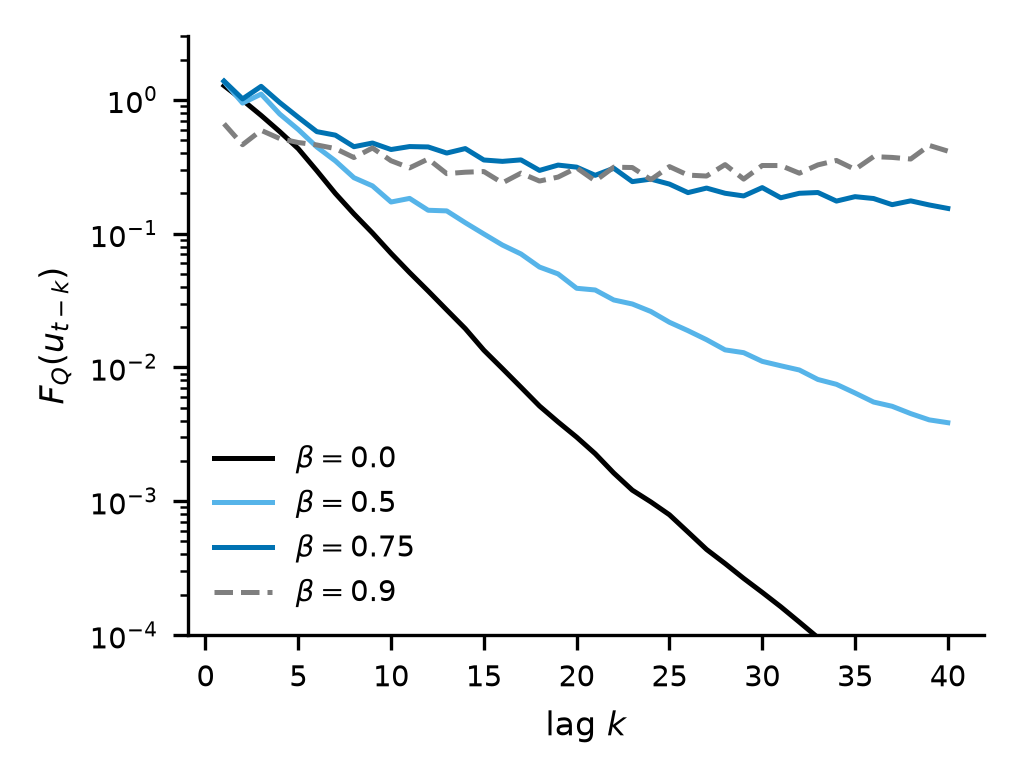}
\caption{\textbf{Input-memory QFI under the conditioned drive.} QFI with respect to the input $k$ samples earlier, averaged over five seeds. The conditioned drive substantially lengthens retention while the injection, metric and access ratios change much less. The $\beta=0.9$ curve is post-edge.}
\label{fig:feedback-memory}
\end{figure}

We decompose the input-memory QFI into four diagnostics: injection $F_Q(u_{t-1})$; retention $F_Q(u_{t-10})/F_Q(u_{t-1})$; a state-space metric ratio $F_Q/\|\partial\rho\|_{\rm HS}^2$; and measurement access $\sum_j(\partial\langle O_j\rangle)^2/F_Q$, with the last two averaged over lags 1--10. Supplementary Table~\ref{tab:feedback} and Supplementary Fig.~\ref{fig:feedback-memory} show that the conditioned drive changes retention far more strongly than the other ratios. From $\beta=0$ to $0.75$, retention grows from $0.056$ to $0.308$ ($5.5\times$), while injection, metric and access change modestly. This is the physical mechanism behind the rise of the novelty factor in the main-text attribution. The envelope derivative through the loop agrees with a centred difference of the fully refitted loss at $\beta=0.5$ to relative $5\times10^{-6}$ (step $10^{-3}$; $8\times10^{-8}$ at step $10^{-4}$).

\begin{table}[!ht]\centering\small
\caption{\textbf{Memory ratios under the conditioned drive at the JC anchor, $\bbeta=\beta e_p$.} $|\bar\lambda_1|$ is the closed-loop slow-mode diagnostic; the frozen open-loop column is retained only as an auxiliary prediction. The $\beta=0.9$ row is post-edge and not converged in cavity cutoff, so its QFI-scale quantities are order-of-magnitude values.}
\label{tab:feedback}
\begin{tabular}{@{}lcccccccccc@{}}\toprule
$\beta$ & NRMSE & loss & injection & retention & metric & access & $|\bar\lambda_1|$ & frozen gain & $\GamQ$ & $\overline{\Var_t}(O_{\ws})$\\\midrule
0    & 0.449 & 11.16 & 1.28 & 0.056 & 3.73 & 0.41 & 0.880 & 0.909 & 0.049 & 16.5\\
0.25 & 0.410 & 9.36  & 1.34 & 0.071 & 4.00 & 0.39 & 0.895 & 0.912 & 0.053 & 39.1\\
0.5  & 0.381 & 8.12  & 1.37 & 0.126 & 4.03 & 0.39 & 0.940 & 0.939 & 0.036 & 87.1\\
0.6  & 0.370 & 7.71  & 1.37 & 0.175 & 3.90 & 0.41 & --    & --    & 0.052 & 128\\
0.7  & 0.357 & 7.41  & 1.38 & 0.253 & 3.74 & 0.43 & --    & --    & 0.184 & 165\\
0.75 & 0.360 & 7.98  & 1.38 & 0.308 & 3.65 & 0.45 & 0.983 & 0.995 & 0.636 & 111\\
0.8  & 0.403 & 9.99  & 1.38 & 0.379 & 3.57 & 0.46 & 0.991 & 1.006 & 1.04  & 20.2\\
0.9  & 0.960 & 51.4  & 0.66 & 0.53  & 5.8  & 0.85 & $>1$  & 1.028 & 0.012 & 1.2\\
\bottomrule\end{tabular}
\end{table}

\subsection*{Delayed-readout check}
The benefit decreases with forecasting horizon. At $\beta=0.7$ the NRMSE gain is $41\%$ at horizon 1, $34\%$ at 2, $20\%$ at 5 and $19\%$ at 10, while horizon 20 worsens. If the dictionary is augmented with the same observables delayed by 5, 10 and 20 samples, the open-loop reservoir already reaches NRMSE $0.164$ at horizon 5 and the conditioned drive gives essentially no further gain ($0.162$). The extra retention supplied by the conditioned drive is therefore redundant once the readout is explicitly given comparable delayed information.

\subsection*{Stopping rule: slope zero, training and held-out optima, dynamical edge}
At each anchor we sweep $\beta$ from zero in the descent direction of the training loss, refit the readout at every gain, and record the training loss, held-out NRMSE, exact derivative $D_\beta\Lc$ (envelope identity, complex step through the loop), the closed-loop slow mode $|\lambda_1(t,\beta)|$ at $n_{\rm samp}$ sampled training times, and the maximum photon number. The quantities are defined in the main-text Methods: $\beta_{\rm slope}$ (zero of $D_\beta\Lc$ at the training minimum; the first zero is listed separately when it differs), $\beta^{\rm train}_{\rm opt}$ and $\beta^{\rm test}_{\rm opt}$ (parabola through the first local minimum of the training loss, respectively the held-out NRMSE, and its two neighbours), $\beta_{\rm edge}$ (first crossing of $\operatorname{median}_t|\lambda_1|=1$), $\beta^{\max}_{\rm edge}$ (with $\max_t$), and $\beta_{\rm nc}$ (maximum photon number twice its $\beta=0$ value). The main-text edge uses the closed-loop diagnostic. The frozen open-loop tangent estimate of Supplementary Table~\ref{tab:feedback} is not the definition of $\beta_{\rm edge}$. Two facts are tautological and serve only as checks of the grid resolution. First, $\beta_{\rm slope}=\beta^{\rm train}_{\rm opt}$ to first order (difference $0.003$ median, $0.006$ maximum over the thirty Mackey--Glass anchors). Second, $\eta_5\to0$ at the zero of $D_\beta\Lc$, because by Eq.~\eqref{eq:s-envelope} $D_\beta\Lc=0$ is $\lo=0$. The non-tautological content is (i) the held-out optimum, (ii) the ordering $\beta^{\rm test}_{\rm opt}<\beta_{\rm edge}$ and (iii) the behaviour of the other links at the zero, where $\BQ$ and $\lp$ stay finite while $\eta_5$ closes. The held-out optimum is computed on the same held-out set that reports the final NRMSE (there is no third split) and is used only as a check on the training-based rule, never for selection.

\emph{Mackey--Glass, thirty one-atom anchors} (Supplementary Table~\ref{tab:stopping-mg}; $n_{\rm samp}=6$, gains $0$--$1$ in steps of $0.025$, then $1.1$--$2.0$). All thirty training and held-out curves are unimodal in the stable range. $\beta_{\rm slope}-\beta^{\rm test}_{\rm opt}$ has median $-0.035$, IQR $[-0.038,-0.032]$ and range $[-0.054,-0.016]$. The offset is small, systematic and independent of the anchor, and $|\beta_{\rm slope}-\beta^{\rm test}_{\rm opt}|\le0.05$ at 29 of 30 anchors. The held-out NRMSE at $\beta_{\rm slope}$ exceeds the best held-out value in the converged range by $0.36\%$ (median regret; IQR $[0.31,0.47]\%$, maximum $0.85\%$). The held-out improvement over $\beta=0$ is $10.1\%$ at $\beta_{\rm slope}$ (median; range $1.5$--$20.1\%$) against $10.6\%$ at the held-out optimum. $\beta^{\rm test}_{\rm opt}<\beta_{\rm edge}$ at 30 of 30 anchors (margin $0.155$ $[0.119,0.177]$, minimum $0.091$). With the stricter $\max_t|\lambda_1|$ criterion, $\beta_{\rm slope}$ still precedes the edge at 26 of 30 anchors (margin $0.095$ median, minimum $-0.018$). At $\beta_{\rm slope}$, $\operatorname{median}_t|\lambda_1|=0.953$ $[0.950,0.956]$ at every anchor. Along $e_\beta$, $\eta_5$ falls from $0.396$ (median at $\beta=0$) to $0.020$ at the gain nearest $\beta_{\rm slope}$ (ratio $0.053$ $[0.023,0.079]$), while $\BQ$ falls only from $82.5$ to $48.4$ (ratio $0.61$ $[0.53,0.65]$) and $\lp$ rises (ratio $2.1$). Seven anchors have more than one zero of $D_\beta\Lc$ before the edge. At all thirty the first zero coincides with the zero at the global pre-collapse training minimum. At the reference anchor the separate scan of main-text Fig.~5a--e ($n_{\rm samp}=10$, gain steps $0.1$, then $0.05$ above $0.6$) gives $\beta_{\rm slope}=0.684$, $\beta^{\rm train}_{\rm opt}=0.682$ and $\beta_{\rm edge}=0.803$, whereas this table ($n_{\rm samp}=6$, steps $0.025$) gives $0.689$, $0.689$ and $0.825$, with $\beta^{\rm test}_{\rm opt}=0.719$. $D_\beta\Lc$ and the training loss are identical in the two scans at common gains, so the differences in $\beta_{\rm slope}$ and $\beta^{\rm train}_{\rm opt}$ come from the gain spacing; added gains at $0.69$, $0.83$ and $0.84$ ($n_{\rm samp}=10$) place them at $0.691$, $0.689$ and $\beta_{\rm edge}$ at $0.833$.

\begin{table}[!ht]\centering\scriptsize
\caption{\textbf{Drive-gain stopping rule, Mackey--Glass, thirty one-atom anchors} ($n_{\rm samp}=6$). Gains: $\beta_{\rm slope}$, $\beta^{\rm train}_{\rm opt}$, $\beta^{\rm test}_{\rm opt}$, $\beta_{\rm edge}$, $\beta^{\max}_{\rm edge}$, $\beta_{\rm nc}$; held-out NRMSE at $\beta=0$, at $\beta_{\rm slope}$ (linear interpolation) and at $\beta^{\rm test}_{\rm opt}$ (parabola value); regret of $\beta_{\rm slope}$ relative to the best held-out value before $\min(\beta_{\rm edge},\beta_{\rm nc})$; $\eta_5$ and $\BQ$ along $e_\beta$ at $\beta=0$ and at the grid gain nearest $\beta_{\rm slope}$. The four anchors off the $5\times5$ sub-grid, $(12,15,11)$, $(15,15,11)$, $(18,15,11)$ and $(18,15,19)$, and the reference anchor $(30,20,10)$ complete the set of thirty.}
\label{tab:stopping-mg}
\resizebox{\textwidth}{!}{\begin{tabular}{@{}rrrcccccccccccccc@{}}\toprule
$g$ & $\epsilon$ & $\kappa_c$ & $\beta_{\rm slope}$ & $\beta^{\rm train}_{\rm opt}$ & $\beta^{\rm test}_{\rm opt}$ & $\beta_{\rm edge}$ & $\beta^{\max}_{\rm edge}$ & $\beta_{\rm nc}$ & NRMSE$(0)$ & NRMSE$(\beta_{\rm slope})$ & NRMSE$(\beta^{\rm test}_{\rm opt})$ & regret (\%) & $\eta_5(0)$ & $\eta_5(\beta_{\rm slope})$ & $\BQ(0)$ & $\BQ(\beta_{\rm slope})$\\\midrule
10 & 15 & 10 & 0.142 & 0.141 & 0.172 & 0.329 & 0.326 & 0.375 & 0.3075 & 0.3029 & 0.3023 & 0.18 & 0.285 & 0.011 & 64.8 & 51.9\\
10 & 15 & 12 & 0.197 & 0.194 & 0.238 & 0.402 & 0.390 & 0.425 & 0.3038 & 0.2974 & 0.2964 & 0.19 & 0.303 & 0.004 & 64.4 & 48.1\\
10 & 15 & 15 & 0.282 & 0.283 & 0.326 & 0.499 & 0.486 & 0.525 & 0.2990 & 0.2898 & 0.2889 & 0.31 & 0.317 & 0.026 & 63.6 & 43.1\\
10 & 15 & 18 & 0.378 & 0.375 & 0.417 & 0.841 & 0.594 & 0.625 & 0.2952 & 0.2827 & 0.2816 & 0.34 & 0.326 & 0.014 & 62.7 & 37.6\\
10 & 15 & 20 & 0.432 & 0.432 & 0.486 & 0.677 & 0.655 & 0.675 & 0.2933 & 0.2787 & 0.2776 & 0.36 & 0.332 & 0.026 & 62.0 & 35.1\\
12 & 15 & 10 & 0.177 & 0.173 & 0.214 & 0.361 & 0.342 & 0.425 & 0.3079 & 0.3006 & 0.2994 & 0.32 & 0.305 & 0.008 & 71.8 & 55.7\\
12 & 15 & 11 & 0.206 & 0.206 & 0.243 & 0.400 & 0.373 & 0.450 & 0.3059 & 0.2972 & 0.2960 & 0.36 & 0.316 & 0.024 & 71.5 & 53.6\\
12 & 15 & 12 & 0.238 & 0.239 & 0.271 & 0.429 & 0.406 & 0.475 & 0.3039 & 0.2939 & 0.2929 & 0.33 & 0.323 & 0.021 & 71.2 & 50.0\\
12 & 15 & 15 & 0.322 & 0.319 & 0.363 & 0.696 & 0.508 & 0.575 & 0.2992 & 0.2856 & 0.2842 & 0.33 & 0.335 & 0.005 & 70.2 & 44.7\\
12 & 15 & 18 & 0.407 & 0.408 & 0.448 & 0.626 & 0.606 & 0.650 & 0.2960 & 0.2790 & 0.2780 & 0.36 & 0.343 & 0.028 & 69.0 & 40.2\\
12 & 15 & 20 & 0.471 & 0.468 & 0.513 & 0.701 & 0.678 & 0.725 & 0.2949 & 0.2751 & 0.2737 & 0.29 & 0.348 & 0.006 & 68.1 & 36.4\\
15 & 15 & 10 & 0.367 & 0.365 & 0.388 & 0.506 & 0.367 & 0.550 & 0.3165 & 0.2916 & 0.2888 & 0.18 & 0.393 & 0.018 & 83.1 & 50.9\\
15 & 15 & 11 & 0.376 & 0.370 & 0.392 & 0.510 & 0.399 & 0.550 & 0.3148 & 0.2889 & 0.2879 & 0.23 & 0.399 & 0.017 & 82.7 & 50.6\\
15 & 15 & 12 & 0.383 & 0.385 & 0.413 & 0.534 & 0.431 & 0.575 & 0.3134 & 0.2866 & 0.2840 & 0.23 & 0.402 & 0.066 & 82.3 & 50.3\\
15 & 15 & 15 & 0.432 & 0.433 & 0.463 & 0.591 & 0.531 & 0.650 & 0.3103 & 0.2805 & 0.2786 & 0.34 & 0.399 & 0.047 & 80.8 & 46.6\\
15 & 15 & 18 & 0.498 & 0.493 & 0.528 & 0.677 & 0.634 & 0.725 & 0.3091 & 0.2760 & 0.2752 & 0.31 & 0.389 & 0.003 & 79.0 & 41.9\\
15 & 15 & 20 & 0.545 & 0.542 & 0.582 & 0.749 & 0.705 & 0.800 & 0.3091 & 0.2739 & 0.2728 & 0.37 & 0.382 & 0.011 & 77.8 & 39.0\\
18 & 15 & 10 & 0.378 & 0.374 & 0.411 & 0.501 & 0.377 & 0.650 & 0.3366 & 0.2891 & 0.2873 & 0.55 & 0.480 & 0.034 & 92.9 & 60.2\\
18 & 15 & 11 & 0.403 & 0.400 & 0.437 & 0.533 & 0.411 & 0.675 & 0.3358 & 0.2874 & 0.2858 & 0.51 & 0.481 & 0.038 & 92.3 & 58.2\\
18 & 15 & 12 & 0.428 & 0.425 & 0.463 & 0.564 & 0.446 & 0.675 & 0.3352 & 0.2862 & 0.2823 & 0.48 & 0.480 & 0.037 & 91.8 & 56.2\\
18 & 15 & 15 & 0.502 & 0.497 & 0.533 & 0.727 & 0.550 & 0.750 & 0.3341 & 0.2841 & 0.2828 & 0.44 & 0.462 & 0.019 & 89.9 & 50.6\\
18 & 15 & 18 & 0.575 & 0.570 & 0.607 & 0.835 & 0.656 & 0.825 & 0.3342 & 0.2833 & 0.2820 & 0.43 & 0.437 & 0.004 & 87.6 & 45.4\\
18 & 15 & 19 & 0.600 & 0.595 & 0.635 & 0.810 & 0.692 & 0.850 & 0.3344 & 0.2831 & 0.2818 & 0.42 & 0.430 & 0.005 & 86.7 & 43.8\\
18 & 15 & 20 & 0.626 & 0.622 & 0.663 & 0.817 & 0.729 & 0.875 & 0.3347 & 0.2829 & 0.2802 & 0.45 & 0.423 & 0.010 & 85.9 & 42.2\\
20 & 15 & 10 & 0.406 & 0.405 & 0.441 & 0.537 & 0.388 & 0.700 & 0.3498 & 0.2973 & 0.2942 & 0.85 & 0.490 & 0.051 & 97.0 & 64.9\\
20 & 15 & 12 & 0.464 & 0.465 & 0.497 & 0.605 & 0.458 & 0.750 & 0.3489 & 0.2963 & 0.2943 & 0.68 & 0.503 & 0.037 & 95.6 & 59.2\\
20 & 15 & 15 & 0.550 & 0.545 & 0.585 & 0.717 & 0.564 & 0.825 & 0.3487 & 0.2966 & 0.2952 & 0.43 & 0.488 & 0.005 & 93.4 & 53.3\\
20 & 15 & 18 & 0.628 & 0.626 & 0.664 & 0.810 & 0.672 & 0.900 & 0.3498 & 0.2975 & 0.2949 & 0.53 & 0.459 & 0.028 & 90.9 & 47.7\\
20 & 15 & 20 & 0.681 & 0.681 & 0.716 & 0.974 & 0.745 & 0.950 & 0.3509 & 0.2976 & 0.2955 & 0.57 & 0.444 & 0.043 & 89.0 & 44.3\\
30 & 20 & 10 & 0.689 & 0.689 & 0.719 & 0.825 & 0.716 & 0.850 & 0.4487 & 0.3587 & 0.3560 & 0.73 & 0.323 & 0.032 & 186.6 & 48.6\\
\midrule
\multicolumn{3}{l}{median} & 0.418 & 0.417 & 0.455 & 0.615 & 0.519 & 0.675 & 0.3141 & 0.2881 & 0.2866 & 0.36 & 0.396 & 0.020 & 82.5 & 48.4\\
\bottomrule\end{tabular}}
\end{table}

\begin{figure}[!ht]
\centering
\includegraphics[width=\textwidth,height=0.6\textheight,keepaspectratio]{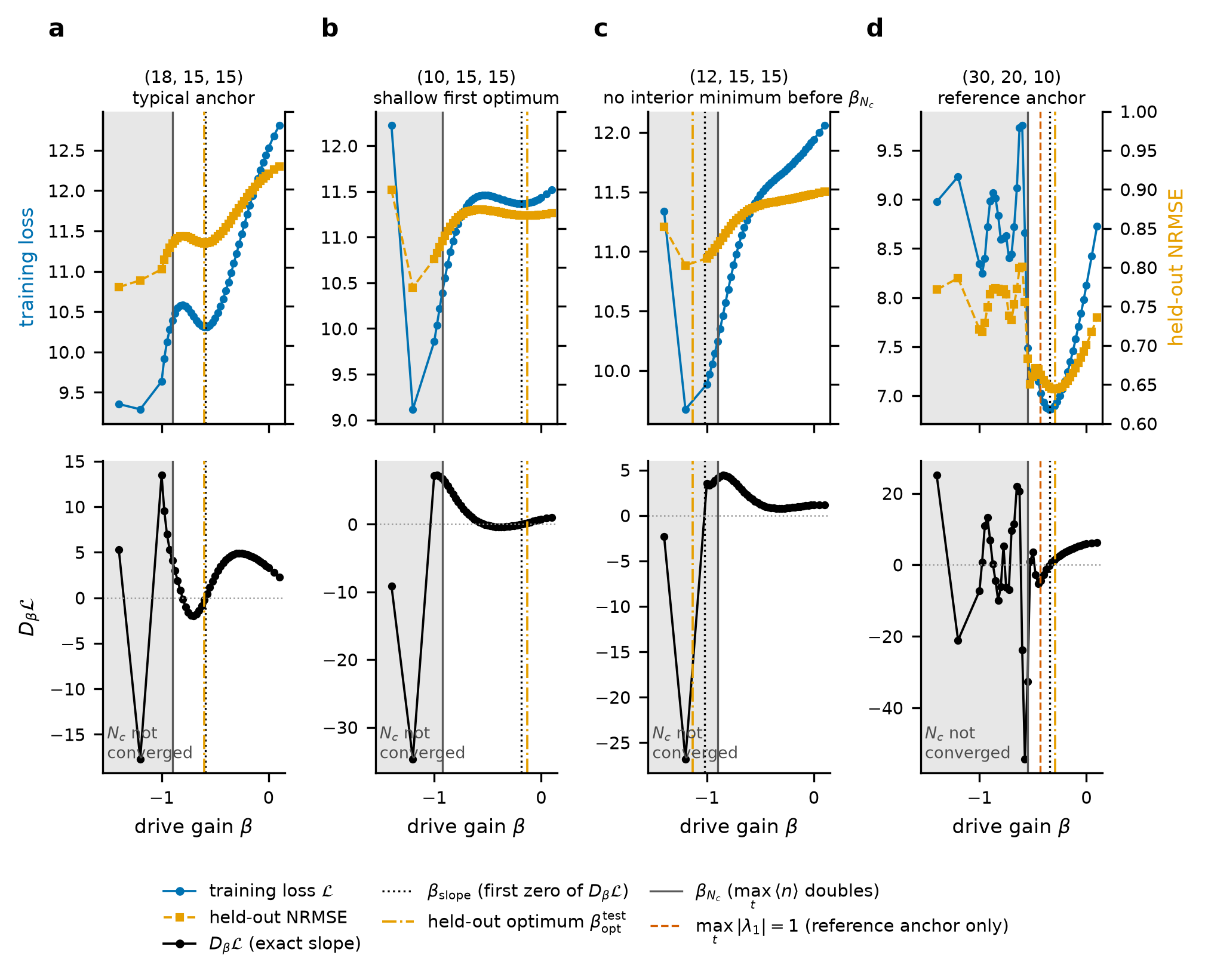}
\caption{\textbf{Conditioned drive on NARMA10: a damped loop whose validity is photon-number limited.} One-atom JC reservoir, $\tilde u_t=u_t+\beta\langle p\rangle_t$, descent direction $\beta<0$, at representative anchors of Supplementary Table~\ref{tab:stopping-narma}. Held-out NRMSE (top) and exact slope $D_\beta\Lc$ (bottom) against the gain, with $\beta_{\rm slope}$ (zero of the slope at the training minimum), the held-out optimum $\beta^{\rm test}_{\rm opt}$ and the $N_c$-convergence marker $\beta_{\rm nc}$ indicated; beyond $\beta_{\rm nc}$ the maximum photon number exceeds twice its $\beta=0$ value, the $N_c=10$ truncation is not converged and the curves are shown for orientation only. $\operatorname{median}_t|\lambda_1|$ stays below one along the whole descent at every anchor. The held-out set is used only for the marked optimum, never for selection.}
\label{fig:narma-feedback}
\end{figure}

\emph{NARMA10, thirty one-atom anchors} (Supplementary Table~\ref{tab:stopping-narma}, Supplementary Fig.~\ref{fig:narma-feedback}; gains $+0.1$, $+0.05$, $0$, $-0.025$ to $-1.0$, then $-1.2$ to $-3.0$). $D_\beta\Lc(0)>0$ at every anchor, so the descent direction is $\beta<0$. Along it the closed loop is damped. $\operatorname{median}_t|\lambda_1|$ decreases from $0.88$ to about $0.80$ and never crosses one at any anchor ($\max_t|\lambda_1|$ crosses only at $(30,20,10)$, $\beta=-0.43$), so there is no edge. Validity is limited instead by photon-number growth. $\beta_{\rm nc}=-0.90$ (median; range $-0.95$ to $-0.55$), and beyond it the maximum photon number reaches $4$--$6$, $N_c=10$ is not converged and the landscape becomes rugged. Twenty-nine of the thirty global grid minima of the held-out NRMSE lie in that region and are not quoted. At the 21 anchors whose $\beta_{\rm slope}$ lies in the converged range, it is within $0.05$ of the held-out optimum at 15 (median $|\Delta|$ $0.011$, IQR $[0.006,0.052]$), and the held-out improvement at $\beta_{\rm slope}$ is $7.8\%$ (median; $0.1$--$10.5\%$) against $8.1\%$ for the best converged grid gain (regret median $0.42\%$, IQR $[0.02,1.68]\%$). At the other nine anchors $\beta_{\rm slope}$ lies beyond $\beta_{\rm nc}$, where the NRMSE is not converged. Two sub-families deviate. At $g=10$ the first held-out optimum is very shallow ($<1\%$ gain) and $\beta_{\rm slope}$ overshoots it by $0.04$--$0.07$ while the held-out NRMSE keeps improving slowly toward $\beta_{\rm nc}$ ($2$--$4\%$ regret). At $g=12$, and at $g=15$ with $\kappa_c\le12$, the training loss has no interior minimum before $\beta_{\rm nc}$ ($\beta_{\rm slope}=-1.02$ to $-1.21$ lies in the unconverged region; 9 of 30 anchors). The chain diagnosis is the same as for Mackey--Glass with the opposite trend of the budget. $\BQ$ grows along the descent (ratio $2.8$ $[2.0,3.9]$ at $\beta_{\rm slope}$ over the 21 converged anchors; $3.7$ over all thirty) while $\eta_5$ falls to $0.011$ ($0.015$ over all thirty).

\begin{table}[!ht]\centering\scriptsize
\caption{\textbf{Drive-gain stopping rule, NARMA10, thirty one-atom anchors} ($n_{\rm samp}=6$; descent direction $\beta<0$; no anchor reaches $\operatorname{median}_t|\lambda_1|=1$). $\beta_{\rm slope}$ is the first zero of $D_\beta\Lc$ and $\beta^{\rm tm}_{\rm slope}$ the zero at the global training minimum of the converged range (``--'': the global training minimum of the converged range lies at its boundary); $n_0$ is the number of zeros before $\beta_{\rm nc}$; NRMSE$^{\rm conv}_{\min}$ is the best held-out value before $\beta_{\rm nc}$. Rows with $\beta_{\rm slope}$ beyond $\beta_{\rm nc}$ (more negative) are in the unconverged region and their NRMSE values are order-of-magnitude only.}
\label{tab:stopping-narma}
\resizebox{\textwidth}{!}{\begin{tabular}{@{}rrrccccccccccccc@{}}\toprule
$g$ & $\epsilon$ & $\kappa_c$ & $\beta_{\rm slope}$ & $\beta^{\rm tm}_{\rm slope}$ & $\beta^{\rm test}_{\rm opt}$ & $\beta_{\rm nc}$ & $n_0$ & NRMSE$(0)$ & NRMSE$(\beta_{\rm slope})$ & NRMSE$^{\rm conv}_{\min}$ & regret (\%) & $\eta_5(0)$ & $\eta_5(\beta_{\rm slope})$ & $\BQ(0)$ & $\BQ(\beta_{\rm slope})$\\\midrule
10 & 15 & 10 & -0.221 & -- & -0.178 & -0.950 & 2 & 0.8818 & 0.8784 & 0.8569 & 2.50 & 0.083 & 0.002 & 90.7 & 53.1\\
10 & 15 & 12 & -0.198 & -- & -0.150 & -0.925 & 2 & 0.8751 & 0.8732 & 0.8538 & 2.28 & 0.073 & 0.001 & 70.8 & 48.1\\
10 & 15 & 15 & -0.183 & -- & -0.127 & -0.925 & 2 & 0.8675 & 0.8666 & 0.8405 & 3.11 & 0.065 & 0.003 & 52.7 & 41.7\\
10 & 15 & 18 & -0.191 & -- & -0.126 & -0.900 & 2 & 0.8615 & 0.8609 & 0.8349 & 3.11 & 0.065 & 0.003 & 41.9 & 36.7\\
10 & 15 & 20 & -0.212 & -- & -0.137 & -0.900 & 2 & 0.8579 & 0.8573 & 0.8280 & 3.54 & 0.068 & 0.003 & 36.9 & 34.4\\
12 & 15 & 10 & -1.030 & -- & -1.131 & -0.925 & 0 & 0.9141 & 0.8345 & 0.8549 & -2.39 & 0.104 & 0.449 & 86.6 & 536.4\\
12 & 15 & 11 & -1.029 & -- & -1.130 & -0.925 & 0 & 0.9097 & 0.8298 & 0.8494 & -2.32 & 0.105 & 0.441 & 76.0 & 531.8\\
12 & 15 & 12 & -1.027 & -- & -1.130 & -0.925 & 0 & 0.9055 & 0.8250 & 0.8441 & -2.26 & 0.106 & 0.433 & 67.6 & 526.8\\
12 & 15 & 15 & -1.023 & -- & -1.131 & -0.900 & 0 & 0.8941 & 0.8107 & 0.8336 & -2.75 & 0.112 & 0.412 & 50.7 & 509.2\\
12 & 15 & 18 & -1.020 & -- & -1.137 & -0.900 & 0 & 0.8840 & 0.7964 & 0.8201 & -2.89 & 0.122 & 0.408 & 40.9 & 487.3\\
12 & 15 & 20 & -1.019 & -- & -1.143 & -0.900 & 0 & 0.8778 & 0.7870 & 0.8121 & -3.08 & 0.131 & 0.419 & 36.4 & 470.3\\
15 & 15 & 10 & -1.206 & -- & -1.190 & -0.900 & 0 & 0.9537 & 0.7846 & 0.8581 & -8.56 & 0.199 & 0.004 & 75.6 & 649.3\\
15 & 15 & 11 & -1.158 & -- & -1.194 & -0.900 & 0 & 0.9484 & 0.7918 & 0.8532 & -7.20 & 0.209 & 0.030 & 68.1 & 659.4\\
15 & 15 & 12 & -1.127 & -- & -1.197 & -0.900 & 0 & 0.9429 & 0.7974 & 0.8483 & -5.99 & 0.220 & 0.066 & 62.0 & 671.4\\
15 & 15 & 15 & -0.815 & -0.815 & -1.204 & -0.900 & 2 & 0.9266 & 0.8334 & 0.8327 & 0.09 & 0.251 & 0.016 & 49.2 & 283.0\\
15 & 15 & 18 & -0.830 & -0.830 & -1.200 & -0.900 & 1 & 0.9106 & 0.8167 & 0.8163 & 0.04 & 0.279 & 0.011 & 41.3 & 270.9\\
15 & 15 & 20 & -0.847 & -0.847 & -0.884 & -0.900 & 1 & 0.9005 & 0.8057 & 0.8055 & 0.03 & 0.295 & 0.008 & 37.6 & 285.9\\
18 & 15 & 10 & -0.589 & -- & -0.588 & -0.900 & 2 & 0.9518 & 0.8621 & 0.8488 & 1.57 & 0.377 & 0.014 & 72.3 & 187.3\\
18 & 15 & 11 & -0.583 & -- & -0.587 & -0.900 & 2 & 0.9456 & 0.8558 & 0.8462 & 1.14 & 0.399 & 0.013 & 66.7 & 174.8\\
18 & 15 & 12 & -0.582 & -- & -0.587 & -0.900 & 2 & 0.9393 & 0.8494 & 0.8435 & 0.71 & 0.417 & 0.012 & 62.1 & 172.5\\
18 & 15 & 15 & -0.591 & -0.591 & -0.601 & -0.900 & 2 & 0.9203 & 0.8309 & 0.8307 & 0.02 & 0.461 & 0.022 & 52.2 & 175.1\\
18 & 15 & 18 & -0.615 & -0.615 & -0.625 & -0.900 & 2 & 0.9021 & 0.8133 & 0.8132 & 0.02 & 0.487 & 0.031 & 45.7 & 176.1\\
18 & 15 & 19 & -0.625 & -0.625 & -0.636 & -0.900 & 2 & 0.8963 & 0.8077 & 0.8077 & 0.00 & 0.493 & 0.001 & 43.9 & 172.3\\
18 & 15 & 20 & -0.636 & -0.636 & -0.647 & -0.900 & 2 & 0.8908 & 0.8024 & 0.8023 & 0.02 & 0.498 & 0.035 & 42.3 & 168.2\\
20 & 15 & 10 & -0.448 & -- & -0.442 & -0.900 & 2 & 0.9096 & 0.8514 & 0.8373 & 1.68 & 0.405 & 0.003 & 75.5 & 154.6\\
20 & 15 & 12 & -0.445 & -- & -0.446 & -0.900 & 2 & 0.8970 & 0.8367 & 0.8320 & 0.56 & 0.458 & 0.009 & 67.2 & 150.0\\
20 & 15 & 15 & -0.462 & -0.462 & -0.467 & -0.900 & 2 & 0.8797 & 0.8167 & 0.8166 & 0.01 & 0.509 & 0.030 & 58.4 & 141.2\\
20 & 15 & 18 & -0.493 & -0.493 & -0.500 & -0.875 & 2 & 0.8644 & 0.7991 & 0.7990 & 0.01 & 0.537 & 0.018 & 51.8 & 146.0\\
20 & 15 & 20 & -0.520 & -0.520 & -0.528 & -0.875 & 2 & 0.8553 & 0.7886 & 0.7885 & 0.01 & 0.547 & 0.015 & 48.0 & 146.2\\
30 & 20 & 10 & -0.344 & -0.344 & -0.292 & -0.550 & 3 & 0.7004 & 0.6464 & 0.6437 & 0.42 & 0.351 & 0.008 & 49.4 & 172.6\\
\midrule
\multicolumn{3}{l}{median} & -0.603 & -0.615 & -0.613 & -0.900 & -- & 0.8987 & 0.8208 & 0.8331 & 0.02 & 0.265 & 0.015 & 52.5 & 174.9\\
\bottomrule\end{tabular}}
\end{table}

\emph{Two-atom reservoir} (reference anchor $(30,20,10)$, distinct couplings $g_i=g\,(1,\tfrac12)$, six features, channel $\langle p\rangle$; Supplementary Table~\ref{tab:stopping-two}, main-text Fig.~5g). The closed-loop tangent map has Liouville dimension $44^2$. Its stationary direction is obtained by a linear solve and checked against the eigenvector at one time ($|\lambda_1|$ identical to $6\times10^{-15}$), and the $\beta=0.5$ point reproduces a separate two-atom scan (loss $5.6717$, NRMSE $0.3245$). For Mackey--Glass the first zero of $D_\beta\Lc$, $\beta=0.249$, is a shallow local minimum of the training loss ($5.886$, rising to $5.940$ at $\beta=0.35$ and then descending to $4.973$ at $0.80$). The held-out NRMSE decreases monotonically through this plateau to its optimum $\beta^{\rm test}_{\rm opt}=0.783$ (NRMSE $0.299$, $-21.7\%$ relative to $\beta=0$), which coincides with the zero at the global pre-collapse training minimum, $\beta=0.779$ ($\Delta=-0.004$). Both precede the median-criterion edge ($0.967$) and the $N_c$ marker ($0.95$), whereas $\beta^{\max}_{\rm edge}=0.574$ lies before the optimum. A rule that stops at the first zero would forfeit $13\%$ in held-out NRMSE ($0.340$ against $0.300$). The main-text rule is therefore phrased as the zero of $D_\beta\Lc$ at the training minimum. For NARMA10 the two-atom loop behaves like the one-atom case, with $\beta_{\rm slope}=-0.109$, $\beta^{\rm train}_{\rm opt}=-0.107$ and $\beta^{\rm test}_{\rm opt}=-0.097$ (held-out improvement $1.4\%$, regret $0.1\%$), no $|\lambda_1|$ crossing and $\beta_{\rm nc}=-0.55$.

\begin{table}[!ht]\centering\small
\caption{\textbf{Two-atom reservoir at $(30,20,10)$.} First zero of $D_\beta\Lc$, zero at the training minimum, training and held-out optima, edges and $N_c$ marker; held-out NRMSE at $\beta=0$, at the first zero, at the zero at the training minimum and at the held-out optimum; regret of the first zero.}
\label{tab:stopping-two}
\resizebox{\textwidth}{!}{\begin{tabular}{@{}lccccccccccccc@{}}\toprule
Task & first zero & zero at train.\ min. & $\beta^{\rm train}_{\rm opt}$ & $\beta^{\rm test}_{\rm opt}$ & $\beta_{\rm edge}$ & $\beta^{\max}_{\rm edge}$ & $\beta_{\rm nc}$ & NRMSE$(0)$ & NRMSE(first zero) & NRMSE(zero at min.) & NRMSE$(\beta^{\rm test}_{\rm opt})$ & regret of first zero (\%)\\\midrule
Mackey--Glass & 0.249 & 0.779 & 0.259 & 0.783 & 0.967 & 0.574 & 0.950 & 0.3818 & 0.3400 & 0.3020 & 0.2990 & 13.22\\
NARMA10 & -0.109 & -0.109 & -0.107 & -0.097 & -- & -- & -0.550 & 0.5688 & 0.5607 & 0.5607 & 0.5601 & 0.11\\
\bottomrule\end{tabular}}
\end{table}

\subsection*{Geometry of the task link: why a larger gain stops helping}
The task factor has a geometric reading that explains why the conditioned drive first helps and then stops helping. Over the $T$ training samples the residual $r$ is the part of the target that the refitted readout still misses, and
\begin{equation}
p_\beta\equiv\PXp X_{e_\beta}\ws
\label{eq:s-pbeta}
\end{equation}
is the novel response, the change of the deployed output per unit gain that a refit of the readout cannot reproduce. Both are vectors in $\mathbb{R}^T$. We define the task angle by
\begin{equation}
\cos\theta_{\rm task}(\beta)=\frac{r^\top p_\beta}{\|r\|\,\|p_\beta\|},\qquad
D_\beta\Lc=-\|r\|\,\|\ws\|\,\lp\,\cos\theta_{\rm task},\qquad
\eta_5=|\cos\theta_{\rm task}|.
\label{eq:s-taskangle}
\end{equation}
The second identity is Eq.~\eqref{eq:s-envelope} with $r^\top X_{e_\beta}\ws=r^\top p_\beta$, which holds because $r\perp\col X$ up to the $10^{-10}$ ridge. Equation~\eqref{eq:s-taskangle} splits the slope into two factors. $\lp$ measures how much new motion the gain produces. $\cos\theta_{\rm task}$ measures how well that motion points along what the task still needs. Raising the gain lowers the loss while $\theta_{\rm task}<90^\circ$, has no first-order effect at $90^\circ$ and raises the loss beyond. ``The task link closes'' therefore means that the novel response has become perpendicular to the residual. For NARMA10 the angle is measured along the descent direction $-e_\beta$, which replaces $p_\beta$ by $-p_\beta$.

Only $\cos\theta_{\rm task}$ can change sign, so $\theta_{\rm task}=90^\circ$ at every interior training optimum. That crossing restates $D_\beta\Lc=0$ and is not evidence. The diagnosis lies in how the angle gets there. Two different mechanisms give the same zero. In the first, the gain stops producing new motion, $\lp\to0$, and the budget is used up. In the second, the motion continues and even grows, but in directions orthogonal to the residual. In the second case the angle can close in two ways. The response can turn away from the task, because the drive produces a different kind of motion. Or the residual can turn: as the gain grows, the refitted readout absorbs the part of the target that the drive has made reachable, and what remains of the target no longer lies along the response.

Because both vectors move, how the closing divides between them depends on the frame, and we report two frames. Write $\hat r=r/\|r\|$ and $\hat p=p_\beta/\|p_\beta\|$. The \emph{path split} accumulates, between consecutive grid gains $k$ and $k+1$, the exact identity
\begin{equation}
\cos\theta_{k+1}-\cos\theta_k=\tfrac12(\hat r_k+\hat r_{k+1})^\top(\hat p_{k+1}-\hat p_k)+\tfrac12(\hat p_k+\hat p_{k+1})^\top(\hat r_{k+1}-\hat r_k).
\label{eq:s-turning}
\end{equation}
Its first term is the turning of the response and its second the turning of the residual. The sums reproduce the change of $\cos\theta_{\rm task}$ to $2\times10^{-16}$. The \emph{frozen-residual split} holds the residual at its $\beta=0$ value $\hat r_0$. It credits the response with $\cos\theta_{\rm task}(0)-\hat r_0^\top\hat p(\beta_{\rm slope})$ and the residual with the rest. Shares below are fractions of the total closing $\cos\theta_{\rm task}(0)$. They can be negative when one vector turns toward the other while the other closes the angle.

All quantities use the grids and conventions of Supplementary Tables~\ref{tab:stopping-mg} and~\ref{tab:stopping-narma} (refit at every gain, complex step through the loop). They reproduce the stored $\eta_5$, $\lp$ and $D_\beta\Lc$ at all 1692 overlapping gains to relative $4\times10^{-7}$. For Mackey--Glass the scan is extended to $\beta_{\rm edge}$, which at $(10,15,18)$ and $(12,15,15)$ lies beyond $\beta_{\rm nc}$. For NARMA10 it ends at the last grid gain within $0.15$ beyond $\beta_{\rm slope}$. Values at $\beta_{\rm slope}$ are linear interpolations, and $\beta_{\rm slope}$ is the value listed in the tables.

\begin{figure}[!ht]
\centering
\includegraphics[width=\textwidth,height=0.6\textheight,keepaspectratio]{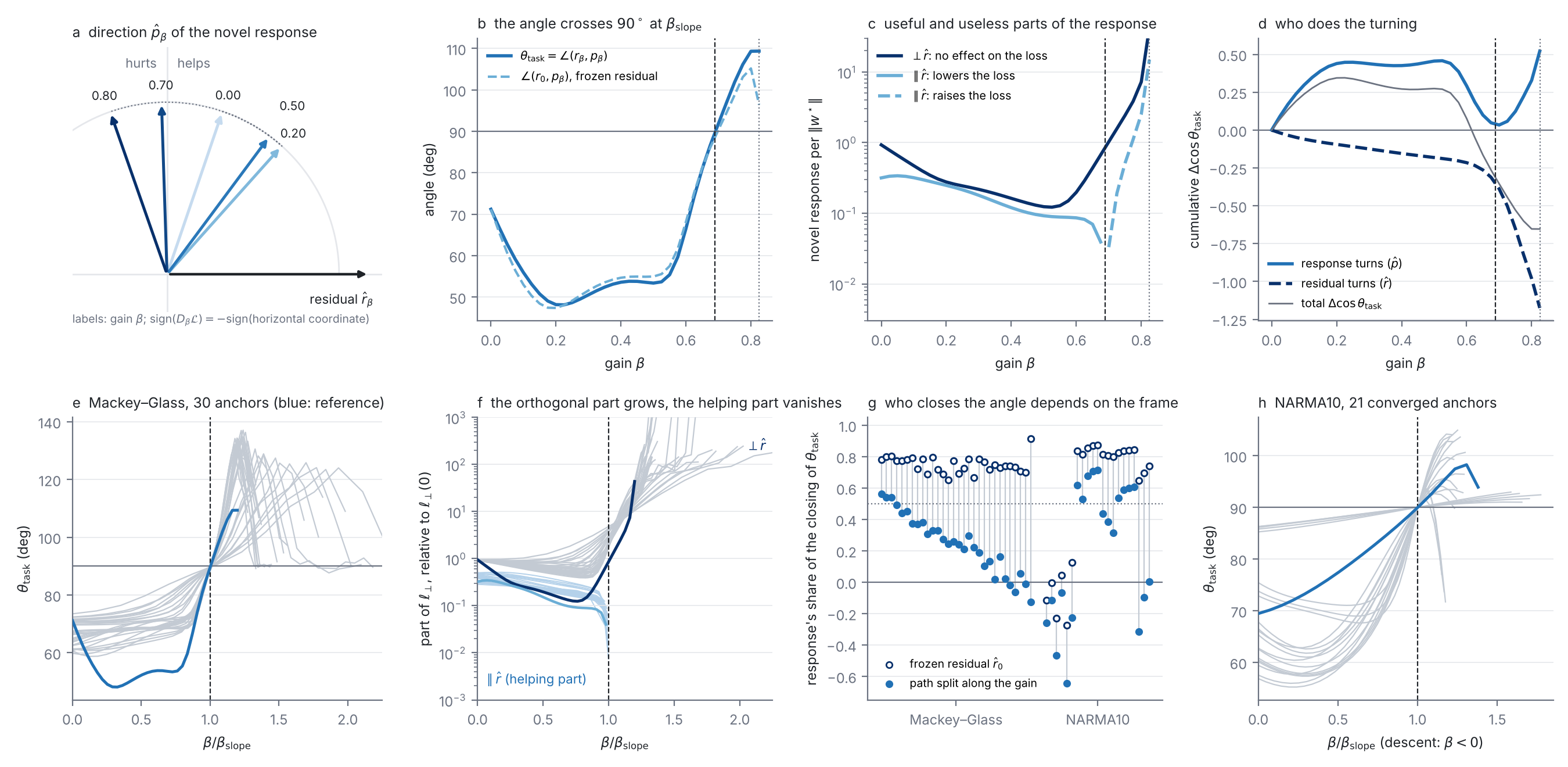}
\caption{\textbf{Geometry of the task link under the conditioned drive.} One-atom JC reservoir, $\tilde u_t=u_t+\beta\langle p\rangle_t$, readout refitted at every gain. \textbf{a}--\textbf{d}, Reference anchor $(30,20,10)$, Mackey--Glass. \textbf{a}, Direction of the novel response $\hat p_\beta$ (Eq.~\eqref{eq:s-pbeta}) in the frame of the current residual, at the labelled gains. The horizontal coordinate is $\cos\theta_{\rm task}$, so $\operatorname{sign}D_\beta\Lc=-\operatorname{sign}\cos\theta_{\rm task}$, and the arrow nearest $\beta_{\rm slope}=0.689$ ($\beta=0.70$) is almost vertical. \textbf{b}, Task angle (Eq.~\eqref{eq:s-taskangle}) and the angle of $\hat p_\beta$ to the residual frozen at $\beta=0$. Vertical dashed line, $\beta_{\rm slope}$; vertical dotted line, $\beta_{\rm edge}$. \textbf{c}, Components of the novel response orthogonal to the residual and along it (the part that changes the loss; dashed after it changes sign). \textbf{d}, Cumulative change of $\cos\theta_{\rm task}$, path split into the turning of the response and of the residual (Eq.~\eqref{eq:s-turning}). \textbf{e}, Task angle at the thirty Mackey--Glass anchors of Supplementary Table~\ref{tab:stopping-mg} against $\beta/\beta_{\rm slope}$, up to $\beta_{\rm edge}$. \textbf{f}, Orthogonal (dark) and helping (light, shown up to $\beta_{\rm slope}$) components of the novel response relative to $\lp(0)$. \textbf{g}, Response's share of the closing of $\theta_{\rm task}$ between $\beta=0$ and $\beta_{\rm slope}$, per anchor, in the frozen-residual split (open) and the path split (filled). Mackey--Glass anchors are ordered by $\beta_{\rm slope}$, NARMA10 anchors by $|\beta_{\rm slope}|$. \textbf{h}, NARMA10 task angle along the descent direction $\beta<0$ at the 21 anchors with $n_0\ge1$, to the last grid gain within $0.15$ beyond $\beta_{\rm slope}$.}
\label{fig:rotation}
\end{figure}

\emph{Reference anchor} (Supplementary Fig.~\ref{fig:rotation}a--d). At $\beta=0$, $\theta_{\rm task}=71.2^\circ$. The drive first improves the alignment, to $48.1^\circ$ at $\beta=0.225$, while $\lp$ shrinks from $0.98$ to a minimum of $0.15$ at $\beta=0.525$. The angle then rises steeply and reaches $90^\circ$ at $\beta_{\rm slope}=0.689$. Meanwhile $\lp$ grows $5.7$-fold from its minimum to $0.86$ ($0.88$ of its $\beta=0$ value), almost entirely orthogonal to the residual. The component along the residual, $\lp\cos\theta_{\rm task}$, falls through zero and changes sign. Past the zero it grows against the residual, and $\theta_{\rm task}=109^\circ$ at the edge. In the path split the response first turns toward the residual and then away. From the best alignment to $\beta_{\rm slope}$ its turning contributes $-0.41$ and the residual's $-0.25$, but over the whole range from $\beta=0$ its net contribution is $+0.04$. Against the frozen residual, $\hat p$ goes from $71^\circ$ to $47^\circ$ and back to $88^\circ$, so the frozen-residual split credits the response with $91\%$. This anchor shows the frame dependence most clearly.

\emph{Thirty Mackey--Glass anchors} (Supplementary Fig.~\ref{fig:rotation}e--g). The reference anchor's early alignment phase is the exception. Elsewhere the alignment improves by at most $2^\circ$, and at 18 of 30 anchors $\theta_{\rm task}$ rises monotonically from its $\beta=0$ value (median $66.7^\circ$, range $59.8$--$73.5^\circ$) to $90^\circ$. Three features recur.

(i) \emph{The response is not exhausted at the zero.} $\lp(\beta_{\rm slope})/\lp(0)=2.1$ $[1.9,3.1]$, range $0.88$ (reference anchor) to $5.0$. At 29 anchors $\lp$ dips to a minimum at about $0.8\,\beta_{\rm slope}$; at $(10,15,10)$ the minimum is at $\beta=0$. From its minimum, $\lp$ grows $4.5$-fold $[3.6,4.7]$ to the zero. The growth is almost entirely orthogonal to the residual (orthogonal part $\times4.7$ from its minimum). The component along the residual decreases monotonically to zero at 27 of 30 anchors.

(ii) \emph{Both vectors turn.} The residual turns by $19^\circ$ $[11,24]^\circ$ between $\beta=0$ and $\beta_{\rm slope}$. The response turns by $82^\circ$ $[81,83]^\circ$ relative to its own $\beta=0$ direction, and its angle to the frozen residual $\hat r_0$ grows from $67^\circ$ to $84^\circ$ $[82,86]^\circ$. The new motion therefore lies mostly in directions of $\mathbb{R}^T$ that were orthogonal both to the old motion and to the task. The division of the closing depends on the frame (Supplementary Fig.~\ref{fig:rotation}g). The frozen-residual split credits the response with $74\%$ $[72,78]\%$ (more than half at all 30 anchors). The path split credits it with $25\%$ $[11,37]\%$ (more than half at 3).

(iii) \emph{Past the zero, the response points against the task.} Between $\beta_{\rm slope}$ and the edge, $\theta_{\rm task}$ peaks at $132^\circ$ $[128,134]^\circ$, about $0.12$ beyond $\beta_{\rm slope}$. At the reference anchor it instead rises to $109^\circ$ at the edge. After the peak the angle falls at 29 of 30 anchors. At 9 it comes back to within $2^\circ$ of $90^\circ$, always beyond $\beta^{\max}_{\rm edge}$, where the loop approaches instability.

\emph{NARMA10} (Supplementary Fig.~\ref{fig:rotation}g,h). The descent is $\beta<0$, and the angle is measured along $-e_\beta$. We use the 21 anchors with $n_0\ge1$ in Supplementary Table~\ref{tab:stopping-narma}. The nine anchors with $n_0=0$, whose $\beta_{\rm slope}$ lies beyond $\beta_{\rm nc}$, are excluded from these statistics and from panel~h. The response is again not exhausted at the zero. $\lp(\beta_{\rm slope})/\lp(0)=1.5$ $[1.1,1.6]$, and up to the zero $\lp$ stays at least at $0.86$ of its $\beta=0$ value at every anchor. The orthogonal part grows $1.7$-fold, while the component along the residual falls to zero. At 16 of these anchors $\BQ$ also grows along the descent (all but $g=10$), so there the closing cannot come from a shrinking budget either. The early alignment phase is more common than for Mackey--Glass (15 of 21 anchors) but small, at most $5^\circ$. Against the frozen residual, the response's angle grows from $66.5^\circ$ (median) to $85.6^\circ$, and the response turns by $75^\circ$ relative to its own $\beta=0$ direction. The frozen-residual split credits the response with $81\%$ $[13,84]\%$ (more than half at 15 anchors), the path split with $38\%$ $[-12,60]\%$ (more than half at 9). At six anchors both frames credit the residual with essentially the whole closing (response share at most $13\%$). These are the five $g=10$ anchors, where the response starts almost perpendicular to the residual ($\theta_{\rm task}(0)=85$--$86^\circ$, $\eta_5\approx0.07$), and the reference anchor. Inside the converged range and within $0.15$ beyond the zero, $\theta_{\rm task}>90^\circ$ at 19 of 21 anchors. At $(15,15,15)$ and $(18,15,10)$ it returns to $90^\circ$ at the second zero of $D_\beta\Lc$ ($n_0=2$).

\emph{Physical reading.} The conditioned drive raises the novel response by lengthening retention (Memory ratios above). Most of this motion is not what the target needs. Already at $\beta=0$, $84\%$ (median, both tasks) of $\|p_\beta\|^2$ is orthogonal to the residual. As the gain grows, both vectors turn. The drive's new motion leaves the original task direction, and the readout absorbs the useful part of what the drive makes reachable, so the residual turns away from the response. The training optimum is the gain at which the remaining useful component reaches zero. Neither the QFI budget nor the novel response is exhausted there. This is consistent with the delayed-readout check: once the readout is given explicitly delayed copies of the features, the extra retention adds nothing. Two qualifications apply. First, $\theta_{\rm task}$ is the angle between two vectors in $\mathbb{R}^{2000}$ that change together with the feature span. It is not a rigid rotation in a fixed plane: $\hat p$ turns by $82^\circ$ (Mackey--Glass) and $75^\circ$ (NARMA10) relative to its own $\beta=0$ direction while its length changes several-fold. Second, no frame-free division of the closing between the response and the residual exists. The two frames above agree that both vectors turn, but they weight the two differently, so we do not attribute the closing to either one alone.

\emph{Caveats.} Past $\beta_{\rm edge}$ or $\beta_{\rm nc}$ the photon number grows and $N_c=10$ is not converged. Losses, NRMSE and $\BQ$ there are order-of-magnitude only and enter none of the reported gains for Mackey--Glass. With $n_{\rm samp}=6$ the median criterion can fire late at isolated anchors (at $(10,15,18)$ it gives $0.841$ although $\max_t|\lambda_1|$ crosses at $0.594$ and the loss collapses at $0.625$), and $\beta^{\max}_{\rm edge}$ lies within $0.02$ of $\beta_{\rm slope}$ at four anchors. Gains with $|\tilde u_t|>10$ abort the propagation and enter only the unstable-gain counts (18 of 30 Mackey--Glass anchors become unstable, but only at $\beta\ge1.6$).

\subsection*{Cutoff checks}
On the stable branch, $N_c=10$ and $N_c=14$ give identical training loss and held-out NRMSE at the reported precision. At $\beta=0.9$, the maximum mean photon number grows from $6.6$ to $11.6$ as $N_c$ increases from 10 to 16, so the state is not converged. Post-edge QFI values are used only at order-of-magnitude level. The poor held-out NRMSE ($0.96$--$0.99$) is already stable across these cutoffs.

\section{Sign-agreement scan of the loss budget, the counter-example anchor and the measurable budget}
\label{sec:sign}
This note reports the sign-agreement scan of the loss budget over the 121-point grid, documents the counter-example anchor $(15,15,15)$ and gives the classical Fisher information of the deployed dictionary as a measurable proxy for the budget. It supports main-text Sec.~2.4 (Table~3 and Fig.~4h), the first limitation stated in the Discussion and the Experimental-estimation paragraph.

The exact decomposition $\Delta\log\lo=\Delta\log\BQ+\sum_k\Delta\log\eta_k$ is a statement about the local slope. To measure how often it predicts the outcome of the finite intervention it motivates, both interventions of main-text Table~3 were applied at every one of the 121 points of the $(g,\kappa_c)$ grid at $\epsilon=15$: the two Fisher-leading SLD eigenprojectors of $\bv_Q^\star$ added to the dictionary at fixed $\bth$ and trajectory, and input profiling with $P\le4$ at fixed $\bth$ (the settings of \ref{sec:input}, twelve iterations per order). At each point the intervention ``on the smallest factor'' is the dictionary intervention where $\eta_3<\eta_4$ (45 points) and profiling where $\eta_4<\eta_3$ (76 points). Supplementary Table~\ref{tab:sign} gives the fractions of points, by type, at which the targeted factor rises, $\lo$ rises, the training loss falls, and the two signs agree; main-text Fig.~4h shows the scatter. Both interventions lower the training loss at all 121 points (for the dictionary intervention because the ridge fit gains two columns, for profiling because the algorithm accepts only loss-lowering steps), so the finite-step sign is not an informative discriminator. The held-out NRMSE also falls at 118 of 121 (dictionary) and 121 of 121 (profiling) points. The informative quantity is whether the local slope $\lo$ along the fixed direction $\bv_Q^\star$ also rises. The intervention aimed at the smallest factor raises it at 65 of 121 points ($54\%$), namely the dictionary intervention at 22 of 45 $\eta_3$-limited points and profiling at 43 of 76 $\eta_4$-limited points. The other intervention raises it at 35 of 121 ($29\%$; profiling at only 2 of 45 $\eta_3$-limited points). The targeted factor itself is not guaranteed to rise. $\Delta\log\eta_4>0$ at 72 of 76 $\eta_4$-limited points under profiling (median $+0.091$), but $\Delta\log\eta_3>0$ at only 30 of 45 $\eta_3$-limited points under the dictionary intervention (median $+0.40$, range $-1.49$ to $+1.84$), because the added projectors change both the readout-used response and the total feature motion. Where the dictionary intervention lowers the slope at an $\eta_3$-limited point (23 of 45), the most negative budget term is $\Delta\log\eta_4$ at 12, $\Delta\log\eta_5$ at 7 and $\Delta\log\eta_3$ at 4 of them. Where profiling lowers the slope at an $\eta_4$-limited point (33 of 76), $\Delta\log\eta_5$ is the most negative term at 31 of them (median $-1.75$; median over the 121 grid points $-0.64$), which means that the profiled input mostly re-aligns the residual with the response instead of changing the span factors. Where profiling raises $\lo$ at an $\eta_4$-limited point (43 cases), $\Delta\log\eta_4>0$ at 41 of them and carries a median $0.59$ $[0.39,0.80]$ of $\Delta\log\lo$. The magnitudes of $\Delta\log\lo$ and $\Delta\log\Lc_\delta$ are uncorrelated for the dictionary intervention (Spearman $0.06$) and positively rank correlated for profiling ($0.84$). For profiling, the largest finite-step improvements occur where the remaining slope along the old direction falls most. Of the 22 $\eta_3$-limited points at which the dictionary intervention raises $\lo$, the one with the largest $\Delta\log\lo$ is $(18,15,14)$ ($+0.51$). There, however, $\eta_3=0.072$ is only marginally below $\eta_4=0.097$, and the rise is carried by $\eta_5$ ($+0.40$) rather than by the targeted $\eta_3$ ($+0.15$). The same holds at $(18,15,15)$, whereas at $\kappa_c=10$--$13$ in that row the $\eta_3$ term is the larger one, and at $(13,15,20)$ and $(13,15,19)$ $\lo$ rises while $\eta_3$ falls. For this reason anchor A of main-text Table~3 is $(17,15,15)$, a point near the counter-example anchor at which the intervention on the smallest factor moves that factor and raises the slope; the nearest such point, $(16,15,15)$, shows the same pattern with a smaller rise ($\Delta\log\eta_3=+0.370$, $\Delta\log\eta_4=-0.450$, $\Delta\log\eta_5=+0.142$, $\Delta\log\lo=+0.113$). At $(17,15,15)$, $\Delta\log\eta_3=+0.473$, $\Delta\log\eta_4=-0.502$, $\Delta\log\eta_5=+0.296$ and $\Delta\log\lo=+0.348$, with training loss $6.231\to4.964$ and held-out NRMSE $0.325\to0.294$. Profiling at the same anchor lowers $\lo$ ($\Delta\log\lo=-2.07$, carried by $\eta_5$) while lowering the training loss $6.23\to5.17$, the same pattern as at the counter-example anchor. Anchor B $(11,15,15)$ was evaluated independently with the full settings, and the two evaluations agree exactly. The two Fisher-leading eigenprojectors of $\bv_Q^\star$ capture a median $0.925$ of the CFI along $\bv_Q^\star$ over the grid. They raise $\eta_1$ along $\bv_Q^\star$ only from $0.346$ to $0.353$ (medians) because they are per-time, direction-specific projectors with small variance weights.

\begin{table}[!ht]\centering\small
\caption{\textbf{Sign-agreement scan over the 121-point grid.} Fractions of grid points at which the intervention on the smallest factor raises the targeted factor, raises the local slope $\lo$, lowers the training loss after the finite step, lowers the held-out NRMSE (reported only), and at which the signs agree ($\Delta\log\lo>0$ while the loss falls); by type of the smallest factor and pooled. The lower block gives the same counts for the intervention aimed at the other factor.}
\label{tab:sign}
\setlength{\tabcolsep}{4pt}
\begin{tabular}{@{}lcccccc@{}}\toprule
Points & $n$ & targeted $\eta_k$ rises & $\lo$ rises & training loss falls & held-out NRMSE falls & signs agree\\\midrule
\multicolumn{7}{@{}l}{\emph{Intervention aimed at the smallest factor}}\\
$\eta_3$-limited (dictionary) & 45 & 30 (67\%) & 22 (49\%) & 45 (100\%) & 45 (100\%) & 22 (49\%)\\
$\eta_4$-limited (profiling) & 76 & 72 (95\%) & 43 (57\%) & 76 (100\%) & 76 (100\%) & 43 (57\%)\\
pooled & 121 & 102 (84\%) & 65 (54\%) & 121 (100\%) & 121 (100\%) & 65 (54\%)\\\midrule
\multicolumn{7}{@{}l}{\emph{Intervention aimed at the other factor}}\\
$\eta_3$-limited (profiling) & 45 & -- & 2 (4\%) & 45 (100\%) & 45 (100\%) & 2 (4\%)\\
$\eta_4$-limited (dictionary) & 76 & -- & 33 (43\%) & 76 (100\%) & 73 (96\%) & 33 (43\%)\\
pooled & 121 & -- & 35 (29\%) & 121 (100\%) & 118 (98\%) & 35 (29\%)\\
\bottomrule\end{tabular}
\end{table}

\subsection*{The counter-example anchor $(15,15,15)$}
The anchor $(15,15,15)$ is $\eta_3$-limited. There the intervention aimed at the smallest factor raises $\eta_3$ by $4.6\times$, but the enlarged feature span changes the residual and reduces $\eta_4$ and $\eta_5$ by more, so that $\lo$ falls by a factor $3.5$ while the training loss improves from $5.68$ to $4.73$. The local slope and the finite-step outcome disagree in sign. Supplementary Table~\ref{tab:oldA} documents this counter-example anchor, which is marked in main-text Fig.~4h.

\begin{table}[!ht]\centering\small
\caption{\textbf{Loss budget at the counter-example anchor} $(g,\epsilon,\kappa_c)=(15,15,15)$ along $\bv_Q^\star$ (one QFI unit before the intervention), before and after adding the two Fisher-leading SLD eigenprojectors to the dictionary at fixed $\bth$ and trajectory, with the log-additive attribution. $\BQ$ after the intervention is evaluated with the changed weights $\omega_t$. Training loss $5.68\to4.73$.}
\label{tab:oldA}
\begin{tabular}{@{}lcc@{}}\toprule
Quantity & before & after\\\midrule
$\BQ$ & 1 & 1.005\\
$\BC$ & 0.308 & 0.318\\
$\normF{X_{\bv}}$ & 0.306 & 0.315\\
$\|X_{\bv}\ws\|/\|\ws\|$ & $5.37\times10^{-3}$ & $2.57\times10^{-2}$\\
$\lp$ & $1.96\times10^{-3}$ & $1.82\times10^{-3}$\\
$\lo$ & $5.55\times10^{-4}$ & $1.59\times10^{-4}$\\\midrule
$\eta_1$ & 0.308 & 0.316\\
$\eta_2$ & 0.992 & 0.992\\
$\eta_3$ & \textbf{0.018} & 0.081\\
$\eta_4$ & 0.365 & 0.071\\
$\eta_5$ & 0.283 & 0.087\\
$\eta_Q$ & $5.55\times10^{-4}$ & $1.58\times10^{-4}$\\\midrule
$\Delta\log\lo$ & -- & $-1.25$\\
$\Delta\log\BQ$ & -- & $+0.005$\\
$\Delta\log\eta_1,\ \Delta\log\eta_2$ & -- & $+0.02$, $0.00$\\
$\Delta\log\eta_3$ & -- & $+1.53$\\
$\Delta\log\eta_4$ & -- & $-1.64$\\
$\Delta\log\eta_5$ & -- & $-1.18$\\
\bottomrule\end{tabular}
\end{table}

\subsection*{Classical Fisher information of the deployed dictionary as a measurable proxy}
\label{sec:cfiproxy}
The measurement factor $\eta_1=\BC/\BQ=\sqrt{\bv^\top\GC\bv/\bv^\top\GQ\bv}$, where $\GC=\sum_t\sum_j\Var_{\rho_t}(O_j)\FC^{(j)}(t)$ is the feature-matched classical Fisher information of the deployed dictionary $\{x,p,\sigma_x,\sigma_y\}$, is the fraction of the quantum budget that the outcome statistics of the deployed measurement already certify. $\BC$ is measurable, and $\BQ\ge\BC$. Supplementary Table~\ref{tab:cfiproxy} gives its distribution over the 121 grid points at $\epsilon=15$ (computed from the $\GQ$ and $\GC$ matrices of the sign scan; the values along $\bv_Q^\star$ and the dominant direction agree with the scan values to all digits) and along the stable branch of the conditioned drive at the reference anchor $(30,20,10)$ (twelve stable gains $\beta\in[0,0.9]$). At the reference anchor without the conditioned drive, $\eta_1$ is $0.325$, $0.483$ and $0.500$ along $e_g$, $e_\epsilon$ and $e_{\kappa}$, $0.384$ along $\bv_Q^\star$, and lies between $0.316$ and $0.522$ over all directions. The measurable budget is therefore a factor $2$--$5$ below the quantum one. The four downstream links involve only measured features and are unchanged.

\begin{table}[!ht]\centering\small
\caption{\textbf{Fraction $\eta_1=\BC/\BQ$ of the quantum budget certified by the classical Fisher information of the deployed dictionary.} Median, interquartile range and range over the 121 grid points $g,\kappa_c\in\{10,\ldots,20\}$, $\epsilon=15$, for the QFI-normalized direction $\bv_Q^\star$, the dominant-QFI direction $\bv_{\max}$, the Euclidean-unit physical axes and the worst and best directions of the $(\GC,\GQ)$ pencil; last row: the stable branch of the conditioned drive at $(30,20,10)$ (twelve gains).}
\label{tab:cfiproxy}
\begin{tabular}{@{}lcccc@{}}\toprule
Direction & median & IQR & min & max\\\midrule
$\bv_Q^\star$ & 0.346 & [0.289, 0.462] & 0.188 & 0.509\\
$\bv_{\max}$ & 0.385 & [0.321, 0.439] & 0.258 & 0.468\\
$e_g$ & 0.328 & [0.269, 0.392] & 0.209 & 0.432\\
$e_\epsilon$ & 0.455 & [0.425, 0.481] & 0.404 & 0.493\\
$e_{\kappa}$ & 0.512 & [0.510, 0.514] & 0.504 & 0.516\\
worst direction & 0.282 & [0.200, 0.369] & 0.157 & 0.420\\
best direction & 0.527 & [0.518, 0.540] & 0.510 & 0.555\\
conditioned-drive branch, $e_\beta$ chain & 0.513 & [0.491, 0.522] & 0.431 & 0.525\\
\bottomrule\end{tabular}
\end{table}

\section{Quantum Fisher information as a design resource: two design gains}
\label{sec:advantage}
This note collects the numerical tests behind the two further design gains of main-text Sec.~2.6, that is, the gains obtained when a design decision is taken on the pre-measurement quantum information (the SLD eigenbasis and the exact QFI-normalized slope) rather than on the classical information of the measured features: the SLD-eigenprojector augmentation of the observable dictionary (main-text Sec.~2.3) and the location of the drive gain from the exact slope (main-text Sec.~2.5). The matched-cost comparison and the stage test that carry the main-text advantage claim are reported in \ref{sec:matched}.

The two tests are the measurement gap and its closure, and the cost of locating a drive-gain optimum from the exact slope. All configurations were selected on training data. Held-out values are reported as checks and never enter a selection. Data and scripts are in the deposited code (directory \texttt{advantage}).

\subsection*{Measurement gap and SLD-eigenprojector augmentation}
At each of the 121 grid points ($\epsilon=15$) the $k$ Fisher-leading eigenprojectors of the SLD along $\bv_Q^\star$ were added to the dictionary $\{x,p,\sigma_x,\sigma_y\}$ at fixed $\bth$ and trajectory (Supplementary Table~\ref{tab:adv-meas}). $\eta_1$ rises at every point but saturates. With the full SLD basis ($k=21$; the 22nd projector is collinear with the bias) $\eta_1'/\eta_1=1.05$--$1.20$ (median $1.08$) at the nine anchors tested, because each projector enters $\BC$ as a two-outcome measurement. The training loss falls everywhere (nested least squares), and the held-out NRMSE falls at 118 of 121 points for $k=2$ (120 for $k=4$). The slope $\lo$ along the fixed base direction rises at 55 of 121 points for both $k$. Along the augmented model's own $\bv_Q^{\star\prime}$ it rises at 83. With the shot model of \ref{sec:shots} ($h=0.5$, $\epsilon_{\rm rel}=0.1$) the ratio $n_{\rm req}({\rm aug})/n_{\rm req}({\rm base})$ along $\bv_Q^\star$ is $1.5$ $[0.83,3.2]$ at a fixed total shot budget split equally over features and $1.05$ $[0.60,2.2]$ with the optimal allocation ($k=2$), and it is $\ge1$ at 76 and 63 of 121 points. The ratio is governed by the change of $D_{\bv}\Lc$ (Spearman $0.95$ with $(D_{\bv}\Lc/D_{\bv}\Lc')^2$), the projector features adding $2\%$ of the shots. Favourable anchors are $(30,20,10)$ ($0.93$/$0.78$) and $(17,15,15)$ ($0.59$/$0.49$). Unfavourable anchors are $(11,15,15)$ ($1.46$/$1.00$) and $(15,15,15)$ ($14.1$/$11.4$). Monte Carlo checks of the variance formula agree to $0.94$--$1.11$ at nine anchors. In the spin reservoir (20 seeds, $(1,1,1)$) the $Z$ basis captures $0.12$ $[0.09,0.14]$ of $\bv_Q^{\star\top}\GQ\bv_Q^\star$, the $J$-SLD basis $0.74$ $[0.54,0.82]$ and the best single-control SLD basis $0.90$ $[0.78,0.97]$. Augmenting $\{Z_i\}$ with $k=2$ $J$-SLD projectors lowers the Mackey--Glass loss by $9.1\%$ $[5.4,17.9]$ and the held-out NRMSE by $0.026$ $[0.013,0.040]$ at 20 of 20 seeds, while $\lo$ along the fixed $\bv_Q^\star$ falls ($\times0.64$, up at 8 of 20). For NARMA10 the loss falls by $0.6\%$ and the held-out NRMSE is unchanged. Replacing $\{Z_i\}$ by projectors is detrimental (loss $\times5.5$). For $k\ge6$ (JC) the design matrix is ill conditioned and $\lo$ collapses, so the saturation statements rest on $\eta_1$ and the loss.

\begin{table}[!ht]\centering\small
\caption{\textbf{SLD-eigenprojector augmentation over the 121 grid points} (median [IQR]; min--max), $k=2$ projectors of $\bv_Q^\star$ added at fixed $\bth$.}
\label{tab:adv-meas}
\begin{tabular}{@{}lcc@{}}\toprule
Quantity & median [IQR] & min--max\\\midrule
$\eta_1$ (base) & $0.346$ $[0.289,0.462]$ & $0.188$--$0.509$\\
$\eta_1'/\eta_1$ & $1.049$ $[1.021,1.072]$ & $1.005$--$1.169$\\
$\lo'/\lo$ (fixed $\bv_Q^\star$) & $0.975$ $[0.684,1.279]$ & $0.030$--$1.764$\\
$\lo'/\lo$ (own $\bv_Q^{\star\prime}$) & $1.37$ $[0.94,1.88]$ & $0.40$--$4.02$\\
$\Delta\Lc/\Lc$ (train) & $-0.093$ $[-0.177,-0.032]$ & $-0.589$--$-0.001$\\
$\Delta$NRMSE (held-out) & $-0.0084$ $[-0.020,-0.0027]$ & $-0.104$--$+0.0004$\\
$n_{\rm req}$ ratio, equal split & $1.51$ $[0.83,3.19]$ & $0.35$--$1217$\\
$n_{\rm req}$ ratio, optimal split & $1.05$ $[0.60,2.22]$ & --\\
\bottomrule\end{tabular}
\end{table}

\subsection*{Locating the drive-gain optimum from the exact slope}
On the stored 30-anchor Mackey--Glass sweeps (\ref{sec:feedback-si}), a root search on $D_\beta\Lc$ was replayed using only stored gains. The search brackets the zero by doubling from $\beta=0$ (step $0.1$) with a loss safeguard (the bracket closes when the training loss rises) and then bisects to the grid step $0.025$. It returns the sweep's $\beta_{\rm slope}$ at 30 of 30 anchors in $9$ $[7,9]$ evaluations (range $5$--$9$; a secant variant $8$ $[6,11]$), against 48 stored gains, $26$ $[21,30]$ gains up to the edge and $18$ $[16,23]$ gains of a fine march to the first sign change. The held-out regret is that of the sweep's $\beta_{\rm slope}$ ($0.36\%$ median). Without the safeguard the doubling step lands beyond the edge at 3 of 30 anchors, returning a spurious zero at one (held-out NRMSE $0.99$) and no sign change at two; with new trajectories it returns a spurious zero at 3 of 10. A golden-section search on the training loss alone needs $13$ $[13,14]$ single-trajectory evaluations. Re-run with new trajectories at ten anchors (tolerance $0.01$), the safeguarded search agreed with the sweep to $|\Delta\beta|\le0.005$ at the eight anchors with a stored sweep, in $11$ $[9.5,11]$ evaluations and $153$ $[131,158]$~s against $1248$ $[1187,1339]$~s for the 48-gain sweep at the same anchors. The golden-section search took $16$ $[14.5,16]$ evaluations and $107$ $[96,111]$~s, because a slope evaluation costs a trajectory and its complex-step derivative ($7.2+7.2$~s) whereas a loss evaluation costs one trajectory. The chain adds $2.3$~s ($16\%$) and $|\lambda_1|$ with six samples $7.1$~s. For NARMA10, 9 of 30 anchors have no zero before $\beta_{\rm nc}$, so the search must be given the truncation limit. In the two-atom reservoir all root rules returned the third zero ($0.779$) rather than the first ($0.249$), by the accident of the step schedule.

\section{Matched-cost comparison with a tuned echo-state network}
\label{sec:matched}
This note reports the matched-cost comparison between the quantum reservoirs designed under the all-parameter protocol and a tuned echo-state network (ESN), referred to in main-text Sec.~2.1, Fig.~6c,d and the Methods subsection ``Matched-cost comparison and stage test'', the stage test that attributes the margin to the QFI natural-gradient stage, the controls of that test (single-block ablation arms and an equal-budget black-box control) and the sensitivity of the comparison to the ESN grid, the ridge, the choice of ESN draw, held-out sampling and the data window. The comparison is on five-step Mackey--Glass forecasting at readout cost $m\in\{4,6,8,12,16,18,24,30\}$. NARMA10 was not run under the all-parameter protocol and is not part of the comparison. All tables of this note are generated by \texttt{code/\allowbreak analysis/\allowbreak build\_benchmark\_validation.py} from the deposited result files (\texttt{data/\allowbreak advantage/\allowbreak performance\_all/}, \texttt{data/\allowbreak advantage/\allowbreak performance\_largem/}, \texttt{data/\allowbreak advantage/\allowbreak performance/}); \texttt{code/\allowbreak run/\allowbreak rebuild\_benchmark\_validation.sh} runs the build. The numerical entries quoted in this note are read from \texttt{supplement\_tables/\allowbreak benchmark\_macros.tex}, written by the same script.

\subsection*{Metrics}
Held-out NRMSE is the root-mean-square error of the fixed readout on the 895 held-out rows divided by the standard deviation of the held-out target; it is the primary quantity and never enters a selection. Training NRMSE is the root-mean-square residual of the ridge fit ($\delta=10^{-10}$) on the 2,000 training rows divided by the standard deviation of the training target ($0.2331$); the training loss $\tfrac12\|r\|^2+\tfrac12\delta\|w\|^2$ on the same rows is the objective of every search stage. $\Delta$ vs ESN $=1-{\rm NRMSE}_{\rm QRC}/\overline{\rm NRMSE}_{\rm ESN}$ with the ESN mean over its 10,000 draws (positive: QRC better); ``draws worse'' is the share of the 10,000 draws whose held-out NRMSE exceeds the QRC value; the QFI-stage gain is $1-{\rm NRMSE}({\rm grid+BO+QFI})/{\rm NRMSE}({\rm grid+BO})$.

\subsection*{Quantum-reservoir protocol}
The reservoir is the driven Jaynes--Cummings model of the main text with $N\in\{1,2,3\}$ atoms, $N_c=10$ ($8$ for three atoms), $V$ virtual nodes per input interval and $m=V(2+2N)$ features, driven through the measurement-conditioned input drive $\tilde u_t=u_t+b+\bbeta^\top X_t$ of main-text Eq.~(15), with $X_t$ the $m$ sub-time samples of the previous interval. The adjustable parameters are the physical parameters $\phi=(\Delta_c,\Delta_{a,i},g_i,\epsilon,\kappa_c,\gamma_i)$ (cavity and atomic detunings, couplings, drive, cavity and atomic decay; $P=3N+3$), the input offset $b$, the $V-1$ interior sampling times $\tau_j$ and the $m$ drive gains $\bbeta$. The time step $dt$, the drive frequency, the signs of $\Delta_c$ and $g_i$, the quadrature of the measured pairs and the initial state were excluded after the checks of Table~\ref{tab:benchmark-exclusions}: each is either redundant with a retained parameter or absorbed by the readout and $\bm{\beta}$. The box is $\Delta_c\in[1,80]$, $\Delta_{a,i}\in[-80,80]$, $g_i\in[2,64]$, $\epsilon\in[2,48]$, $\kappa_c\in[1,48]$, $\gamma_i\in[0.1,24]$, $|b|\le\sigma_u$, $0.02\,dt<\tau_1<\cdots<\tau_{V-1}<0.98\,dt$; admissibility requires $|\tilde u_t|\le10$ on the closed loop and $\max_t\langle n\rangle\le3$.

\emph{Layouts.} Two configurations were run at $m=12$ (one atom with $V=3$; two atoms with $V=2$) and at $m=16$ (one atom with $V=4$; three atoms with $V=2$). The reported layout per $m$ is the one with the lower training loss at the end of the QFI stage; the other is listed in Table~\ref{tab:benchmark-matched} as well.

\emph{Reference point and sensitivity screen.} Each configuration was centred on a reference point $\theta^\star$: for $m\le18$ the fixed-control winner of the earlier comparison (its anchor and optimized drive vector), for $m=24$ and $30$ the end point of the chain on the controls $(g_i,\epsilon,\kappa_c)$ and the drive vector described below. At $\theta^\star$ every parameter was scanned one at a time over its box (nine points, log-spaced for rates, readout refitted, screening loss over the first 2,100 samples); a parameter would have been dropped only if the relative range of the admissible losses stayed below $10^{-2}$ in every configuration with the same atom number. No parameter met the condition (Table~\ref{tab:benchmark-sensitivity}), so all $d=P+1+(V-1)+m$ parameters entered the search: $d=\rDimMin$ to $\rDimMax$ (\rDimList; Table~\ref{tab:benchmark-matched}).

\emph{Rough grid.} A sub-sampled factorial of 2,000--4,000 points (fixed seed) around $\theta^\star$: three levels per rate parameter $\{\phi^\star/3,\phi^\star,3\phi^\star\}$ clipped to the box, three levels per detuning $\{\Delta/2,\Delta,2\Delta\}$ with a random sign, $b\in\{-\tfrac12,0,\tfrac12\}\sigma_u$, three sampling-time profiles (uniform, early, late) and a structured drive vector (one amplitude per feature class, three sub-time profiles) inside a box calibrated at $\phi^\star$ to keep at least half of the points admissible under the photon cap. The reference point is grid point 0. The best admissible point seeds the Bayesian stage (Table~\ref{tab:benchmark-resources}).

\emph{Bayesian optimization.} Phase 1: Gaussian-process expected improvement in the structured drive-vector coordinates inside $\pm$ one grid spacing around the seed (log-scaled rates, signed detunings in units of $\Delta^\star$, $b$ in units of $\sigma_u$, ordered sampling-time fractions). Phase 2: trust-region Bayesian optimization with Thompson sampling on the full space, including all $m$ components of $\bm{\beta}$. Stop: no relative improvement above $10^{-3}$ of the best training loss over 40 consecutive evaluations (counter restarted at the phase change) or 600 evaluations. Every configuration stopped on the stall rule after \rBoEvals{} evaluations, except $m=8$, whose trust region had contracted below its floor (Table~\ref{tab:benchmark-resources}). At $m=24$ and $30$ the grid and Bayesian-optimization stages of Table~\ref{tab:stage} searched the drive vector with the controls $(g_i,\epsilon,\kappa_c)$ held at $\theta^\star$, the training-selected end point of earlier natural-gradient runs from local starts (natural gradient on the controls alternating with BFGS on the drive vector); the QFI stage of that search (the controls-and-drive-vector search) was followed by the all-parameter grid, BO and QFI stage described here, started from its end point. The reported stage gain at these $m$ therefore spans both QFI stages; the gain of the all-parameter stage over its own start is the upper block of Table~\ref{tab:ablation}.

\emph{QFI natural-gradient stage.} From the BO optimum, rounds of up to three natural-gradient steps on the physical parameters alternate with up to eight BFGS steps on $z=(\bbeta,b,\tau)$. A natural-gradient step uses the closed-loop complex-step gradient $g$ of the training loss and the trajectory QFI metric $\GQ=\sum_t\omega_tF_Q(t)$ summed over the training sub-steps, moves along $-\GQ^{+}g/\GamQ$ with a nominal step of $0.1\sqrt{T}$ in the metric, halves the step until the refitted training loss decreases and projects the candidate into the box; the BFGS steps use exact adjoint gradients for $(\bbeta,b)$ and complex-step gradients for $\tau$ with Armijo backtracking, $b$ kept in $[-\sigma_u,\sigma_u]$ and the $\tau_j$ kept ordered. The stage stops after three consecutive rounds with a relative training-loss decrease below $10^{-4}$ (``converged'') or at the wall budget (``budget''): 6~h for $m\le18$ and 12~h for $m\ge24$, extended by another budget while the last round's relative decrease exceeded $10^{-3}$, up to 36~h. When an end point sat on a box face with an outward descent direction, that face was widened ($\times2$ for rates, $\pm40$ for detunings) and the arm continued; Table~\ref{tab:benchmark-constraints} lists the widened boxes, the rounds, wall times, last-round decreases and the constraints active at every end point. \rPhoton\rFaces

\subsection*{Echo-state-network protocol}
The ESN with $m$ nodes has $m$ trained readout weights plus a bias, the same ridge and the same input information as the reservoir. Its spectral radius $\rEsnLevelsRadius$, leak $\rEsnLevelsLeak$, input scale $\rEsnLevelsScale$ and bias $\rEsnLevelsBias$ were tuned over the \rEsnNSettings{} settings of this grid, with the reservoir matrix and the input matrix and bias drawn at random for every draw. Stage 1 evaluated all \rEsnNSettings{} settings with 1,000 draws each and ranked them by the mean training loss over the draws; stage 2 re-evaluated the five leaders with 10,000 draws each and selected the setting with the lowest mean training loss. The selected settings (spectral radius, leak, input scale, bias) are \rEsnSettingsList. No selected setting lies on a radius or input-scale edge of the grid. The leak sits at the grid maximum $1.0$, the physical maximum of a leaky integrator, at $m=$ \rEsnLeakMaxMs, and at the grid minimum $0.1$ at $m=$ \rEsnLeakMinMs; at the latter a smaller leak level might still gain at the per-cent level (the held-out means of the five stage-1 leaders at these $m$ lie within $\rExtTopSpreadMin$--$\rExtTopSpreadMax$ of each other). The comparison reports the held-out NRMSE mean and standard deviation over the 10,000 draws at the selected setting and the share of draws worse than the reservoir (Table~\ref{tab:benchmark-esn}). Held-out data never entered the selection. The grid is a strict superset of the \rEsnNSettingsOrig-setting grid (spectral radius $\ge0.5$, input scale $\ge0.1$) of an earlier build; the extended run reproduces the deposited draws of the original selected settings to $|\Delta{\rm NRMSE}|=\rEsnRepro$, and the original run reproduced the ten seeds of the earlier ten-seed comparison to $|\Delta{\rm NRMSE}|\le8\times10^{-7}$. The two grids are compared in the subsection `Sensitivity of the comparison'. Averaging the performance of a fixed hyperparameter setting over random reservoir and input matrices is the standard ESN convention; the tuning of the reservoir's physical parameters has no analogue in this convention (see Caveats).

\subsection*{Results}
Table~\ref{tab:benchmark-matched} lists every configuration, Table~\ref{tab:stage} the stage test for the reported layout per $m$, and Fig.~6c,d of the main text plot both. At the QFI end point the reservoir's held-out NRMSE is below the ESN mean at every $m$, by $\rGainMin\%$ ($m=\rGainMinM$) to $\rGainMax\%$ ($m=\rGainMaxM$), and below at least $\rShareFloor\%$ of the 10,000 ESN draws at every $m$ ($\rShareMin\%$ at $m=\rShareMinM$\rShareOthers). The margin is produced at the last stage. After grid + BO the reservoir is ahead of the ESN mean by \rBoAheadRange\% at \rBoAheadMs{} and \rBoBehind. The QFI stage lowers the held-out NRMSE at every $m$: by \rStageSmall{} at $m=$ \rStageSmallMs, where the grid + BO point is already close to the end point, and by \rStageLargeRange\% from $m=12$ on (\rStageLargeList). The training loss falls by \rTrainLossSmall\% at $m\le8$ and \rTrainLossLarge\% at $m\ge12$, in the same order as the held-out values. At $m=30$ the stage turns a deficit of $\rBoGainB\%$ against the ESN mean into a margin of $\rQfiGainB\%$. At $m=12$ and $16$ the layout with more atoms has the lower training loss and is the reported one; the one-atom layouts also beat the ESN mean (\rAltGains) and also owe most of their margin to the QFI stage (\rAltStage{} reductions).

\begin{table}[!ht]\centering\scriptsize
\caption{\textbf{All-parameter protocol against the tuned ESN on Mackey--Glass (train / held-out NRMSE).} One row per configuration; the reported layout per $m$ is the one with the lower training loss at the QFI stage ($m$ marked $\checkmark$). $d=(3+3N)+1+(V-1)+m$ is the number of optimized parameters (physical, offset, sampling times, drive gains). Status: stop reason (rounds, wall time; `r.' = rounds). Training NRMSE = RMSE of the ridge fit on the 2,000 training rows divided by the standard deviation of the training target (0.2331); held-out NRMSE on the 895 held-out rows. $\Delta$ vs ESN $=1-{\rm NRMSE}_{\rm QRC}/\overline{\rm NRMSE}_{\rm ESN}$ (held-out, QFI stage); `draws worse' = share of the 10,000 ESN draws with a higher held-out NRMSE than the QRC. Active constraints at the QFI end point: photon cap $\max_t\langle n\rangle\le3$, box faces, $|\tilde u_t|\le10$. At $m=24$ and $30$ the grid and BO stages were run on the controls $(g_i,\epsilon,\kappa_c)$ and the drive vector, the QFI stage on all parameters.}
\label{tab:benchmark-matched}
\setlength{\tabcolsep}{3.5pt}
\resizebox{\textwidth}{!}{\begin{tabular}{@{}llrrrrrrll@{}}\toprule
$m$ & $N$, $V$, $N_c$ & $d$ & grid & grid + BO & grid + BO + QFI & $\Delta$ vs ESN & draws worse & active & status\\\midrule
4$\checkmark$ & 1, 1, 10 & 11 & 0.1781 / 0.1820 & 0.1743 / 0.1781 & 0.1724 / \textbf{0.1760} & +30.0\% & 100.0\% & $\langle n\rangle$ cap & converged (4 r., 0.6 h)\\
6$\checkmark$ & 2, 1, 10 & 16 & 0.1695 / 0.1716 & 0.1649 / 0.1673 & 0.1597 / \textbf{0.1616} & +24.9\% & 99.5\% & $\langle n\rangle$ cap & converged (5 r., 2.6 h)\\
8$\checkmark$ & 1, 2, 10 & 16 & 0.1462 / 0.1504 & 0.1294 / 0.1345 & 0.1219 / \textbf{0.1269} & +34.0\% & 100.0\% & box face: $\gamma_1$ & converged (39 r., 9.9 h)\\
12$\checkmark$ & 2, 2, 10 & 23 & 0.0957 / 0.0982 & 0.0818 / 0.0858 & 0.0563 / \textbf{0.0608} & +62.4\% & 100.0\% & -- & budget (47 r., 18.1 h)\\
12 & 1, 3, 10 & 21 & 0.1263 / 0.1316 & 0.1215 / 0.1270 & 0.0718 / \textbf{0.0754} & +53.4\% & 100.0\% & -- & budget (51 r., 12.0 h)\\
16$\checkmark$ & 3, 2, 8 & 30 & 0.0853 / 0.0898 & 0.0638 / 0.0658 & 0.0371 / \textbf{0.0400} & +69.4\% & 100.0\% & -- & budget (15 r., 18.1 h)\\
16 & 1, 4, 10 & 26 & 0.1221 / 0.1237 & 0.0927 / 0.0942 & 0.0622 / \textbf{0.0647} & +50.5\% & 100.0\% & -- & budget (44 r., 12.1 h)\\
18$\checkmark$ & 2, 3, 10 & 30 & 0.0916 / 0.0970 & 0.0769 / 0.0806 & 0.0419 / \textbf{0.0449} & +60.0\% & 100.0\% & -- & budget (28 r., 12.0 h)\\
24$\checkmark$ & 2, 4, 10 & 37 & 0.0738 / 0.0798 & 0.0593 / 0.0636 & 0.0368 / \textbf{0.0394} & +51.3\% & 100.0\% & -- & budget (21 r., 12.2 h)\\
30$\checkmark$ & 2, 5, 10 & 44 & 0.0843 / 0.0916 & 0.0761 / 0.0816 & 0.0369 / \textbf{0.0397} & +35.1\% & 99.9\% & -- & converged (9 r., 8.6 h)\\
\bottomrule\end{tabular}}
\end{table}

\begin{table}[!ht]\centering\scriptsize
\caption{\textbf{Stage test: held-out NRMSE after grid, after grid + BO and after the QFI natural-gradient stage (reported layout per $m$).} QFI-stage gain $=1-{\rm NRMSE}({\rm grid+BO+QFI})/{\rm NRMSE}({\rm grid+BO})$ on held-out NRMSE (train NRMSE in brackets). $\Delta$ vs ESN and `draws worse' as in Table~\ref{tab:benchmark-matched}, given for the grid + BO point and for the QFI end point. At $m=24$ and $30$ the grid and BO stages were run on the controls $(g_i,\epsilon,\kappa_c)$ and the drive vector, the QFI stage on all parameters.}
\label{tab:stage}
\setlength{\tabcolsep}{3.5pt}
\resizebox{\textwidth}{!}{\begin{tabular}{@{}llrrrrrrrl@{}}\toprule
$m$ & layout & grid & grid + BO & $\Delta$ vs ESN & draws worse & grid + BO + QFI & $\Delta$ vs ESN & draws worse & QFI-stage gain (train)\\\midrule
4 & 1 atom, $V=1$ & 0.1820 & 0.1781 & +29.2\% & 100.0\% & 0.1760 & +30.0\% & 100.0\% & +1.2\% (+1.1\%)\\
6 & 2 atoms, $V=1$ & 0.1716 & 0.1673 & +22.2\% & 99.2\% & 0.1616 & +24.9\% & 99.5\% & +3.4\% (+3.1\%)\\
8 & 1 atom, $V=2$ & 0.1504 & 0.1345 & +30.0\% & 99.8\% & 0.1269 & +34.0\% & 100.0\% & +5.7\% (+5.8\%)\\
12 & 2 atoms, $V=2$ & 0.0982 & 0.0858 & +46.9\% & 100.0\% & 0.0608 & +62.4\% & 100.0\% & +29.1\% (+31.1\%)\\
16 & 3 atoms, $V=2$ & 0.0898 & 0.0658 & +49.7\% & 100.0\% & 0.0400 & +69.4\% & 100.0\% & +39.2\% (+41.9\%)\\
18 & 2 atoms, $V=3$ & 0.0970 & 0.0806 & +28.1\% & 98.6\% & 0.0449 & +60.0\% & 100.0\% & +44.3\% (+45.5\%)\\
24 & 2 atoms, $V=4$ & 0.0798 & 0.0636 & +21.2\% & 95.1\% & 0.0394 & +51.3\% & 100.0\% & +38.1\% (+38.0\%)\\
30 & 2 atoms, $V=5$ & 0.0916 & 0.0816 & -33.6\% & 3.0\% & 0.0397 & +35.1\% & 99.9\% & +51.4\% (+51.5\%)\\
\bottomrule\end{tabular}}
\end{table}

\begin{table}[!ht]\centering\scriptsize
\caption{\textbf{Tuned echo-state network with 10,000 random draws per $m$ (Mackey--Glass, held-out NRMSE).} Setting = (spectral radius, leak, input scale, bias) with the lowest mean training loss among the five stage-1 leaders of the 810-setting grid: radius $\{0.1, 0.2, 0.3, 0.5, 0.7, 0.9, 0.95, 0.99, 1.1\}$, leak $\{0.1, 0.3, 0.5, 0.7, 1.0\}$, input scale $\{0.005, 0.01, 0.02, 0.05, 0.1, 0.3, 0.5, 1.0, 2.0\}$, bias $\{0, 1\}$; statistics over the 10,000 draws of the reservoir and input matrices at that setting. Train = mean training NRMSE. Edge = grid level of the selected setting that lies on an edge of the grid (leak $1.0$ is the maximum of a leaky integrator, leak $0.1$ the smallest level offered); no selected setting lies on a radius or input-scale edge.}
\label{tab:benchmark-esn}
\begin{tabular}{@{}rlrrrrrrrrl@{}}\toprule
$m$ & setting & train & mean & std & median & min & 1\% & 99\% & max & edge\\\midrule
4 & $(0.7, 1.0, 0.02, 0)$ & 0.2568 & 0.2516 & 0.0418 & 0.2424 & 0.1824 & 0.1942 & 0.3938 & 0.4854 & leak $=1.0$ (max)\\
6 & $(0.9, 1.0, 0.02, 0)$ & 0.2172 & 0.2151 & 0.0287 & 0.2094 & 0.1416 & 0.1682 & 0.3078 & 0.3941 & leak $=1.0$ (max)\\
8 & $(0.9, 1.0, 0.05, 0)$ & 0.1911 & 0.1921 & 0.0238 & 0.1892 & 0.1192 & 0.1450 & 0.2677 & 0.3875 & leak $=1.0$ (max)\\
12 & $(0.9, 1.0, 0.05, 0)$ & 0.1614 & 0.1618 & 0.0193 & 0.1655 & 0.0949 & 0.1143 & 0.2022 & 0.2582 & leak $=1.0$ (max)\\
16 & $(0.2, 0.1, 1.0, 0)$ & 0.1307 & 0.1307 & 0.0219 & 0.1282 & 0.0701 & 0.0908 & 0.1914 & 0.2385 & leak $=0.1$ (min)\\
18 & $(0.2, 0.1, 1.0, 0)$ & 0.1115 & 0.1121 & 0.0173 & 0.1105 & 0.0651 & 0.0794 & 0.1611 & 0.2122 & leak $=0.1$ (min)\\
24 & $(0.3, 0.1, 1.0, 1)$ & 0.0775 & 0.0808 & 0.0118 & 0.0795 & 0.0428 & 0.0578 & 0.1137 & 0.1446 & leak $=0.1$ (min)\\
30 & $(0.7, 0.3, 0.5, 0)$ & 0.0565 & 0.0611 & 0.0093 & 0.0598 & 0.0364 & 0.0442 & 0.0880 & 0.1121 & none\\
\bottomrule\end{tabular}
\end{table}

\subsection*{Controls for the stage test}
The QFI stage bundles natural-gradient steps on the physical parameters, which use the QFI metric, with BFGS steps on the classical parameters $(\bbeta,b,\tau)$, which use exact adjoint gradients only. Three pieces of evidence separate the two contributions (Tables~\ref{tab:ablation}--\ref{tab:ablation-share}; \texttt{code/\allowbreak scripts/\allowbreak adv\_all\_joint.py} with \texttt{--freeze\_phi} or \texttt{--freeze\_z}, \texttt{adv\_all\_bo\_equal.py}, \texttt{adv\_all\_stages\_abl.py}). (i) Two single-block arms start from exactly the point at which the reported stage starts, under its constraints, admissibility rules, evaluation accounting, budgets and stopping rules: arm Z freezes the physical parameters and runs the BFGS blocks alone (classical gradient only); arm $\Theta$ freezes $(\bbeta,b,\tau)$ and runs the natural-gradient rounds alone. (ii) An equal-budget black-box control, BO$_{\rm eq}$, continues the protocol's trust-region Bayesian optimization from the same point for exactly the number of trajectory evaluations the reported stage used (\rCtrlEvalsC), with the same training-loss objective and box, no stall stop (a collapsed trust region is re-opened around the incumbent) and the held-out NRMSE evaluated once at the end. (iii) The round records of the reported arm give a sequential split of its $\log$-loss decrease into the two blocks. No held-out information enters any arm or control; the arms Z and $\Theta$ used \rCtrlEvalsZ{} and \rCtrlEvalsTheta{} trajectory evaluations. At $m=24$ and $30$ the test is run twice: from the start of the final all-parameter stage (upper block of Table~\ref{tab:ablation}) and from the BO optimum of the controls-and-drive-vector search, which is the start of the stage reported in Table~\ref{tab:stage} (lower block: Z$'$ = BFGS on $\bm{\beta}$ with the controls frozen, BO$'_{\rm eq}$ = trust-region BO on the controls and $\bm{\beta}$ for the evaluation count of that stage, QFI$'$ = the QFI stage of that search).

\emph{Reading.} Held-out NRMSE of the combined stage divided by that of the better classical control (Z or BO$_{\rm eq}$): \rCtrlRatioBestList. At $m=12$, $16$ and $18$ the combined stage ends \rCtrlBelowMid{} below the better control (\rCtrlBestMid) and below each single block: C/Z $=$ \rCtrlRatioZList; C/$\Theta$ $=$ \rCtrlRatioThetaList; C/BO$_{\rm eq}$ $=$ \rCtrlRatioBOList. Held-out gains over the start, Z / $\Theta$ / BO$_{\rm eq}$ / C: \rCtrlGainsMid. The natural-gradient block alone recovers \rThetaRecoveryMid{} of the combined gain at these $m$; the two blocks together exceed either one, and the sequential split attributes \rSeqQngShareMidRange\% of the combined arm's $\log$-loss decrease to the natural-gradient steps at $m=12$--$18$ (\rSeqQngShares; Table~\ref{tab:ablation-share}). At $m\le8$ the better control ends \rCtrlAboveSmall{} \emph{below} the combined stage (\rCtrlBestSmall; gains Z / $\Theta$ / BO$_{\rm eq}$ / C: \rCtrlGainsSmall): the stage gain at these $m$ is matched or exceeded by classical search from the same point, and the photon cap is active at the Z, $\Theta$ and C end points at $m=$ \rCtrlPhotonMs{} ($\max_t\langle n\rangle=3.000$; the equal-budget control ends at \rCtrlPhotonBO). At $m=24$ and $30$ the start of the upper block is the end point of the controls-and-drive-vector search, whose QFI stage had already refined the controls, so the upper block measures only the incremental all-parameter stage (training-loss reduction \rBlockOneCLossGainA\% and \rBlockOneCLossGainB\%, against \rMidCLossGainRange\% at $m=12$--$18$; C/best $=\rBlockOneRatioBestA$ and $\rBlockOneRatioBestB$; gains Z / $\Theta$ / BO$_{\rm eq}$ / C: \rCtrlGainsLargeBlockOne) and does not test the stage of Table~\ref{tab:stage}. The lower block does. From the BO optimum of that search (held-out $\rPrevStartHeldA$ at $m=24$, $\rPrevStartHeldB$ at $m=30$) the classical-gradient arm Z$'$ reaches $\rPrevZpHeldA$ at $m=24$, the same held-out NRMSE as the QFI stage of that search ($\rPrevQfiHeldA$; training loss $\rPrevZpLossA$ against $\rPrevQfiLossA$), so the natural-gradient and BFGS contributions are not separable by this test at $m=24$; the equal-budget control reaches $\rPrevBOHeldA$ (QFI$'$/BO$'_{\rm eq}=\rPrevRatioQfiBOA$), and the final all-parameter point ends $\rPrevCBelowA\%$ below the better control (C/best $=\rPrevRatioCBestA$). At $m=30$ Z$'$ stops at $\rPrevZpHeldB$ and the equal-budget control at $\rPrevBOHeldB$, against $\rPrevQfiHeldB$ for the QFI stage (QFI$'$/Z$'=\rPrevRatioQfiZpB$, QFI$'$/BO$'_{\rm eq}=\rPrevRatioQfiBOB$; C/best $=\rPrevRatioCBestB$, $\rPrevCBelowB\%$ below). The $\Theta$ arms at $m=$ \rCtrlThetaBudgetMs{} stopped at their wall budget and are upper bounds on what the natural-gradient block alone reaches. Fig.~6d of the main text plots, per $m$, the held-out reduction of the better classical control from the same start as the plotted stage gain (upper block at $m\le18$, lower block at $m=24$ and $30$).

\begin{table}[!ht]\centering\scriptsize
\caption{\textbf{Controls for the stage test: single-block ablation arms and an equal-budget black-box control against the reported QFI stage (held-out NRMSE; gain over the start point in brackets).} All arms start from the point at which the reported QFI stage starts (the grid + BO optimum; training loss / held-out NRMSE in the `start' column) under the stage's constraints, admissibility rules, evaluation accounting, budgets and stopping rules. Z: physical parameters frozen, BFGS with the exact adjoint gradient on $(\bm{\beta},b,\tau)$ (classical-gradient only). $\Theta$: $(\bm{\beta},b,\tau)$ frozen, natural-gradient rounds on the physical parameters (QNG only). BO$_{\rm eq}$: trust-region Bayesian optimization of the protocol's second phase continued from the same point for exactly the number of trajectory evaluations the reported stage used (training-loss objective; held-out NRMSE evaluated once at the end). C: the reported combined stage. Ratios are held-out NRMSE of C over that of the control ($<1$: the combined stage ends lower); `best' is the lower of the two classical controls Z and BO$_{\rm eq}$. Lower block ($m=24$, $30$): the same test from the BO optimum of the controls-and-drive-vector search, the start of the stage reported in Table~\ref{tab:stage}: Z$'$ = BFGS on $\bm{\beta}$ with the controls frozen, BO$'_{\rm eq}$ = trust-region BO on the controls and $\bm{\beta}$ for the evaluation count of that stage, QFI$'$ = the QFI stage of that search (natural gradient on the controls alternating with BFGS on $\bm{\beta}$), C = the final all-parameter end point. Selection and stopping use the training loss only.}
\label{tab:ablation}
\setlength{\tabcolsep}{3.5pt}
\resizebox{\textwidth}{!}{\begin{tabular}{@{}rlrrrrrrrrr@{}}\toprule
$m$ & $N$, $V$ & start: loss / held-out & Z & $\Theta$ & BO$_{\rm eq}$ & C (reported) & C/Z & C/$\Theta$ & C/BO$_{\rm eq}$ & C/best\\\midrule
4 & 1, 1 & 1.65031 / 0.1781 & 0.1737 (+2.5\%) & 0.1760 (+1.2\%) & 0.1720 (+3.4\%) & \textbf{0.1760} (+1.2\%) & 1.014 & 1.000 & 1.024 & 1.024\\
6 & 2, 1 & 1.47667 / 0.1673 & 0.1633 (+2.4\%) & 0.1619 (+3.2\%) & 0.1575 (+5.9\%) & \textbf{0.1616} (+3.4\%) & 0.989 & 0.998 & 1.026 & 1.026\\
8 & 1, 2 & 0.91567 / 0.1345 & 0.1300 (+3.4\%) & 0.1344 (+0.0\%) & 0.1136 (+15.5\%) & \textbf{0.1269} (+5.7\%) & 0.976 & 0.944 & 1.117 & 1.117\\
12 & 2, 2 & 0.39535 / 0.0858 & 0.0711 (+17.2\%) & 0.0791 (+7.9\%) & 0.0702 (+18.2\%) & \textbf{0.0608} (+29.1\%) & 0.856 & 0.769 & 0.867 & 0.867\\
16 & 3, 2 & 0.22108 / 0.0658 & 0.0533 (+19.0\%) & 0.0577 (+12.3\%) & 0.0485 (+26.3\%) & \textbf{0.0400} (+39.2\%) & 0.751 & 0.693 & 0.825 & 0.825\\
18 & 2, 3 & 0.33265 / 0.0806 & 0.0671 (+16.8\%) & 0.0561 (+30.5\%) & 0.0563 (+30.2\%) & \textbf{0.0449} (+44.3\%) & 0.669 & 0.800 & 0.797 & 0.797\\
24 & 2, 4 & 0.08925 / 0.0416 & 0.0403 (+3.3\%) & 0.0413 (+0.8\%) & 0.0416 (+0.0\%) & \textbf{0.0394} (+5.5\%) & 0.977 & 0.953 & 0.945 & 0.977\\
30 & 2, 5 & 0.09258 / 0.0417 & 0.0405 (+2.8\%) & 0.0417 (+0.0\%) & 0.0417 (+0.0\%) & \textbf{0.0397} (+4.8\%) & 0.980 & 0.952 & 0.952 & 0.980\\
\midrule
$m$ & $N$, $V$ & start: previous BO optimum & Z$'$ & QFI$'$ & BO$'_{\rm eq}$ & C (final) & QFI$'$/Z$'$ & QFI$'$/BO$'_{\rm eq}$ & C/Z$'$ & C/best\\\midrule
24 & 2, 4 & 0.19511 / 0.0636 & 0.0417 (+34.5\%) & 0.0416 (+34.6\%) & 0.0599 (+5.9\%) & \textbf{0.0394} (+38.1\%) & 1.000 & 0.696 & 0.945 & 0.945\\
30 & 2, 5 & 0.33450 / 0.0816 & 0.0766 (+6.2\%) & 0.0417 (+49.0\%) & 0.0717 (+12.2\%) & \textbf{0.0397} (+51.4\%) & 0.544 & 0.581 & 0.518 & 0.553\\
\bottomrule\end{tabular}}
\end{table}

\begin{table}[!ht]\centering\scriptsize
\caption{\textbf{Controls for the stage test: end-point training loss, stop reason, rounds, accepted steps, trajectory evaluations and wall time of every arm.} Arms as in Table~\ref{tab:ablation}. Stop: `conv.' = three consecutive rounds with a relative training-loss decrease below $10^{-4}$; `stalled' = two rounds without an accepted step; `budget' = wall budget reached (6~h for $m\le18$, 12~h for $m\ge24$, extended by one budget while the last round's relative decrease exceeded $10^{-3}$, at most 18~h); `evals' = the control's evaluation count reached. Steps = accepted natural-gradient steps on the physical parameters / accepted BFGS steps on $(\bm{\beta},b,\tau)$; for BO$_{\rm eq}$ the number of trust-region restarts. Evaluations = serial-equivalent trajectory count of the protocol's accounting; wall times of the parallel per-round code (2--13 workers) are not comparable across arms. QFI$'$: the reported stage of the controls-and-drive-vector search; its evaluation count set the budget of BO$'_{\rm eq}$.}
\label{tab:ablation-runs}
\setlength{\tabcolsep}{3pt}
\resizebox{\textwidth}{!}{\begin{tabular}{@{}rlrlrrrrrlrrrrrlrrrrrlrrrr@{}}\toprule
& & \multicolumn{6}{c}{Z} & \multicolumn{6}{c}{$\Theta$} & \multicolumn{6}{c}{BO$_{\rm eq}$} & \multicolumn{6}{c}{C (reported)}\\
\cmidrule(lr){3-8}\cmidrule(lr){9-14}\cmidrule(lr){15-20}\cmidrule(lr){21-26}
$m$ & $N$, $V$ & loss & stop & rounds & steps & evals & wall (h) & loss & stop & rounds & steps & evals & wall (h) & loss & stop & rounds & steps & evals & wall (h) & loss & stop & rounds & steps & evals & wall (h)\\\midrule
4 & 1, 1 & 1.56317 & stalled & 3 & 0/8 & 138 & 0.2 & 1.61426 & conv. & 5 & 10/0 & 291 & 0.4 & 1.52705 & evals & -- & 4 rst. & 371 & 0.3 & 1.61430 & conv. & 4 & 4/8 & 371 & 0.6\\
6 & 2, 1 & 1.40658 & conv. & 6 & 0/21 & 461 & 1.6 & 1.39158 & conv. & 6 & 16/0 & 423 & 0.6 & 1.33421 & evals & -- & 5 rst. & 614 & 1.2 & 1.38583 & conv. & 5 & 6/16 & 614 & 2.6\\
8 & 1, 2 & 0.86261 & conv. & 10 & 0/36 & 242 & 0.9 & 0.91452 & conv. & 5 & 15/0 & 359 & 0.6 & 0.64010 & evals & -- & 19 rst. & 2754 & 3.5 & 0.81672 & conv. & 39 & 117/78 & 2754 & 9.9\\
12 & 2, 2 & 0.25471 & conv. & 22 & 0/118 & 709 & 5.0 & 0.31529 & budget & 81 & 241/0 & 3828 & 6.0 & 0.24273 & evals & -- & 20 rst. & 3255 & 5.2 & 0.17325 & budget & 47 & 137/146 & 3255 & 18.1\\
16 & 3, 2 & 0.13490 & conv. & 8 & 0/40 & 243 & 4.0 & 0.16621 & budget & 53 & 157/0 & 2781 & 12.0 & 0.12717 & evals & -- & 5 rst. & 1176 & 5.5 & 0.07466 & budget & 15 & 44/71 & 1176 & 18.1\\
18 & 2, 3 & 0.22869 & conv. & 7 & 0/39 & 275 & 2.4 & 0.16180 & budget & 59 & 172/0 & 2429 & 6.0 & 0.16082 & evals & -- & 6 rst. & 1618 & 4.1 & 0.09856 & budget & 28 & 82/49 & 1618 & 12.0\\
24 & 2, 4 & 0.08047 & conv. & 6 & 0/26 & 241 & 1.6 & 0.08846 & conv. & 46 & 138/0 & 2549 & 5.4 & 0.08925 & evals & -- & 7 rst. & 1416 & 4.1 & 0.07764 & budget & 21 & 62/27 & 1416 & 12.2\\
30 & 2, 5 & 0.08400 & conv. & 5 & 0/23 & 275 & 1.8 & 0.09257 & conv. & 3 & 9/0 & 209 & 0.7 & 0.09258 & evals & -- & 4 rst. & 970 & 3.2 & 0.07941 & conv. & 9 & 27/40 & 970 & 8.6\\
\midrule
& & \multicolumn{6}{c}{Z$'$} & \multicolumn{6}{c}{QFI$'$} & \multicolumn{6}{c}{BO$'_{\rm eq}$} & \multicolumn{6}{c}{C (final)}\\
\cmidrule(lr){3-8}\cmidrule(lr){9-14}\cmidrule(lr){15-20}\cmidrule(lr){21-26}
24 & 2, 4 & 0.09376 & conv. & 17 & 0/102 & 862 & 5.3 & 0.08925 & rep. & -- & -- & 1080 & -- & 0.17098 & evals & -- & 6 rst. & 1080 & 2.3 & 0.07764 & budget & 21 & 62/27 & 1416 & 12.2\\
30 & 2, 5 & 0.29410 & conv. & 9 & 0/42 & 418 & 2.7 & 0.09258 & rep. & -- & -- & 951 & -- & 0.24749 & evals & -- & 3 rst. & 951 & 2.1 & 0.07941 & conv. & 9 & 27/40 & 970 & 8.6\\
\bottomrule\end{tabular}}
\end{table}

\begin{table}[!ht]\centering\scriptsize
\caption{\textbf{Sequential attribution of the reported QFI stage's training-loss decrease to its two blocks.} For every round of the reported arm C the decrease of $\log L$ is split into the part achieved by the accepted natural-gradient steps on the physical parameters (QNG) and the part achieved by the BFGS block on $(\bm{\beta},b,\tau)$ that followed; shares are relative to the total $\log(L_{\rm start}/L_{\rm end})$ of the arm. The split is sequential (each block acts on the point the other left), so it is an attribution, not an independent measurement; the ablation of Table~\ref{tab:ablation} is.}
\label{tab:ablation-share}
\begin{tabular}{@{}rlrrrrrr@{}}\toprule
$m$ & $N$, $V$, $N_c$ & rounds & accepted QNG steps & accepted BFGS steps & $\log(L_{\rm start}/L_{\rm end})$ & QNG share & BFGS share\\\midrule
4 & 1, 1, 10 & 4 & 4 & 8 & 0.022 & 93.9\% & 6.1\%\\
6 & 2, 1, 10 & 5 & 6 & 16 & 0.063 & 88.7\% & 11.3\%\\
8 & 1, 2, 10 & 39 & 117 & 78 & 0.114 & 23.3\% & 76.7\%\\
12 & 2, 2, 10 & 47 & 137 & 146 & 0.825 & 34.5\% & 65.5\%\\
16 & 3, 2, 8 & 15 & 44 & 71 & 1.086 & 43.7\% & 56.3\%\\
18 & 2, 3, 10 & 28 & 82 & 49 & 1.216 & 46.6\% & 53.4\%\\
24 & 2, 4, 10 & 21 & 62 & 27 & 0.139 & 9.7\% & 90.3\%\\
30 & 2, 5, 10 & 9 & 27 & 40 & 0.153 & 29.6\% & 70.4\%\\
\bottomrule\end{tabular}
\end{table}

\subsection*{Sensitivity of the comparison}
Five checks of the comparison (\texttt{code/\allowbreak scripts/\allowbreak adv\_perf\_esn\_ext.py}, \texttt{adv\_ridge\_tune.py}, \texttt{adv\_bootstrap.py}, \texttt{adv\_fresh\_segments.py}; Tables~\ref{tab:esn-ext}--\ref{tab:fresh}). Every selection or tuning decision below uses the 2,000 training rows only.

\emph{Extended hyperparameter grid} (Table~\ref{tab:esn-ext}). Against the original \rEsnNSettingsOrig-setting grid, the \rEsnNSettings-setting grid lowers the draw-averaged ESN held-out NRMSE by \rExtEsnChanges{} and leaves it unchanged at $m=$ \rExtEsnUnchangedMs{} (at $m=30$ the original setting is selected again; at $m=24$ a different setting with a mean $0.09\%$ lower); the stage-1 rank of the original selected setting on the extended grid is \rExtRankOrig. The reservoir's margin over the ESN mean becomes \rExtGainRange\% (\rOrigGainRange\% on the original grid) and the share of draws worse than the reservoir stays at least \rExtShareFloor\% (\rOrigShareFloor\%); the lowest single draw among the 10,000 lies below the reservoir at \rExtMinBelowMs. Table~\ref{tab:benchmark-matched}, Table~\ref{tab:stage}, the tables below and Fig.~6c,d of the main text use the extended grid.

\emph{Ridge on both sides} (Table~\ref{tab:ridge}). Training-internal validation picks $\delta=10^{-10}$, the smallest value offered, for the reservoir readout at \rRidgeQrcTen{} $m$ (\rRidgeQrcOther, where the validation curve is flat to four digits below $10^{-6}$) and for \rRidgeEsnShare\% of the ESN draws at every $m$; the tuned held-out values differ from the fixed-$\delta$ values by at most \rRidgeMaxChange{} and the readout norms do not move. The validation NRMSE rises monotonically above $\delta\approx10^{-6}$ on both sides, so a finer ladder would not find an interior optimum; the large readout norms are a conditioning property of the feature sets and do not cost held-out accuracy. The shot counts $n^\star$ of Table~\ref{tab:benchmark-shots} therefore cannot be lowered by regularizing the readout under the training-only rule.

\emph{Training-selected draw} (Table~\ref{tab:esn-selected}). The draw with the lowest training loss among the 10,000 draws of the selected setting ranks \rSelRankRange{} of 10,000 on held-out data ($r=$ \rSelRRange{} between training loss and held-out NRMSE across the draws), so the selection is effective and the draw is a legitimate deployed network, chosen with 10,000 training runs. Against it the reservoir loses at \rSelBeat, ties at \rSelTie{} and wins at \rSelLose{} (margins $1-{\rm NRMSE}_{\rm QRC}/{\rm NRMSE}_{\rm draw}$; paired bootstrap intervals in Table~\ref{tab:esn-selected}); on the original grid the margins were \rSelOrig. The reservoir's advantage therefore holds against the draw-averaged network at every $m$ from $4$ to $30$ and against the training-selected draw at $m=4$ and $m=12$--$24$; at $m=6$, $8$ and $30$ the training-selected draw is within $15\%$ of the reservoir and ahead of it or level.

\emph{Bootstrap} (Table~\ref{tab:bootstrap}). The 95\% moving-block intervals of the reservoir's held-out NRMSE have half-widths of \rBsHalfWidth\% of the point value, an effective held-out sample of order $10^2$. Because the intervals are paired on time blocks, the differences are much tighter: the stage gain is positive in \rBsSharePos, with intervals \rBsStageSmall, \rBsStageMid, and \rBsStageLarge{} for the final all-parameter stage over its own start. The margin over the ESN mean has intervals from $\rBsMarginMin$ ($m=\rBsMarginMinM$) to $\rBsMarginMax$ ($m=\rBsMarginMaxM$). These intervals describe the time-sampling uncertainty of the one held-out series under one optimization path; the fresh windows below are the independent check.

\emph{Fresh windows} (Table~\ref{tab:fresh}). With the design frozen and only the readout refitted, the reservoir's held-out NRMSE on five later windows of the Mackey--Glass trajectory lies \rFreshQrcRange{} from the manuscript value (SD over windows \rFreshQrcSd\% of the mean); the manuscript window is the worst or second-worst of the six at \rFreshKZeroWorst. The ESN mean on the same windows moves by up to \rFreshEsnMove\%. The margins over the ESN mean on the window means are \rFreshMargins{} for $m=4,\ldots,30$, and the stage gain persists on every window (\rFreshStageSmall{} at $m\le8$, \rFreshStageMid{} at $m=12$--$18$, \rFreshStageLarge{} for the final stage at $m=24$ and $30$). Two qualifications: the windows are later segments of the same deterministic trajectory, not independent realizations; and the $m=4$ and $6$ end points exceed the photon cap \rFreshExcluded{} ($\max_t\langle n\rangle=$ \rFreshMaxnRange), a consequence of their having been driven to the cap on the training series, so the admissibility statement of the protocol holds on the training series.

\begin{table}[!ht]\centering\scriptsize
\caption{\textbf{Extended against original ESN hyperparameter grid.} Original grid: 300 settings (radius $\{0.5,0.7,0.9,0.95,0.99,1.1\}$, leak $\{0.1,0.3,0.5,0.7,1.0\}$, input scale $\{0.1,0.3,0.5,1.0,2.0\}$, bias on/off); extended grid: 810 settings (Table~\ref{tab:benchmark-esn}), a strict superset, same draws, update rule, readout and split; both selected in two stages on the mean training loss (all settings $\times$ 1,000 draws, then the five leaders $\times$ 10,000 draws). Mean $\pm$ SD over the 10,000 draws of the selected setting; `worse' = share of the draws with a higher held-out NRMSE than the reservoir's QFI end point; $\Delta$ = margin of the reservoir over the ESN mean; rank = stage-1 rank of the original selected setting on the extended grid; edge as in Table~\ref{tab:benchmark-esn}. Table~\ref{tab:benchmark-matched}, Table~\ref{tab:stage} and Fig.~6c,d of the main text use the extended grid.}
\label{tab:esn-ext}
\setlength{\tabcolsep}{3.5pt}
\resizebox{\textwidth}{!}{\begin{tabular}{@{}rrlrrrlrrrrl@{}}\toprule
$m$ & QRC & original setting & mean $\pm$ SD & worse & $\Delta$ & extended setting & mean $\pm$ SD & worse & $\Delta$ & rank & edge\\\midrule
4 & 0.1760 & $(0.7, 1.0, 0.1, 0)$ & $0.2998\pm0.0617$ & 99.99\% & +41.3\% & $(0.7, 1.0, 0.02, 0)$ & $0.2516\pm0.0418$ & 100.00\% & +30.0\% & 32 & leak $=1.0$ (max)\\
6 & 0.1616 & $(0.5, 1.0, 0.1, 0)$ & $0.2273\pm0.0192$ & 100.00\% & +28.9\% & $(0.9, 1.0, 0.02, 0)$ & $0.2151\pm0.0287$ & 99.54\% & +24.9\% & 11 & leak $=1.0$ (max)\\
8 & 0.1269 & $(0.7, 1.0, 0.1, 0)$ & $0.1986\pm0.0196$ & 100.00\% & +36.1\% & $(0.9, 1.0, 0.05, 0)$ & $0.1921\pm0.0238$ & 99.97\% & +34.0\% & 6 & leak $=1.0$ (max)\\
12 & 0.0608 & $(0.9, 1.0, 0.1, 0)$ & $0.1634\pm0.0207$ & 100.00\% & +62.8\% & $(0.9, 1.0, 0.05, 0)$ & $0.1618\pm0.0193$ & 100.00\% & +62.4\% & 2 & leak $=1.0$ (max)\\
16 & 0.0400 & $(0.5, 0.3, 0.5, 0)$ & $0.1397\pm0.0184$ & 100.00\% & +71.4\% & $(0.2, 0.1, 1.0, 0)$ & $0.1307\pm0.0219$ & 100.00\% & +69.4\% & 2 & leak $=0.1$ (min)\\
18 & 0.0449 & $(0.5, 0.3, 0.5, 0)$ & $0.1225\pm0.0170$ & 100.00\% & +63.4\% & $(0.2, 0.1, 1.0, 0)$ & $0.1121\pm0.0173$ & 100.00\% & +60.0\% & 4 & leak $=0.1$ (min)\\
24 & 0.0394 & $(0.5, 0.3, 0.5, 0)$ & $0.0808\pm0.0127$ & 100.00\% & +51.3\% & $(0.3, 0.1, 1.0, 1)$ & $0.0808\pm0.0118$ & 100.00\% & +51.3\% & 1 & leak $=0.1$ (min)\\
30 & 0.0397 & $(0.7, 0.3, 0.5, 0)$ & $0.0611\pm0.0093$ & 99.93\% & +35.1\% & $(0.7, 0.3, 0.5, 0)$ (same) & $0.0611\pm0.0093$ & 99.93\% & +35.1\% & 1 & none\\
\bottomrule\end{tabular}}
\end{table}

\begin{table}[!ht]\centering\scriptsize
\caption{\textbf{Training-selected ESN draw.} Among the 10,000 draws of the selected setting, the draw with the lowest training loss (held-out data never used); its held-out NRMSE, its held-out rank among the 10,000, the mean held-out NRMSE of the ten best-by-training draws and the Pearson correlation $r$ between training loss and held-out NRMSE across the draws, for the extended and the original grid. Margin = $1-{\rm NRMSE}_{\rm QRC}/{\rm NRMSE}_{\rm draw}$ (negative: the selected draw beats the reservoir); the 95\% interval and the share of resamples in which the reservoir is ahead come from the paired moving-block bootstrap of Table~\ref{tab:bootstrap}. The selected draw is a best-of-10,000 statistic: the ESN mean describes the hyperparameter setting, the selected draw a deployed network chosen with 10,000 training runs.}
\label{tab:esn-selected}
\setlength{\tabcolsep}{3.5pt}
\resizebox{\textwidth}{!}{\begin{tabular}{@{}rrrrrrrrrrrr@{}}\toprule
& & \multicolumn{7}{c}{extended grid} & \multicolumn{3}{c}{original grid}\\\cmidrule(lr){3-9}\cmidrule(lr){10-12}
$m$ & QRC & ESN mean & selected draw & rank & 10-best mean & $r$ & margin [95\% CI] & QRC ahead & selected draw & rank & margin\\\midrule
4 & 0.1760 & 0.2516 & 0.1861 & 6 & 0.1872 & 0.953 & +5.4\% [+3.4\%, +7.2\%] & 100.0\% & 0.1731 & 1 & -1.7\%\\
6 & 0.1616 & 0.2151 & 0.1416 & 1 & 0.1490 & 0.972 & -14.2\% [-20.9\%, -9.7\%] & 0.0\% & 0.1749 & 1 & +7.6\%\\
8 & 0.1269 & 0.1921 & 0.1192 & 1 & 0.1287 & 0.978 & -6.4\% [-12.2\%, -4.7\%] & 0.0\% & 0.1424 & 2 & +10.9\%\\
12 & 0.0608 & 0.1618 & 0.0995 & 4 & 0.1027 & 0.982 & +38.9\% [+34.6\%, +41.3\%] & 100.0\% & 0.0918 & 1 & +33.7\%\\
16 & 0.0400 & 0.1307 & 0.0772 & 4 & 0.0775 & 0.962 & +48.2\% [+39.5\%, +52.1\%] & 100.0\% & 0.0729 & 2 & +45.1\%\\
18 & 0.0449 & 0.1121 & 0.0680 & 4 & 0.0718 & 0.951 & +34.0\% [+27.8\%, +39.8\%] & 100.0\% & 0.0615 & 1 & +27.1\%\\
24 & 0.0394 & 0.0808 & 0.0474 & 4 & 0.0507 & 0.946 & +17.0\% [+4.7\%, +28.5\%] & 99.9\% & 0.0507 & 4 & +22.4\%\\
30 & 0.0397 & 0.0611 & 0.0376 & 2 & 0.0395 & 0.971 & -5.5\% [-26.1\%, +9.8\%] & 22.8\% & 0.0376 & 2 & -5.5\%\\
\bottomrule\end{tabular}}
\end{table}

\begin{table}[!ht]\centering\scriptsize
\caption{\textbf{Ridge tuning on both sides by training-internal validation.} Fit on training rows 1--1600, validate on rows 1601--2000, $\delta\in\{10^{-10}, 10^{-8}, 10^{-6}, 10^{-4}, 10^{-2}, 10^{0}\}$ chosen by validation NRMSE, refit on all 2,000 rows, evaluated once on the 895 held-out rows. ESN: extended-grid selected setting, all 10,000 draws (share of draws choosing each $\delta$; held-out mean $\pm$ SD at the fixed $\delta=10^{-10}$ and at the tuned $\delta$). QRC: reported end point (validation NRMSE per $\delta$; chosen $\delta$; held-out NRMSE fixed $\to$ tuned). $\|w\|$ = mean readout norm, fixed $\to$ tuned.}
\label{tab:ridge}
\setlength{\tabcolsep}{3.5pt}
\resizebox{\textwidth}{!}{\begin{tabular}{@{}rrrlrlrrl@{}}\toprule
& \multicolumn{4}{c}{ESN (10,000 draws)} & \multicolumn{4}{c}{QRC end point}\\\cmidrule(lr){2-5}\cmidrule(lr){6-9}
$m$ & fixed & tuned & share $\delta=10^{-10}/10^{-8}/10^{-6}/\ge10^{-4}$ & $\|w\|$ & validation NRMSE per $\delta$ & chosen $\delta$ & fixed $\to$ tuned & $\|w\|$\\\midrule
4 & $0.2516\pm0.0418$ & $0.2516\pm0.0418$ & 99.3 / 0.5 / 0.1 / 0.0\% & 44908 $\to$ 44900 & 0.1636, 0.1636, 0.1636, 0.1756, 0.3471, 0.7722 & $10^{-8}$ & 0.1760 $\to$ 0.1760 & 144 $\to$ 144\\
6 & $0.2151\pm0.0287$ & $0.2151\pm0.0288$ & 97.0 / 2.9 / 0.1 / 0.0\% & 39909 $\to$ 39858 & 0.1548, 0.1548, 0.1545, 0.1852, 0.3421, 0.6811 & $10^{-6}$ & 0.1616 $\to$ 0.1618 & 248 $\to$ 244\\
8 & $0.1921\pm0.0238$ & $0.1922\pm0.0238$ & 96.1 / 3.8 / 0.1 / 0.0\% & 25887 $\to$ 25854 & 0.1141, 0.1383, 0.2471, 0.3431, 0.4648, 0.7502 & $10^{-10}$ & 0.1269 $\to$ 0.1269 & 13600 $\to$ 13600\\
12 & $0.1618\pm0.0193$ & $0.1618\pm0.0193$ & 99.7 / 0.3 / 0.0 / 0.0\% & 34347 $\to$ 34342 & 0.0514, 0.0563, 0.1147, 0.1833, 0.2887, 0.5020 & $10^{-10}$ & 0.0608 $\to$ 0.0608 & 4510 $\to$ 4510\\
16 & $0.1307\pm0.0219$ & $0.1307\pm0.0219$ & 100.0 / 0.0 / 0.0 / 0.0\% & 26897 $\to$ 26893 & 0.0325, 0.0329, 0.0398, 0.1027, 0.1865, 0.4290 & $10^{-10}$ & 0.0400 $\to$ 0.0400 & 894 $\to$ 894\\
18 & $0.1121\pm0.0173$ & $0.1121\pm0.0173$ & 100.0 / 0.0 / 0.0 / 0.0\% & 30389 $\to$ 30389 & 0.0406, 0.0476, 0.0796, 0.1432, 0.2226, 0.4715 & $10^{-10}$ & 0.0449 $\to$ 0.0449 & 7827 $\to$ 7827\\
24 & $0.0808\pm0.0118$ & $0.0808\pm0.0118$ & 100.0 / 0.0 / 0.0 / 0.0\% & 20865 $\to$ 20865 & 0.0349, 0.0435, 0.0535, 0.0757, 0.2486, 0.4447 & $10^{-10}$ & 0.0394 $\to$ 0.0394 & 9150 $\to$ 9150\\
30 & $0.0611\pm0.0093$ & $0.0611\pm0.0093$ & 99.9 / 0.1 / 0.0 / 0.0\% & 11083 $\to$ 11082 & 0.0357, 0.0478, 0.0584, 0.0890, 0.2717, 0.4344 & $10^{-10}$ & 0.0397 $\to$ 0.0397 & 10351 $\to$ 10351\\
\bottomrule\end{tabular}}
\end{table}

\begin{table}[!ht]\centering\scriptsize
\caption{\textbf{Moving-block bootstrap of the held-out residuals: 95\% percentile intervals for the cells of Table~1 of the main text and for the paired stage gain.} Blocks of 50 consecutive held-out rows, resampled with replacement to the 895 held-out rows, 2,000 resamples; NRMSE of a resample $=\sqrt{\overline{r^2}}/\sigma_y$ over the resampled rows; the same block indices are used for every cell of a given $m$, so the ratios (stage gain, margins) are paired. Cells: QRC QFI end point and grid + BO start point (stored features, readout refit at $\delta=10^{-10}$); ESN mean = mean over the 10,000 draws of the per-draw resampled NRMSE (extended-grid setting); ESN selected draw = the training-selected draw of Table~\ref{tab:esn-selected}. `$>0$' = share of resamples with a positive stage gain. The share of ESN draws worse than the reservoir (Table~\ref{tab:stage}) is a statistic over draws, not over held-out rows, and is not resampled. The intervals describe the time-sampling uncertainty of the one held-out series; series-to-series variation is checked in Table~\ref{tab:fresh}. $^{\S}$: at $m=24$ and $30$ the grid + BO cell is the start point of the final all-parameter stage (the end point of the controls-and-drive-vector search, whose features are stored), not the grid + BO value of Table~\ref{tab:stage}; the stage-gain interval at these $m$ is that of the final stage over its own start.}
\label{tab:bootstrap}
\setlength{\tabcolsep}{3.5pt}
\resizebox{\textwidth}{!}{\begin{tabular}{@{}rllllrlll@{}}\toprule
$m$ & QRC QFI end point & grid + BO start & ESN mean & ESN selected draw & stage gain & $>0$ & $\Delta$ vs ESN mean & $\Delta$ vs selected draw\\\midrule
4 & 0.1760 [0.1620, 0.2005] & 0.1781 [0.1639, 0.2029] & 0.2516 [0.2332, 0.2826] & 0.1861 [0.1714, 0.2106] & +1.2\% [+1.1\%, +1.2\%] & 100.0\% & +30.0\% [+28.3\%, +31.5\%] & +5.4\% [+3.4\%, +7.2\%]\\
6 & 0.1616 [0.1492, 0.1842] & 0.1673 [0.1544, 0.1908] & 0.2151 [0.1996, 0.2416] & 0.1416 [0.1292, 0.1597] & +3.4\% [+2.9\%, +3.9\%] & 100.0\% & +24.9\% [+22.5\%, +26.7\%] & -14.2\% [-20.9\%, -9.7\%]\\
8 & 0.1269 [0.1173, 0.1420] & 0.1345 [0.1241, 0.1501] & 0.1921 [0.1780, 0.2170] & 0.1192 [0.1086, 0.1301] & +5.7\% [+4.4\%, +6.6\%] & 100.0\% & +34.0\% [+31.9\%, +36.3\%] & -6.4\% [-12.2\%, -4.7\%]\\
12 & 0.0608 [0.0547, 0.0695] & 0.0858 [0.0794, 0.0940] & 0.1618 [0.1496, 0.1829] & 0.0995 [0.0913, 0.1088] & +29.1\% [+24.2\%, +33.1\%] & 100.0\% & +62.4\% [+60.5\%, +64.9\%] & +38.9\% [+34.6\%, +41.3\%]\\
16 & 0.0400 [0.0370, 0.0453] & 0.0658 [0.0608, 0.0751] & 0.1307 [0.1209, 0.1477] & 0.0772 [0.0692, 0.0839] & +39.2\% [+34.1\%, +44.4\%] & 100.0\% & +69.4\% [+66.3\%, +72.1\%] & +48.2\% [+39.5\%, +52.1\%]\\
18 & 0.0449 [0.0411, 0.0509] & 0.0806 [0.0749, 0.0907] & 0.1121 [0.1038, 0.1267] & 0.0680 [0.0618, 0.0777] & +44.3\% [+40.7\%, +47.3\%] & 100.0\% & +60.0\% [+56.2\%, +63.4\%] & +34.0\% [+27.8\%, +39.8\%]\\
24 & 0.0394 [0.0341, 0.0475] & 0.0416 [0.0366, 0.0495]$^{\S}$ & 0.0808 [0.0738, 0.0902] & 0.0474 [0.0421, 0.0563] & +5.5\% [+2.8\%, +8.6\%]$^{\S}$ & 100.0\% & +51.3\% [+44.7\%, +56.2\%] & +17.0\% [+4.7\%, +28.5\%]\\
30 & 0.0397 [0.0342, 0.0477] & 0.0417 [0.0367, 0.0491]$^{\S}$ & 0.0611 [0.0551, 0.0676] & 0.0376 [0.0339, 0.0426] & +4.8\% [+1.4\%, +8.0\%]$^{\S}$ & 99.7\% & +35.1\% [+25.2\%, +41.5\%] & -5.5\% [-26.1\%, +9.8\%]\\
\bottomrule\end{tabular}}
\end{table}

\begin{table}[!ht]\centering\scriptsize
\caption{\textbf{Fresh Mackey--Glass windows.} Five later 3,000-sample windows of the same trajectory (transient $1{,}000+500k$ samples discarded, $k=1\ldots5$; $k=0$ is the manuscript series), same split (100 washout, 2,000 training, 895 held-out). QRC: the reported end point of each $m$ and the chain's grid + BO start point, propagated once per window with the design frozen and only the readout refitted ($\delta=10^{-10}$); ESN: the extended-grid selected setting, 100 draws per window. Mean $\pm$ SD over the five windows (ESN: SD of the per-window means; the within-window SD over draws is given separately); $\Delta$ = margin of the reservoir over the ESN mean, both averaged over the retained windows; stage gain = gain of the QFI end point over the grid + BO point on the same means. $^{\P}$: windows on which the end point exceeds the photon cap $\max_t\langle n\rangle\le3$ (the $m=4$ and $6$ end points were driven to the cap on the training series) are excluded from the QRC means; $n$ = windows retained. $^{\S}$: at $m=24$ and $30$ the grid + BO point is the start of the final all-parameter stage, as in Table~\ref{tab:bootstrap}.}
\label{tab:fresh}
\setlength{\tabcolsep}{3.5pt}
\resizebox{\textwidth}{!}{\begin{tabular}{@{}rrlrlrlrrr@{}}\toprule
& \multicolumn{3}{c}{QRC QFI end point} & \multicolumn{2}{c}{grid + BO start} & \multicolumn{3}{c}{ESN} & \\\cmidrule(lr){2-4}\cmidrule(lr){5-6}\cmidrule(lr){7-9}
$m$ & $k=0$ & windows 1--5 ($n$) & $k=0$ & windows 1--5 & $k=0$ & windows 1--5 & within-window SD & $\Delta$ & stage gain\\\midrule
4 & 0.1760 & $0.1740\pm0.0011$ (3)$^{\P}$ & 0.1781 & $0.1761\pm0.0011$ & 0.2558 & $0.2558\pm0.0025$ & 0.0465 & +31.6\% & +1.2\%\\
6 & 0.1616 & $0.1595\pm0.0014$ (3)$^{\P}$ & 0.1673 & $0.1651\pm0.0012$ & 0.2163 & $0.2152\pm0.0018$ & 0.0228 & +25.5\% & +3.4\%\\
8 & 0.1269 & $0.1245\pm0.0006$ (5) & 0.1345 & $0.1320\pm0.0007$ & 0.1892 & $0.1874\pm0.0015$ & 0.0214 & +33.5\% & +5.7\%\\
12 & 0.0608 & $0.0588\pm0.0008$ (5) & 0.0858 & $0.0841\pm0.0007$ & 0.1622 & $0.1601\pm0.0012$ & 0.0179 & +63.3\% & +30.1\%\\
16 & 0.0400 & $0.0385\pm0.0009$ (5) & 0.0658 & $0.0650\pm0.0004$ & 0.1311 & $0.1291\pm0.0049$ & 0.0237 & +70.2\% & +40.9\%\\
18 & 0.0449 & $0.0435\pm0.0006$ (5) & 0.0806 & $0.0791\pm0.0006$ & 0.1124 & $0.1097\pm0.0046$ & 0.0181 & +60.4\% & +45.0\%\\
24 & 0.0394 & $0.0387\pm0.0005$ (5) & 0.0416$^{\S}$ & $0.0411\pm0.0007$ & 0.0798 & $0.0764\pm0.0045$ & 0.0122 & +49.3\% & +5.8\%\\
30 & 0.0397 & $0.0389\pm0.0006$ (5) & 0.0417$^{\S}$ & $0.0411\pm0.0007$ & 0.0612 & $0.0578\pm0.0018$ & 0.0104 & +32.6\% & +5.2\%\\
\bottomrule\end{tabular}}
\end{table}

\subsection*{Saturation of the black-box stages}
Two observations show that the QFI stage did not merely continue a search that had been stopped early. First, every Bayesian stage ended on its stall rule (no relative improvement above $10^{-3}$ in 40 consecutive evaluations) or, at $m=8$, on a collapsed trust region. Second, a wide scrambled-Sobol search over the full joint box of the \rSecNcfg{} reported configurations with $m\ge\rSecMmin$ (log-uniform rates, uniform detunings, $b$, sampling-time fractions and a structured drive vector; not part of the protocol) evaluated \rSecPoints{} points per configuration and stayed far above the chain optima: its best admissible training loss is \rSecRatioBo{} times the grid + BO loss and \rSecRatioQfi{} times the loss at the QFI end point (Table~\ref{tab:benchmark-secondary}). The black-box stages had therefore located the operating region, and the natural-gradient stage moved inside it: the physical parameters moved by \rPhiMove\% of their norm (Table~\ref{tab:benchmark-constraints}).

\begin{table}[!ht]\centering\scriptsize
\caption{\textbf{Secondary wide random search (Sobol points over the full parameter box, all parameters).} Points evaluated so far, admissible points (photon cap and stability), best admissible training loss, against the training loss of the chain's grid + BO optimum and of the QFI end point (same loss $\tfrac12\|r\|^2+\tfrac12\delta\|w\|^2$ on the 2,000 training rows; the ratio search/chain in brackets).  The search is not part of the protocol; it tests whether the black-box stages missed a better basin; point counts are those at the close of the study. Configurations without a row were not given a secondary search.}
\label{tab:benchmark-secondary}
\begin{tabular}{@{}rlrrrrr@{}}\toprule
$m$ & layout & points & admissible & best loss (search) & loss grid + BO & loss QFI end point\\\midrule
12 & 2, $V=2$ & 14,013 & 6,943 & 0.5423 & 0.3954 ($\times1.4$) & 0.1732 ($\times3.1$)\\
16 & 3, $V=2$ & 18,698 & 10,147 & 0.3808 & 0.2211 ($\times1.7$) & 0.0747 ($\times5.1$)\\
18 & 2, $V=3$ & 16,765 & 6,954 & 0.4252 & 0.3327 ($\times1.3$) & 0.0986 ($\times4.3$)\\
24 & 2, $V=4$ & 13,872 & 4,837 & 0.3547 & 0.1951 ($\times1.8$) & 0.0776 ($\times4.6$)\\
30 & 2, $V=5$ & 16,506 & 3,745 & 0.3984 & 0.3345 ($\times1.2$) & 0.0794 ($\times5.0$)\\
\bottomrule\end{tabular}
\end{table}

\begin{table}[!ht]\centering\scriptsize
\caption{\textbf{QFI natural-gradient stage: budgets, box widening and constraints at the end point (all configurations).} Rounds = completed rounds of (three natural-gradient steps on the physical parameters, eight BFGS steps on $(\bm{\beta},b,\tau)$); wall = elapsed time of the arm; `last rel.' = relative training-loss decrease of the last round (extension rule: continue while $>10^{-3}$); widened = parameters whose box was widened after an end point sat on a face ($\times2$ for rates, $\pm40$ for detunings); $\max\langle n\rangle$ against the cap 3; $|\tilde u|_{\max}$ against the cap 10; $\|\Delta\phi\|/\|\phi_0\|$ = relative move of the physical parameters from the BO optimum.}
\label{tab:benchmark-constraints}
\begin{tabular}{@{}rlrrrlrrrl@{}}\toprule
$m$ & layout & rounds & wall (h) & last rel. & widened & $\max\langle n\rangle$ & $|\tilde u|_{\max}$ & $\|\Delta\phi\|/\|\phi_0\|$ & status\\\midrule
4 & 1, $V=1$ & 4 & 0.6 & $0$ & none & 3.000 & 5.85 & 0.016 & converged\\
6 & 2, $V=1$ & 5 & 2.6 & $0$ & none & 3.000 & 6.53 & 0.057 & converged\\
8 & 1, $V=2$ & 39 & 9.9 & $2.2\times10^{-5}$ & $\gamma_1\in[0.0125,24]$ & 0.168 & 1.12 & 0.020 & converged\\
12 & 2, $V=2$ & 47 & 18.1 & $2.7\times10^{-4}$ & none & 0.464 & 2.60 & 0.164 & budget\\
12 & 1, $V=3$ & 51 & 12.0 & $1.6\times10^{-4}$ & $\Delta_c\in[1,160]$, $g_1\in[2,128]$ & 0.305 & 3.73 & 0.128 & budget\\
16 & 3, $V=2$ & 15 & 18.1 & $5.6\times10^{-3}$ & none & 0.801 & 1.73 & 0.131 & budget\\
16 & 1, $V=4$ & 44 & 12.1 & $1.0\times10^{-4}$ & $\Delta_c\in[1,160]$ & 0.476 & 1.17 & 0.035 & budget\\
18 & 2, $V=3$ & 28 & 12.0 & $8.2\times10^{-5}$ & none & 1.009 & 3.24 & 0.259 & budget\\
24 & 2, $V=4$ & 21 & 12.2 & $5.6\times10^{-5}$ & none & 1.448 & 4.03 & 0.012 & budget\\
30 & 2, $V=5$ & 9 & 8.6 & $6.4\times10^{-5}$ & none & 0.962 & 2.47 & 0.031 & converged\\
\bottomrule\end{tabular}
\end{table}

\begin{table}[!ht]\centering\scriptsize
\caption{\textbf{Sensitivity screen.} For every physical parameter of every configuration, the relative change of the training loss over an admissible one-parameter scan around the reference point ($\delta_{\rm rel}$; `adm.' = number of admissible scan points). A parameter would have been dropped only if $\delta_{\rm rel}<0.01$ in every configuration with the same atom number (scans with fewer than two admissible points count as `keep'); no parameter met this condition, so all $P$ physical parameters (plus $b$, the $V-1$ interior sampling times and the $m$ drive gains) entered the grid, BO and QFI stages. `Single admissible point': parameters whose scan left the reference point as the only admissible point under the photon cap, kept by the rule.}
\label{tab:benchmark-sensitivity}
\begin{tabular}{@{}rlrrlll@{}}\toprule
$m$ & layout & $P$ & dropped & $\min\delta_{\rm rel}$ (parameter) & $\max\delta_{\rm rel}$ (parameter) & single admissible point\\\midrule
4 & 1, $V=1$ & 6 & 0 & 0.60 ($\gamma_1$) & 8.5 ($\epsilon$) & $g_1$, $\Delta_{a,1}$\\
6 & 2, $V=1$ & 9 & 0 & 0.01 ($\gamma_2$) & 7.8 ($\epsilon$) & $\Delta_{a,2}$, $g_1$, $g_2$, $\Delta_{a,1}$\\
8 & 1, $V=2$ & 6 & 0 & 0.20 ($\gamma_1$) & 13.7 ($\Delta_{a,1}$) & --\\
12 & 2, $V=2$ & 9 & 0 & 0.33 ($\gamma_2$) & 28.9 ($\Delta_{a,1}$) & --\\
12 & 1, $V=3$ & 6 & 0 & 0.50 ($\epsilon$) & 20.7 ($\Delta_{a,1}$) & --\\
16 & 3, $V=2$ & 12 & 0 & 0.60 ($\gamma_2$) & 8.4 ($\Delta_{a,2}$) & --\\
16 & 1, $V=4$ & 6 & 0 & 1.08 ($\Delta_c$) & 28.0 ($\Delta_{a,1}$) & --\\
18 & 2, $V=3$ & 9 & 0 & 1.32 ($\kappa_c$) & 9.3 ($\Delta_{a,1}$) & --\\
24 & 2, $V=4$ & 9 & 0 & 0.83 ($\Delta_c$) & 47.4 ($\kappa_c$) & $\Delta_{a,2}$, $\Delta_{a,1}$\\
30 & 2, $V=5$ & 9 & 0 & 0.19 ($\kappa_c$) & 20.7 ($\gamma_2$) & $\Delta_{a,2}$, $g_1$, $\Delta_{a,1}$\\
\bottomrule\end{tabular}
\end{table}

\begin{table}[!ht]\centering\scriptsize
\caption{\textbf{Excluded parameters: checks at a two-atom, $V=3$ reference point ($N_c=8$, 700 samples).} Relative change of the training loss under the transformation; a pass confirms that the parameter is redundant with those retained and was therefore not optimised.}
\label{tab:benchmark-exclusions}
\begin{tabular}{@{}lp{0.56\textwidth}rl@{}}\toprule
excluded parameter / check & reason & rel. change of loss & pass\\\midrule
time step $dt$ & rescaling all rates by $s=2.5$ with $dt\to dt/s$ leaves every propagator unchanged (features agree to $4\times10^{-14}$); $dt$ is redundant with the overall rate scale & $5.5\times10^{-8}$ & yes\\
global sign flip & $(\Delta_c,\Delta_{a,i},g_i)\to-(\Delta_c,\Delta_{a,i},g_i)$ is the complex-conjugation symmetry of the Lindbladian; $p$ and $\sigma_{y,i}$ change sign, compensated by $\bm{\beta}$ and the readout, so $\Delta_c>0$ without loss of generality & $0$ & yes\\
individual $g_i\to-g_i$ & a $\pi$ rotation of atom $i$ about $z$; $\sigma_{x,i},\sigma_{y,i}$ change sign, compensated by $\bm{\beta}$ and the readout, so $g_i>0$ & $0$ & yes\\
quadrature rotation & a rotation (angle 0.7) of the measured pairs $(x,p)$, $(\sigma_{x,i},\sigma_{y,i})$ is a fixed invertible linear map of the feature vector; the readout span is invariant and $vR^{\top}$ reproduces the closed loop & $8.0\times10^{-9}$ & yes\\
initial state & the 100-sample washout erases the initial state (3 initial states compared) & $1.2\times10^{-7}$ & yes\\
drive frequency & fixed by the rotating frame; a change is a shift of the detunings $\Delta_c$, $\Delta_{a,i}$, which are optimised & -- & yes\\
gradients & adjoint $(\bm{\beta},b)$ and complex-step $\tau$ gradients against complex-step and central differences (relative) & $\le2.5\times10^{-6}$ & yes\\
\bottomrule\end{tabular}
\end{table}

\subsection*{Closed-loop checks of the end points}
The diagnostics of the drive-gain study (main-text Sec.~2.5, \ref{sec:feedback-si}) were applied to every end point (Supplementary Table~\ref{tab:loopcheck}). Every end point reproduces its stored training loss and held-out NRMSE. (i) \emph{Contraction.} The closed-loop tangent map of main-text Eq.~(16), generalized to the multiplexed drive with the loop state $(\rho_t,\bbeta^\top X_t)$ (Methods), has $\mathrm{median}_t|\lambda_1|=0.903$--$0.956$ at the eight reported end points; individual sampled times reach $1.020$ ($m=16$) and $1.015$ ($m=24$). By the criterion of Sec.~2.5, $\mathrm{median}_t|\lambda_1|=1$, every design lies before the dynamical edge. The drive moves the loop towards the edge: the closed-loop median exceeds the open-loop one at every $m$ except $8$. For $V=1$ and $\bbeta=\beta e_p$ the implementation reproduces the single-channel computation of Sec.~2.5 to a relative $2\times10^{-8}$. (ii) \emph{Fading memory and start-up.} Trajectories started from three other states (a thermal cavity state with $\bar n=0.3$ and maximally mixed atoms, a coherent state with $\alpha=0.5$, and a random pure state with at most two photons) converge to the reported trajectory at every end point. With the washout extended to $500$ samples of the same Mackey--Glass trajectory, all eight designs reproduce their held-out NRMSE to within $0.04\%$ from all four starts, and every training and held-out row satisfies $|\tilde u_t|\le10$ and $\langle n\rangle\le3$ (largest values $3.24$ and $0.96$; Supplementary Table~\ref{tab:loopcheck}). At the standard $100$-sample washout, the designs at $m=4$, $6$, $8$, $16$, $18$ and $24$ already reproduce their held-out NRMSE to within $0.02\%$ from all four starts. With the drive gains ramped linearly from zero over the first $50$ input intervals, every design from $m=8$ to $m=30$ reproduces its held-out NRMSE to within $0.13\%$ at the standard washout from all four starts and satisfies both bounds at every step, washout included ($\max_t|\tilde u_t|=3.24$, $\max_t\langle n\rangle=1.56$). Code: \texttt{code/\allowbreak scripts/\allowbreak loop\_check.py}, \texttt{code/\allowbreak scripts/\allowbreak startup.py} and \texttt{code/\allowbreak analysis/\allowbreak loop\_check\_table.py}; data: \texttt{data/\allowbreak advantage/\allowbreak loop\_check/}.

\begin{table}[!ht]\centering\small\setlength{\tabcolsep}{4pt}
\caption{\textbf{Closed-loop checks of the reported end points.} Held-out NRMSE of each end point (main-text Table~1). $|\lambda_1|$: largest eigenvalue modulus of the closed-loop tangent map on traceless perturbations (Methods), at ten training times, with the loop closed and with $\bbeta$ decoupled (open). Four starts: held-out NRMSE with the readout refitted, for trajectories started from the standard initial state, a thermal cavity state ($\bar n=0.3$) with maximally mixed atoms, a coherent state ($\alpha=0.5$) and a random pure state with at most two photons, each preceded by 400 further samples of the same Mackey--Glass trajectory (washout $500$); $\max|\tilde u_t|$ and $\max\langle n\rangle$ over the training and held-out rows of these runs.}
\label{tab:loopcheck}
\begin{tabular}{@{}rlccccccc@{}}\toprule
& & & \multicolumn{3}{c}{$|\lambda_1|$} & \multicolumn{3}{c}{four starts, washout 500}\\\cmidrule(lr){4-6}\cmidrule(l){7-9}
$m$ & $N$, $V$ & held-out & median & max & open median & held-out & $\max|\tilde u_t|$ & $\max\langle n\rangle$\\\midrule
4 & 1, 1 & 0.1760 & 0.914 & 0.916 & 0.843 & 0.1760 & 0.69 & 0.08\\
6 & 2, 1 & 0.1616 & 0.921 & 0.927 & 0.894 & 0.1616 & 0.74 & 0.08\\
8 & 1, 2 & 0.1269 & 0.907 & 0.908 & 0.952 & 0.1269 & 0.81 & 0.00\\
12 & 2, 2 & 0.0608 & 0.912 & 0.923 & 0.850 & 0.0608 & 2.60 & 0.24\\
16 & 3, 2 & 0.0400 & 0.948 & 1.020 & 0.903 & 0.0400 & 1.73 & 0.80\\
18 & 2, 3 & 0.0449 & 0.956 & 0.993 & 0.840 & 0.0449 & 3.24 & 0.26\\
24 & 2, 4 & 0.0394 & 0.903 & 1.015 & 0.846 & 0.0394 & 2.10 & 0.94\\
30 & 2, 5 & 0.0397 & 0.930 & 0.946 & 0.860 & 0.0397 & 2.43 & 0.96\\
\bottomrule\end{tabular}
\end{table}

\subsection*{Cutoff check and quadratic readout}
Every finished end point was re-evaluated at $N_c+2$ with the same parameters and drive vector (Table~\ref{tab:benchmark-resources}). The held-out NRMSE changes by at most \rNcRest{} relatively in every case except $m=\rNcWorstM$ (\rNcWorst); \rNcPending. For the record, a $2m$-feature readout on $[X,X^2]$ at the same end points lowers the held-out NRMSE further at every finished $m$; the matched-cost comparison uses the linear readout on both sides.

\begin{table}[!ht]\centering\scriptsize
\caption{\textbf{Evaluations per stage, cutoff check and quadratic readout (reported layouts).} Grid = rough-grid trajectories (admissible in brackets); BO = evaluations of the Bayesian stage with its stop reason; QFI = trajectories of the natural-gradient stage. $N_c+2$: relative change of the held-out NRMSE when the end point is re-evaluated at $N_c+2$. Quadratic: held-out NRMSE of a $2m$-feature readout on $[X,X^2]$ (same ridge and split) at the same end point (reported for the record; not part of the matched-cost comparison). `--': closing tasks pending for a running arm.}
\label{tab:benchmark-resources}
\begin{tabular}{@{}rlrrlrrr@{}}\toprule
$m$ & layout & grid (adm.) & BO & BO stop & QFI evals & $N_c+2$ rel. change & quadratic\\\midrule
4 & 1, $V=1$ & 2,000 (915) & 114 & stalled & 371 & $1.6\times10^{-7}$ & 0.1714\\
6 & 2, $V=1$ & 4,000 (1,987) & 114 & stalled & 614 & $8.1\times10^{-6}$ & 0.1495\\
8 & 1, $V=2$ & 2,000 (748) & 343 & converged & 2754 & $7.0\times10^{-9}$ & 0.1212\\
12 & 2, $V=2$ & 4,000 (2,130) & 253 & stalled & 3255 & $2.1\times10^{-7}$ & 0.0507\\
16 & 3, $V=2$ & 2,500 (1,205) & 193 & stalled & 1176 & $8.1\times10^{-2}$ & 0.0361\\
18 & 2, $V=3$ & 4,000 (1,601) & 197 & stalled & 1618 & $1.6\times10^{-3}$ & 0.0375\\
24 & 2, $V=4$ & 4,000 (1,485) & 126 & stalled & 1416 & $3.8\times10^{-5}$ & 0.0296\\
30 & 2, $V=5$ & 4,000 (993) & 130 & stalled & 970 & $3.6\times10^{-5}$ & 0.0240\\
\bottomrule\end{tabular}
\end{table}

\subsection*{Finite-shot budget}
The shot model adds independent Gaussian noise of variance $\Var_t(O_j)/n$ to every feature and time sample, the diagonal of the multinomial covariance of a projective measurement of $O_j$, and either refits the readout on the noisy training features (median over ten noise seeds) or keeps the noiseless readout fixed, in which case the held-out NRMSE is $\sqrt{{\rm NRMSE}^2+S/(n\Var y)}$ with $S=\overline{\sum_j(\partial\hat y_t/\partial X_{tj})^2\Var_t(O_j)}$. Table~\ref{tab:benchmark-shots} gives both for the reported end points. At $10^4$, $10^5$ and $10^6$ shots per feature and time sample the refitted reservoir has held-out NRMSE \rShotFour, \rShotFive{} and \rShotSix, \rShotAbove. The margin over the ESN mean appears at $n^\star_{\rm refit}=\rNstarRefitMin$ ($m=\rNstarRefitMinM$) to $\rNstarRefitMax$ ($m=\rNstarRefitMaxM$) shots with the refitted readout, and at $\rNstarFixedMin$ to $\rNstarFixedMax$ shots with the fixed readout, the difference being the readout norm $\|w\|$ of $10^2$--$10^4$ that the fixed readout carries onto the noise. The advantage is therefore a statement about expectation values, or about shot counts of $\rNstarDecades$ per feature and time sample.

\begin{table}[!ht]\centering\scriptsize
\caption{\textbf{Finite-shot cost of the reported QRC end points (Mackey--Glass, held-out NRMSE).} Gaussian noise of variance $\Var_t(O_j)/n$ on every feature and time sample, readout refitted (ridge $\delta=10^{-10}$), median over ten noise seeds, for $n=10^4$, $10^5$, $10^6$ shots; $n^\star_{\rm refit}$ = shots at which the refitted median falls below the ESN mean (bisection in $\log_{10}n$ between decades); $n^\star_{\rm fixed}$ = fixed-readout shot count at which $\sqrt{{\rm NRMSE}^2+S/(n\,\Var y)}$ equals the ESN mean, $S=\overline{\sum_j(\partial\hat y_t/\partial X_{tj})^2\Var_t(O_j)}$; $\|w\|$ = readout norm.}
\label{tab:benchmark-shots}
\begin{tabular}{@{}rlrrrrrrr@{}}\toprule
$m$ & layout & noiseless & ESN mean & $10^4$ & $10^5$ & $10^6$ & $n^\star_{\rm refit}$ & $n^\star_{\rm fixed}$ ($\|w\|$)\\\midrule
4 & 1, $V=1$ & 0.1760 & 0.2516 & 0.631 & 0.425 & 0.329 & $7.2\times10^{6}$ & $1.1\times10^{7}$ (144)\\
6 & 2, $V=1$ & 0.1616 & 0.2151 & 0.561 & 0.404 & 0.326 & $3.4\times10^{7}$ & $5.3\times10^{7}$ (248)\\
8 & 1, $V=2$ & 0.1269 & 0.1921 & 0.710 & 0.583 & 0.403 & $2.1\times10^{10}$ & $1.5\times10^{11}$ (13600)\\
12 & 2, $V=2$ & 0.0608 & 0.1618 & 0.412 & 0.297 & 0.250 & $2.0\times10^{8}$ & $1.6\times10^{10}$ (4510)\\
16 & 3, $V=2$ & 0.0400 & 0.1307 & 0.413 & 0.262 & 0.166 & $6.6\times10^{6}$ & $1.2\times10^{9}$ (894)\\
18 & 2, $V=3$ & 0.0449 & 0.1121 & 0.362 & 0.232 & 0.198 & $9.0\times10^{8}$ & $1.0\times10^{11}$ (7828)\\
24 & 2, $V=4$ & 0.0394 & 0.0808 & 0.393 & 0.295 & 0.224 & $6.7\times10^{7}$ & $2.6\times10^{11}$ (9155)\\
30 & 2, $V=5$ & 0.0397 & 0.0611 & 0.394 & 0.312 & 0.254 & $1.6\times10^{9}$ & $7.6\times10^{11}$ (10350)\\
\bottomrule\end{tabular}
\end{table}

\subsection*{Caveats}
(i) The comparison uses noiseless expectation values; the finite-shot cost above bounds its reach, and the ridge check shows that the shot counts cannot be lowered by regularizing the readout. (ii) The QFI metric requires the density matrix and its parameter derivatives, which are available in simulation and at design time; the deployed reservoir uses only the measured features. (iii) The reservoir's physical parameters, input offset, sampling times and drive gains are optimized, whereas the ESN follows the standard convention of tuning four hyperparameters and averaging over random reservoir and input matrices; the readout cost is matched exactly, the reservoir beats at least $\rExtShareFloor\%$ of the individual draws, not only their mean, and it beats the network's training-selected draw at $m=4$ and $12$--$24$, but not at $m=6$ and $8$, and ties it at $m=30$ (subsection `Sensitivity of the comparison'). (iv) The QFI stages that stopped at their wall budget ($m=$ \rBudgetList, Table~\ref{tab:benchmark-constraints}) did so with last-round relative decreases of $6\times10^{-5}$--$3\times10^{-4}$, except the three-atom run at $m=16$, stopped at $18$ h with a last-round decrease of $6\times10^{-3}$; their end points are upper bounds on the reservoir's loss\rInterimCaveat. (v) The Bayesian and natural-gradient stages use the same admissibility rules and the same training-loss objective; the natural-gradient stage additionally uses exact gradients and the QFI metric, and the two controls of the subsection `Controls for the stage test' (exact-gradient descent on the classical parameters alone; black-box search for the stage's own evaluation count) place the stage \rCtrlBelowMidRange\% below the better control at $m=12$--$18$ and \rCtrlAboveSmallRange\% above it at $m\le8$, where the stage gain is within the reach of classical search. (vi) The held-out series is one 895-row window; the paired bootstrap and the five fresh windows bound its sampling uncertainty, and the fresh windows are later segments of the same trajectory, not independent realizations. Data: \texttt{data/\allowbreak advantage/\allowbreak performance\_all/}, \texttt{data/\allowbreak advantage/\allowbreak performance\_largem/}, \texttt{data/\allowbreak advantage/\allowbreak performance/}; scripts \texttt{code/\allowbreak scripts/\allowbreak adv\_all\_*.py}, \texttt{code/\allowbreak scripts/\allowbreak adv\_perf\_esn\_ext.py}, \texttt{adv\_ridge\_tune.py}, \texttt{adv\_bootstrap.py}, \texttt{adv\_fresh\_segments.py}, \texttt{code/\allowbreak analysis/\allowbreak build\_benchmark\_validation.py}.

\section{Alternative matching by physical elements}
\label{sec:altmatch}
Main-text Sec.~2.1 matches the echo-state network to the quantum reservoir by readout cost, the convention of the field (\ref{sec:baseline-survey}). This note reports an alternative matching at $m=30$, by physical elements. The $m=30$ reservoir has three dynamical elements, two atoms and one cavity mode, and is compared with networks of three coupled nodes read out through their three node states. Unlike the network of Sec.~2.1, every parameter of these networks is trained.

\emph{Networks.} The nodes update as $x_{t+1}=(1-a)x_t+a\tanh(Ax_t+B\tilde u_t+c)$ from $x_0=0$, with $\tilde u_t=u_t+b+\bm{k}^\top x_t$, $A\in\mathbb{R}^{3\times3}$, a leak $a\in[10^{-3},1]$ and an input offset $b$. Four variants are trained. In the first, the input enters every node ($B\in\mathbb{R}^3$, $c=0$) and gains $\bm{k}\in\mathbb{R}^3$ on the node states give the network an input conditioned on its own state, the state feedback of Ref.~\cite{SBEhlersNN2025}; this is an ad hoc addition to the echo-state network, included so that the network has every kind of freedom the quantum reservoir has ($9+3+1+1+3=17$ parameters). The second is the echo-state network in its original form, with node biases $c\in\mathbb{R}^3$ and no feedback ($\bm{k}=0$; again $17$ parameters). In the third and fourth, the input and offset enter only the node that stands for the cavity mode, $B=(B_1,0,0)^\top$, as in the reservoir, whose drive acts on the cavity and reaches the atoms through the coupling; the third has state feedback ($15$ parameters), the fourth has none ($12$). The readout is the three node states plus a bias (four weights, against $31$ for the reservoir), with the same ridge ($\delta=10^{-10}$), washout, split and admissibility bound $|\tilde u_t|\le10$ (imposed as a penalty) as the quantum reservoir.

\emph{Search.} For each variant, a quasi-random screen of $16{,}384$ parameter vectors ($8{,}192$ for the second) is followed by L-BFGS-B with exact gradients (JAX, double precision, at most $5{,}000$ iterations) from the $512$ ($256$) best screened points and $6{,}656$ ($6{,}912$) random starts, $7{,}168$ starts per variant and $28{,}672$ in all, on the ridge training loss of the quantum reservoir plus the penalty. A search counts as saturated when doubling the number of random starts lowers the best training loss by less than $10^{-3}$ relatively; all four searches meet this, each reaching its best training loss within the first half of its random starts. The four searches use $2.2\times10^{8}$ forward-pass equivalents ($4.8$--$6.6\times10^{7}$ per variant), against $5{,}100$ trajectories for the grid, BO and QFI-aware stages of the quantum reservoir (\ref{sec:matched}).

\emph{Selection and evaluation.} Two rules were fixed before the runs. Rule A takes the lowest training loss. Rule B ranks the 200 lowest training losses whose design has fading memory on the training inputs (trajectories from four initial states agree to $10^{-6}$ over the last 100 training rows) by a training-internal validation NRMSE (readout fitted on the first $1{,}600$ training rows, error on the last $400$). The selected design is evaluated once on the held-out rows, with a paired moving-block bootstrap (blocks of $50$, $2{,}000$ resamples) against the stored reservoir features, and on $20$ fresh windows, later segments of the same Mackey--Glass trajectory offset by multiples of $500$ samples, with the design frozen and the readout refitted on both sides.

\emph{Result} (Supplementary Table~\ref{tab:altmatch}). The reservoir is ahead of every variant under both rules, on the held-out rows and on each of the $20$ fresh windows. Under rule B the networks with input on every node reach held-out NRMSEs of $0.0685$ (state feedback) and $0.0673$ (node biases) against $0.0397$ for the reservoir, margins of $42.1\%$ ($95\%$ interval $33.2$--$50.3\%$) and $41.1\%$ ($31.5$--$49.1\%$); on the fresh windows they average $0.075\pm0.012$ and $0.086\pm0.031$, against $0.0385\pm0.0013$ for the reservoir. With the input on the cavity node only, as in the reservoir, both rules select the same design, at $0.1079$ with state feedback and $0.1085$ without, margins of $63.2\%$ and $63.5\%$ (intervals $56.6$--$68.4\%$ and $56.6$--$68.6\%$), and the fresh windows give $0.104\pm0.004$. For the networks with input on every node, rule A selects designs without fading memory, whose held-out errors ($0.0694$ and $0.0567$; margins $42.9\%$ and $30.0\%$) do not carry over to the fresh windows ($0.53\pm0.18$ and $0.39\pm0.23$); rule B was included for this case. The reservoir, with $44$ system parameters and $30$ readouts against $12$--$17$ and $3$, has a held-out error $30$--$64\%$ below that of every trained three-node network under either rule. Code: \texttt{code/\allowbreak matched\_esn/} (\texttt{esn\_model.py}, \texttt{run\_starts.py}, \texttt{collect.py}, \texttt{selftest.py}, \texttt{make\_si\_table.py}); results: \texttt{data/\allowbreak advantage/\allowbreak altmatch/}.

\begin{table}[!ht]\centering\scriptsize\setlength{\tabcolsep}{3pt}
\caption{\textbf{Alternative matching by physical elements ($m=30$, Mackey--Glass, five-step forecasting).} The quantum reservoir (two atoms and one cavity mode, $V=5$) against three-node networks, read out through their three node states, whose every parameter is trained (\ref{sec:altmatch}). All nodes, cavity node: input and offset on every node or on the cavity node only; feedback: state feedback; biases: node biases without feedback, the original form of the network. Readouts: 30 for the reservoir, 3 for the networks. Rule A: lowest training loss; rule B: among the 200 best, fading memory required, lowest training-internal validation NRMSE. Margin: $1-{\rm NRMSE}_{\rm QRC}/{\rm NRMSE}_{\rm net}$ on the held-out rows with a 95\% paired moving-block bootstrap interval (positive: reservoir better). Fresh: mean $\pm$ standard deviation over 20 later windows with the design frozen and the readout refitted, and the number of windows on which the reservoir is better. Every search met the saturation criterion.}
\label{tab:altmatch}
\resizebox{\textwidth}{!}{\begin{tabular}{@{}lcccccccccc@{}}\toprule
Model & Params & Starts (feasible) & Rule & Fading memory & Train NRMSE & Held-out & Margin [95\% CI] & Fresh & Ahead\\\midrule
Quantum reservoir & 44 & -- & -- & -- & 0.0369 & 0.0397 & -- & 0.0385 $\pm$ 0.0013 & --\\\midrule
All nodes, feedback & 17 & 7,168 (6,981) & A & no & 0.0649 & 0.0694 & $+$42.9\% [$+$34.9\%, $+$49.9\%] & 0.5322 $\pm$ 0.1843 & 20/20\\
 & 17 & 7,168 (6,981) & B & yes & 0.0711 & 0.0685 & $+$42.1\% [$+$33.2\%, $+$50.3\%] & 0.0747 $\pm$ 0.0124 & 20/20\\
All nodes, biases & 17 & 7,168 (7,152) & A & no & 0.0498 & 0.0567 & $+$30.0\% [$+$22.4\%, $+$38.3\%] & 0.3870 $\pm$ 0.2298 & 20/20\\
 & 17 & 7,168 (7,152) & B & yes & 0.0658 & 0.0673 & $+$41.1\% [$+$31.5\%, $+$49.1\%] & 0.0864 $\pm$ 0.0305 & 20/20\\
Cavity node, feedback & 15 & 7,168 (6,695) & A, B & yes & 0.1035 & 0.1079 & $+$63.2\% [$+$56.6\%, $+$68.4\%] & 0.1040 $\pm$ 0.0039 & 20/20\\
Cavity node & 12 & 7,168 (6,721) & A, B & yes & 0.1036 & 0.1085 & $+$63.5\% [$+$56.6\%, $+$68.6\%] & 0.1042 $\pm$ 0.0042 & 20/20\\
\bottomrule\end{tabular}}
\end{table}

\section{How classical baselines are set in the quantum-reservoir literature}
\label{sec:baseline-survey}
This note surveys how published quantum-reservoir studies choose and tune the classical model they compare against, and places the two comparisons of this work (main-text Sec.~2.1 with \ref{sec:matched}, and \ref{sec:altmatch}) among them. It supports the statement in main-text Sec.~2.1 that the echo-state-network (ESN) baseline follows the standard convention and is tuned more extensively than the classical baselines of the field.

\emph{Scope and extraction.} We read 66 studies from 2017 to 2026 in four groups: foundational and theoretical work, oscillator and cavity platforms, hardware demonstrations, and studies that examine the fairness of classical baselines. For each we recorded the classical comparator, the quantity held equal, what was tuned on the classical side and how, the reported statistic (single draw, mean or best over draws), the resources given only to the quantum side, and the reported outcome. Every candidate found by the citation search was checked: of 41 candidates beyond the first 46 studies, 16 were already in the set or listed twice, 5 contain no quantum model, and 20 were read and added. Twenty-eight studies have no classical comparator; Supplementary Table~\ref{tab:baseline-survey} lists the 38 that do. Several studies were read in their arXiv versions and values shown only in figures were not digitized; the deposited survey table records the source of every entry.

\emph{What is held equal.} Fifteen of the 38 comparisons match the number of readout features, giving the ESN as many nodes as the quantum reservoir has measured observables, virtual nodes included \cite{SBFujiiNakajima2017,SBNakajimaetal2019,SBKobayashietal2024,SBZhuetal2025PRR,SBZhuetal2025npjQI,SBKubotaetal2023,SBHuetal2023}. Six compare against the classical limit or a dephased version of the same system \cite{SBGoviaetal2021,SBKalfusetal2022,SBBravoetal2022,SBKornjacaetal2024,SBPalaciosetal2024,SBSpagnoloetal2022}, which isolates the quantum contribution but uses a classical system with two to a few state variables. Three match an ESN of $2^n$ nodes to $n$ qubits whose $2^n$ outcome probabilities are read out exactly \cite{SBPfefferetal2023,SBAhmedetal2024,SBSenanianetal2024}, an assumption of infinite shots; four match nodes to oscillators, qubits or detectors \cite{SBNokkalaetal2021,SBXiaetal2023,SBEhlersetal2025,SBConnertyetal2025}; four apply the same readout to the raw input or use another criterion \cite{SBSuzukietal2022,SBHouetal2026,SBPrietoGarciaetal2026,SBWangetal2026}; five hold nothing equal \cite{SBKutvonenetal2020,SBHuetal2024,SBHamhoumetal2025,SBMlikaetal2023,SBPandey2026c}. Only one study counts trainable physical parameters, and it compares two quantum models \cite{SBDudasetal2026}.

\emph{What is tuned.} On the classical side, twelve studies search a grid or random sample of one to four ESN hyperparameters while keeping the reservoir and input matrices random; ten use an untuned or default network, including both studies of the few-atom cavity reservoir closest to ours, which average an untuned ReLU ESN over 1,000 random draws \cite{SBZhuetal2025PRR,SBZhuetal2025npjQI}; eight do not report any tuning; two compare against trained classical models that are not reservoirs, a memoryless classifier \cite{SBHuetal2024} and a feed-forward network with all 1,153 weights trained \cite{SBHuetal2023}. Six tune the classical side with the same procedure as the quantum side \cite{SBGoviaetal2021,SBKalfusetal2022,SBPfefferetal2023,SBKornjacaetal2024,SBPandey2026a,SBXiaetal2023}; the most extensive of these configures the coupling matrix of a six-node ESN with the genetic algorithm used for the quantum reservoir and keeps its input matrix random \cite{SBXiaetal2023}. The reported classical statistic is almost always the mean or median over random reservoir draws, over at most 1,000 draws \cite{SBZhuetal2025npjQI,SBEhlersetal2025}.

\emph{The prevailing convention.} The most common baseline is therefore a random ESN with as many nodes as the quantum reservoir has readout features, its hyperparameters at defaults or chosen on a small grid, and its performance taken as the mean over random draws of its matrices. Averaging over draws is the defining convention of reservoir computing, in which only the readout is trained and the reservoir is characterized by its ensemble \cite{SBLukosevicius2009}.

\emph{Design elements on the quantum side only.} The quantum model frequently has design elements with no counterpart in the classical comparator: virtual nodes or multiplexed instances that multiply the number of features \cite{SBFujiiNakajima2017,SBNakajimaetal2019,SBCarlesetal2026,SBHouetal2026,SBKubotaetal2023}, quadratic, covariance or correlator readouts \cite{SBNokkalaetal2021,SBZhuetal2025PRR,SBZhuetal2025npjQI,SBSenanianetal2024,SBPandey2026a}, an input drive conditioned on measured readouts, called feedback in those papers \cite{SBKobayashietal2024,SBZhuetal2025npjQI,SBPaparelleetal2026,SBPandey2026a,SBWangetal2026}, and optimized or scanned physical parameters while the classical matrices stay random \cite{SBFujiiNakajima2017,SBZhuetal2025npjQI,SBAbbasMaksymov2024}. Where the classical side is tuned with the same care, reported quantum margins shrink or reverse in several studies: a degree-2 polynomial model reaches an NRMSE of $0.363$ against $0.726$ for a tuned correlator reservoir, and a tuning-matched ESN beats a quantum reservoir with a conditioned drive in 92\% of paired runs \cite{SBPandey2026a}; a nonlinear vector autoregression beats a gate-based quantum reservoir \cite{SBHamhoumetal2025}; the comparison of Ref.~\cite{SBAhmedetal2024} is split between tasks. A recent review asks that every claim of matching state what was held equal, including trainable parameters and search budget \cite{SBTariqetal2026}.

\emph{Where the baselines of this work stand.} The ESN of main-text Sec.~2.1 follows the prevailing convention and is more stringent than the surveyed baselines in five respects. (i) The readout cost is matched exactly, with the same ridge, washout, split and bias column and no direct input term on either side. (ii) Four hyperparameters are chosen among 810 settings on the mean training loss over 1,000 random draws each, and the five leaders are re-evaluated over 10,000 draws, ten times the largest number of draws in the surveyed comparisons. (iii) The full distribution over the 10,000 draws is reported, not only its mean: the reservoir is below 99.5--100\% of the individual draws at every $m$ (\ref{sec:matched}). (iv) Selection uses training data only, and the held-out window, five further windows, a block bootstrap, the ridge and an extended grid are reported as sensitivity checks (\ref{sec:matched}). (v) Beyond the convention, \ref{sec:altmatch} reports an alternative matching in which three-node networks are matched to the quantum reservoir by physical elements and every one of their parameters is trained with exact gradients under selection rules fixed in advance; the reservoir's held-out error is $30$--$64\%$ below that of each of the four network variants, and it is lower on every one of $20$ fresh windows. None of the surveyed studies trains every parameter of a classical reservoir; the closest are the genetic-algorithm configuration of the ESN coupling matrix in Ref.~\cite{SBXiaetal2023}, with the input matrix random, and the equal-budget tuning of two ESN hyperparameters in Ref.~\cite{SBPandey2026a}.

\begingroup\scriptsize\setlength{\tabcolsep}{2pt}\renewcommand{\arraystretch}{1.1}
\begin{longtable}{@{}p{0.95in}p{0.8in}p{0.9in}p{0.78in}p{0.78in}p{0.9in}p{0.92in}@{}}
\caption{\textbf{Classical baselines in the quantum-reservoir literature.} The 38 surveyed studies with a classical comparator (of 66 read; 2017--2026), in order of publication, and the two comparisons of this work. ``Matched quantity'': what the study holds equal between the quantum and the classical model. ``Classical tuning'': what was optimized on the classical side. ``Quantum-side design absent on the classical side'': design elements of the quantum model with no counterpart in the classical comparator. Entries are condensed from the papers and their supplements (arXiv versions where marked in the deposited survey table).}\label{tab:baseline-survey}\\
\toprule
Study & Quantum system & Classical comparator & Matched quantity & Classical tuning & Quantum-side design absent on the classical side & Reported outcome\\
\midrule
\endfirsthead
\toprule
Study & Quantum system & Classical comparator & Matched quantity & Classical tuning & Quantum-side design absent on the classical side & Reported outcome\\
\midrule
\endhead
\bottomrule
\endlastfoot
Fujii \& Nakajima 2017~\cite{SBFujiiNakajima2017} & 5--7-qubit spin reservoir & tanh ESN, 10--500 nodes; linear regression & none fixed; ESN size swept & spectral radius only & virtual nodes ($V\le50$); scanned physical parameters & 5--7 qubits comparable to 100--500-node ESNs\\
Nakajima et al. 2019~\cite{SBNakajimaetal2019} & multiplexed spin reservoirs & tanh ESN, 5--300 nodes & implicit readout count & grid (input scale $\times$ radius) & virtual nodes and spatial multiplexing & ESN lines as reference; no explicit claim\\
Kutvonen et al. 2020~\cite{SBKutvonenetal2020} & spin reservoir & ARIMA, LSTM & none & untuned (LSTM trained once) & lagged readouts, virtual nodes & QRC better than LSTM (qualitative)\\
Govia et al. 2021~\cite{SBGoviaetal2021} & single Kerr oscillator & classical Kerr oscillator & classical limit; same draws and observable & equal (ridge search, same draws) & nonlinear quantum readout & quantum errors orders of magnitude lower\\
Nokkala et al. 2021~\cite{SBNokkalaetal2021} & 8 Gaussian oscillators & 8-neuron ESN & node count = oscillators & single untuned draw & $O(N^2)$ covariance readout & quantum higher information-processing capacity\\
Bravo et al. 2022~\cite{SBBravoetal2022} & Rydberg qRNN & classical stochastic RNN & classical limit (analytic) & not trained & readout time optimized & quantum tasks infeasible classically (analytic)\\
Kalfus et al. 2022~\cite{SBKalfusetal2022} & single qudit/oscillator & Duffing oscillator & classical limit; same observable & best over sweeps, both sides & Hilbert dimension grows & quantum lower error plateau\\
Pfeffer et al. 2022~\cite{SBPfefferetal2022} & 7--9-qubit circuits & leaky ESN, up to 512 nodes & $N=$ $$2^n$$ (readout = Hilbert dim) & grid & exact $$2^n$$ probabilities & similar accuracy at 4$\times$ smaller size\\
Suzuki et al. 2022~\cite{SBSuzukietal2022} & IBM hardware, 8--12 qubits & linear regression; tanh ESN 2--50 & none; equivalent size scanned & radius grid, 100 draws & none noted & beats LR on NARMA2/5, loses on NARMA10\\
Spagnolo et al. 2022~\cite{SBSpagnoloetal2022} & photonic quantum-memristor reservoir, 3 photons in 9 modes (simulated) & same optics with coherent-state input & classical limit; same optics and readout & untuned (same training as quantum side) & single-photon encoding & 95\% (quantum) vs 71\% (coherent-state) test accuracy\\
Dudas et al. 2023~\cite{SBDudasetal2023} & coupled quantum oscillators & static ReLU reservoir; spin-torque oscillators & measured neurons = measured states & not reported & all basis states contribute & 99.7\% with 16 states vs $\ge$ 24--40 classical neurons\\
Pfeffer et al. 2023~\cite{SBPfefferetal2023} & hybrid circuits & leaky ESN, up to 2,048 & $N=$ $$2^n$$ & equal grid search & exact $$2^n$$ probabilities & similar accuracy at 4--8$\times$ smaller size\\
Kubota et al. 2023~\cite{SBKubotaetal2023} & 4-qubit noise-induced reservoirs, spatially multiplexed & tanh ESN, 5--520 nodes & readout count (computational nodes) & spectral radius per task; mean of 10 draws & instances selected from 1,024 noise configurations & 520-node QRC matches a 110-node ESN (NARMA2)\\
Xia et al. 2023~\cite{SBXiaetal2023} & 6-qubit Ising reservoir configured by a genetic algorithm & tanh ESN, 6 nodes (60 and 120 in SI) & node count = qubits & coupling matrix set by the same genetic algorithm; input matrix random & 3 Pauli readouts per qubit at V sub-intervals & QRC training loss 0.000028 vs 0.1 for the 6-node ESN\\
Hu et al. 2023~\cite{SBHuetal2023} & 6-qubit IBM device, 64 basis probabilities & trained 64-node feed-forward network; product-state circuit & feature count (64) & all 1,153 network weights trained & none (classical side noiseless) & only infinite-shot quantum results approach the trained network\\
Mlika et al. 2023~\cite{SBMlikaetal2023} & 4-qubit Ising reservoir (simulator and IBM device) & 500-unit ESN; LSTM & none (classical models larger) & coarse grid (not tuned to optimality) & none noted & QRC lowest MSE in all six cases\\
Ahmed et al. 2024~\cite{SBAhmedetal2024} & 5--12-qubit circuits & leaky ESN, $N=$ $$2^n$$ & Hilbert dimension & grid + Bayesian optimization & exact probabilities, classical leaky memory & ESN better on Lorenz-63; QRC better on Lorenz-96\\
Hu et al. 2024~\cite{SBHuetal2024} & 7-qubit IBM device and simulation & rounding; logistic regression & none & trained (no hyperparameters) & baseline memoryless & QRC beats logistic regression\\
Kobayashi et al. 2024~\cite{SBKobayashietal2024} & 8-qubit reservoir & leaky ESN, 8/50/100 nodes & readout count (8 + 1) & grid (radius $\times$ leak) & measurement-conditioned drive & QRC beats 8-node ESN\\
Kornjača et al. 2024~\cite{SBKornjacaetal2024} & neutral-atom array & classical-spin limit; SVM; NN & classical limit; input dimension & same procedure as QRC & multi-time correlator features & mixed; reaches classical benchmarks on some tasks\\
Palacios et al. 2024~\cite{SBPalaciosetal2024} & 5-qubit Ising reservoir & dephased same system & classical limit (noise) & n/a & none & capacity carried by classical correlations under noise\\
Senanian et al. 2024~\cite{SBSenanianetal2024} & microwave qubit--cavity & leaky ReLU ESN; linear classifier & Hilbert-dimension proxy & swept per width & higher-order moment readout & matches $\sim$32--64-dim ESN depending on shots\\
Abbas \& Maksymov 2024~\cite{SBAbbasMaksymov2024} & driven atom-cavity system, continuous measurement & ESN (standard code) & neuron count (16 Fock-state features) & fixed defaults & measurement rate tuned per task & 16 quantum neurons comparable to an ESN with more than 500 neurons (visual)\\
Zhu et al. 2025 (PRR)~\cite{SBZhuetal2025PRR} & few-atom cavity & ReLU ESN & readout count & untuned, mean of 1,000 & quadratic readout & QRC below ESN at equal features\\
Zhu et al. 2025 (npj QI)~\cite{SBZhuetal2025npjQI} & few-atom cavity, measurement-conditioned drive & ReLU ESN, no feedback & readout count & untuned, mean of 1,000 & trained drive gains, quadratic readout & 51\% lower NRMSE at equal neurons\\
Hamhoum et al. 2025~\cite{SBHamhoumetal2025} & multi-qubit QRC & NVAR, ESN, LSTM (imported) & none & best of source study & different data pipelines & NVAR beats QRC\\
Ehlers et al. 2025~\cite{SBEhlersetal2025} & stochastic qubit reservoir network & deterministic ESN with the same random matrices & detector count; same random draws & untuned; mean of 100--1,000 draws & $$2^n$$ outcome probabilities read out & stochastic network ahead with exact probabilities; ESN ahead at finite statistics from 4--7 detectors\\
Wang et al. 2025~\cite{SBWangetal2025} & quantum next-generation reservoir, optical implementation & NG-RC with 28 features (literature, not re-run) & feature-count equivalence & not reported & quantum features reproduce NG-RC monomials & longer prediction claimed; no classical numbers\\
Connerty et al. 2025~\cite{SBConnertyetal2025} & gate-based quantum ESN on IBM hardware & ESN; linear regression & nodes = readout qubits & not reported & best of several seeds; regularization tuned on test error & small gain over the ESN; no advantage claimed\\
Carles et al. 2026~\cite{SBCarlesetal2026} & transmon--resonator experiment & imported classical results & readout count & not reported & time multiplexing; post hoc feature choice & 8 quantum vs $\ge$ 25 classical features\\
Dudas et al. 2026~\cite{SBDudasetal2026} & same hardware as QRC and QNN & illustrative MLP & trainable parameters (quantum vs quantum) & not reported & n/a & trained QNN needs fewer observables\\
Hou et al. 2026~\cite{SBHouetal2026} & 9-spin liquid-state NMR & ESN 500--10,000; classical spin limit & none; classical limit & grid per size & 653 multiplexed features & comparable to ESN(1,000)--ESN(10,000)\\
Paparelle et al. 2026~\cite{SBPaparelleetal2026} & CV optical reservoir & leaky ESN; LSTM & modes = neurons; equal readout parameters & grid & covariance features, conditioned drive, multiplexing & mixed by task\\
Pandey 2026a~\cite{SBPandey2026a} & simulated Ising QRC $\pm$ conditioned drive & Poly2 ridge; tuning-matched ESN & feature dimension and tuning budget & as careful as quantum & correlator readout; conditioned drive & classical matches or beats QRC\\
Pandey 2026b~\cite{SBPandey2026b} & simulated QRC & ESN with $2^q$ units & readout count & not reported & none noted & QRC better on most Lorenz-96 cells\\
Pandey 2026c~\cite{SBPandey2026c} & 5-qubit random circuit, 32 basis probabilities & 500-neuron ESN & none (sizes differ) & fixed hyperparameters & none (ESN favoured) & ESN test MSE 2.08 vs 3.2 for the QRC (Lorenz)\\
Prieto-Garcia et al. 2026~\cite{SBPrietoGarciaetal2026} & two coupled nonlinear bosonic modes (simulated) & likelihood-ratio tests; same readout on raw input & same readout on raw input & none & dynamical reservoir; drive amplitude set per task & QRC better on short series, marginally worse on long\\
Wang et al. 2026~\cite{SBWangetal2026} & integrated photonic squeezed-light reservoir (experiment) & logistic regression on raw inputs & same readout on raw input & none & photon-correlation features; measurement-conditioned drive & 98.3\% vs 33.3\% test accuracy (static XOR)\\
\midrule
This work, Sec.~2.1 and \ref{sec:matched} & 1--3-atom cavity with conditioned drive & tanh ESN, $m$ nodes, random matrices & readout count $m$ (exact), same ridge, split and washout & 810-setting grid on the mean training loss over 1,000 draws; five leaders over 10,000 draws & optimized physical parameters, offset, sampling times and drive gains (ESN in its standard form) & QRC below the ESN mean by 25--69\% and below 99.5--100\% of the draws\\
This work, \ref{sec:altmatch} (alternative matching) & 2-atom cavity with conditioned drive, $m=30$ & four 3-node networks, all 12--17 parameters trained & physical elements (2 atoms + 1 mode = 3 nodes) & 7,168 exact-gradient starts per network, selection rules fixed in advance & 44 system parameters against 12--17; 30 readouts against 3 & QRC 30--64\% below every network; ahead on all 20 fresh windows (Supplementary Table~\ref{tab:altmatch})\\
\end{longtable}
\endgroup